\documentclass[%
 aip,floatfix,
 amsmath,amssymb,
 reprint,%
]{revtex4-1}

\usepackage{graphicx}
\usepackage{dcolumn}
\usepackage{upgreek}
\usepackage{bm}
\usepackage[utf8]{inputenc}
\usepackage[T1]{fontenc}
\usepackage{mathptmx}
\usepackage{etoolbox}
\usepackage{multirow}
\usepackage{hyperref}
\makeatletter
\def\@email#1#2{%
 \endgroup
 \patchcmd{\titleblock@produce}
  {\frontmatter@RRAPformat}
  {\frontmatter@RRAPformat{\produce@RRAP{*#1\href{mailto:#2}{#2}}}\frontmatter@RRAPformat}
  {}{}
}%
\makeatother
\begin{document}

\preprint{AIP/123-QED}

\title{Invited Article: Micro-fabricated components for cold atom sensors}
\author{J. P. McGilligan}
\email[Authors to whom correspondence should be addressed:]{james.mcgilligan@strath.ac.uk}
\affiliation{SUPA and Department of Physics, University of Strathclyde, Glasgow, G4 0NG, United Kingdom}
\author{K. Gallacher}
\affiliation{James Watt School of Engineering, University of Glasgow, Glasgow, G12 8LT, United Kingdom}
\author{P. F. Griffin}
\affiliation{SUPA and Department of Physics, University of Strathclyde, Glasgow, G4 0NG, United Kingdom}
\author{D. J. Paul}
\affiliation{James Watt School of Engineering, University of Glasgow, Glasgow, G12 8LT, United Kingdom}
\author{A.\ S.\ Arnold}
\affiliation{SUPA and Department of Physics, University of Strathclyde, Glasgow, G4 0NG, United Kingdom}
\author{E. Riis}
\affiliation{SUPA and Department of Physics, University of Strathclyde, Glasgow, G4 0NG, United Kingdom}

\date{\today}

\begin{abstract}
Laser cooled atoms have proven transformative for precision metrology, playing a pivotal role in state-of-the-art clocks and interferometers, and having the potential to provide a step-change in our modern technological capabilities. To successfully explore their full potential, laser cooling platforms must be translated from the laboratory environment and into portable, compact quantum sensors for deployment in practical applications. This transition requires the amalgamation of a wide range of components and expertise if an unambiguously chip-scale cold atom sensor is to be realized. We present recent developments in cold-atom sensor  miniaturization, focusing on key components that enable laser cooling on the chip-scale. The design, fabrication and impact of the components on sensor scalability and performance will be discussed with an outlook to the next generation of chip-scale cold atom devices. 

\end{abstract}

\maketitle

\section{Introduction}

From its first proposal in 1957 \cite{schawlow} and realization in 1960 \cite{MAIMAN1960}, the laser has revolutionized the capabilities of experimental physics, with a profound impact on atomic spectroscopy and quantum sensors, where laser cooled atoms lie at their core. Within a decade of the laser's realization it was used to demonstrate the acceleration and trapping of particles by radiation pressure \cite{ashkin, HANSCH197568, wineland}, laying the foundation for the laser cooling of ions and neutral atoms below the milliKelvin regime \cite{chu}.

Today, laser cooled atoms are central to modern precision measurements, as their slow speed and low temperature enable long interrogation times in unperturbed atomic samples, making possible orders of magnitude more accurate and precision measurements in metrology \cite{ludlow, ADAMS19971}. As such, cold atom technology lies at the core of interferometers, used for inertial sensing \cite{aidanring,landragin,inertial}, gravimetry \cite{Poli,bidel} and accelerometry \cite{garrido}. Additionally, cold atoms are exploited in primary standard fountain atomic clocks, where the SI second is defined by the 9,192,631,770~Hz frequency separation of the ground-state energy levels of Cs, extracted from a cold-atom fountain measurement \cite{Bize_2005}. The state-of-the-art optical atomic clocks laser cool alkaline earth atoms, such as Yb and Sr, confined in optical lattices to realise a measurement uncertainty of the atomic frequency below\cite{ludlownist, McGrew2018, hanschreview, ludlowreview}  $10^{-19}$. The capability to isolate individual cold atoms in optical lattice sites and dipole traps is also utilized for the quantum gate functions at the core of neutral atom quantum computers \cite{Kaufman2015, Wu2019, Picken_2018}. Additionally, laser cooling atoms below the recoil-limit has enabled experimental measurement of the phase transitions to Bose-Einstein condensates (BEC) \cite{ketterle, BECCU, Aveline2020} and degenerate Fermi gases \cite{Greiner2002}, placing cold atoms at the forefront of fundamental research in quantum simulation \cite{kuhr1, kuhr2} and quantum fluids. The high state of coherence that can be attained from these quantum gases are at the core of atom lasers \cite{atomlaser} and atomic lithography \cite{atomlithography}, with the use of delta-kick collimation \cite{arnold2002} leading to record low temperatures \cite{Deppner2021} and increased coherence.
 
 However, with high-performance cold atom sensors largely remaining bound to laboratory environments due to core-component size, weight and power (SWaP) issues, a large emphasis has been placed on global institutions to tackle these issues through the development of compact cold atom sensors \cite{Rushton2014, SALOMONreview, garrido, UKHub, muquans, ColdQuanta}.

In the past decade significant effort has been made to miniaturise atomic sensors to a scale where they can make the largest technological and economic impact \cite{Kitching2018}. This has been driven by the tantalising prospect of providing compact yet adequate primary measurement capability to the end-user. This motivation has led to a revolution in the micro-fabrication and miniaturization of micro-electro-mechanical-systems (MEMS) vapor cells, semiconductor diode lasers and methods for atomic deposition that have enabled the realization of fully incorporated, chip-scale atomic packages \cite{Knappe2004,moreland}. Although this progress has been made in the miniaturization of thermal atom metrological devices \cite{boudotmemscell,Hasegawa2011_buffer_gas,Hummon:18}, the 
vapor cells at the heart of this technology typically require the inclusion of buffer gasses or wall coatings to increase the light-atom interaction time and reduce relaxations from wall collisions. Ultimately, these inclusions degrade the stability of the sensor due to temperature dependent pressure shifts from the buffer gas, or the wall coatings degrading over time \cite{Kitching2018}. However, many of the limitations placed on the performance of hot atomic gasses can be circumvented by using laser cooled atoms \cite{Rushton2014}.

\subsection{Laser cooling}

The workhorse of cold-atom experiments is the magneto-optical trap (MOT) \cite{Raab}. This system utilises a balanced optical radiation force to reduce the momentum of thermal atoms through photon scattering in a spatially localized trap provided by the light in conjunction with a magnetic quadrupole field. The radiation force used to reduce the momentum spread of the atoms is primarily due to light absorption. If an atom absorbs a photon, then the photon energy will mostly be converted to the internal energy of the atom, exciting the electronic state. The momentum kick, $p=mv=\hbar k$, imparted during this process induces a recoil in the direction of the incident light. Since the direction of the spontaneously emitted photon is spatially isotropic the momentum kick due to emission averages out to zero. Therefore, the net momentum imparted on the atom from the absorption-emission process is in the direction of the incident light.

Since the atoms under interrogation have a finite velocity, we must also consider the Doppler effect's role in the cooling process. If a single atom is travelling towards resonant cooling light, then it will be blue-shifted in frequency from the atomic resonance. Hence, if the incident light frequency is red-detuned below an atomic cycling transition then the number of absorbed photons, and hence recoil momentum kicks, are maximized for atoms propagating anti-parallel to the cooling light. If we simplify this to the case of a two-level atom with velocity $\vec{v}$ then the time-averaged force experienced by the atom from the interaction with the light from the $i^\textrm{th}$ laser beam (with wave-vector $\vec{k}_i$) is described by \cite{ADAMS19971}:
\begin{equation}
    \vec{F}_i=\frac{d\vec{p}_i}{dt}=R(I_i,\Delta_i)\hbar\vec{k}_i,
\end{equation}
where  $R(I_i,\Delta_i)=\frac{\Gamma}{2}\frac{I_i/I_\textrm{S}}{1+I_\textrm{T}/I_\textrm{S}+4{\Delta_i}^2/\Gamma^2}$ is the photon scattering rate of the atom, $I_i,$ $I_\textrm{S}$ and $I_\textrm{T}$ are the incident, saturation and total intensities for the transition being addressed, and  $\Delta_i=\Delta-\vec{k}_i\cdot\vec{v}$ is the Doppler-shifted detuning from resonance. The natural linewidth of the excited state is $\Gamma=1/\tau$, where $\tau$ is the excited state lifetime.

If the atom is positioned between two counter-propagating laser beams, derived from the same laser with the same intensity, then the atom will experience a velocity-dependent acceleration. In this scenario, the Doppler effect breaks the symmetry of absorption between the two beams, where a balanced force from both laser beams is only achieved when the atom is at rest \cite{2Dcooling}. While the atoms are not actually in thermal equilibrium, the spread of the atomic velocity is Gaussian and conventionally described by a temperature, corresponding to that of a Maxwell Boltzmann distribution with the particular velocity spread. It is because of the reduction in atomic temperature with laser scattering events that we refer to the process as laser cooling. In the limit of low intensity, the achievable temperature from the stochastic nature of the absorption and spontaneous emission processes is restricted by the theoretical Doppler temperature limit \cite{Lett:89, Ungar:89}, 
\begin{equation}
    T_\textrm{D}=\frac{\hbar\Gamma}{2k_{\rm{B}}}
\end{equation}
where $k_{\rm{B}}$ is the Boltzmann constant. For e.g.\ $^{87}$Rb and the $780\,$nm transition with $\tau=26.2\,$ns \cite{Steck} ($\Gamma=2\pi\times6.1\,$MHz) this corresponds to  $T_\textrm{D}=146~\upmu$K. 

While the Doppler limit theoretically restricts the ensemble temperature to the hundreds of $\upmu$K range for the alkalis, experimental laser cooling has overcome this limitation through sub-Doppler cooling mechanisms \cite{molasses1,molasses2,Weiss:89}. These cooling mechanisms arise from the complicated atomic level structure of the alkali atom and the interference pattern formed in the cooling light, such that the achievable temperature, that is routinely realized, is that of a distribution with only a few photon momenta. This results in temperatures in the single-digit micro-Kelvin regime being commonly realized, and with additional experimental techniques, can be pushed into the nano-Kelvin regime.

\begin{figure}[t]
\centering
\includegraphics[width=0.48\textwidth]{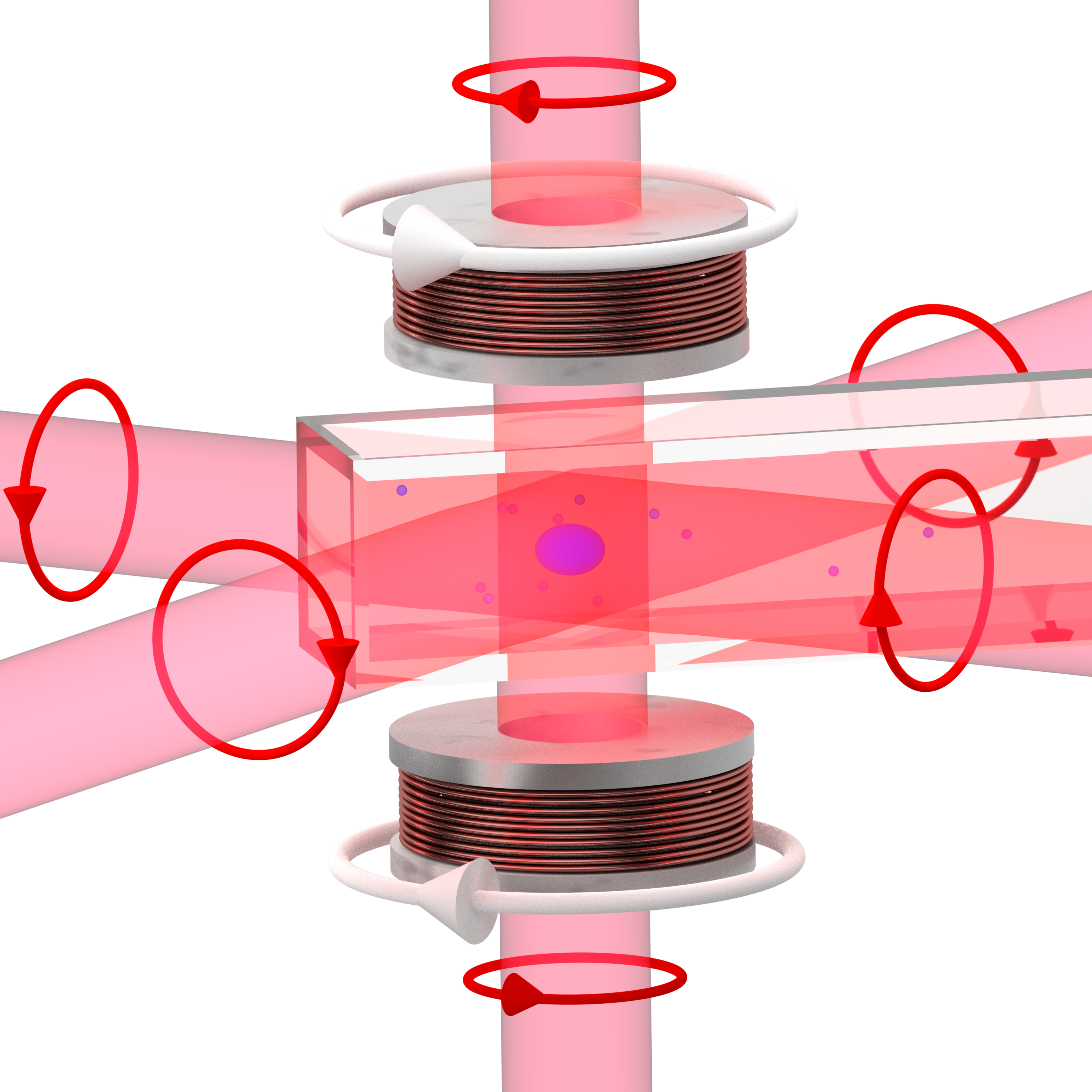}
\caption{\label{MOT} Illustration of the standard 6-beam MOT. Light polarization is indicated with red arrows. Magnetic trap coil current orientation is indicated with white arrows. An ultra-high-vacuum glass cuvette is shown, with the alkali density and MOT emphasized in blue within the vacuum.}
\end{figure}

The lower limit of the dimensional scaling of laser cooling follows $N_\textrm{b} \geqslant N_\textrm{d}+1$, where $N_\textrm{b}$ and $N_\textrm{d}$ refer to the number of beams and dimensions, respectively. Thus laser cooling in three dimensions requires at least four beams. Due to its simplicity and convenience of alignment, the typical formation of the MOT has been constructed around six counter-propagating laser fields in three dimensions, as shown in Fig.~\ref{MOT}, to satisfy a balanced radiation force $\sum_{i=1}^{N_\textrm{b}}I_i\, \vec{k}_i=0$, where $I_i$ is the laser intensity from the $i^\textrm{th}$ beam .

Although the cooling process in 3 dimensions produces a velocity-dependent force, the scattering force is not position-dependent without the application of appropriate circular beam polarizations and a gradient magnetic field, completing the magneto-optical trap (MOT) \cite{Raab,monroe}. The addition of the quadrupole magnetic field through the overlapping laser volume, produced from e.g.\ an anti-Helmholtz coil pair, induces a splitting of the atomic sub-levels due to the Zeeman effect. This gives rise to a complex optical pumping scheme, where each beam's local light polarization critically affects its  scattering rate.

The experimental parameters required for laser cooling of $^{87}$Rb are outlined in Table~\ref{table1}.  Cooling light ideally has a laser linewidth, and locked absolute frequency stability, below the natural linewidth of the cooling transition \cite{wieman}. In the case of $^{87}$Rb, the cycling transition exists between the $5S_{1/2}\,F=2\leftrightarrow5P_{3/2}\,F'=3$ states, with a natural linewidth of $\Gamma=2\pi\times$6.1~MHz. As such, the incident light would be preferred with a linewidth of $\leqslant$1~MHz, locked red-detuned from the cycling transition between 1$\Gamma$-2$\Gamma$. While the cycling transition is required for atomic cooling, an off-resonant decay route from the excited $F'=2$ to ground $F=1$ state depletes the atoms from the cycle. While this is a relatively weak decay channel, over many cycles the atoms will be shelved in the non-addressed ground-state. To correct the loss and bring these atoms back into the cooling cycle, re-pumping light resonantly addresses e.g.\ the $F=1\rightarrow F'=2$ transition, rapidly returning atoms to the $F=2$ state and back onto the cycling transition. While the example covers the case of Rb, a variety of re-pumping wavelengths is commonly required to address dark or metastable states in the more complicated electronic structure of alkaline earth atoms or molecules. 

The output cooling beams are circularly polarized with a $\lambda/4$ wave-plate to maximise scattering through the magnetically sensitive stretched states. The quadrupole magnetic field, $\vec{B}_\textrm{q}=B_1\{-x/2,-y/2,z\}$, is often provided from an anti-Helmholtz coil pair, with the optimum gradient field providing maximum scattering at the extremes of the beam overlap volume radius, $r$. As such, the approximate optimum axial gradient is $B_1\sim-\Delta/(\mu_\textrm{B} g_{F\textrm{e}} r)$, where $\mu_\textrm{B}=1.4$~MHz/G is the Bohr magneton and $g_{F\textrm{e}}$ is the excited state Land\'e $g$ factor \cite{McGilliganphase, aidaninprep,wieman1992,aidanthesis}. In the case of $^{87}$Rb atoms driven by circularly  polarized light with an average beam radius of $1\,$cm (so $-\Delta \sim 2\Gamma$), and $g_{F\textrm{e}}=2/3$, this yields $B_1\sim13\,$G/cm. The application of this field is capable of spatially shifting the resonance of these magnetically sensitive sub-levels, such that a laser detuned below the non-degenerate excited state resonance will have an increased probability of being brought into resonance with these sub-levels at larger magnetic fields. This creates the position dependent scattering rate that provides a restoring force, which combined with the cooling realizes a MOT.

The optical configuration and coil orientation described until this point are built around an ultra-high-vacuum (UHV) chamber, an essential requirement in laser cooling application to reduce energetic collision with background gases as well as provide a suitable alkali density \cite{wieman,wieman1992}. Typically machined from non-magnetic stainless steel or titanium, the main vacuum chamber contains windows or a glass cuvette for optical access. Initially evacuated by a mechanical pumping rig, the chamber pressure is maintained within the range of $P<10^{-7}\,$mbar with active pumping mechanisms such as an ion-pump. The final component required within the vacuum chamber is a mechanism to release the atomic element for laser cooling. In the case of the alkali elements, this can be achieved from resistively heated dispensers \cite{grifflaserheating, amdpaper, fastAMD} and ampoule sources \cite{Kitching2018} to achieve a sufficient alkali vapor density in the range of $10^8-10^{10}$~atoms/cm$^3$ at UHV pressures.

\begin{table}[t] 
\caption{$^{87}$Rb Chip-scale Cooling Challenge} 
\centering 
\begin{tabular}{c c c} 
\hline\hline 
Parameter & Optimum & References  \\ [1ex]
\hline 
Atom number & 10$^6$-10$^7$ & \cite{Rushton2014,Hoth13} \\
Laser linewidth (MHz) & $<$1$\Gamma$ & \cite{wieman}\\
Laser power (mW) & 30 &  \cite{wieman,monroe, McGilliganphase}\\
Re-pumping power ($\%$) & 5$\%$ & \cite{wieman,McGilligan2017} \\ 
Absolute frequency stability (MHz) & 0.5~$\Gamma$ & \cite{wieman} \\ 
Detuning range (MHz) & 4$\Gamma$ & \cite{monroe, Weiss:89, Lett:89} \\ 
Total background pressure (mbar) & $\leqslant10^{-7}$ & \cite{pressurecriteria,landragin, monroe}   \\ 
Alkali density (atoms/cm$^3$) & 10$^8$-10$^{10}$ & \cite{FLEMMING1997269}  \\ 
Gradient field (G/cm) & 12 & \cite{monroe} \\ [1ex] 
\hline 
\end{tabular} 
\label{table1} 
\end{table} 

This review article will look at the recent developments in the miniaturization of cold-atom systems, focusing on the fabrication of components that enable mass production and the reduction of the cold-atom system to the chip-scale. Section~\ref{PIC} will discuss the current state-of-the-art of photonic integrated packages, comprising narrow linewidth laser sources, waveguides, beam-splitters and optical control components required for cold-atom sensor packages. Section~\ref{waveref} highlights the options available for chip-scale wavelength references, providing an adequate frequency stability for laser cooling while maintaining a low SWaP. The fabrication techniques used in vapor cell manufacturing are outlined and their compatibility with atomic sensors is reviewed. In Section~\ref{atomdepo} we discuss clean and mass producible atomic sources, compatible with chip-scale cold-atom packages, with an insight to compact solutions to vacuum pressure longevity and alkali density regulation. Section~\ref{lasercooling} provides an overview to micro-fabricated optical components that miniaturize laser cooling packages, such as pyramidal and grating magneto-optical traps, as well as the impact of meta-surface lenses. In Section~\ref{UHV} we discuss the recent demonstrations of micro-fabricated vacuum cells, as well as suitable materials and fabrication methods. Section~\ref{vacuumpumps} highlights the potential solutions for maintaining UHV in a chip-scale package, through topics in micro ion-pumps and non-evaporable getters, with a focus on vacuum isolation technique's suitable to cold-atom apparatus. In Section~\ref{magneticfield} we discuss planar coil solutions that have been demonstrated to reduce the size of the coil arrangement required in the MOT. Finally, Section~\ref{conclusion} summarises the amalgamation of this technology and our outlook to where this technology will go next.

\section{Photonic integrated circuits}
\label{PIC}
An essential building block in atomic sensors and many quantum technologies are the laser sources and photonic components that enable cooling and probing on the atomic scale. The recent drive for portable cold-atom sensors has led to a number of macroscopic, miniaturized optical packages being demonstrated for ground-based sensors \cite{Bongs2019, LEE2021106698}, portable drone-mounted apparatus \cite{atoms10010032}, and space applications \cite{Strangfeld:22, Elliott2018,Frye2021, Dinkelaker:17,Luvsandamdin:14}, where SWaP are of critical importance. While this technology has demonstrated robustness and scalability, the reduction of the apparatus to the chip-scale would greatly increase the rigidity and SWaP constraints for in-field deployment. 

In recent years significant gains have been made in the miniaturization of photonic devices to facilitate the growing need for chip-scale quantum technologies. In this section we will highlight the recent advancements to photonic components that directly benefit the SWaP of the laser cooling apparatus.

\subsection{On-chip laser solutions}
The advent of the laser realized a step change in the capabilities of atomic sensors, providing a previously unobtainable coherence and linewidth to the atom-light interaction. Since early demonstrations of particle trapping and the realization of the spontaneous and dipole force from laser light at Bell labs in the 1970's \cite{ashkin, ashkin1}, the laser has become a key instrument in atomic physics, providing global access to a plethora of applications such as optical lattices \cite{katori}, non-linear dynamics \cite{kippenberg}, frequency metrology \cite{Heavner_2005}, and importantly, laser cooling \cite{Raab}. 

The stringent requirements placed on the optical power and laser linewidth for laser cooling have hindered a simplistic route for on-chip cold atom light sources. In the past, laboratory experiments have instead focused on cost-effective methods to meeting this criteria, largely driving diode based laser sources \cite{diodelaserwieman} such as the extended-cavity diode-laser (ECDL) \cite{diodelaserhollberg}. This now well established configuration provides frequency selective feedback through an extended, external cavity, typically using a grating in a Littrow set-up \cite{ecdlmooradian,Ricci1995,ecdlarnold,Cook2012}. Alternatively, compact ECDL designs have been demonstrated with frequency selective interference filters to improve mechanical robustness and sensitivity to misalignment \cite{BAILLARD2006609}.

Since both the free-spectral range (FSR=$\frac{c}{2l}$) and linewidth of the laser determined from the Schawlow-Townes limit are inversely proportional to the cavity length $l$, a trade-off arises between the spacing of adjacent cavity modes of the laser and the free scanning range that is useful for atomic spectroscopy. However, in the case of a $2\,$cm extended cavity, a semiconductor diode laser can provide a mode spacing of $\sim8\,$GHz with a corresponding linewidth less than $1\,$MHz on second timescales and enough power for even large beam alkali metal MOTs \cite{Daffurn2021}. It should be noted that while the ECDL provides a cost-effective solution for laser cooling, more precise applications in atomic spectroscopy with cold atoms require significantly larger and more complex laser systems \cite{ludlowreview}.

With this being said, the critical nature of the laser in portable atomic sensors has provided a drive to bring this component to the chip-scale, realizing technology such as the vertical-cavity surface-emitting laser (VCSEL) \cite{vcsel1}. Formed in a semiconductor stack, the VCSEL is composed of a thin gain medium containing multiple quantum wells, sandwiched between two distributed Bragg reflector mirrors. With application of a small current, guided through the semiconductor stack to the active gain region, the VCSEL emits light orthogonal to the wafer surface. VCSELs have provided a dramatic reduction in SWaP for a chip-scale atomic clock (CSAC) by eliminating the atomic discharge lamp and heating elements ($\sim2\,$W) with a source that requires less than $5\,$mW of power.\cite{Serkland_06} Utilizing a coherent population trapping architecture also removed the requirement for microwave syntheses ($\sim1\,$W), where the VCSEL can instead be directly modulated at half of the hyperfine splitting frequency.\cite{Serkland_07} This resulted in a CSAC with a volume of $9.5\,$mm$^3$, power consumption $<75\,$mW and fractional frequency instability of 2.5 $\times$ 10$^{-10}$ at one second.\cite{Knappe2004} 

While these light sources have found application in thermal atom spectroscopy for atomic clocks and magnetometers, the short cavity length on the order of a few $\mu$m, limits the typical device linewidth to $\sim50$~MHz.\cite{Iga_18} This limits the performance due to the frequency noise of the laser being directly converted to an amplitude noise on a transmitted beam through an atomic vapour\cite{Mikhailov_10} and makes them also unsuitable for laser cooling.\cite{vcselkitching} 

One technique to improve both power and linewidth is to utilize an external cavity arrangement (VECSEL) that has recently demonstrated sub kHz performance.\cite{Moriya_20} However, this significantly increases the size due to the requirement of a high finesse external cavity and limits the potential for chip-scale integration.  

A second noteworthy chip-scale laser source is the distributed feedback laser (DFB).\cite{Scifres_1974} These edge-emitting semiconductor-based lasers can offer the same tunability as an ECDL but at a fraction of the size (typically a few $\upmu$m wide and mm long). The DFB is composed of a periodically structured gain medium, acting as a Bragg reflector at the wavelength of interest while remaining fabricated on-chip. Compared to a Fabry Perot semiconductor laser this provides much larger mode hop free tuning ranges ($\sim$ nm) and side mode suppression ratios (> 40 dB). Typically these have been composed of AlGaInAs/InP layers for emission at telecommunication wavelengths\cite{Gaetano:19} where frequency doubling can be utilized to lock to an Rb absorption line for accurate wavelength division multiplication schemes.\cite{Poulin_1997}

More recently, DFB lasers composed of GaAs/AlGaAs can provide gain bandwidths that directly cover the transition lines of Rb and Cs atoms. These have demonstrated linewidths narrow enough for saturated absorption spectroscopy.\cite{Kraft_05,Ligeret_08} Due to the larger cavity lengths, sub-MHz linewidths can be readily obtained with optical powers suitable for laser cooling.\cite{DiGaetano:20}

\subsection{On-chip laser integration strategies}

\begin{figure}[t]
\centering
\includegraphics[width=0.48\textwidth]{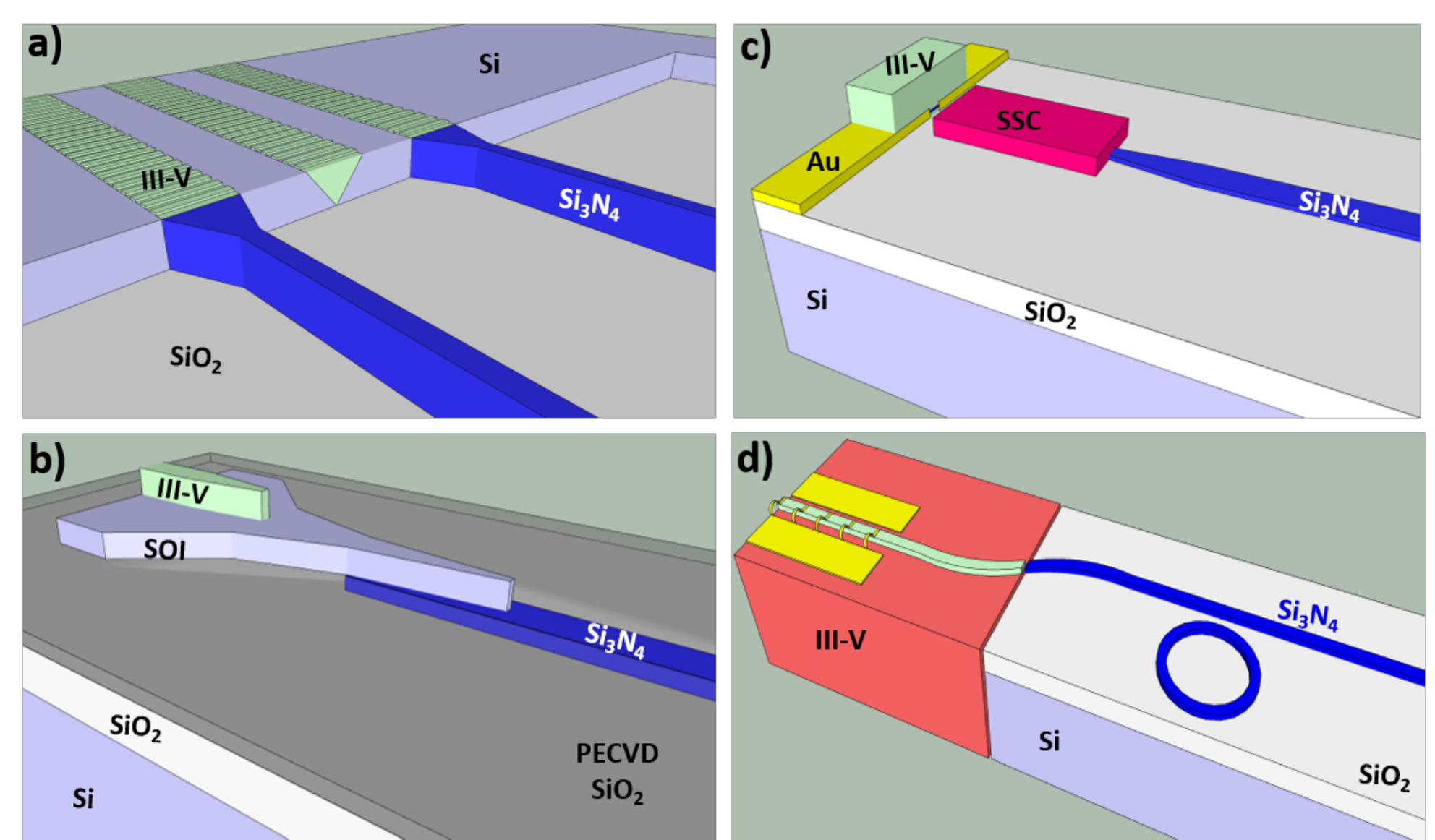}
\caption{\label{DFB_int} A schematic diagram of possible strategies for integrating a DFB narrow-linewidth laser with Si$_3$N$_4$ waveguides. (a) III-V material selectively grown on defined Si trenches. (b) DFB laser flip chipped or wafer bonded directly on top of SOI waveguides and evanescently coupled using inverse tapers. (c) DFB laser flip chipped and butt coupled using a spot-size converter. (d) Butt-coupling of two discrete chips containing the III-V laser and Si$_3$N$_4$ waveguides.}
\end{figure}

Due to the low linewidths, a DFB edge-emitting laser is an ideal candidate for integration with photonic integrated circuits for miniaturizing cold atomic devices. Integration could be achieved via several approaches (see Fig. \ref{DFB_int}) such as monolithic, where the III-V gain material is selectively grown directly on top of Si trenches to overcome the large lattice mismatch.\cite{Wang_15} Heterogeneous integration is where the III-V can be wafer or flip-chip bonded directly on top of waveguides and coupled evanescently.\cite{Xie:19} However, due to the large refractive index difference between the III-V laser structure and Si$_3$N$_4$ waveguides, inversely tapered silicon on insulator (SOI) waveguides are usually required as an intermediate. This approach has recently demonstrated an on-chip laser with a $4\,$kHz linewidth when the Si$_3$N$_4$ was used as the extended laser cavity.\cite{Xiang_20} 

An alternative strategy is required, however, below the absorption band edge of Si where the Rb and Cs atomic optical transitions reside. This could be achieved by butt-coupling the flip-chipped laser and the waveguides. To enable efficient coupling the mode field diameter of the laser can be matched by using a spot size converter\cite{Ohira_10,Gallacher_19} or inverse taper structure.\cite{Romero-Garcia_14} An alternative would be to butt-couple two discrete III-V and Si$_3$N$_4$ waveguide chips. This hybrid approach allows the III-V to operate as a semiconductor optical amplifier (SOA) and the Si$_3$N$_4$ waveguides provide feedback via a high finesse grating or ring resonator. The advantage being the effective cavity length can be greatly increased due to the significant lower propagation losses of Si$_3$N$_4$.\cite{Boller_20} This has recently resulted in the demonstration of a $1\,$Hz linewidth laser.\cite{Jin_21}

\subsection{Photonic circuits}

Photon routing and manipulation is a necessity for a fully integrated on-chip cold-atom sensor. Replacing discrete optical components with their waveguide equivalent will enable photonic integrated circuits (PIC) that significantly reduce the SWaP and cost (SWaP-C).

PICs specifically based on silicon photonics can leverage the huge investments made in CMOS foundries. SOI, which is a commercialized platform for telecommunication applications, cannot operate below wavelengths of $1\,\upmu$m. Therefore, alternative materials with low material absorption, suitable refractive index for guiding and compatibility with laser integration are required.\cite{Blumenthal_20} Potential candidates include tantalum pentoxide (Ta$_2$O$_3$),\cite{Belt_17} aluminium nitride (AlN),\cite{Lu_18} aluminium oxide (Al$_2$O$_3$),\cite{Aslan_10} and silicon nitride (Si$_3$N$_4$).\cite{Sinclair_20} 

These platforms have demonstrated low propagation losses at visible and near-infrared (NIR) wavelengths. With a comparable refractive index, the main distinguishing feature is the short wavelength cut-off. In this regard, Al$_2$O$_3$ has demonstrated low loss down to 370 nm.\cite{West_19} However, for Cs and Rb atomic devices, Si$_3$N$_4$ is an attractive option since it is the more mature platform with process design kits (PDK) accessible from several foundries.\cite{Munoz_19} In addition, several high-performance devices at telecommunication wavelengths have been demonstrated, such as modulators\cite{Chang_17} switches,\cite{Joo_18, Zheng_19} polarization control,\cite{Sun16, Zhan21} grating couplers,\cite{Sacher_14,Mak_18} optical isolators,\cite{Yan_20,Tian_21} optical frequency combs (OFC),\cite{Okawachi_11,Marin-Palomo_17,Stern_18} and integrated lasers.\cite{Zhan21} The difficulty is to create a similar library of components for cold atom PICs. 

 One of the most challenging devices to replicate at shorter wavelengths is the OFC. OFCs are essential for an optical atomic clock since they can act as a gear to convert from the THz to GHz frequency domain for electronic measurement. Apart from increased losses due to Rayleigh scattering ($\sim \lambda^{-4}$), the bulk material dispersion of Si$_3$N$_4$ has limited octave-spanning comb generation to above $900\,$nm wavelength.\cite{Okawachi_14} Ideally, the OFC should be broadly spanning for self-reference schemes\cite{Brasch_17,Pasquazi_18} while generating lines at the atomic transitions for stabilization. A potential strategy includes utilizing other non-linear processes such as second or third harmonic generation, which have demonstrated additional comb lines at 770 and 520$\,$nm wavelengths, respectively.\cite{Miller_14,Wang_16} A further challenge for integration is that previous demonstrations of self-referenced combs have produced $\sim$ THz repetition rates, therefore requiring multiple combs for the frequency conversion.\cite{Newman:19}    
 
 Switches and modulators on the Si$_3$N$_4$ platform have mainly been demonstrated by utilizing thermo-optic phase shifters in combination with interferometric devices such as a Mach-Zehnder.\cite{Joo_18} Compared with SOI that can take advantage of the plasma dispersion effect, this has limited both speed (< $\mu$s) and power consumption (10s of mWs). In addition, these devices usually have a much larger footprint due to the small thermal-optic coefficient of Si$_3$N$_4$.\cite{Arbabi_13} For switching a beam on and off to interrogate a cold atom cloud, these limitations are acceptable since only millisecond timescales are required. The main challenge is achieving an on-off extinction ratio that exceeds 60 dB. 
 
 Interferometric-based approaches will most likely require cascading and compensating for imperfections in beam splitting.\cite{Wilkes_16} To date, the highest extinction ratios have been demonstrated with MEMS-based switches where coupling between two vertically displaced SOI waveguides was achieved by electrostatic actuation.\cite{Nagai_18} This type of design also benefits from being non-blocking with the light routed to a different plane rather than being reflected but is still limited in switching speed. To achieve higher modulation and switching rates then materials such as graphene,\cite{Shiramin_17} lead zirconate titanate (PZT)\cite{Alexander_18} and lithium niobate (LiNb0$_3$)\cite{Jin_16} can be integrated. These provide large electro-optic coefficients that can enable GHz modulation rates on Si$_3$N$_4$. The challenge is to reduce the associated insertion loss and increase extinction ratios.

 Optical isolators are another critical component since they prevent reflected light from de-stabilizing the laser. Optical non-reciprocity has traditionally been achieved in bulk devices using magneto-optical (MO) materials that induce a Faraday rotation. Integrating MO materials with waveguides has proven to be a significant challenge.\cite{Stadler_14} This was only recently achieved on the SOI platform by direct bonding of a cerium substituted yttrium iron garnet (Ce:YIG) onto a Mach Zehnder interferometer. This demonstrated a peak isolation ratio of 21 dB with an 8 dB insertion loss at 1550 nm.\cite{Shoji_08} 
 
 Further work capitalized on Ce:YIG integration by combining with a ring resonator to miniaturize the footprint,\cite{Dai_12} and create polarization diverse isolators.\cite{Zhang_19} Ce:YIG was subsequently integrated with Si$_3$N$_4$ waveguides, where high performance was obtained (isolation ratio of 28 dB with a 1 dB insertion loss).\cite{Yan_20} Disadvantages of MO materials are their incompatibility with CMOS foundries and the requirement of a permanent magnet. In addition, utilizing Ce:YIG below 1$\,\upmu$m wavelength might prove difficult due to the reduced optical transparency\cite{Goto_12} but this is potentially compensated by a larger Faraday rotation.\cite{Onbasli_16}  
 
 To overcome these bottlenecks, magnetic-free schemes have been demonstrated through optical non-linearities,\cite{Hua_16} Brillouin scattering,\cite{Kittlaus_18} and spatio-temporal modulation using surface acoustic waves.\cite{Kittlaus_21,Tian_21} These are all interesting approaches but some effort is required to reach an isolation ratio of 60 dB with low insertion loss.
 
\begin{figure}[t]
\centering
\includegraphics[width=0.45\textwidth]{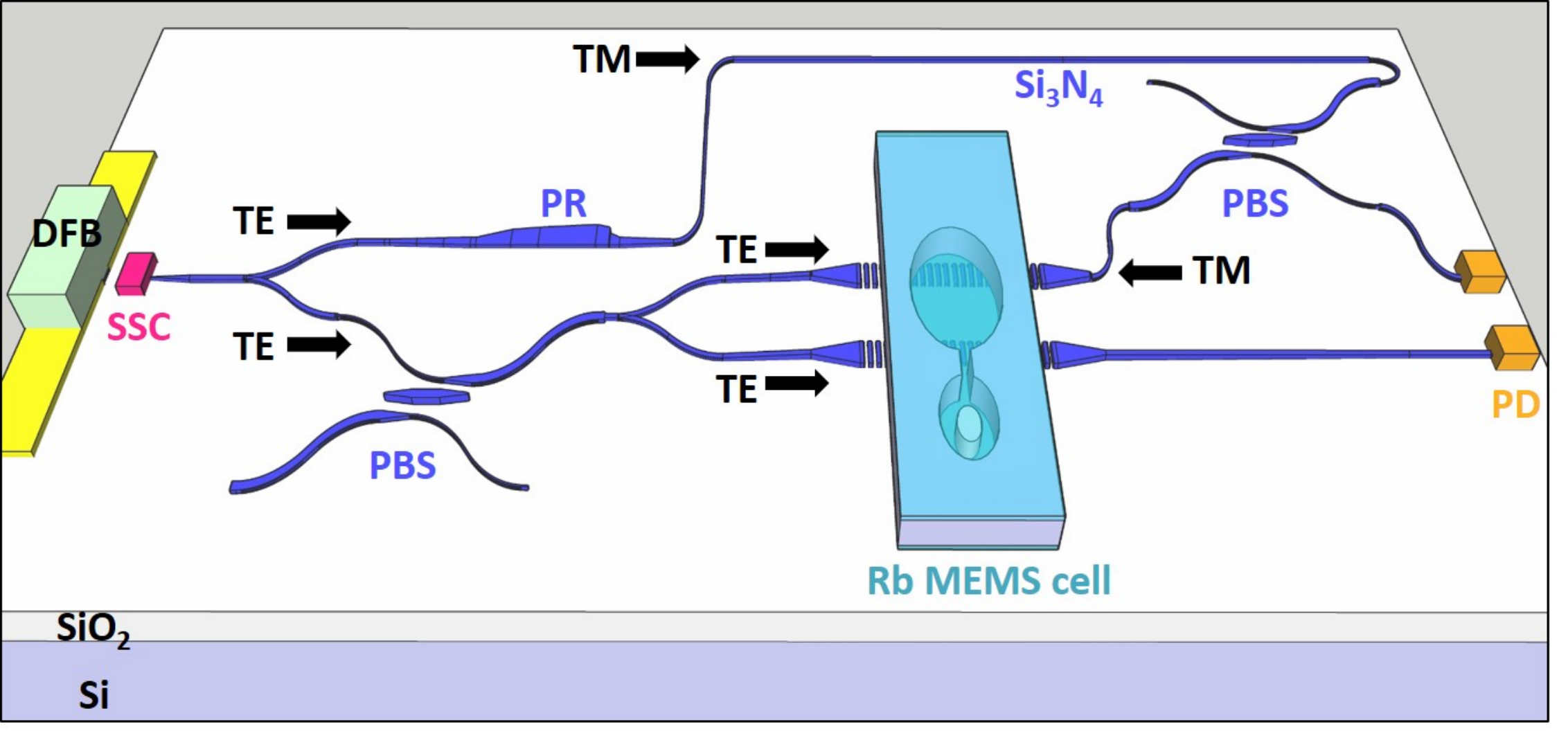}
\caption{\label{SAS_PIC} A schematic diagram of a photonic integrated circuit for saturated absorption spectroscopy of rubidium atoms on-chip using Si$_3$N$_4$ waveguides. The key components include a Si$_3$N$_4$ waveguide polarization rotator (PR) and a polarization beam splitter (PBS), distributed feedback laser (DFB), Rb vapor MEMS cell, and photodetectors (PD).}
\end{figure}

 High-performance polarization components such as rotators and beam-splitters have been difficult to realize with Si$_3$N$_4$ due to the limited geometric birefringence that can be induced compared to SOI devices.\cite{Dai18} Recently, however, a Si$_3$N$_4$ waveguide polarization rotator and polarization beam splitter were demonstrated below telecommunication wavelengths. These devices provide a polarization extinction ratio close to 30 dB at the D2 transition of Rb.\cite{Gallacher_22} The polarization rotator is based on the mode evolution approach using adiabatic tapers, whereas the polarization beam splitter utilizes a cascaded tapered asymmetric directional coupler for TE and TM mode separation. Si$_3$N$_4$ waveguides have recently demonstrated saturated absorption spectroscopy of Rb atoms using antiresonant reflecting optical waveguides,\cite{Yang2007} extreme mode-converting apodized gratings,\cite{Hummon:18} and atomic-cladding waveguides.\cite{Zektzer21} The next integration step will include the circuitry for generating the required counter-propagating pump and probe beams on-chip. Polarization control would allow the pump power to be tuned and filtered from returning to the laser (see Fig. \ref{SAS_PIC})    

\subsection{Apodized grating couplers}
\begin{figure}[!b]
\centering
\includegraphics[width=0.45\textwidth]{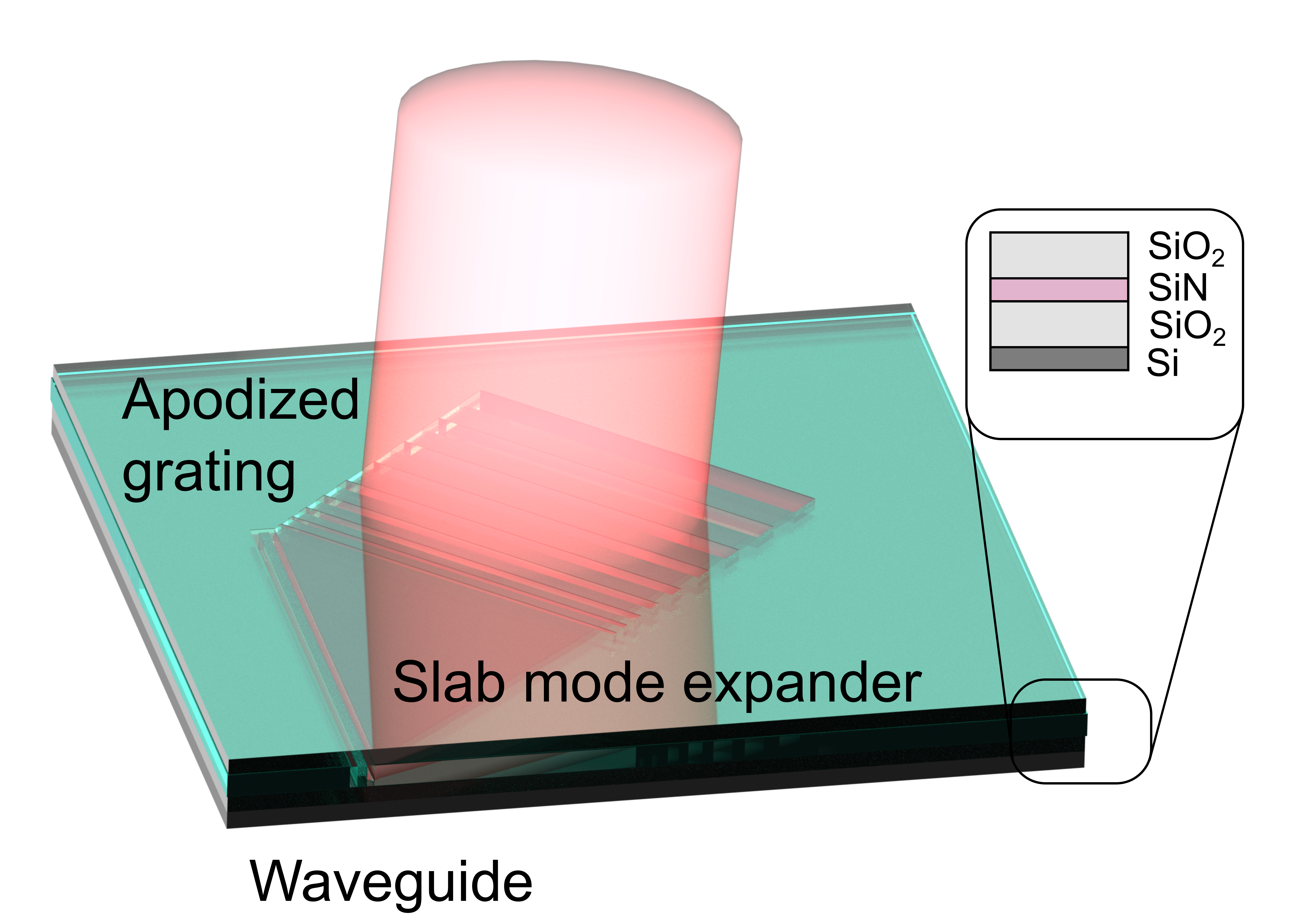}
\caption{\label{PICfig} Schematic diagram of a photonic chip integrated extreme mode converter, based on work demonstrated in Ref.~\cite{Kim2018}. The input waveguide is evanescently coupled to the slab mode expander which is out-coupled through the spatially modulated grating structure with a gradiented duty-cycle and period to provide a Gaussian profile in free-space.}
\end{figure}
With viable solutions for on-chip laser sources and waveguide based photonic circuits, a key direction of recent integrated photonics has been on the efficient out-coupling from waveguide to free-space for controllable atom-light interactions. One such solution is the periodic modulation of the waveguide structure to diffract light from the photonic chip into free-space while remaining micro-fabricated and low volume \cite{Kim2018}. Similar to the waveguides previously discussed, the grating out-coupler is formed of a period stack of SiO$_2$ and SiN, layered upon a silicon substrate. 

However, due to the exponential loss of light through a uniformly spaced grating structure in the waveguide, the out-coupled light would have an exponential intensity profile. Recent work has shown that a Gaussian profile can instead be achieved with the use of an apodized grating structure in the photonic chip \cite{Kim2018}. Additionally, a key recent advancement that facilitates wider beam modes for applications in atomic sensors is the implementation of a slab mode expander \cite{Kim2018}, as illustrated in the photonic chip schematic in Fig.~\ref{PICfig}. By tuning the evanescent coupling between the light confined in the photonic waveguide and the slab mode expander via the waveguide-to-slab-mode gap size, a one dimension Gaussian profile can be formed. When connecting the slab mode expander to an apodized grating chip (with appropriately tuned period and duty cycle), S. Kim $et. al.$ demonstrated a Gaussian profile in the out-coupled radiation with a modal area of 160$\,\upmu$m$\times160\,\upmu$m in early demonstrations \cite{Kim2018}. 

While these initial beam waist demonstrations are not directly applicable to laser cooling, more recent research has pushed the achievable beam area \cite{Yulaev:21}. Additionally, the successful amalgamation of the photonic integrated chip with pre-existing cold atom apparatus, such as the grating magneto-optical trap (discussed in Sect.~\ref{lasercooling}), provides a clear route to the device incorporation into chip-scale cold atom sensors \cite{McGehee_2021}.

\section{Micro-fabricated wavelength reference}
\label{waveref}
Laser referencing to an atomic wavelength is widely implemented in the measurement of magnetic fields \cite{Hunter:18}, time \cite{Newman:19}, rotation \cite{donleyrotation} and length \cite{Hummon:18}, finding application in navigation, geological surveying, medicine, communication and finance. Atomic transitions corresponding to electronic excitation can be measured from atomic vapor using spectroscopic techniques \cite{RevModPhys.89.035002}. While a range of sub--Doppler spectroscopic techniques are suitable for atomic referencing, so--called saturated absorption spectroscopy (Sat-Spec) remains one of the most commonly used methods due to its simplicity and robustness against environmental fluctuations \cite{diodelaserwieman,hughes,Siddons_2008}.

\begin{figure}[t]
\centering
\includegraphics[width=0.4\textwidth]{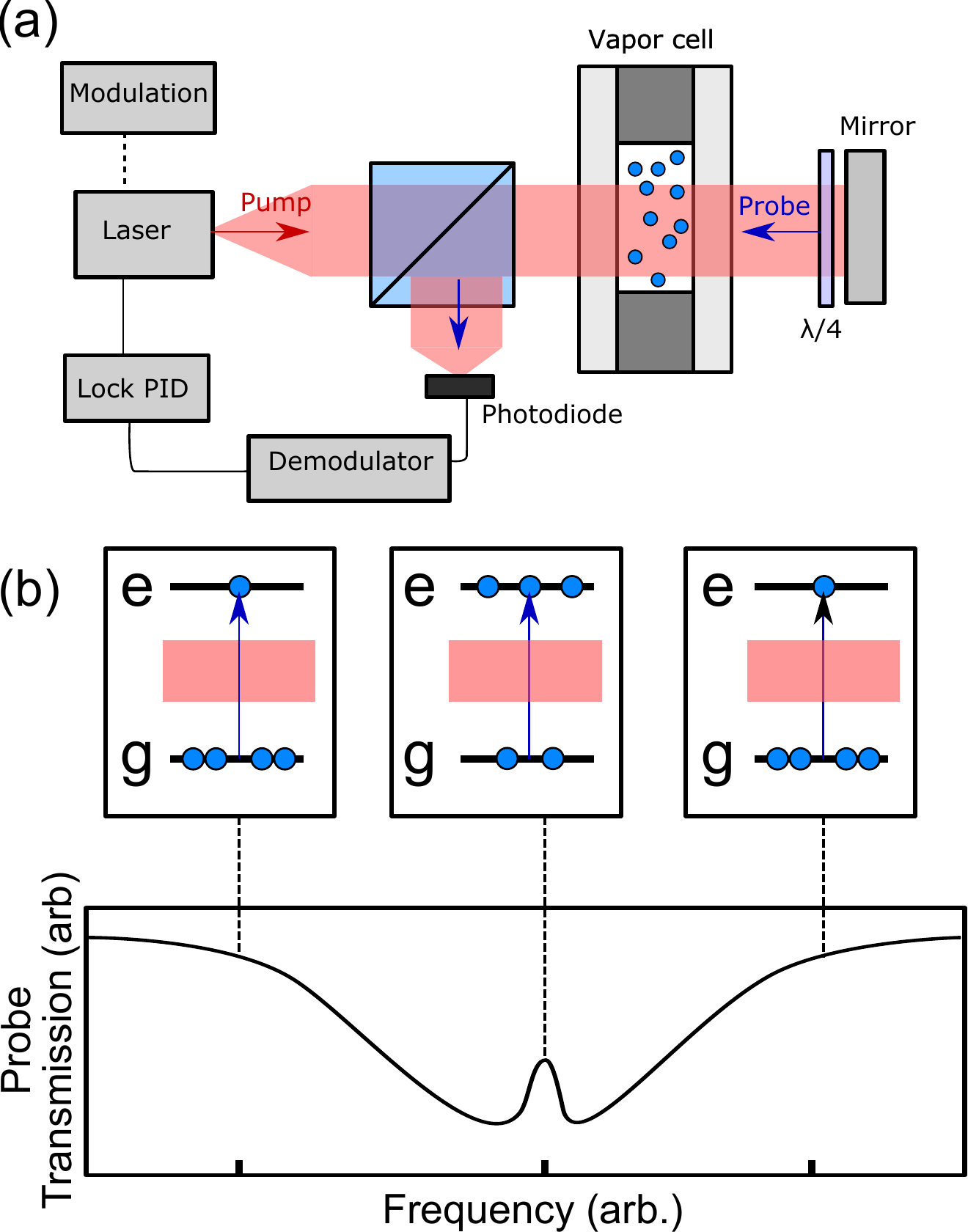}
\caption{\label{satspec} (a): Atomic wavelength reference based on saturated absorption spectroscopy. Laser light is retro-reflected through an atomic vapor cell, with the transmission measured on a photodiode. A modulation frequency, injected to the incident light, is demodulated from the system and sent into lock electronics. The feedback correction signal is coupled back to the laser wavelength. (b): The optical pumping experienced by the probe light as a function of frequency, where the central feature is the atomic transition. When on resonance, optical pumping between the ground (g) and excited (e) state is saturated by the pump beam, such that less light is absorbed by the probe, increasing the photodiode transmission, providing a narrow lock feature.}
\end{figure}

An illustration of the Sat-Spec process is shown in Fig.~\ref{satspec} (a). Here, a single wavelength laser is expanded and passed through a polarizing beam splitter (PBS) for intensity control and polarization purity. The output propagates through an atomic vapor cell, and is retro reflected with the opposite handedness of linear polarization, such that it exits the opposite port of the PBS. Following the double pass, the light transmission from the cell is measured on a photodiode. Since the atomic gas possess a wide range of velocities, $\vec{v}$, the frequency of the interrogating laser light is Doppler-shifted from resonance, such that the Doppler shifted angular frequency is $\omega_{\rm Doppler}=\omega\mp\vec{k}\cdot\vec{v}$, where $\omega$ is the unperturbed resonance angular frequency. While a single laser frequency is used to provide the pump and probe configuration required for Sat-Spec, the anti-parallel direction of the beams introduces a preferential absorption of the atom, depending on the direction of the atom's motion. As such, both lasers, in general, will not simultaneously interact with a single atom with motion along the laser axis. If the atom has zero velocity along the beam, then the frequency of both beams will be the same in the atoms rest frame and both beams will address the same atom. Since the cyclic atomic transition can then be saturated by the pump beam, there is a decreased absorption of photons from the probe beam, effectively increasing the laser intensity at the photodiode. The atomic absorption as a function of the laser frequency can then be converted to a voltage signal through the photodiode. An illustration of the pumping process and corresponding absorption profile of a simplified two-level atom is shown in Fig.~\ref{satspec} (b). 

While the Sat-Spec process provides atomic absorption spectra corresponding to the natural atomic resonances, these absorption peaks do not provide a frequency-dependent error signal with zero-crossing in the voltage that can enable frequency feedback to a servo loop for laser wavelength stabilization. To overcome this, a modulation frequency can be mixed with the laser frequency, such that a lock-in amplifier can demodulate this frequency to provide a lockable error signal that has a zero crossing in voltage axis. Sat-Spec systems in glass-blown vapor cells have demonstrated a frequency stability of $3\times10^{-12}\,\tau^{-1/2}$, where the short-term limitation was attributed to the frequency modulation of the laser diode for locking electronics \cite{AFFOLDERBACH2005291,Ye:96}. 

\subsection{Vapor cell fabrication}
In several laboratory based experiments, the alkali source under interrogation is typically contained within a commercially--off--the--shelf (CoTS) 5-10-cm long glass-blown cell, to provide a sufficient absorption of resonant light within the alkali vapor. The glass blown cells are manufactured by heating glass, typically borosilicates, to the working point of the material where the required cell geometry can be mechanically pressed into the malleable glass \cite{pawel}. These cells are connected to a larger glass vacuum cell and pumping apparatus, required to provide a sufficient pressure and alkali density, provided from an ampoule source that is directed into the cell by localized heating of the chamber. The glass-blown stem connecting the larger vacuum to the cell is then re-shaped with a blow torch to isolate the cell \cite{pawel,Kitching2018}, discussed in more detail in Sect.~\ref{cellclosure}. The manufacturing time, and difficulty in cell-to-cell reproducibility of the inner vacuum environment, have limited the capabilities and scalability of glass-blown macro-scale cells. While macro-scale vapor cells are limited in production scalability, micro-fabricated glass-blown cell structures have been demonstrated with anodically bonded silicon-glass structures \cite{EKLUND2008175}. This routine enables the mass production of 900~$\mu$m diameter micro spheres, blown from 100~$\mu$m thick borosilicate glass at 850~$^{\circ}$C. However, the room temperature absorption in these micro-spheres is significantly reduced compared to the standard macro-cell, degrading the performance of atomic sensors based on this technology. Additionally, the short radius of curvature of the optical surfaces provides significant focusing of the probing light.

MEMS technologies have pioneered the advancement of chip-scale atomic devices such as clocks, magnetometers, gyroscopes and wavelength references \cite{Knappe2004,kitchingmag,donleyrotation,Hummon:18}. MEMS cells consist simply of hermetically sealed silicon frames, micro-fabricated to achieve the designed geometry, size and shape required for the cell. A commonly used method for silicon processing is reactive-ion-etching (RIE), and its supplementary deep-reactive-ion-etching (DRIE) \cite{RIE,DRIE}. This method of silicon etching is based largely around ionic bombardment of an exposed silicon surface to remove material that is not coated in a protective layer. Such protective layers are easily deposited and patterned upon the silicon substrate with photo-lithography, enabling the user to well define narrow structures within the etching process \cite{VLADIMIRSKY1999205}. The primary advantage of this etching technique is the ability to achieve anisotropic etching without the need to rely upon the Miller indices of the crystalline structure of the silicon. Secondly, with etching rates for silicon on the order of 50~$\mu$m/min, the DRIE process is significantly faster than wet etch processing \cite{DRIE}. A drawback to the use of ion etching techniques is the poor surface quality that results from the duty cycle of the bombardment process. Although, this can largely be overcome in post processing with wet etch solutions such as KOH or TMAH \cite{wetetch}.

Wet-etching provides an alternative route to silicon processing with the added benefit of simplicity and cost. This process relies on the solubility of silicon, typically using alkali hydroxide with a sufficiently high pH to etch along the crystal planes of the substrate. While silicon possesses a diamond-cubic lattice structure, the plane selection is dependent upon the etch rate of the reactant solution, as well as the concentration and temperature during etching \cite{wetetch}. For example, commonly used KOH solution will attack the $\{100\}$ and $\{110\}$ planes $\times$400 and $\times$600 faster than the $\{111\}$ plane respectively. As such, it becomes relatively easy to reveal the $\{111\}$ silicon plane from the wet-etch processing, at an angle of 54.74$^{\circ}$. Since the etch follows well defined Miller indices through the crystal lattice, optically smooth walls are regularly achieved with this process. However, with etching times for the $\{100\}$ plane $\sim1~\mu$m/min, the processing times for deep features are slow, with a 1~mm etch requiring $\sim$17 hours of etching \cite{wetetch}.

While these techniques are well suited to etching features on the nano-scale, fabrication on the millimetre-scale can be achieved with more direct approaches such as mechanical cutting. While traditional machine milling offers a simple alternative to bulk material processing, its application with brittle materials, such as silicon, may reduce the yield from wafer cuts and hinder the achievable surface quality due to damage \cite{siliconmachining}. Alternatively, laser micro-machining offers a cheap and versatile method for the processing of a wide range of materials, including silicon \cite{lasermachining}. Unlike the other methods outlined for silicon processing, laser machining is a thermal ablation process, using very short optical pulses to vaporize material without heat damage to the surrounding structure \cite{lasercuttingreview}. Importantly, this method of fabrication is not limited in etch depth by dry etch phenomena or wet-etch processing times, providing a simple means to thick silicon fabrication. The success of this method has led to its use in micron-feature drilling, patterning, and the dicing of silicon \cite{laserdicing, lasercuttingreview}. More recently, water jet cutting of silicon has been shown as a suitable approach in the fabrication of deep silicon cells \cite{dyer}. Using a high-pressure jet with a water and abrasive mixture, this process can cut 6~mm deep into the silicon, with a 1.5~mm minimum feature size demonstrated in the first generation. The reduced processing costs and fabrication simplicity make this an attractive routine for the generation of silicon-based vapor cells for atomic spectroscopy. Once the silicon frame has been processed to the desired cell geometry, micro-electro processes are used to form hermetic seals for vacuum encapsulation, such as anodic bonding.

\subsubsection{Anodic bonding}
\begin{figure}[!t]
\centering
\includegraphics[width=0.45\textwidth]{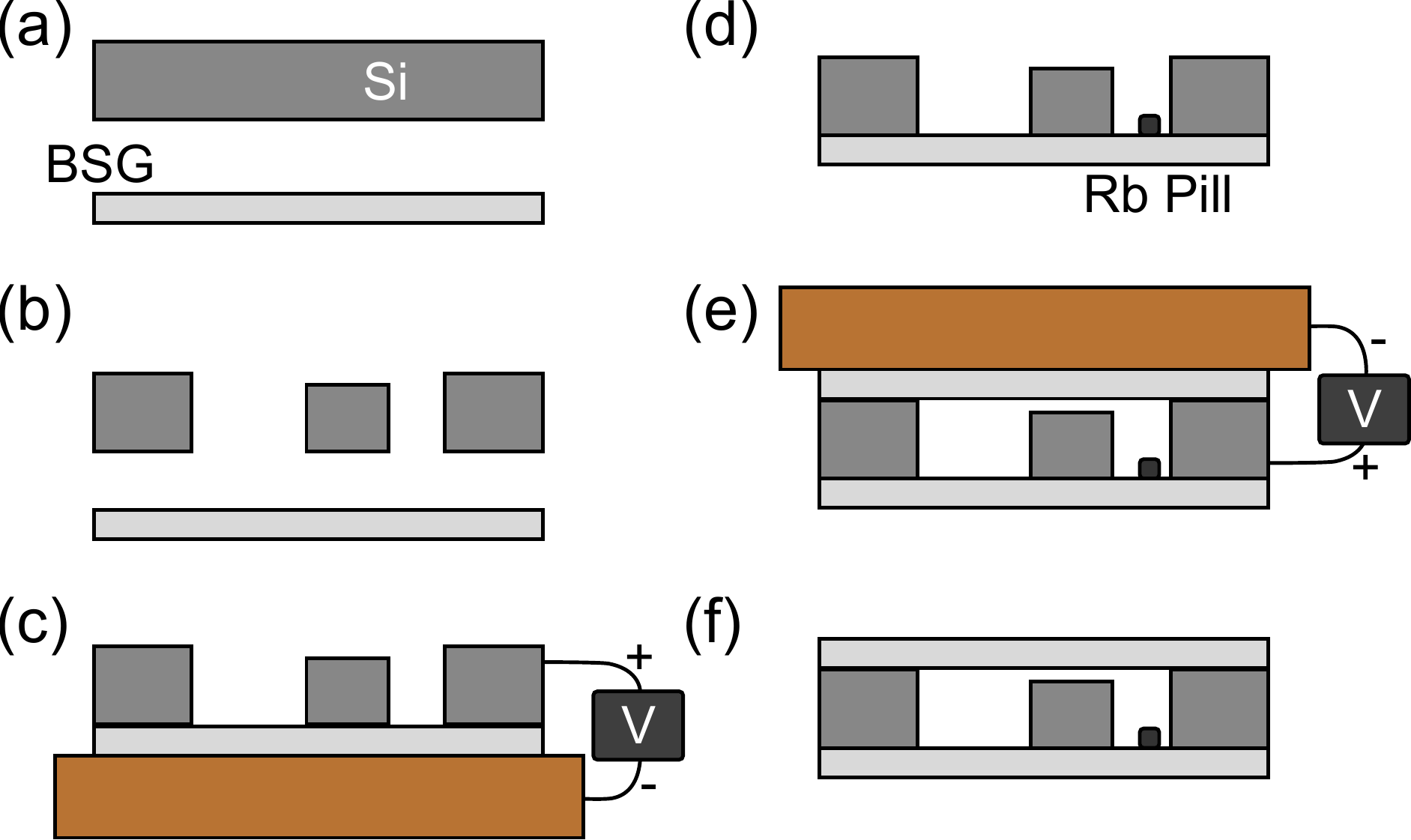}
\caption{\label{anodicbond} Simplified anodic bonding sequence between silicon and borosilicate glass (BSG) wafers, adapted from Ref.~[\cite{vincentthesis}]. (a): The silicon and glass substrate prior to processing. (b): A DRIE is used to etch the main cell, channel and pill cavity. (c): The silicon and glass wafer are brought into contact with a heated surface electrode holding the glass wafer. The bonding voltage is applied across the substrates. (d): The alkali pill source is located in the dedicated pill cavity before the upper borosilicate glass wafer and silicon surface are brought into contact. (e): Second anodic bond under vacuum for cell encapsulation. (f): The final, hermetically sealed cell with a controlled vacuum environment and internal alkali pill source.}
\end{figure}

Anodic bonding is a widely-used core technology to hermetically bond the surfaces of hard glass and polished silicon frames that have low enough surface roughness and large ionic mobility within the glass for electrostatic attraction of the two materials \cite{wallis,Kitching2018}. The anodic bonding process, highlighted in Fig.~\ref{anodicbond}, consists simply of a high-voltage power supply and heat source. The heat source is used to control the glass temperature and hence the ionic mobility of Na$^{+}$ ions within the glass, with standard bonding being generally carried out between 200-400$^{\circ}$C \cite{Kitching2018, Rushton2014}. Low-temperature anodic bonding is required for MEMS vapor cells that are hindered by vapor diffusion into the glass cell windows at high temperatures. Reducing the bond temperature also alleviates the risk of damage to integrated components and circuitry \cite{heatinganodic}. Such low-temperature anodic bonding of borosilicate for MEMS vapor technology has been demonstrated lower than 200$^{\circ}$C \cite{pawel}. 

The high-voltage power supply applies a positive voltage to the silicon frame via a surface platen or point electrode. In the example highlighted in Fig.~\ref{anodicbond} a simple two chamber vapor cell structure is fabricated. Initially the silicon is coated with a thin SiO$_2$ layer, to which photolithography can be used to spatially expose regions of the silicon surface for etching. The exposed silicon regions can then be etched with DRIE, to give structure to the cell geometry, as seen in Fig.~\ref{anodicbond} (b). Following the dry etch, the remaining SiO$_2$ is chemically stripped, and the silicon surface is cleaned for anodic bonding. Next, the glass and silicon wafer surfaces are brought into contact, with the glass typically held on a heated surface electrode, highlighted in Fig.~\ref{anodicbond} (c). The voltage supply ground is connected to the surface electrode, providing a negative charge to the glass, which is now sandwiched between the negatively charged platen and positively charged frame. The platens are heated to a steady-state temperature of $\sim300^{\circ}$C prior to the voltage being applied. The electric field then attracts the glass impurities such as Na$^{+}$ ions to the cathode, leaving behind the relatively immobile oxygen anions in a now negatively charged space. This negatively charged region forms at the interface with a positively charged silicon layer, which, when the spacing between both layers is on the order of $\mu$m, permits the electrostatic force to pull together these two materials and form SiO$_2$ at the interface. However, in order for the electrostatic attraction to enable the anodic bond, the bonding surfaces must have a RMS surface roughness $<50$~nm \cite{anodicbondtextbook}. When both surfaces meet the surface roughness and cleanliness criteria, the anodic bonding process is able to form a hermetic seal for vacuum encapsulation of atomic vapor cells. Typically, a failed bond will show interference fringes between the glass and silicon, highlighting an air gap between the substrates. A notable drawback to this bond is the release of O$_2$ into the vacuum cell at this stage \cite{anodicbondtextbook}. While the quantity of released oxygen is a function of the bonding parameters and wafer composition, J. A. Rushton $\textit{et. al.}$ estimate between 10$^{13}-10^{14}$ molecules per millimeter inner bonding circumference \cite{Rushton2014}. Methods to overcome the released O$_2$ content in anodically bonded cells are outlined in Sect.~\ref{deposition}.

Following the first bond, the alkali source is deposited into the glass-silicon frame, as shown in Fig.~\ref{anodicbond} (d). The alkali pill is deposited into a dedicated cavity, fluidly connected to the main chamber with micro-channels, discussed in detail in Sect.~\ref{deposition}. The second anodic bond is then carried out inside a controlled vacuum environment, where the total background pressure can be pumped to achieve a suitable environment prior to encapsulating the cell as seen in Fig.~\ref{anodicbond} (e) and (f). The simplicity in cell fabrication that can be achieved with this approach is emphasized by the wide implementation of anodically bonded technology in vapor cell fabrication \cite{Kitching2018}, accelerometers \cite{anodicaccelerometer}, gas sensors \cite{anodicgas}, and micro-pumps \cite{anodicpump} among other applications.

\subsubsection{Fusion bonding}
Fusion bonding enables the hermetic sealing of two highly polished wafers brought into contact at room temperature via the Van Der Walls attraction. Since the Van Der Walls attraction is short-ranged, the wafers have a significantly higher requirement of the surface smoothness and cleanliness compared to the previously discussed bonding method, with a typical RMS surface roughness $\leqslant 1$~nm required with silicon bonding \cite{Cui2008,directbond2}. The fusion bond, also referred to as direct bonding, can be initially achieved at room temperature. However, this initial bond between the wafers is relatively weak compared to other bonding methods, such that the wafers typically require heat treatment between 800-1200$^{\circ}$C to convert the Van Der Walls bond to a much stronger covalent bond \cite{Cui2008}. Alternatively, research has looked at the direct formation of covalent bonds between silicon wafers when brought into contact under UHV conditions, where the purity of the wafer surface can be maintained without the build up of contaminants or oxidization layers that would otherwise degrade the bond \cite{directbondUHV}. While the processing of the silicon wafers requires significantly more effort than an anodically bonded stack, fusion bonding remains an advantageous method for future cold atom system fabrication due to the lack of contaminant release during bonding and the ability to achieve hermetic bonds between adjacent wafers at room temperature.

\subsubsection{Thermo-compression bonding}

\begin{figure}[!t]
\centering
\includegraphics[width=0.4\textwidth]{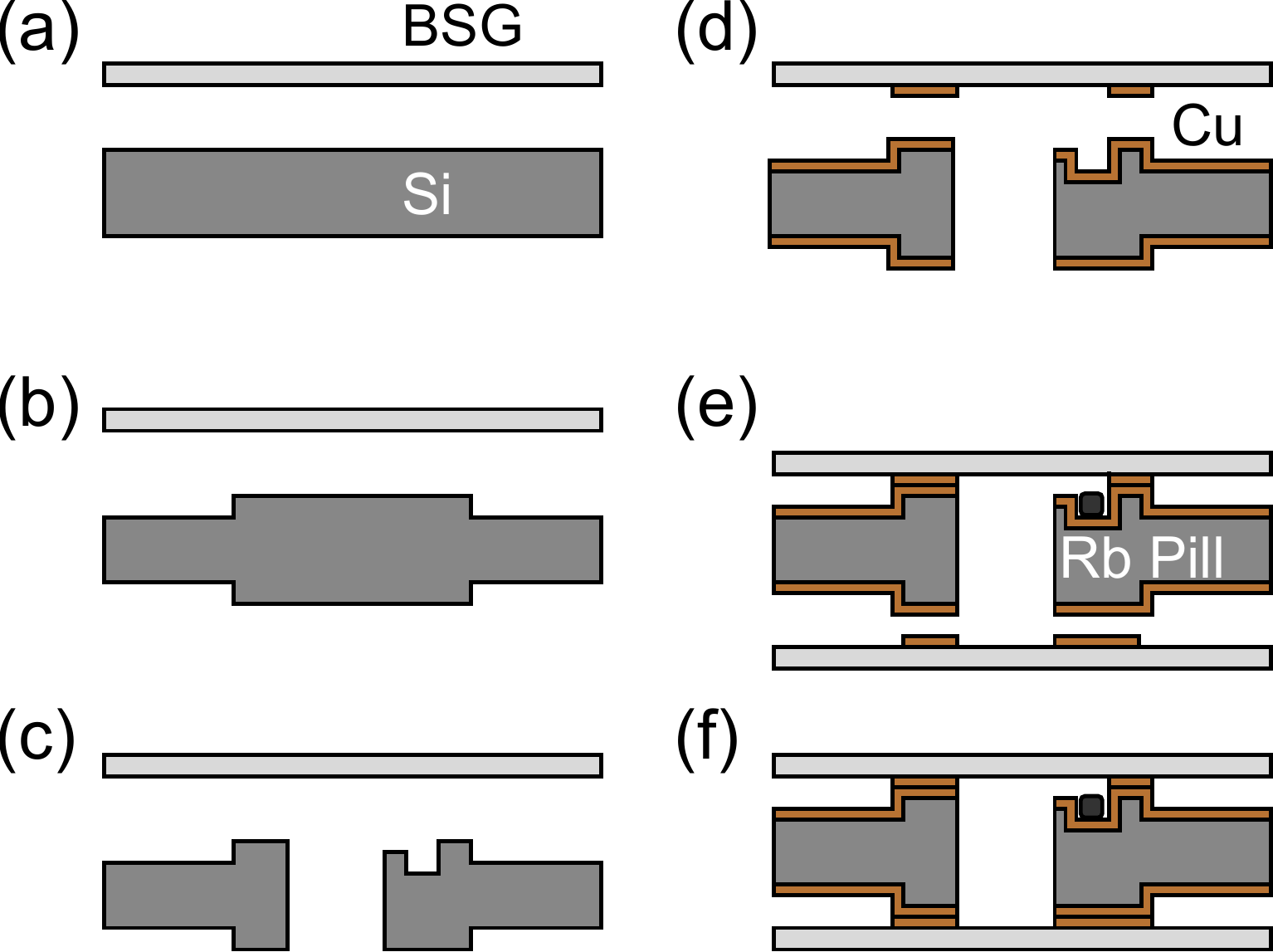}
\caption{\label{cucu} A simplified Cu-Cu thermo-compression bonding sequence between silicon and borosilicate glass (BSG) wafers, adapted from Ref.~[\cite{karlencucu}]. (a): The silicon and glass substrate prior to processing. (b): A shallow DRIE is used to emphasize the bonding regions on both sides of the silicon. (c): DRIE of main chamber, channels and cavities. (d): Sputter deposition of a 50~nm Ta adhesion layer, followed by a sputtered 500~nm thick Cu layer on the bonding surfaces of the silicon. The deposition of Ta and Cu is also applied to the glass wafers, with the addition of spatial selection of the coating with use of lift-off stripping. (e): The alkali pill source is added to the cell cavity before the Cu-Cu thermo-compression bond between the upper borosilicate wafer and Si surface. (f): Cu-Cu thermo-compression bonding of the second borosilicate wafer to the silicon surface, encapsulating the cell vacuum environment.}
\end{figure}

Thermo-compression bonding is a diffusion based process capable of forming hermetic seals between two metal coated surfaces, using metals such as Cu, Au or Al. When pressed together under heat and force, the atoms within the metal coated surfaces diffuse in to the lattice structure of the in-contact metal, while remaining below the melting point of the metal. This method has recently been demonstrated for the bonding of silicon and glass in atomic vapor cells, with the process outlined in Fig.~\ref{cucu} (a)-(f), as described in Ref.~[\cite{karlencucu}].

The first step of the process uses a shallow DRIE to allow contact between the wafers only at the required bonding regions. The inner cell cavity is then etched with DRIE, including the pill cavity and micro channels. Next, the bonding surfaces of the silicon and glass surfaces are sputter coated in a 50~nm Ta adhesion layer, followed by a 50~nm thick Cu layer for thermo-compression. The Cu coating area of the glass wafer is spatially selected with lift-off stripping from non-bonding surface regions. The bonding surfaces are then pressed together under force and high temperature to form a hermetic seal. Finally, the same procedure is used for the bonding of the lower glass wafer to the silicon body, following the deposition of the alkali source.

The ability to surface coat materials with a suitable bonding metal, such as Cu-Cu, and then form a hermetic seal from thermo-compression widens the available materials that can then be incorporated into the cell fabrication process. Additionally, this method does not release oxygen into the cell during bonding, a key advantage over anodic bonding. However, a key limitation to this technique is sensitivity to the total thickness variation of the surface deposited metal, such that critical deposition control is required to coat the bonding surfaces. Critically, the bonding time is significantly longer than anodic bonding since this is a diffusion driven process \cite{karlenthesis}. Finally, this process has been noted to have a more stringent requirement on surface cleanliness than anodic bonding, restricting the fabrication process to a higher level of clean-room environment \cite{karlencucu}. While this method uses bonding temperatures comparable to anodic bonding, the lack of the electrostatic force that exists in the anodic bond has been noted to require an increased force of $\sim14$~kN between the wafers during thermo-compression \cite{karlencucu}. Finally, additional vacuum preparation stages are required to remove the copper-oxide layer formation from the bonding surface prior to thermo-compression \cite{karlencucu,karlenthesis}.

\subsubsection{Eutectic bonding}
As demonstrated with thermo-compression bonding, the hermetic sealing of wafer stacks can be achieved via the inclusion of an intermediate adhesion layer. On such intermediate layer bonding technique is the eutectic bond that is possible between materials that can form a eutectic alloy, which melts at a substantially lower temperature than the individual materials melting point \cite{Rushton2014, eutectic}. 

For silicon wafers, both Al and Au are capable of forming such a eutectic alloy. In the case of Au, a thin Ti adhesion layer is first deposited on the glass wafer that will be bonded to the silicon. Subsequent deposition of $\sim$1~$\mu$m of Au onto the adhesion layer, followed by bringing the wafers into contact and heating above the eutectic temperature, $362^{\circ}$C for Au-Si, forms a strong bond between the wafers.

This enables the bonding between silicon and glass wafers where eutectic alloy forming metal is coated on the glass bonding region. While Au is reactive with some alkali elements and is potentially less favourable as a result, Si-Al alloys could be implemented in component bonding for cold-atom systems. With this being said, the Al coated wafer will require careful processing prior to bonding to avoid a oxidization layer on the bonding interface \cite{karlenthesis}. 

\subsection{Vapor cell geometry}

\begin{figure}[!t]
\centering
\includegraphics[width=0.41\textwidth]{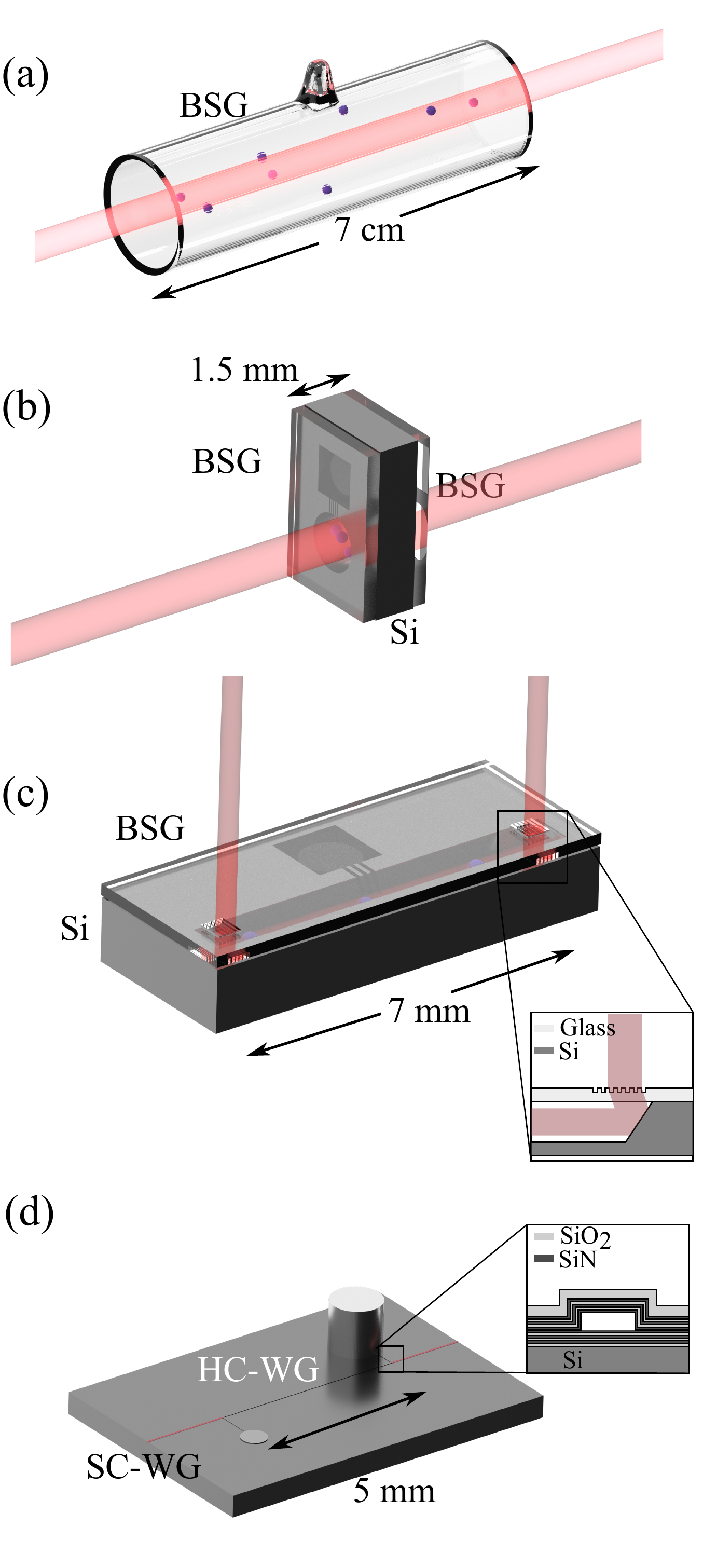}
\caption{\label{geometry} Cell geometry options for micro-fabricated atomic wavelength references. (a): Glass blown borosilicate glass cell with a 7~cm absorption path length. (b): Line of sight glass-silicon-glass MEMS cell, as demonstrated in Ref.~[\cite{Hasegawa2011_buffer_gas}] with a 1.5~mm absorption path length. (c): Wet-etched, elongated glass-silicon cell, as demonstrated in Ref.~[\cite{chutani}] with a 7~mm absorption path length. (d): Hollow core waveguide, based on the anti-resonant reflecting optical waveguides (ARROW) shown in Ref.~[\cite{hollowcoreschmidt}]. The light is routed through a solid-core waveguide (SC-WG) before entering the hollow-core waveguide (HC-WG) connected to the alkali reservoir. The inset shows the cross-section of the HC-WG dielectric layer stacks.}
\end{figure}

Glass blown, borosilicate cells are commonly used at the core of laboratory atomic wavelength references, due to the 7~cm absorption path length available from the manufacturing of the glass, shown in Fig.~\ref{geometry} (a). However, the glass blowing process does not provide a simple route to miniaturization nor reproducible vacuum conditions. In the past decade, significant effort has been made to miniaturize atomic wavelength references through the development of micro-fabricated alkali vapor cells \cite{moreland}. The simplicity of manufacturing from DRIE, wet-etching or mechanical cutting to achieve a simple line of sight glass-silicon-glass cell geometry, has paved the way for MEMS cells to be implemented at the core of chip-scale atomic clocks and commercial atomic products \cite{Knappe2004}. An example of a two-chamber MEMS cell with a 1.5~mm absorption path length is shown in Fig.~\ref{geometry} (b), based on the work described in Ref.~[\cite{Hasegawa2011_buffer_gas}]. However, these miniaturized packages are typically limited in performance by a poor signal-to-noise ratio resulting from the short absorption path length in straight lined silicon cells \cite{Kitching2018}, that result from technical difficulties in deep etches or etching times for dry and wet etch processes respectively \cite{CHUTANIdrie,P_tremand_2012}. Although some performance gains can be reclaimed by increasing the vapor density via external heating of the cell, the heating apparatus requires additional circuitry, heaters and electrical power consumption, degrading the SWaP footprint of the device. Recent literature has overcome the performance of standard sub-Doppler spectroscopy of alkali atoms in line of sight cells by demonstrating an optical wavelength reference in a Rb MEMS cell using two-photon spectroscopy to demonstrate an instability of 1.8$\times10^{-13}\,\tau^{-1/2}$ \cite{newman2021highperformance}. 

Alternatively, performance gains can be achieved from a longer absorption path length, as described by the Beer-Lambert law $I(x)=I_0\exp{(-\alpha x)}$, where $I(x)$ is the intensity of the light at position $x$, $I_0$ is the incident beam intensity, and $\alpha$ is the absorption coefficient; dependent on the frequency of the incident light and the temperature of the atomic medium. Hence, the poor signal-to-noise ratio from the line of sight cells can be overcome with a longer absorption path length. With this in mind, alternative geometries in silicon fabrication have been explored to increase the effective cell path length while remaining micro-fabricated. A notable example is the anisotropic wet etch of the $\{111\}$ crystallographic plane at 54.7$^{\circ}$ to route the incident light through an elongated silicon cell geometry \cite{chutani}, as illustrated in Fig.~\ref{geometry} (c). In this case, a 650~$\mu$m deep cavity was wet-etched into the 1.5~mm thick silicon frame to achieve a 7~mm long path length through the long axis of the silicon. Since incident orthogonal light will not be routed parallel to the silicon surface by reflection from the 54$^{\circ}$ walls, a diffraction grating is etched into the upper glass wafer to compensate the incident light angle. The addition of the transmissive grating provides a routine for simple alignment of the incident light. However, the inclusion of the additional fabrication stages and critical wafer alignment increases the production complexity and cost. 

To aid the device scalability, H. Nishino $\textit{et. al.}$ have demonstrated the capability to wet-etch the $\{$100$\}$ plane of a silicon wafer to achieve 45$^{\circ}$ optically smooth walls as a method for light routing in MEMS vapor cells \cite{nishino}. In order to realize the 45$^{\circ}$ silicon surfaces, the wafer is cut-off at 9.74$^{\circ}$ towards the $\{011\}$ plane, such that KOH wet etching reveals a $54.74^{\circ}-9.74^{\circ}=45^{\circ}$ wall with respect to the $\{100\}$ crystal plane. However, light scatter from the silicon wall roughness post etching reduced the transmitted light by half of the theoretically expected value. Additionally, the shallow etched reflective walls restrict the incident beam width and atomic absorption within the cell. Secondary to angled cleaving of the silicon cut, KOH/isopropanol mixtures have been demonstrated to reveal slower etch planes other than the $\{$100$\}$, capable of realizing 45$^{\circ}$ silicon features \cite{Backlund_1992, Rola2014}. This etching process is highly dependent on the solution mix ratio, temperature and etch time, restricting the process to a short window where $45^{\circ}$ silicon walls are attainable. More recent work on elongated vapor cell fabrication has stepped away from wet-etch processing in favour of mechanically cut, and anodically bonded glass mirrors \cite{Nishino:212}. Formed from glass dicing, the 45$^{\circ}$ reflectors are coated in multi-layers of dielectric coatings and anodically bonded in position once aligned within the cell to provide an increase absorption path length in micro-fabricated cells.  

Hollow core waveguides offer a simple method for an increased absorption path length while remaining micro-fabricated \cite{hollowcore, hollowcoreschmidt, Yin:04}, with an example schematic from Ref.~[\cite{Yang2007}] shown in Fig.~\ref{geometry} (d). Using similar principles of operation to previously demonstrated hollow core fibres \cite{Perrella:13, Perellahollowcore}, the hollow core waveguides use stacked dielectric coatings to confine light and atoms within a single mode photonic structure, providing a competitive absorption path length with a scale significantly smaller than micro-fabricated vapor cells \cite{Yang2007}. 

The hollow core, anti-resonant reflecting optical waveguides (ARROW) have been fabricated in silicon with a 5~mm absorption path length, coupled to a solid-core waveguide at either end for light coupling, and connected to a Rb reservoir to provide vapor content. When operated at 70$^{\circ}$C, the ARROW system has demonstrated spectroscopy with a signal-to-noise ratio that is comparable to that measured from a standard glass-blown cell \cite{Yang2007}. The silicon structure of the ARROW waveguide facilitates the monolithic integration of this technology with other silicon components for device amalgamation into a chip-scale laser cooling platform. Since the light is guided through the waveguide with minimal loss, the structural geometry can vary from direct path channels to the use of meandering waveguides to increase the absorption path length over a relatively small device area \cite{csmc}. The meandering waveguide approach demonstrated a cell length of 14~cm while having a cross sectional area of $\approx1\times0.5$~mm$^2$. While this particular demonstration was machined into aluminium, it is an interesting prospect for future silicon based waveguide adaptations.

As well as hollow-core waveguides, significant efforts have been made in the miniaturization of atomic wavelength referencing via the evanescent interaction from a micro-fabricated waveguide \cite{Stern2013}. The narrow, solid core waveguide enables a strong atom-light interaction light confined within the SiN structure, while the transparency of the core material allows a spectroscopic measurement from the alkali vapor cladding surrounding the waveguide. These narrow waveguides are capable of providing saturated spectroscopy of Rb using nano-watts of optical power in an absorption path length of 1.5~mm. Subsequently, meandering SiN waveguides were fabricated with a 17~mm long path length, demonstrated for applications in the near-IR \cite{Stern2017} and telecommunications \cite{hummonstern} wavelength ranges. This technology provides a significantly reduced SWaP, while remaining fully micro-fabricated and capable of systems integration for on-chip laser locking. 

\section{Atomic deposition and regulation}
\label{atomdepo}
The purity of the alkali vapor content in cold-atom experiments is of critical importance to reduce contaminant collisions and vacuum degradation. As such, the alkali is typically provided by an ampoule source or resistively heated alkali-metal dispensers (AMDs) due to the low level of background gas release during sourcing \cite{McGilligan2017, c-adams, grifflaserheating}. While these methods have proven useful in the past, the inclusion of electrical feedthroughs to activate AMDs, as well as the difficultly of handling alkali ampoules in air, make these unattractive options when reducing the cold-atom apparatus down to the chip-scale. Alternatively, existing literature has shown the compatibility of azide and chloride compounds with MEMS cells due to their dispensing simplicity during cell fabrication \cite{kitching2002, moreland}. After dispensing an aqueous mix of the azide solution, the compound is decomposed under UV light to form 2RbN$_3\rightarrow$2Rb$+$3N$_2$. While the natural abundance of N$_2$ can prove favourable for applications in vapor cell clocks and magnetometers, the presence of the buffer gas would be detrimental in a cold-atom system. A recent experimental technique outlined by D. G. Bopp~$\textit{et. al.}$ has shown that a combination of wafer displacement and localized alkali condensation prior to bonding could enable azide to be used in a manner that could potentially be free from N$_2$ buffer gas \cite{Bopp_2020}.

In this section we outline recent research into cold-atom compatible atomic sources and density regulators that would remain suitable with micro-fabricated components and simple fabrication routines.

\subsection{Alkali sourcing}
\label{micropill}
\begin{figure*}[t!]
\centering
\includegraphics[width=0.8\textwidth]{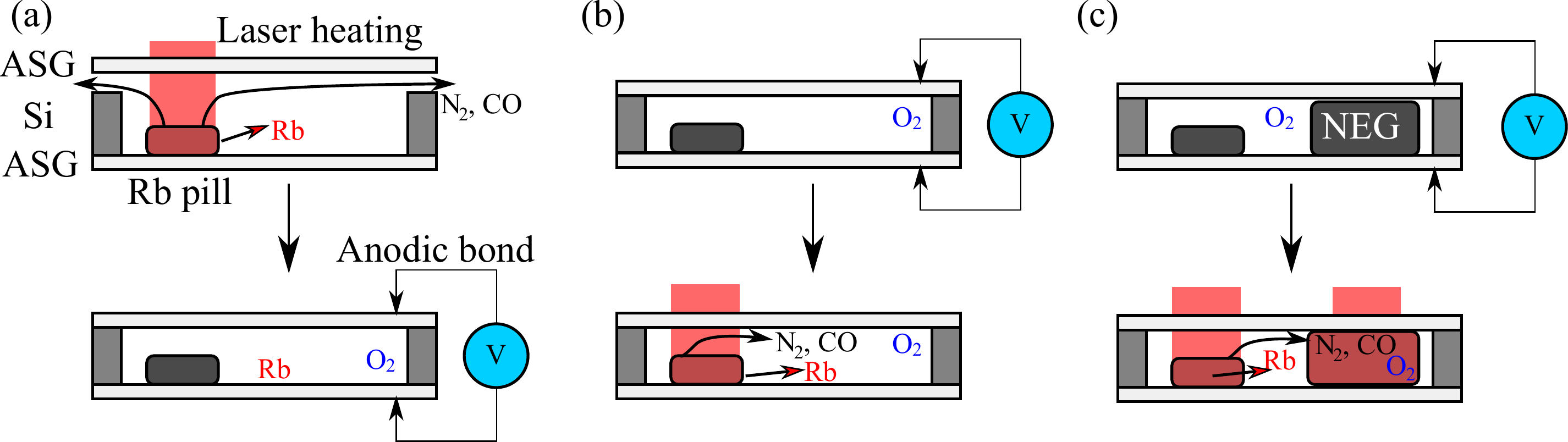} 
\caption{\label{deposition} Rubidium cell deposition options. (a): Pill is laser heated before the cell is hermetically sealed. Temperature gradients between the external vacuum and MEMS cell walls can then be used to evacuate the non-Rb contaminants (N$_2$,CO in this example) while condensing the metal Rb within the cell. Anodic bonding then seals the cell closed with a Rb vapor density. The anodic bonding process provides an additional O$_2$ contaminant into the cell. (b): The cell is anodically bonded for a hermetic seal prior to pill activation. Once sealed, the pill is laser heated, releasing both the Rb and non-Rb elements into the vapor. (c): A non-evaporable getter (NEG) and pill combination are deposited into the cell prior to cell sealing. Both the pill and NEG are laser heated, such that the non-Rb and O$_2$, produced during the bonding process, diffuse into the NEG bulk and form stable chemical compounds that are irreversible to diffuse back out of the bulk. This leaves a a pure Rb vapor density behind in the cell.}
\end{figure*}
A simple and effective approach to atomic sourcing is the commercially available solid alkali pill, capable of in-air handling and external activation with no significantly reactive atomic species produced \cite{gorecki}. Based on a chemical reaction of 2Rb$_2$CrO$_4$+ZrAl$_2\rightarrow$2Rb+Cr$_2$O$_3$+Al$_2$O$_3$+3ZrO$_2$, the pill source provides a means to release a substantial alkali vapor density with minimal reactant by-products \cite{boudotpill}. The reduced complexity of incorporating the pill into the pre-bonded cell, as well as the ability to avoid an electrical feed-through in favour of laser heating \cite{grifflaserheating}, make this approach a promising candidate for chip-scale sensors \cite{mcgilliganSPIE}. Potential activation routes are highlighted in Fig.~\ref{deposition} (a)-(c). The first option is the activation of the Rb pill prior to cell closure. The pill is placed inside the pre-bonded glass-silicon cavity and laser heated. During activation the cell is kept at a lower temperature than the surrounding vacuum to enable Rb condensation within the cell, whilst the non-Rb contaminants (N$_2$ and CO) are pumped away by the larger vacuum. This approach is similar to that outlined in Refs.~[\cite{Bopp_2020,Nishino:21}]. The upper glass is then brought into contact with the Si wafer and bonded. If anodic bonding is used for this stage, then an unspecified amount of O$_2$ will be released from the edges of the bonding interface, where free oxygen atoms do not join with the Si substrate \cite{HENMI1994243,Rushton2014}. This unwanted oxygen content will rapidly deplete the neutral Rb vapor density by forming Rb$_2$O. It is worth noting that this activation method risks contamination of the bonding surface, which can potentially impact the yield and hermeticity of the cell. To avoid such contamination, the pill can be activated post-bonding, as shown in Fig.~\ref{deposition} (b). However, in this scenario there is no clear route to remove the non-Rb elements released during pill activation, which will ultimately limit the vacuum pressure and device stability. A simple route to overcome these issues, shown in Fig.~\ref{deposition} (c), is the addition of a non-evaporable getter (NEG) pill to sorb the unwanted contaminants from pill activation and the oxygen released during bonding. This enables post-bonding activation while providing a simple route to alkali deposition on the wafer-scale. 

A notable drawback to the pill dispenser is the required silicon `real-estate' for pill inclusion. Necessary to be held firmly in place for activation and ensuring the pill does encroach on the optical access, the pills are typically located in a dedicated cavity, connected to the main cell via micro-channels etched in the silicon \cite{Hasegawa2011_buffer_gas}. These micro-channels have been demonstrated for chevron \cite{Newman:19}, curved \cite{karlencucu} and line-of-sight \cite{boudotlineofsightchannel} geometries, with the predominant purpose being a fluid connection for vapor density to the main cell, while minimising contaminant spray during activation and particle mobility that could hinder the optical access \cite{karlencucu}. Recent work has advanced the use of narrow silicon etched channels between the source and cell to form collimated thermal beams in micro-fabricated cells \cite{Li2019}.

Recently, the pill composition has been simplified to a paste form, providing a method for simple wafer level deposition using more conventional liquid dispensing techniques \cite{paste}. Most importantly, the deposition of the paste enables fine control over the size of the deposited amount, such that the paste can be placed directly into the main cell without the necessity for an additional pill cavity. Authors V. Maurice $\textit{et.~al.}$ demonstrated that cells fabricated with the paste and subsequently laser activated could achieve a clock fractional frequency stability of $1\times10^{-11}$ at one day integration time. It is likely that the advantages demonstrated with the alkali paste deposition may supersede the pill dispensers in the future. However, it remains that due to the simplicity of pill inclusion in chip-scale apparatus, alkali sourcing from solid state pills is now widespread in vapor cell fabrication \cite{boudotpill,Hasegawa2011_buffer_gas,Hummon:18,Newman:19,BoppStern}, with recent industrial transfer of vapor cell technology that incorporate an alkali pill source \cite{VICARINI201899}.

Alternative atomic sources exist in the form of the recently demonstrated graphite reservoirs \cite{AFRL}. Highly oriented pyrolytic graphite (HOPG) is an attractive candidate for atomic sourcing due to its species selective intercalation of alkali atoms, enabling the diffusion of Rb into the bulk of the graphite when heated. It has been shown that common vacuum impurities such as nitrogen and hydrogen do not intercalate into the graphite bulk. This enables the HOPG to effectively filter the vacuum environment to produce a clean alkali source that could meet the needs of cold-atom applications. Atomic sourcing with HOPG bulk has been recently demonstrated in a laser cooling apparatus for a clean alternative to alkali sourcing in compact systems \cite{AFRL}. In this scenario, the HOPG was loaded with Rb by submergence in a ampoule source under heat, to encourage alkali diffusion into the graphite reservoir \cite{AFRL}. Once the HOPG was cooled down, the chamber was evacuated of the alkali vapor. Localized heating of the HOPG then encourages diffusion of the alkali metal back out of the graphite and into the vacuum vapor for clean vacuum sourcing. Since the HOPG is not vacuum encapsulated, particular care needs to be taken to avoid oxidization.

\subsection{Alkali density regulators}
While a clean atomic source is essential to aid the longevity of an isolated vacuum system, the ability to regulate and potentially recycle the alkali density is crucial to increase the device usable lifetime, as well as for stabilizing the impact of temperature dependent density shifts \cite{Kang:19}, fluctuating atom numbers \cite{mcgilliganAIB} and variations in the optical density of the interrogating light \cite{microdisks}. In larger vacuum apparatus, density control has been demonstrated with pulsed alkali-metal dispensers \cite{grifflaserheating,fastAMD,fastamd2}, or light-induced-atomic-desorption (LIAD) \cite{Bogi:09,Marmugi:12,Torralbo-Campo2015}. In recent years, LIAD has been demonstrated in parallel with a MEMS vapor cell as a non-thermal solution to increase the vapor density for a 2~mm absorption path length \cite{MEMSLIAD}. While the apparatus requires an additional light source in the ultra-violet, a 395~nm light emitting diode (LED) provides an efficient and low-SWaP solution to atomic desorption \cite{MOTLIAD}.

An interesting development related to density regulation from a graphite source is the alkali-ion battery (AIB). The AIB utilises electrode plated electrolyte connected to a graphitic reservoir as a voltage controllable source of neutral alkali atoms \cite{kang-roper}. The reversible electrochemical dissociation/recombination at the electrode-electrolyte interface enables sourcing and sinking of the alkali content from the reservoir and vacuum environment, permitting a micro-fabricated, solid-state solution to alkali recycling and density regulation. The sinking process relies on the adsorption of Rb from the vacuum vapor. This Rb then diffuses on the upper device surface to the electrode interface, where electrochemical disassociation occurs. The Rb$^+$ is then conducted through the Rb-$\beta''$-alumina to the lower electrode where electrochemical recombination occurs to neutralise the Rb that can then diffuse into the heated graphite reservoir for storage. When the voltage across the electrodes is reversed, so to is this process, with the addition of a final surface evaporation stage to transport the Rb atoms from the device upper surface into the vapor phase.

The capabilities of the AIB have demonstrated the sourcing of a MOT directly from the device internal graphite reservoir \cite{Kang:19}, with sourcing and sinking time constants on the order of 1$\,$s \cite{mcgilliganAIB}. Interestingly, the graphite reservoir within the AIB has been shown to survive atmospheric exposure. The lack of oxidization of the Rb with the graphite reservoir provide a simple route to the fabrication of pre-loaded AIB's that can be implemented in vacuum chambers and load MOTs directly from this source without the necessity for additional alkali sources to load the AIB once in vacuum. If coupled to a small vacuum volume, the AIB could both potentially extend the vacuum lifetime by providing a clean alkali source and simultaneously enable a wider field of deployment by regulating the density against environmental temperature fluctuations. 

A simple approach to density regulation in a MEMS vapor cell is the previously demonstrated inclusion of Au micro-disks for controllable, local vapor condensation \cite{microdisks}. Patterned onto the borosilicate glass wafer with a 50~nm Au layer over a 10~nm Ti adhesion layer, 8 micro disks of 100~$\mu$m diameter are placed around the circumference of the MEMS cell. The authors demonstrated that when used in conjunction with a Rb micro-pill dispenser and heating the cell temperature to 180~$^{\circ}$C for 1 hour, the alkali would favour condensing on the Au disks, providing a vacuum window clean from alkali vapor condensation \cite{karlenthesis}. While this first demonstration has shown the initial condensation of the alkali vapor into the solid state with external temperature control on a 1 hour time scale, future generations could incorporate local temperature control at the Au micro-disks for real-time vapor density regulation.

\section{Chip-scale optical components}
\label{lasercooling}
A sizable volume of the cold-atom apparatus is due to the large number of optical components that are essential for control of the optical alignment, polarization and beam shaping. In recent years there has been a drive to bring a number of these components down to the chip-scale to facilitate a platform for compact laser cooling. This section will highlight key advancements that have been made in the micro-fabrication of optical components for cold-atom systems.

\subsection{Pyramidal MOT}
The first component to be discussed is the pyramidal MOT (PMOT) \cite{Lee:96}. Using a pyramidal mirror structure, these systems are capable of reducing the traditional 6-beam MOT down to a single laser beam, incident upon the mirror structure to form the additional beams required from the mirror reflections. Unlike the 6 beam MOT, where the polarizations of the three counter-propagating pairs must be set to ensure the coil axis has the opposite circular handedness to the other axes, the pyramidal mirror MOT realizes all the necessary polarizations and $\textit{k}$ vectors required to achieve a MOT. First demonstrated with Al machined mirror structures \cite{Lee:96}, the pyramidal MOT held an estimated $10^7$ atoms within the mirror structure when illuminated with a 2.1~cm diameter beam and 6.5~mW of optical power. Since then, the pyramid MOT has been widely implemented in cold atom applications, including cold atom flux sources \cite{footpyramid}, gravimetry \cite{landraginpyramid,mueller,dosSantos2016} and atomic clocks \cite{Bowden2019}.

Although this provides a significant reduction on the scalability of critical optics for laser cooling, the pyramidal system is not free from flaws. The optical overlap volume resulting from the reflected mirror orders and single incident beam is 6 times smaller than would be expected from a traditional 6-beam MOT with the same incident beam waists \cite{Lee:96}. Additionally, the cold atom sample forming within the conical structure hinders imaging and probing of the cold atom sample, made more difficult by the reflected MOT images and scattered light.

Many of the issues limiting the incorporation of the pyramidal MOT can be circumvented by raising the angle of reflection into a tetrahedral MOT configuration \cite{Shimizu91}. This improved geometry has been demonstrated in the tetrahedral mirror MOT \cite{Vangeleyn:09}. Here the optical overlap volume exists above the surface of the mirror structure, such that there is available optical access for imaging and interrogation of the atoms, but critically this enables the mirrors to be held outside of vacuum for simplicity of alignment and optimization. While the tetrahedral pyramid MOT has circumvented a number of restrictions from the original MOT, it remains a device that requires hand assembly and careful alignment of bespoke components. 

\begin{figure}[!t]
\centering
\includegraphics[width=0.48\textwidth]{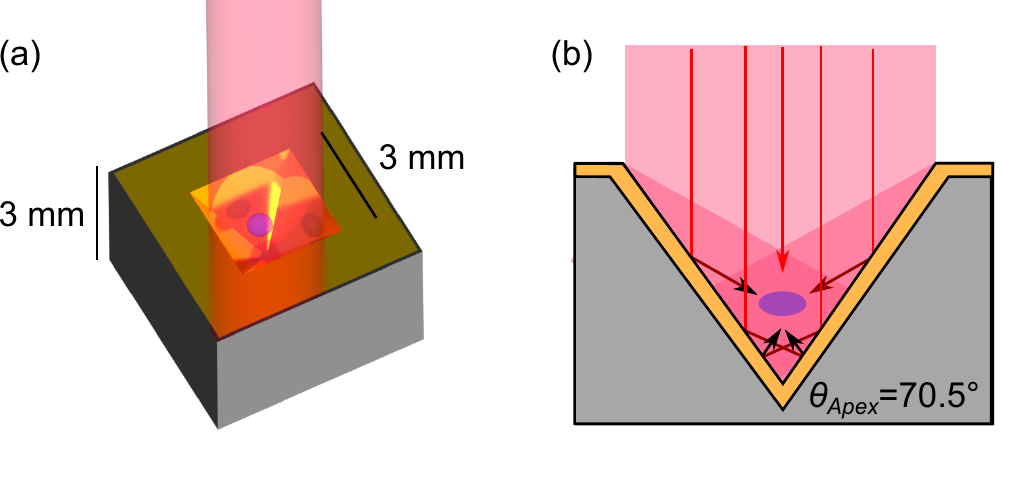}
\caption{\label{pyramid} (a): A micro-fabricated pyramid MOT chip, etched into 3~mm thick silicon and coated in gold, adapted from Ref.~[\cite{Pollock_2011}]. A MOT and reflected images of the MOT can be seen due to the position of the trap within the pyramid and the reflectivity of the gold. (b): Dominant reflection contribution to laser cooling in the micro-fabricated pyramid with a 70.5$^{\circ}$ apex angle.}
\end{figure}

Contrary to this, the scalability of the pyramidal structure has been reduced to a micro-fabricated component, by wet etching silicon to form reflective optical structures capable of trapping small atom clouds \cite{Pollock_2011,trumpke}, illustrated in Fig.~\ref{pyramid} (a). As has been discussed in previous sections, the wet etching process for $\{100\}$ silicon will follow the crystal plane at 54.74$^{\circ}$, such that the apex angle of the silicon etched pyramid is restricted to $(90^{\circ}-54.74^{\circ})\times2=70.5^{\circ}$, shown in Fig.~\ref{pyramid} (b). Additionally, due to etch rates for KOH on the order of $\sim1~\mu$m/min, a 3~mm deep pyramid structure would require a 50~hour etch period, placing a limitation on the pyramid size due to time constraints and costs within clean-room facilities. Additionally, the long etch times can degrade the surface quality of the reflector wall and hence the optical quality of the reflected beam. As is outlined by S. Pollock $\textit{et. al.}$, an apex angle results in three main reflection types, where the dominant contribution to laser cooling, type 1 reflections, are highlighted in Fig.~\ref{pyramid} (b). The overlap volume of the PMOT from the type 1 reflections forms deep within the silicon substrate, restricting the imaging axis. Importantly, the position of the overlap volume also frustrates the ability to differentiate the true PMOT from the image reflections on the angled walls that are in close proximity to the cold atoms \cite{Pollock_2011}.

\subsection{Grating MOT}
While the pyramid MOT has provided a clear route to micro-fabricated cold-atom components, the atom number that is achievable, hindered by the attainable etch depth, remains unfavourable for atomic sensors. A novel approach that aided the miniaturization further was the reduction of the angled mirrors to a planar diffraction grating, to demonstrate the first grating magneto-optical trap \cite{Vangeleyn:10}. Rather than utilize the reflections from the angled mirrors of the pyramid MOT, the grating MOT diffracts light at an angle relative to the grating period, dictated by the Bragg condition $n\lambda=d\sin(\theta)$, where $n$ is the order of diffraction, $d$ is the grating period, and $\theta$ is the angle of diffraction with respect to the normal. A key difference to the tetrahedral and pyramidal MOTs, the diffraction process results in a compression of the diffracted beam waist, $W_1$, to produce an effectively larger intensity in the diffracted order $I_1$, relative to the incident beam waist, $W_{In}$ and intensity, $I_{In}$, such that $I_1/I_{In}=\eta W_{In}/ W_{1}=\eta\sec(\theta)$, where $\eta$ is the diffraction efficiency. To satisfy a good radiation pressure balance for optical molasses the diffraction efficiency should have the value $\eta=1/N$, where $N$ is the number of diffracted beams contributing to the cooling process. In its first demonstration, the grating MOT utilized $\times$3 blazed gratings with $d=1200$~nm and $12\times12$~mm$^2$ surface area, positioned around the grating plane at $\frac{2\pi}{3}$~rad relative to each other. The incident beam is expanded to a $\approx 20$~mm diameter to equally fill the three grating surfaces, with an ideal radial balance of $\eta_{theory}=\frac{1}{3}=33\%$ being achieved when including reflection losses, to achieve a first order diffraction efficiency of $\eta_{1^{st}}=32\%$. This initial proof-of-principle atomic trap was capable of bringing 10$^5$ $^{87}$Rb atoms down to 30~$\mu$K, proving highly advantageous to the miniaturization of cold-atom sensor platforms.

\begin{figure}[!t]
\centering
\includegraphics[width=0.48\textwidth]{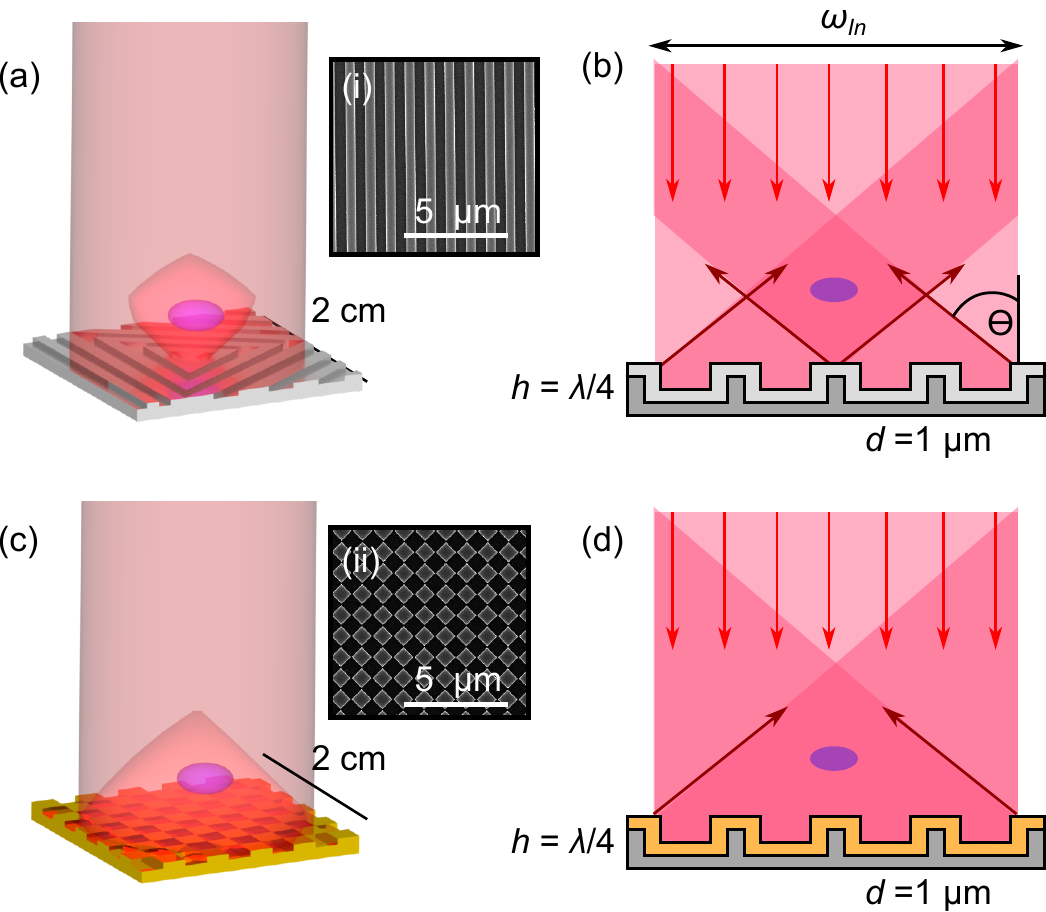}
\caption{\label{grating} Surface patterned grating chips and corresponding overlap volumes, adapted from Ref.~[\cite{Nshii2013,McGilligan:16}]. (a): 1-D Tri grating chip, with 3-distinct segmented regions that meet at the chip centre. The e-beam grating pattern is written with a 1~$\mu$m period and is coated with 100~nm of Al. The optical overlap volume is shown above the grating surface. (b): Illustration of the grating optical overlap volume for the linear segmented chips. The intersection point of the grating segments are represented at the centre of the 1D illustration, pinning the point of symmetry for grating alignment with the incident beam. (c): 2-D Checkerboard grating chip, with a uniform etched pattern across the full surface. The e-beam grating pattern is written with a 1~$\mu$m period and is coated with 80~nm of Au. The optical overlap volume is shown above the grating surface. (d): Illustration of the grating optical overlap volume in the checkerboard geometry. The holographic structure of the grating results in a overlap volume that is largest at the grating surface and has no central point of symmetry. Scanning-electron-microscope images of both chips are provided in inset (i) and (ii).}
\end{figure}

To further drive the miniaturization and scalability of this technology, the footprint of the grating MOT was reduced further by micro-fabricating the optical components into silicon \cite{Nshii2013}. The micro-fabricated grating chips were patterned with e-beam lithography, and etched with RIE to a depth of $\lambda/4$ to destructively interfere the zeroth diffraction order. The binary grating profiles were etched with periods between 1000-1400~nm with 50:50 duty cycles (etched:unetched ratio) \cite{Nshii2013,Cotter2016}. Following etching, the chips reflective metals, typically Al and Au, are evaporated onto the surface at 100~nm thickness. Two of the predominantly used grating chip geometries, are shown in Fig.~\ref{grating}.

The 1-D binary grating structure, illustrated in Fig.~\ref{grating} (a) is made of three distinct segments, rotated around the chip by $\frac{2\pi}{3}$, as was shown with the macroscopic blazed grating MOT. The second chip, Fig.~\ref{grating} (b), uses a 2-D structure, composed of two binary, linear gratings on orthogonal planes superimposed on top of each other to form a 'checkerboard' design. The profile of the grating is shown in scanning electron microscope images in inset (i) and (ii). The formation of the optical overlap volumes for these chips are illustrated in Fig.~\ref{grating} (b) and (d) for the Tri and checkerboard respectively. 

A key difference in the overlap volumes between the chips is the point of symmetry that exists in the segmented Tri grating, requiring critical alignment from the incident beam to this point, forming the centre of the optical overlap volume above this location. Contrary to this, the checkerboard possesses no point of critical alignment since the grating profile is uniform across the full chip surface. This results in the optical overlap volume of the checkerboard extending to, and being largest at the chip surface, whereas the overlap volume of the Tri is instead largest at a height $H=\frac{W_{In}}{4\tan{\theta}}$. 

In the case of both grating geometries, chips fabricated with a 2~cm$\times$2~cm area provide optical overlap volumes on the order of 1~cm$^3$, greatly exceeding the achievable overlap volume of previously micro-fabricated pyramidal MOTs on the order of 0.01~cm$^3$. It is worth noting that the trapped atom number measured in the micro-fabricated PMOT and GMOT systems scale differently with the optical overlap volume, with the PMOT scaling with $N\propto V^2$~ \cite{Pollock_2011} and the GMOT scaling as $N\propto V^{1.2}$~ \cite{Nshii2013,Bregazzi}. The exponential term with the PMOT changes to $V^{1.2}$ for incident beam diameters larger than $\approx$7~mm. However, the micro-fabricated PMOT devices were restricted well below this, for reasons previously highlighted. 

Although the micro-fabricated PMOT has a better scaling with the optical overlap volume, trap volumes larger than 0.01~cm$^3$ have not been achieved experimentally. This is primarily due to the long fabrication times and difficulty maintaining an optically smooth surface over such a deep wet etch without resist breakdown. 

As a result of the large optical overlap volume that can be achieved with the micro-fabricated GMOT, 10$^8$ $^{87}$Rb atoms have been trapped \cite{McGilliganphase}, with experimental demonstrations showing laser cooling down to 3~$\mu$K\cite{McGilligan2017}. The success of this component at miniaturizing cold atom systems has led to GMOT chips being used in clocks\cite{Elvin:19}, ion sources\cite{franssen}, pressure sensors\cite{McGilligan2017,eckelpressure}, gradiometer arrays\cite{McGilligan2017}, high flux atomic sources\cite{imhofgmot}, and interferometers \cite{lee2021coldatom}. Additionally, GMOTs have now been used for laser cooling a number of atomic species including Rb\cite{Nshii2013}, Li \cite{eckelbarker} and more recently Sr\cite{eckelsr,bondza2021twocolor}.

\subsection{Mirror MOT}
With an optical configuration closer to the standard 6-beam MOT, the mirror MOT (MMOT) uses an in-vacuum reflector to form an optical overlap volume close the the chip surface, such that only 4 incident beams are required, with the additional 2 provided by the reflected orders \cite{mirrorMOT1}. As a result, the optical overlap volume is reduced by $\approx$1/2, with the optimum atom number found a few mm from the substrate surface. While this degrades the achievable atom number, the MOT formation in close proximity to a substrate surface is advantageous for applications that require atomic interaction with surface deposited wires, such as chip traps \cite{mirrorMOT2}. 

A common configuration used for the MMOT is for two axes of incident light reflecting at 45$^{\circ}$ with respect to the substrate surface. To ensure the correct polarizations in such an optical configuration, the quadrupole field orientation is aligned along the axis of one of the 45$^{\circ}$ beams. Alternatively, a U wire configuration can be etched in the substrate surface to realize a near ideal quadrupole field to greatly simplify the apparatus required for the MMOT \cite{mirrorMOT3}.

Recently, more novel geometries for MMOTs have been explored that enable cooling from a single optical access vacuum window \cite{Roy2018}. However, the non-trivial alignment reduces the potential for this approach to be successfully implemented in future chip-scale cold atom sensors. Additionally, the large vacuum window diameter required for 4 beams to enter degrades the SWaP when compared to the PMOT and GMOT configurations.

\subsection{Micro-fabricated lenses}
\begin{figure}[!b]
\centering
\includegraphics[width=0.4\textwidth]{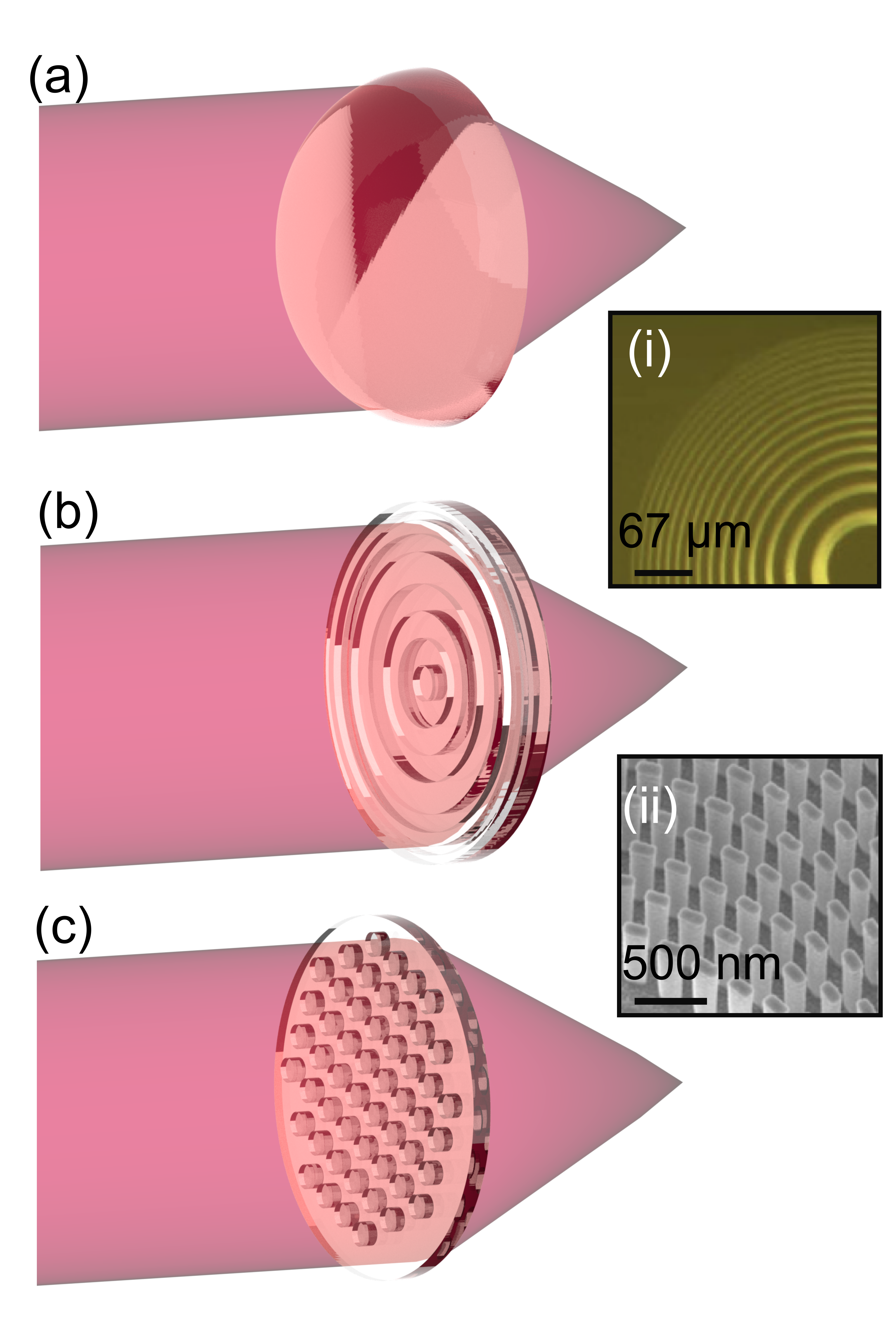}
\caption{\label{microoptics} Illustration of optical elements for focusing light. (a): Refractive optical lens. (b): Diffractive optical element. Inset (i): Microscope image of a transmissive Fresnel lens, reprinted from Scientific Reports $\textbf{6}$, 25348 (2016). (c): Meta surface lens. Inset (ii): Scanning electron microscope image of a dielectric silicon pillar based metasurface lens, reprinted from New Journal of Physics $\textbf{23}$, 013021 (2021).}.
\end{figure}
Optical elements are of critical importance in atomic sensors for beam alignment, shaping and polarization. While refractive optics, illustrated in Fig.~\ref{microoptics} (a), are widely implemented in cold-atom experiments, the physical size of standard mirrors and lenses can constrain the sensor scalability. The miniaturization of refractive optics to diffractive optical elements (DOEs), illustrated with a Fresnel zone plate in Fig.~\ref{microoptics} (b), has provided a routine to further device simplification and component mass production of lenses, gratings and mirrors. In this scenario, the lens curvature is brought down onto a planar surface and replaced with a ramped binary pattern to spatially modulate the amplitude or phase of the incident light. The transmitted light will diffract at an angle relative to the local period of the surface geometry, such that arbitrary patterns can be generated following propagation of the beam. As such, the focal point is strongly tied to the incident wavelength. The step to DOEs has been demonstrated in compact atomic sensors for cold-atom waveguides \cite{Henderson:20}, atom embedded quantum optical elements \cite{BoppStern}, chip-scale alignment \cite{chutani}, and grating magneto-optical traps \cite{Nshii2013}.

In recent years, meta-surface lenses, illustrated in Fig.~\ref{microoptics} (c), have grown in popularity for implementation in atomic experiments. While these meta materials still rely on diffraction, the phase accumulation is not a direct product of path length difference between the peak and trough of the etched material. Instead, the phase of the meta-surface component is induced by the micro-structures etched on the substrate surface, which act as local phase retarders \cite{Engelberg2020}. This pillar-orientation dependent phase shift on the incident light enables the generation of custom phase in the output beam from the meta-surface. Additionally, unlike a standard DOE, the meta-lens is composed of sub-wavelength features, positioned in a sub-wavelength quasi-periodic pattern. Typically formed from silicon and glass, the nano-features can be positioned such that the phase accumulation permits beam shaping \cite{McGehee_2021,beammetashaping}, polarization control \cite{metapolarization,metapolarization2}, angle of diffraction \cite{metasurface}, focusing and multi focal point generation \cite{metamultifocus}. 

Importantly, these properties have been demonstrated with meta-surface devices in cold atom experiments. Highlighted in Ref~[\cite{McGehee_2021}] and shown in Fig.~\ref{microoptics} inset (ii), W. McGehee $\textit{et.~al.}$ fabricate a planar dielectric metasurface lens, based off of Pancharatnam-Berry phase accumulation, to expand and shape the beam profile of the cooling light for optimal coupling to a GMOT chip. In other recent work, a meta surface chip has been used to simplify the 6-beam MOT optics \cite{metasurface}, and generate optical tweezer arrays \cite{regalmeta,regalmeta2}.

\section{Ultra-high vacuum cells}
\label{UHV}
As previously outlined, laser cooling relies on ultra-high vacuum environment to avoid thermal collisions and reactions with contaminant gasses. In the past this has been achieved primarily with off the shelf, modular and generic components; constructed with metal chamber bodies, power feedthroughs, atomic sources, active pumping mechanisms and optical access with flange adapted windows or glass cuvettes. These inclusions have previously restricted the vacuum apparatus to the litre scale, placing a dominant constraint upon the miniaturization of cold-atom sensors. Recent work has looked at methods to circumvent the scalability of this apparatus through innovative developments and transitional technologies, such as MEMS components, ceramic bodies and machined titanium cores. 

\subsection{Chamber body}
Recent investigations into the miniaturization of cold-atom packages have diverged through a number of potential solutions to the ideal component selection for vacuum encapsulation. Ranging from micro-fabricated silicon frames, to 3-D printed chamber cores, the benefits and limitations of the proposed vacuum solutions are evaluated with an outlook to remaining chip-scale and mass producible.

\subsubsection{Machined metal}

In the context of cold-atom UHV cells, the two primary challenges that require addressing are the scalability, and ensuring that the vacuum longevity meets the requirements of deployable cold-atom sensors. Standard laboratory based UHV chambers are typically constructed from stainless steel bodies, with appropriate optical access for cooling and probing \cite{wieman}. These standard machined metal chambers have been shown to offer a clear route to chip-scale platforms with reduced vacuum volumes for portable sensors \cite{hothlittle,prestage}. Due to its low out-gassing and permeability, while remaining machinable and tough, titanium is an ideal vacuum body for UHV applications. Although stainless steel is also helium impermeable and a competitive candidate for chamber construction, titanium possess a significantly reduced hydrogen out-gassing rate \cite{takeda}.

Recent work has demonstrated a cubic titanium chamber, machined with $\sim$41~mm width and total vacuum volume of 70~mL, for laser cooling applications \cite{hothlittle}. Optical access is provided from sapphire windows, braised into titanium frames and laser-welded onto the chamber body. The scalability of machined titanium has also demonstrated a 1~cm$^3$ vacuum chamber for ion trapping applications \cite{prestage}. This package again utilizes sapphire windows for optical access and makes use of high temperature co-fired ceramic (HTCC) electrical feed-throughs. The 11~mm$\times$12~mm$\times6$~mm vacuum dimensions of the chamber are comparable to the optimum silicon laser cooling dimensions. However, the machining and adhesion process involved in the chamber encapsulation do not currently meet the needs of mass production nor parallel fabrication.

\subsubsection{Silicon}
Chip-scale thermal atom sensors have driven the integration of photonic and MEMS components to enable semi-conductor foundry-level scalability in vapor cell processing. The amalgamation of micro-engineering and atomic physics has greatly reduced the volume and cost of vapor cell fabrication, while providing a means to mass production and simplified manufacturability. 

\begin{figure}[t]
\centering
\includegraphics[width=0.45\textwidth]{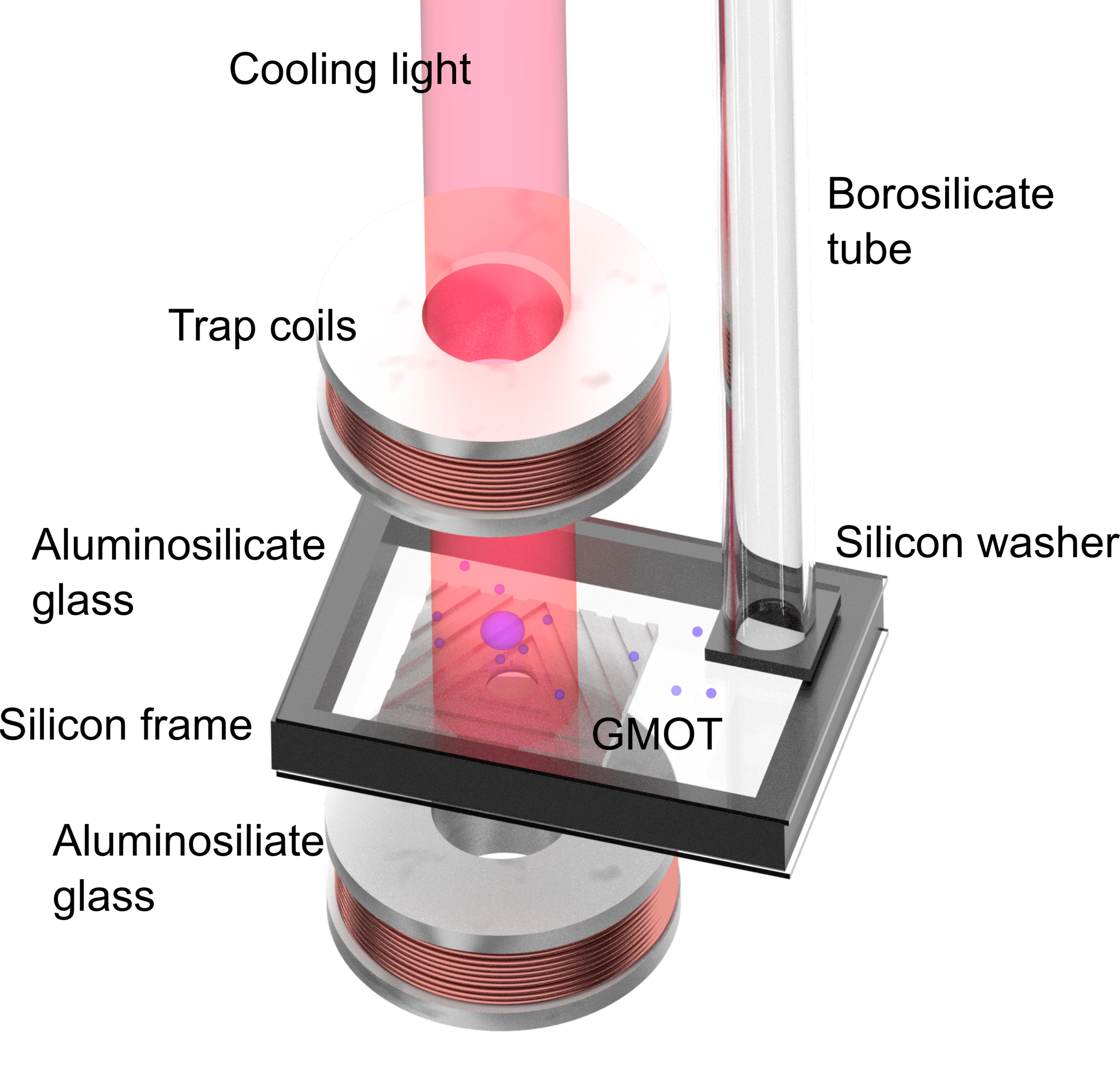}
\caption{\label{MEMSGMOT} Micro-fabricated UHV cell composed of a glass-silicon-glass stack. Upper glass is drilled and anodically bonded to a silicon washer and borosilicate glass tube. The cell is aligned with a GMOT chip and trap coils to demonstrate a chip-scale laser cooling platform. Image adapted from Ref.~[\cite{McGilligan2020}]}
\end{figure}

While the pre-exisiting hot atom vapor cells are not suitable for sustaining cold-atom samples, the fabrication process can be easily adapted to meet the needs of laser cooling, as was first shown by J. P. McGilligan $\textit{et. al.}$ \cite{McGilligan2020}. In this paper, the authors demonstrated the ability to laser cool $10^5$ atoms in a micro-fabricated vapor cell, based on a glass-silicon-glass anodically bonded stack, shown in Fig.~\ref{MEMSGMOT}. To achieve UHV, the upper glass wafer was drilled and the hole was bonded to a silicon washer and borosilicate glass tube. The glass tube was adhered at the opposite end to a standard vacuum flange, connected to a UHV chamber and ion-pump. The cell was initially evacuated to a pressure of 10$^{-8}$~mbar using a turbo/roughing pump combination. Following this initial pumping stage, the larger pumping apparatus was disconnected and the vacuum pressure was sustained by an ion pump.

While this micro-fabricated cell has remained connected to a larger vacuum apparatus, simple methods of cell isolation (outlined in later sections) remain available to reduce the vacuum volume and improve the scalability of the device. The initial demonstration with the micro-fabricated cell made use of a 6-beam MOT, requiring careful alignment of the cell due to the silicon frame restricting the optical access. To simplify the platform further, the authors coupled the cell with pre-existing GMOT technology, for a fully micro-fabricated laser cooling system \cite{McGilligan2020}.

Although the amalgamation of this technology facilitates a reduced scalability of the cooling system, the reduced optical access from the silicon frame hindered standard fluorescence imaging. The difficulty in imaging was found to be caused surface scatter from the glass and grating surfaces when viewing the system at a non-orthogonal angle.

More recent research has utilized a central hole in the grating chip as an imaging axis for absorption imaging. This central region plays a minimal role in the optical overlap volume from the grating chip, thus having little impact on the achievable atom number. However, the authors found that the 3~mm thick silicon frames restrict the optical overlap volume that the atoms experience, reducing the achievable trapped atom number \cite{Bregazzi}.

\subsubsection{Ceramic}
An advantageous material for the fabrication of cold atom chambers are low-temperature co-fired ceramics (LTCC). Commonly used for vacuum sealing \cite{LTCCcell} and a component in the body of optical cavities, the LTCC is non-reactive with alkali atoms and capable of sustaining UHV. 

Recently, a ceramic body for a cold-atom vacuum cell was demonstrated, with preliminary helium permeation rates proving negligible for time scales over a year \cite{burrow2021}. While LTCC cannot be scaled as easily as silicon fabrication, it has an advantage of simple manufacturing with less restriction on the cell dimensions over silicon. As such, the ceramic body could be formed into a cubic geometry with 32~mm sides to permit optical window adhesion on each axis. The bonding of the ceramic body to vacuum windows and copper pinch off for vacuum pumping will likely involve multiple step processes at high temperatures to ensure hermetic sealing. However, the form-customization of the ceramic body provides a key advantage over alternative materials for the fabrication of bespoke vacuum bodies, making this material a promising candidate for the construction of future cold-atom sensors.

\subsubsection{Additive manufacturing}

An exciting recent development in the miniaturization of UHV equipment is the fabrication of metal alloy chambers through additive manufacturing \cite{COOPER2021,PRXQuantum}. Using a laser powder bed fusion technique, the 3-D printing of AlSi10Mg has been demonstrated for the fabrication of custom UHV chamber bodies, with a 70$\%$ reduced mass to an equivalent metal machined chamber. The achievable surface roughness of the 3-D printed chamber walls is critical to minimise virtual leaks and ensure hermeticity from traditional knife-edge sealing to the chamber walls. However, optical and electron microscopy of the 3-D printed AlSi10Mg chamber have revealed a surface roughness of 5.3$\pm$0.1~$\mu$m, which combined with mass spectrometry measurements to quantify the absence of out-gassing from the chamber walls, results in a material that remains suitable for UHV applications \cite{COOPER2021}. 

The 3-D printed chamber was manufactured with standard conflat ports, enabling the body to be later connected to vacuum windows and other standard vacuum components. The suitability of this technology to atomic sensors has been emphasized by the production of 2.5$\times10^8$ $^{85}$Rb atoms inside the 3-D printed chamber when connected to an ion pump through the conflat flanges in the chamber body.

While the first demonstrations of this developing technology have remained on the litre volume scale, the ability to rapidly prototype new geometries and sizes could enable this fabrication method to realize chambers with milli-litre vacuum volumes for chip-scale apparatus. 

\subsection{Vacuum windows}

While the miniaturization of the chamber body sets the scalability of the vacuum cell, the chamber must incorporate optical access for laser cooling and measurement of the atomic signal. However, the inclusion of optical access must have a limited degradation of the chamber performance. Ideally, the windows would (1) remain impermeable to noble gasses, (2) meet the requirements of a mass producible bonding method for adhesion to the chamber body, and (3) have a suitable transmission at the atomic wavelength while not being birefringent. 

\subsubsection{Borosilicate glass}

The manufacturability and size of atomic vapor cells are made possible with the pioneering anodic bonding of silicon and hard glass. One such hard glass, historically used for glass blown vapor cells due to the relatively low softening point, borosilicate glass (BSG) is now widely used at the wafer level for cell fabrication with its commercially available Pyrex and SCHOTT Borofloat.

Made popular by its high optical transmission across the ultra-violet, visible and infra-red, as well as possessing a well matched coefficient of thermal expansion (CTE) to silicon, BSG has been demonstrated in a number of chip-scale atomic vapor cells \cite{Bopp_2020,BoppStern,Kitching2018,Knappe2004,Hasegawa2011_buffer_gas,boudotmemscell}. Typically made up of 81$\%$SiO$_2$+13$\%$B$_2$O$_3$+4$\%$Na$_2$O/K$_2$O+2$\%$Al$_2$O$_3$, the BSG wafers have an ideal chemically composition for anodic bonding \cite{anodicbondtextbook}. However, the significant helium permeation rate of $K_{Pyrex}=3\times10^{-10}$~cm$^2$/s at 96$^{\circ}$C measured in BSG fabricated cells \cite{dellis2016lowHe} greatly limits the transfer of this material from thermal atom sensors to a cold atom platform, without a sufficient means of pumping the permeated helium \cite{Rushton2014}.

\subsubsection{Aluminosilicate glass}

Another hard glass, with a CTE well matched with silicon for bonding, is aluminosilicate glass (ASG). As well as sharing many of the promising properties of BSG, recent literature has highlighted that ASG exhibits a significantly reduced helium permeation rate of $K_{ASG}=1.4\times10^{-12}$~cm$^2$/s at 91$^{\circ}$C \cite{dellis2016lowHe}, attributed to the $\sim20\%$Al$_2$O$_3$ content within the glass \cite{anodicbondtextbook, Rushton2014}. The low permeation rate makes ASG an attractive candidate for the development of MEMS cells with reduced long-term frequency drifts and UHV cell longevity. 

Among the existing options for ASG, the SD2 material produced by Hoya has been used for successful cell fabrication and demonstrated with laser cooling applications \cite{McGilligan2020, Bregazzi}. This work indicated that the alkali ion content within a 700~$\mu$m thick glass wafer is sufficient for anodic bonding at temperatures around 300~$^{\circ}$C for an applied voltage of 800~V across the silicon-glass interface. However, low-temperature anodic bonding is favourable to avoid atomic diffusion into the windows at higher temperatures, which would reduce the vapor density and increase the opacity of the cell window. Importantly, a lower bonding temperature also alleviates the risk of damage to integrated components and circuitry \cite{heatinganodic}. Recent studies have investigated the use of lithium doped ASG to increase the ionic content for a sufficient charge transfer at a lower temperature to achieve a hermetic seal between the glass and silicon substrates \cite{SHOJI199895}. Such doped ASG has been anodically bonded for UHV MEMS cells at temperatures as low as 150~$^{\circ}$C \cite{McGilligan2020}. Interestingly, the authors note that the increased ion content within the doped glass produced a $\times$6 larger initial bonding current and $\times$8 faster bonding time at 300~$^{\circ}$C compared to BSG under the same conditions.

While this glass has a significantly reduced helium permeation rate when compared to borosilicate, the remaining rate of permeation places a limit on the vacuum longevity on the order of 1 year \cite{dellis2016lowHe,Rushton2014}. To extend the vacuum longevity, additional optical coatings, such as graphene \cite{graphene} and Al$_2$O$_3$ \cite{KarlenALD} could be used to reduce the diffusion rates of alkali and noble gasses through the glass \cite{WOETZEL2013158}. 

\subsubsection{Sapphire}

Another optical material that is a suitable candidate for integration into chip-scale sensor platforms is sapphire. Attractive properties of sapphire include a broad optical transparency, resistance to alkali diffusion, and an expected low-to-no helium diffusion \cite{wiemancoatings}. However, sapphire is a notoriously difficult material for incorporation into chip-scale devices due to the hard crystalline structure hindering its compatibility with common etching methods. Importantly, sapphire does not meet the 3 objectives we have outlined for idealized vacuum windows, being that it is has a birefringent crystalline structure. However, the uniaxial birefringence of sapphire can be overcome by machining the bulk such that a single optical axis will be unaffected, enabling its inclusion for optical applications with a single $k$ vector along the material axis. With this being said, sapphire has been used to demonstrate cm- \cite{sapphirecm} and mm-sized \cite{sapphiremm} atomic vapor cells. However, due to the relatively large mismatched CTE with silicon, and low alkali ion content, sapphire bonding to silicon with anodic or thermo-compression bonding is not possible \cite{karlenthesis}. Instead, sapphire cells are typically composed solely of sapphire, with a hermetic seal provided from a Au-Au thermo-compression bond, deposited in atomically thin layers (12~nm) to reduce the formation of alkali-Au alloys \cite{sapphiremm}. 

Sapphire vacuum windows have also been used in a range of cold atom and ion applications \cite{hothlittle,prestage}. Due to the incompatibility with wafer level bonding techniques, the sapphire windows require being brazed into titanium sleeves and adhered to the vacuum bodies with a laser weld. Unlike the adhesion to silicon, the conjunction of sapphire and titanium joints do not break under thermal stress due to a good match between the CTE of the two materials.

As well as existing as a bulk window material, the amorphous form of sapphire, Al$_2$O$_3$ is commonly implemented cell coating for reduced alkali diffusion in atomic vapor cells \cite{KarlenALD, WOETZEL2013158}. A number of chip-scale vapor cells have been fabricated for thermal atom applications with Al$_2$O$_3$ deposited on the inner walls by atomic-layer or molecular-vapor deposition \cite{karlencucu}.

\subsubsection{Ceramic}
Among the available optical materials for vacuum construction are the transparent ceramics such as spinel, MgAl$_2$O$_3$ \cite{spinel, Rushton2014}. With a wavelength dependent optical transparency that is comparable to sapphire \cite{spinel2019}, spinel could potentially be integrated into a broad range of atomic species based sensors. The authors note that while there was no literature on the helium permeation rates of spinel that they are aware of, its chemical composition could indicate a helium permeation rate that is also comparable to sapphire. The optical properties of MgAl$_2$O$_3$ ceramics have led recent research to develop optically smooth vacuum windows based on this technology \cite{ceramicreview}.

\subsubsection{Silicon}

To circumvent significant efforts being placed on a reduced helium permeable optical material, certain crystaline materials that meet the outlined requirements for optical windows exist with no measured helium permeation, such as silicon carbide. Silicon carbide, SiC, possesses a number of interesting properties that support its integration into atomic sensors such as its compatibility with anodic bonding for wafer level manufacturing \cite{SiC,SiCbonding}. The attractive transmission a 780~nm ($\sim93~\%$) and favourable thermal behaviour of SiC have been utilized for in-vacuum, transparent atom chip experiments with cold Rb atoms \cite{SiCchiptrap}. However, the narrow optical transmission range of SiC would restrict such cells from expanding to other alkali metal based sensors. 

As well as there being particular interest in SiC for vacuum optical access, recent work has looked at the fabrication of silicon nitride, SiN, windows \cite{SiNwindow}. Following the nitride deposition, the silicon wafer is wet etched straight through to reveal the 50~nm thick SiN membrane on the opposite surface. While commonly used as a waveguide material for atomic sensors \cite{hollowcoreschmidt,Stern2013}, SiN membrane windows have been used in vacuum window formation \cite{SiNvacuum}, with polycrystalline cubic silicon nitride, c-Si$_3$N$_4$, being demonstrated with a wide optical transparency and material toughness comparable to diamond \cite{c-SiN}. Importantly, SiN is compatible with anodic bonding, and its simple deposition from low-pressure chemical vapor deposition (LPCVD) favours integration with silicon frame based cells \cite{SiNbonding}.

\section{Vacuum pumps}
\label{vacuumpumps}

With the vacuum bonding, components and materials selected, particular care must be taken in the selection of a chip-scale compatible vacuum pump. While certain vacuum pumps remain suitable to macro-size vacuum systems, many have not translated the boundary to a low SWaP version for chip-scale systems. One such example is a titanium sublimation pump, where a filament wire is heated to T$>$1000$^{\circ}$C to sputter titanium particles into the vacuum \cite{micropumps1}. The highly reactive Ti then reacts with the vacuum contaminants to remove them from the vacuum. However, the power consumption for heating, combined with the high temperature not being suitable with many potentially incorporated technologies, the sublimation pump remains unattractive to chip-scale platforms. This section will evaluate the scalability of technology that has been demonstrated in micro-scale components.

\subsection{Ion pumps}

\begin{figure*}[!t]
\centering
\includegraphics[width=0.92\textwidth]{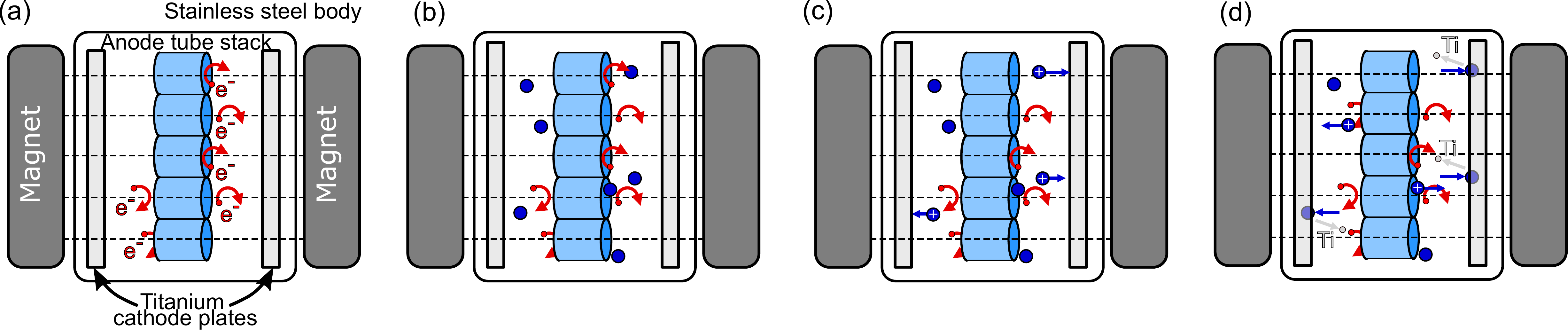}
\caption{\label{ionpump} Illustration of the components and working of a typical ion pump. The ion pump is formed of an anode tube stack, placed between two titanium cathode plates. The cathode and anode are vacuum encapsulated in a stainless steel body, with two permanent ferrite magnets placed immediately outside the vacuum. (a): At an starting pressure of 10$^{-3}$~mbar, the ion pump voltage will draw free electrons to the anode. The magnetic field through the body causes electrons to orbit around the anode tube. (b): As foreign gas species enter the chamber there is a high probability of colliding with the energetic free electrons and ionizing the gas in the process. (c): The ionized gas is now drawn to the cathode plates, while their ejected electron contributes to further collisions. (d): The ionized gas collides with the titanium cathode plate and chemically sorbed. Additionally, the collision sputters titanium atoms into the vacuum, acting as an additional vacuum getter.}
\end{figure*}

A commonly used device for the maintenance of UHV pressures is the ion pump. A schematic of the ion pump composition and description of its pumping mechanism is provided in Fig.~\ref{ionpump} (a)-(d). Encapsulated under vacuum in a stainless steel body, the ion pump can operate with a background pressure on the order of 10$^{-3}$~mbar, achieved with a mechanical roughing and turbo pump. Mounted externally to the vacuum encapsulated body are two ferrite permanent magnets. At its core, the ion pump is formed of an anode tube stack at the device centre, placed between two titanium cathode plates. An applied voltage across the cathode and anode will attract free electrons to the anode voltage potential. However, the strong magnetic field, provided from the permanent magnets induces the electrons into a circular orbit around the anode tube. This increases the mean-free-path of the electron, increasing the probability of colliding with other atomic species within the vacuum, highlighted in blue within Fig.~\ref{ionpump} (b). These collisions ionize the contaminant atomic species, producing an additional free electron, while the ionized atom is fired towards the titanium cathodes, as shown in Fig.~\ref{ionpump} (c). Upon collision with the titanium plate, the ion embeds into the plate, removing itself from the vacuum vapor while sputtering titanium into the vacuum, shown in Fig.~\ref{ionpump} (d). The sputtering of titanium into the chamber acts as a secondary pumping mechanism by chemical reaction with contaminant gas and chamber coating. The current drawn across the anode and cathode is then calibrated as a gauge of the total vacuum pressure at the pump.

\subsection{Micro ion pumps}
While standard ion pump volumes are typically on the order of 1000~cm$^3$, compact pumps have been demonstrated in 2.5~cm$^3$ packages \cite{sputterpump}. More recently, micro-pumps with a total volume of 0.08~cm$^3$ have been shown to achieve a self-pumping limit of $3\times10^{-7}$~mbar in a 25~cm$^3$ chamber \cite{dziuban}. Based on a Penning cell architecture, the pump is micro-fabricated into a glass-silicon stack, and anodically bonded for a hermetic seal. Permanent magnets are placed outside either cathode to increase the mean free path of electron collisions within the vacuum. The compatibility of fabrication techniques to scale down the micro-pump to these extremes while maintaining pressures in the 10$^{-7}$~mbar range is extremely advantageous to chip-scale cold-atom platforms. Interestingly, the authors noted that the deflection of the upper silicon cathode membrane could be used to infer the pressure reading from within the pump. While this is only used for low vacuum pressures, it can be combined with a ionization sensor to provide an increased pressure read-out range \cite{deflector}.

However, the application of miniaturized ion pumps in cold-atom systems is made difficult for two main reasons. Firstly, reducing the size of the cathode and anode components limits the mean free path of the free electrons, in turn reducing the pumping efficiency \cite{AUDI1987629}. Secondly, as the volume of the vacuum apparatus is reduced, the ion pump is placed closer to the science cell, where the permanent magnets perturb the cold-atom sample, degrading the performance of the atomic sensor \cite{boudotmcgilligan}. With the large magnetic field of the ion pump being a dominant constraint to the precision of the atomic sensor, it would be preferred if this could be intermittently removed during precision measurements by using an electromagnet in place of the permanent magnets. However, since the ion pump requires a magnetic field on the order $\sim$10~T, the required power consumption restricts this option for compact metrology \cite{sputterpump}. 

To overcome the impact of the large magnetic field, recent research has focused on the demonstration of novel, chip-scale ion pump architectures that do not require a magnetic field \cite{Basu_2016}. To mitigate the necessity of a magnetic field, the electron source is instead provided by field emission from an electrostatic plate. Charged plates are then used to manipulate the trajectory of the free electrons around a pump cavity to increase their effective mean free path and probability of an ionizing gas collision. The field emission cavity demonstrated by A. Basu $\textit{et al.}$ achieved vacuum pressures as low as 10$^{-9}$~mbar in a 25~cm$^3$ vacuum chamber, showing a clear compatibility with on-chip cold-atom systems. Providing a clear means of pumping vacuum contaminants, including helium, the miniaturization of the ion-pump is a clear favourite for vacuum maintenance in the next generation of chip-scale cold atom sensors \cite{micropumps1}. 

\subsection{Non-evaporable getters}

While the miniaturization of the ion pump has an optimistic outlook for the active pumping of UHV systems, alternative solutions exist that possess a significantly reduced size-weight-and-power, such as non-evaporable getters (NEGs), mentioned previously in Sect.~\ref{micropill}. The commercially available NEG pills (SAES ST172/WHC/4-2/100) and getter tubes (SAES ST172/HI/7.5-7) are composed of a Zr powder and a SAES manufactured metal alloy (ST707) containing a blend of Zr, V and Fe \cite{BENVENUTI200157, BENVENUTI1999219} to form a porous bulk with a large effective surface area for vacuum pumping. A thin oxide layer remains on the NEG surface after the manufacturing process, enabling the pill to be handled in air and inserted within the vacuum environment prior to encapsulation. The deposition simplicity is complemented by the ease of NEG activation, achievable by laser \cite{grifflaserheating, boudotmcgilligan} or resistive heating \cite{hothlittle} to a temperature of $\sim$900$^{\circ}$C for 10 minutes. Once the activation temperature has been reached, the NEG can sorb vacuum elements such as CO, CO$_2$, O$_2$ and N$_2$ into its bulk, where they undergo chemical reactions to form stable compounds such as ZrC, ZrN and ZrO$_2$ \cite{scherer}. The decomposition of these newly formed compounds does not occur below the activation temperature, such that their pumping can be regarded as irreversible. On the contrary, H$_2$ can diffuse into the NEG bulk, but does not undergo a irreversible chemical reaction, such that its sorption process is reversible. Common vacuum elements such as water vapor are pumped through the disassociation of H$_2$O$\rightarrow$H$_2$+O$_2$ on the NEG surface. Additionally, the NEG pills do not pump alkali from the vapor, making them compatible for inclusion in cold-atom sensors. However, a key weakness of the NEG pill is the inability to pump He from the vacuum, such that particular care must be taken to restrict the He permeation to the inner vacuum environment \cite{boudotmcgilligan}.  

The attractive properties of NEGs have been utilized in cold atom experiments as a solution to passive pumping in standard glass-cell \cite{scherer}, chip-scale \cite{boudotmcgilligan}, ceramic \cite{burrow2021} and compact titanium platforms \cite{hothlittle}. In these experiments, both the larger resistively heated and micro-pill NEGs demonstrated the need to purge the oxide layer at high temperature, noting a significant degradation of the vacuum pressure by $\sim$2 orders of magnitude. Following the initial cleaning of the surface, the getters were found to contribute to the vacuum pumping. In the case of the micro-pills, the authors described a MOT survival time that improved by 5 orders of magnitude following the laser heating of 4 pill pumps \cite{boudotmcgilligan}. The authors noted that the MOT survival time provided from the NEG could potentially be longer than was demonstrated, due to the MOT survival time being limited by He permeation in the experimental apparatus. While NEG pills are best suited to chip-scale apparatus, where the inclusion of electrical feed-throughs are not favourable, other compact systems can benefit from the larger resistively heated NEG tubes, where activation is simpler and a larger pumping rate can be attained.

Recent work on passive pumping in a compact vacuum cell has demonstrated that the MOT can survive timescales exceeding 200 days in an isolated vacuum cell with only the resistively heated NEG for pumping \cite{hothlittle}. In this scenario, the initial UHV pressures within the cell were achieved with a standard vacuum and ion pump combination prior to cell closure. Following cell isolation, a total vacuum pressure, measured from the MOT loading time, of $\sim10^{-7}$~mbar was sustained by the NEG pill over 200 days. Unlike the micro ion pump, the NEG has no direct readout to the end-user of the inner vacuum pressure. However, as has been highlighted in literature, the MOT rise time\cite{hothlittle,burrow2021, griffwindow,boudotmcgilligan} or magnetic trap lifetime \cite{monroe,McGilligan2017} of the cold atoms can be used to back out the total background pressure from the vacuum system, circumventing the necessity for further component inclusion.

\subsection{Vacuum isolation}

\begin{figure}[!b]
\centering
\includegraphics[width=0.45\textwidth]{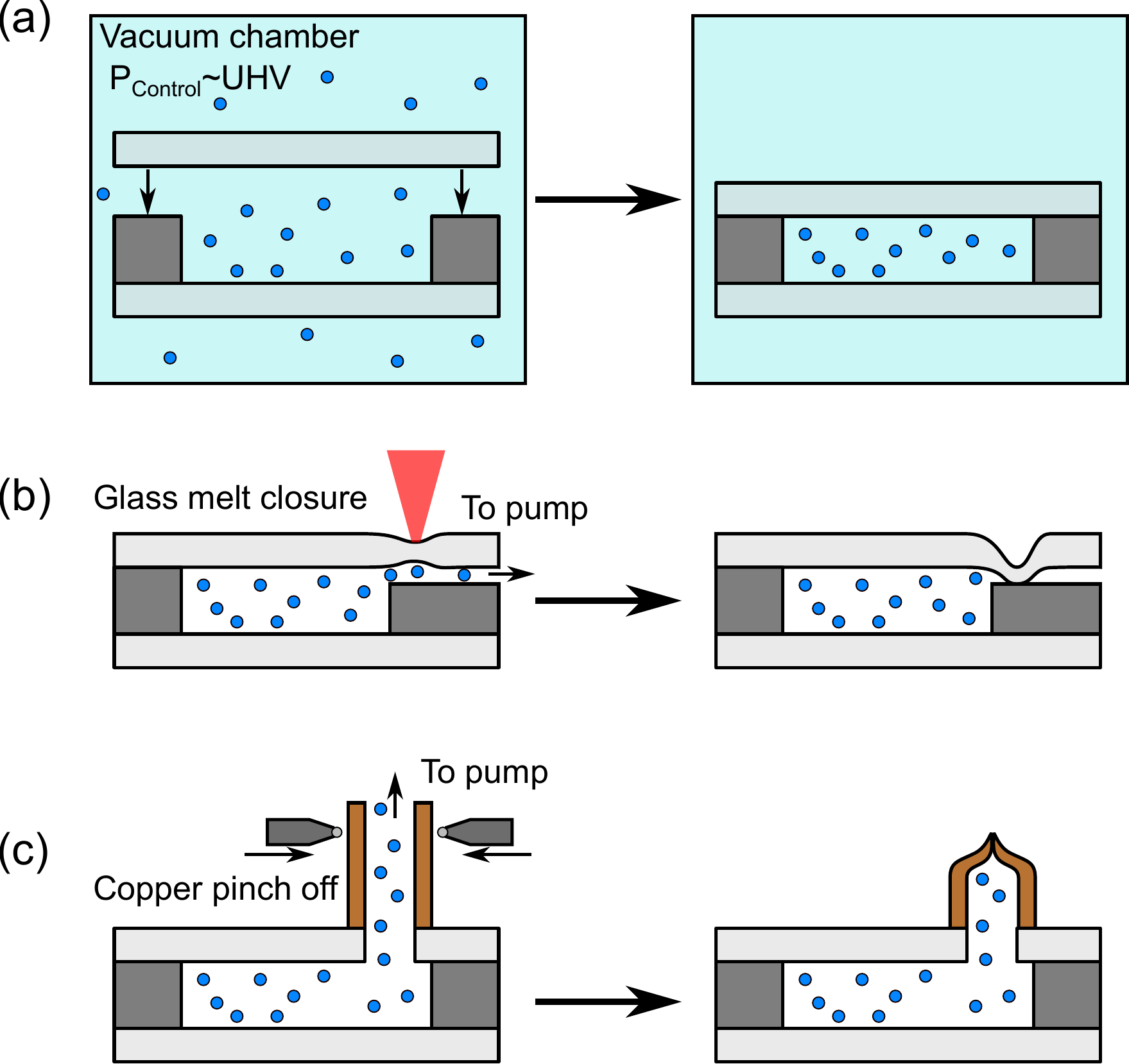}
\caption{\label{cellclosure} Illustrated methods for vacuum closure and isolation. (a): Cell bonding within a UHV environment, such as a bonding station or Bell chamber with UHV capabilities. (b): Internal cell plug to isolate the cell from an external pumping station. The example given is that highlighted in Ref.~[\cite{vincentthesis}] for the laser heating of a glass membrane to provide a hermetic seal between the glass and silicon substrates. (c): Mechanical cold weld sealing of a copper pinch off tube, adhered to the cell with a fluid connection, to isolate the cell from an external pumping station once the required UHV pressures have been achieved within the cell.}
\end{figure}

While the discussion thus far has focuses on the maintenance of UHV pressures, the method for achieving an initial pressure that is suitable for laser cooling has yet to be discussed. Here, we will review the previously used methods to achieve UHV pressures in compact cold atom systems, outlining the applicability to chip-scale integration. The three main experimental techniques are highlighted in Fig.~\ref{cellclosure} and discussed below.

\subsubsection{In-vacuum encapsulation}

A well established technique for vacuum encapsulation in micro-fabricated vapor cells is the fabrication of the cell within a controlled vacuum environment, such as a bell chamber or commercially available bonding station \cite{Bopp_2020,BoppStern, Hummon:18,Newman:19,karlencucu}. The wide-spread implementation of this method for vapor cell fabrication is due to the total pressure and partial buffer gas pressure that can be well controlled within the inner chamber environment \cite{microspheres, pawel, Kitching2018}. With an in-situ pressure gauge, the chamber pressure can be well tracked at the onset of the cell sealing process. After cell closure, the bonding chamber can be vented to atmosphere, with the previous vacuum pressure theoretically retained within the now hermetically sealed cell.

While this method can be adapted for cold-atom systems, a number of issues have thus far limited its suitability to the application. Firstly, if anodic bonding were to be used to close the cell, the inner pressure of the cell would be degraded from the measured control pressure by the O$_2$ released into the inner cell environment as a by-product of the bonding process \cite{Rushton2014}. Additionally, the issue of released O$_2$, combined with the difficulty in clean atomic deposition with anodically bonded cells \cite{Bopp_2020} as outlined in Sect.~\ref{atomdepo}, makes control of the alkali vapor density within the cell post-bonding difficult to pre-establish as being suitable for the formation of a MOT. 

However, if instead of anodic bonding, a more suitable technique, such as thermo-compression bonding, were used for this encapsulation stage, then the cell inner vacuum parameters are more likely to be that of the bonding chamber initial conditions, with the increased likelihood of compatibility with laser cooling.

\subsubsection{Vacuum seal}

A second method outlined in Fig.~\ref{cellclosure} (b) is the routine of sealing a cell post pumping. This method involves the cell initially having a direct connection to a vacuum pump, where the inner vacuum dynamics can be well characterized and optimized to the end-user application. Once the vacuum characteristics meet the user requirements, a seal is applied to isolate the cell from the larger pump apparatus. A key advantage to this approach is the ability to implement laser cooling prior to cell isolation and separation, such that the temporal vacuum pressure can be tracked to optimise the process, ensuring critical information can be extracted for feedback to the isolation process.

A well known process that utilizes this approach is the `hot-torch' sealing of glass-blown atomic vapor cells. For this process, a number of macro-scale glass vapor cells would be glass blown in borosilicate stems, connected internally to a ampoule reservoir of the selected alkali. The localized heating of the ampoule and glass is used to dictate the vapor pressure in each connected glass-blown cell by carefully directing the alkali source down the stems and into the designated cell volumes. Once the required pressure and density was sufficient within the cell, a hot torch would be used to bring the borosilicate stem above the glass working temperature ($\sim$1270$^{\circ}$C \cite{borofloat33}), where it can be mechanically pinched, to close and separate the individual cells from the larger system \cite{glassmelt, pawelmicromachines}. While this technique has had widespread success in the manufacturing of glass-blown atomic reference cells \cite{thorlabs, PGB, Sacher}, its methodology has not been directly applied to the fabrication of MEMS vapor cells. 

A notable micro-fabricated routine that has been inspired by this classical approach is the glass membrane closure, outlined by V. Maurice \cite{vincentthesis, meltcellarxiv}. Here, the author locally heats a thin glass membrane of the MEMS cell upper window with a CO$_2$ laser to soften and deflect the glass. The pressure differential between the cell inner pressure and outer atmospheric pressure encourages the glass to sag and close a 40~$\mu$m diameter channel etched into the silicon frame. This procedure has been demonstrated to isolate the micro-fabricated cell from the rest of the substrate, which can then be diced without evidence of a substantial vacuum leak. The transfer of this technology to a larger vacuum conductance channel for cold atom systems would enable laser cooling demonstrations in the cell prior to isolation, which has been shown to have success in UHV characterization for passive pumping \cite{McGilligan2020}.

\subsubsection{Cold weld}

A widely used method for the hermetic isolation of UHV vacuum components, shown in Fig.~\ref{cellclosure} (c), is the cold weld sealing of copper tubes \cite{atomchippinchoff,prestage,nasapinchoff}. The technique of copper pinch-off involves the compression of a hollow copper stem under vacuum, where the inner surfaces are clean from oxide build up, such that when pressure is applied between the surfaces, the individual atoms can diffuse into the opposing wall, forming a hermetic seal that can sustain the inner vacuum at UHV pressures \cite{coldweld}.

In recent years this method has been used in the manufacturing of compact cold atom systems, initially pumped down by a larger vacuum apparatus prior to pinch off and device isolation \cite{burrow2021,hothlittle,Bregazzi}. The advantages provided by the isolation of the system after pre-establishing laser cooling within the cells have enabled these systems to extract critical information to the longevity of the instruments and optimise the performance of passive vacuum pumping. 

Unlike the methods for vacuum sealing previously discussed, such as the glass melt technique, the ability to close the vacuum via copper pinch off requires a larger footprint for the inclusion of the copper stem stem. Additionally, the inclusion of a copper stem is not easily adapted to micro-fabricated components. While previous chip-scale vacuum cells have included a copper stem adhered to the cell surface using vacuum compatible epoxy \cite{Bregazzi}, the longevity of the cell may ultimately be limited by the outgassing of the epoxy over time. Instead, if the cell surface is coated with an evaporated layer of suitably smooth copper, the copper stem could be adhered with Cu-Cu thermo-compression bonding. This would provide the ability to later pinch off the cell without the potentially degrading the inner vacuum from out-gassing. 

\section{Magnetic field generation}
\label{magneticfield}
\subsection{Planar coil solutions}
A remaining component that is essential for the realization of an unambiguously chip-scale cold-atom sensor is a micro-fabricated solution to the generation of a quadrupole magnetic field. The need for a low-power and compact solution to trapping geometries for magnetic fields has been explored with additive manufacturing of novel coil structures for portable quantum technology \cite{Saint2018}. While this process of fabrication has demonstrated the ability to generate a suitable gradient magnetic field for laser cooling in a MOT, the device requires in-vacuum implementation that greatly restricts the available optical access for cooling and interrogation of the atomic ensemble. Importantly, the demonstrated generation of the device does not meet the requirements of chip-scale components and would not have a direct compatibility with existing technology. 

An alternative approach that has been used for the implementation of current carrying wires in chip-scale devices are lithographically structured coil arrays. Surface deposited wires have been implemented in atomic sensors for applications ranging from static RF magnetic fields \cite{microfabcoilkitching, Kitching2018}, magnetic guides for atomic traps \cite{garrido,jakobreichelcoil}, cell heaters \cite{coilheater}, and atomic chip-trap geometries \cite{baumgartner,Sewell_2010,mirrorMOT1,mirrorMOT2}.  
\begin{figure}[t]
\centering
\includegraphics[width=0.45\textwidth]{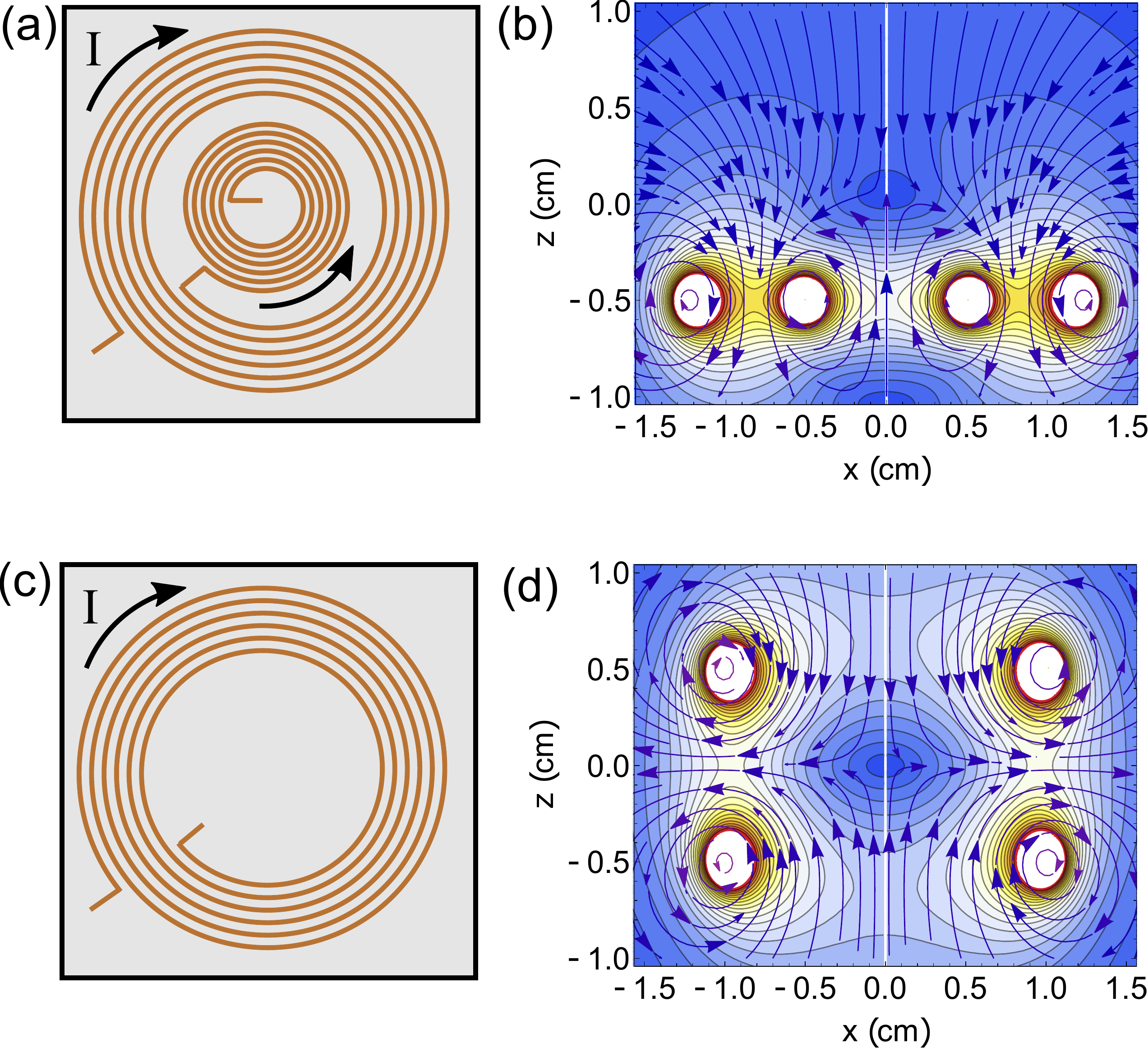}
\caption{\label{coildesign} (a): Illustration of the planar coil design for a quadrupole magnetic field used for laser cooling in Ref.~[\cite{chen2021planar}]. (b): Magnetic field simulation for the coil design shown in (a), where the outer coil has 13 turns and a mean radius of 12.2~mm. The inner coil, with opposite current polarity has 13 turns and a mean radius of 5.2~mm. A current of 0.9~A is used for this simulation. (c): Planar coil design for a standard anti-Helmholtz pair, formed on to separate chips and separated by $d$ the diameter of the coil. (d): Magnetic field simulation for the coil design shown in (b), where each upper and lower coil in the anti-Helmholtz pair have 13 turns and a current of 0.9~A.}
\end{figure}

Recent work has used printed circuit board (PCB) with surface deposited copper wires, using a thickness, width and period of 70~$\mu$m, 200~$\mu$m and 200~$\mu$m to achieve a $\sim$10~G/cm gradient trapping field with an applied current of 0.9~A \cite{chen2021planar}. The compatibility of this technology with other chip-scale quantum devices was emphasized by the integration with a GMOT chip, demonstrating a planar stacked cooling and magnetic field generating chip capable of confining 2$\times10^4$ $^{87}$Rb atoms in a MOT. While this is an advantageous study to address the miniaturization of the coil configuration for chip-scale cold atom instruments, the authors elected to use a single planar coil chip, with a small and large coil of opposite current polarity to generate the gradient magnetic field, shown in Fig.~\ref{coildesign} (a) and (b). Ultimately the reduced saddle potential from the gradient field will reduce the achievable number of atoms that can be trapped in the reduced trapping region. 

Alternatively, this technology could be used in a standard anti-Helmholtz configuration if used in conjunction with a GMOT-MEMS cell, where the vertical separation between the planar coil stacks can be reduced to the millimeter scale, enabling low power consumption and a reduced footprint for device amalgamation. An illustration of the coil and generated field lines are shown in Fig.~\ref{coildesign}.

\section{Outlook}

Laser cooled atoms have revolutionized our capabilities in precision measurement, redefined the second, paved the way to the creation of new states of matter, and driven research in fundamental physics. At the core of this review article we have examined the potential for cold-atom systems to reach out beyond the laboratory environment where the precision and stability of ultra-cold ensembles can have the largest social and economic impact by facilitating a step-change in our technological capabilities. 
\label{conclusion}

\begin{acknowledgments}
The authors would like to thank A. Bregazzi, S. Dyer, R. Boudot and D. Burt for useful comments on the manuscript. The authors acknowledge funding from Defence Security and Technology Laboratory, Engineering and Physical Sciences Research Council (EP/T001046/1), and Defence and Security Accelerator. J. P. M. gratefully acknowledges funding from a Royal Academy of Engineering Research Fellowship. This collaborative work was carried out within the International Network for Micro-fabricated Atomic Quantum Sensors (INMAQS) network.
\subsection{Conflict of interest}
This article has been submitted to Review of Scientific Instruments.
The authors declare they have no competing interest. Approved for Public Release, Distribution Unlimited.
\end{acknowledgments}

\section*{Data Availability Statement}
The data that support the findings of this study are available from the corresponding author upon reasonable request.

\bibliography{Ref}

\providecommand{\noopsort}[1]{}\providecommand{\singleletter}[1]{#1}%
\begin{thebibliography}{316}%
\makeatletter
\providecommand \@ifxundefined [1]{%
 \@ifx{#1\undefined}
}%
\providecommand \@ifnum [1]{%
 \ifnum #1\expandafter \@firstoftwo
 \else \expandafter \@secondoftwo
 \fi
}%
\providecommand \@ifx [1]{%
 \ifx #1\expandafter \@firstoftwo
 \else \expandafter \@secondoftwo
 \fi
}%
\providecommand \natexlab [1]{#1}%
\providecommand \enquote  [1]{``#1''}%
\providecommand \bibnamefont  [1]{#1}%
\providecommand \bibfnamefont [1]{#1}%
\providecommand \citenamefont [1]{#1}%
\providecommand \href@noop [0]{\@secondoftwo}%
\providecommand \href [0]{\begingroup \@sanitize@url \@href}%
\providecommand \@href[1]{\@@startlink{#1}\@@href}%
\providecommand \@@href[1]{\endgroup#1\@@endlink}%
\providecommand \@sanitize@url [0]{\catcode `\\12\catcode `\$12\catcode
  `\&12\catcode `\#12\catcode `\^12\catcode `\_12\catcode `\%12\relax}%
\providecommand \@@startlink[1]{}%
\providecommand \@@endlink[0]{}%
\providecommand \url  [0]{\begingroup\@sanitize@url \@url }%
\providecommand \@url [1]{\endgroup\@href {#1}{\urlprefix }}%
\providecommand \urlprefix  [0]{URL }%
\providecommand \Eprint [0]{\href }%
\providecommand \doibase [0]{https://doi.org/}%
\providecommand \selectlanguage [0]{\@gobble}%
\providecommand \bibinfo  [0]{\@secondoftwo}%
\providecommand \bibfield  [0]{\@secondoftwo}%
\providecommand \translation [1]{[#1]}%
\providecommand \BibitemOpen [0]{}%
\providecommand \bibitemStop [0]{}%
\providecommand \bibitemNoStop [0]{.\EOS\space}%
\providecommand \EOS [0]{\spacefactor3000\relax}%
\providecommand \BibitemShut  [1]{\csname bibitem#1\endcsname}%
\let\auto@bib@innerbib\@empty
\bibitem [{\citenamefont {Schawlow}\ and\ \citenamefont
  {Townes}(1958)}]{schawlow}%
  \BibitemOpen
  \bibfield  {author} {\bibinfo {author} {\bibfnamefont {A.~L.}\ \bibnamefont
  {Schawlow}}\ and\ \bibinfo {author} {\bibfnamefont {C.~H.}\ \bibnamefont
  {Townes}},\ }\href {https://doi.org/10.1103/PhysRev.112.1940} {\bibfield
  {journal} {\bibinfo  {journal} {Phys. Rev.}\ }\textbf {\bibinfo {volume}
  {112}},\ \bibinfo {pages} {1940} (\bibinfo {year} {1958})}\BibitemShut
  {NoStop}%
\bibitem [{\citenamefont {Maiman}(1960)}]{MAIMAN1960}%
  \BibitemOpen
  \bibfield  {author} {\bibinfo {author} {\bibfnamefont {T.~H.}\ \bibnamefont
  {Maiman}},\ }\href {https://doi.org/10.1038/187493a0} {\bibfield  {journal}
  {\bibinfo  {journal} {Nature}\ }\textbf {\bibinfo {volume} {187}},\ \bibinfo
  {pages} {493} (\bibinfo {year} {1960})}\BibitemShut {NoStop}%
\bibitem [{\citenamefont {Ashkin}(1970)}]{ashkin}%
  \BibitemOpen
  \bibfield  {author} {\bibinfo {author} {\bibfnamefont {A.}~\bibnamefont
  {Ashkin}},\ }\href {https://doi.org/10.1103/PhysRevLett.24.156} {\bibfield
  {journal} {\bibinfo  {journal} {Phys. Rev. Lett.}\ }\textbf {\bibinfo
  {volume} {24}},\ \bibinfo {pages} {156} (\bibinfo {year} {1970})}\BibitemShut
  {NoStop}%
\bibitem [{\citenamefont {Hänsch}\ and\ \citenamefont
  {Schawlow}(1975)}]{HANSCH197568}%
  \BibitemOpen
  \bibfield  {author} {\bibinfo {author} {\bibfnamefont {T.}~\bibnamefont
  {Hänsch}}\ and\ \bibinfo {author} {\bibfnamefont {A.}~\bibnamefont
  {Schawlow}},\ }\href
  {https://doi.org/https://doi.org/10.1016/0030-4018(75)90159-5} {\bibfield
  {journal} {\bibinfo  {journal} {Optics Communications}\ }\textbf {\bibinfo
  {volume} {13}},\ \bibinfo {pages} {68} (\bibinfo {year} {1975})}\BibitemShut
  {NoStop}%
\bibitem [{\citenamefont {Wineland}\ \emph {et~al.}(1978)\citenamefont
  {Wineland}, \citenamefont {Drullinger},\ and\ \citenamefont
  {Walls}}]{wineland}%
  \BibitemOpen
  \bibfield  {author} {\bibinfo {author} {\bibfnamefont {D.~J.}\ \bibnamefont
  {Wineland}}, \bibinfo {author} {\bibfnamefont {R.~E.}\ \bibnamefont
  {Drullinger}},\ and\ \bibinfo {author} {\bibfnamefont {F.~L.}\ \bibnamefont
  {Walls}},\ }\href {https://doi.org/10.1103/PhysRevLett.40.1639} {\bibfield
  {journal} {\bibinfo  {journal} {Phys. Rev. Lett.}\ }\textbf {\bibinfo
  {volume} {40}},\ \bibinfo {pages} {1639} (\bibinfo {year}
  {1978})}\BibitemShut {NoStop}%
\bibitem [{\citenamefont {Chu}\ \emph {et~al.}(1985)\citenamefont {Chu},
  \citenamefont {Hollberg}, \citenamefont {Bjorkholm}, \citenamefont {Cable},\
  and\ \citenamefont {Ashkin}}]{chu}%
  \BibitemOpen
  \bibfield  {author} {\bibinfo {author} {\bibfnamefont {S.}~\bibnamefont
  {Chu}}, \bibinfo {author} {\bibfnamefont {L.}~\bibnamefont {Hollberg}},
  \bibinfo {author} {\bibfnamefont {J.~E.}\ \bibnamefont {Bjorkholm}}, \bibinfo
  {author} {\bibfnamefont {A.}~\bibnamefont {Cable}},\ and\ \bibinfo {author}
  {\bibfnamefont {A.}~\bibnamefont {Ashkin}},\ }\href
  {https://doi.org/10.1103/PhysRevLett.55.48} {\bibfield  {journal} {\bibinfo
  {journal} {Phys. Rev. Lett.}\ }\textbf {\bibinfo {volume} {55}},\ \bibinfo
  {pages} {48} (\bibinfo {year} {1985})}\BibitemShut {NoStop}%
\bibitem [{\citenamefont {Schioppo}\ \emph {et~al.}(2017)\citenamefont
  {Schioppo}, \citenamefont {Brown}, \citenamefont {McGrew}, \citenamefont
  {Hinkley}, \citenamefont {Fasano}, \citenamefont {Beloy}, \citenamefont
  {Yoon}, \citenamefont {Milani}, \citenamefont {Nicolodi}, \citenamefont
  {Sherman}, \citenamefont {Phillips}, \citenamefont {Oates},\ and\
  \citenamefont {Ludlow}}]{ludlow}%
  \BibitemOpen
  \bibfield  {author} {\bibinfo {author} {\bibfnamefont {M.}~\bibnamefont
  {Schioppo}}, \bibinfo {author} {\bibfnamefont {R.~C.}\ \bibnamefont {Brown}},
  \bibinfo {author} {\bibfnamefont {W.~F.}\ \bibnamefont {McGrew}}, \bibinfo
  {author} {\bibfnamefont {N.}~\bibnamefont {Hinkley}}, \bibinfo {author}
  {\bibfnamefont {R.~J.}\ \bibnamefont {Fasano}}, \bibinfo {author}
  {\bibfnamefont {K.}~\bibnamefont {Beloy}}, \bibinfo {author} {\bibfnamefont
  {T.~H.}\ \bibnamefont {Yoon}}, \bibinfo {author} {\bibfnamefont
  {G.}~\bibnamefont {Milani}}, \bibinfo {author} {\bibfnamefont
  {D.}~\bibnamefont {Nicolodi}}, \bibinfo {author} {\bibfnamefont {J.~A.}\
  \bibnamefont {Sherman}}, \bibinfo {author} {\bibfnamefont {N.~B.}\
  \bibnamefont {Phillips}}, \bibinfo {author} {\bibfnamefont {C.~W.}\
  \bibnamefont {Oates}},\ and\ \bibinfo {author} {\bibfnamefont {A.~D.}\
  \bibnamefont {Ludlow}},\ }\href {https://doi.org/10.1038/nphoton.2016.231}
  {\bibfield  {journal} {\bibinfo  {journal} {Nature Photonics}\ }\textbf
  {\bibinfo {volume} {11}},\ \bibinfo {pages} {48} (\bibinfo {year}
  {2017})}\BibitemShut {NoStop}%
\bibitem [{\citenamefont {Adams}\ and\ \citenamefont
  {Riis}(1997)}]{ADAMS19971}%
  \BibitemOpen
  \bibfield  {author} {\bibinfo {author} {\bibfnamefont {C.}~\bibnamefont
  {Adams}}\ and\ \bibinfo {author} {\bibfnamefont {E.}~\bibnamefont {Riis}},\
  }\href {https://doi.org/https://doi.org/10.1016/S0079-6727(96)00006-7}
  {\bibfield  {journal} {\bibinfo  {journal} {Progress in Quantum Electronics}\
  }\textbf {\bibinfo {volume} {21}},\ \bibinfo {pages} {1} (\bibinfo {year}
  {1997})}\BibitemShut {NoStop}%
\bibitem [{\citenamefont {Zawadzki}\ \emph {et~al.}(2010)\citenamefont
  {Zawadzki}, \citenamefont {Griffin}, \citenamefont {Riis},\ and\
  \citenamefont {Arnold}}]{aidanring}%
  \BibitemOpen
  \bibfield  {author} {\bibinfo {author} {\bibfnamefont {M.~E.}\ \bibnamefont
  {Zawadzki}}, \bibinfo {author} {\bibfnamefont {P.~F.}\ \bibnamefont
  {Griffin}}, \bibinfo {author} {\bibfnamefont {E.}~\bibnamefont {Riis}},\ and\
  \bibinfo {author} {\bibfnamefont {A.~S.}\ \bibnamefont {Arnold}},\ }\href
  {https://doi.org/10.1103/PhysRevA.81.043608} {\bibfield  {journal} {\bibinfo
  {journal} {Phys. Rev. A}\ }\textbf {\bibinfo {volume} {81}},\ \bibinfo
  {pages} {043608} (\bibinfo {year} {2010})}\BibitemShut {NoStop}%
\bibitem [{\citenamefont {Geiger}\ \emph {et~al.}(2020)\citenamefont {Geiger},
  \citenamefont {Landragin}, \citenamefont {Merlet},\ and\ \citenamefont
  {Pereira Dos~Santos}}]{landragin}%
  \BibitemOpen
  \bibfield  {author} {\bibinfo {author} {\bibfnamefont {R.}~\bibnamefont
  {Geiger}}, \bibinfo {author} {\bibfnamefont {A.}~\bibnamefont {Landragin}},
  \bibinfo {author} {\bibfnamefont {S.}~\bibnamefont {Merlet}},\ and\ \bibinfo
  {author} {\bibfnamefont {F.}~\bibnamefont {Pereira Dos~Santos}},\ }\href
  {https://doi.org/10.1116/5.0009093} {\bibfield  {journal} {\bibinfo
  {journal} {AVS Quantum Science}\ }\textbf {\bibinfo {volume} {2}},\ \bibinfo
  {pages} {024702} (\bibinfo {year} {2020})}\BibitemShut {NoStop}%
\bibitem [{\citenamefont {Nelson}\ \emph {et~al.}(2020)\citenamefont {Nelson},
  \citenamefont {Fertig}, \citenamefont {Hamilton}, \citenamefont {Brown},
  \citenamefont {Estey}, \citenamefont {Müller},\ and\ \citenamefont
  {Compton}}]{inertial}%
  \BibitemOpen
  \bibfield  {author} {\bibinfo {author} {\bibfnamefont {K.~D.}\ \bibnamefont
  {Nelson}}, \bibinfo {author} {\bibfnamefont {C.~D.}\ \bibnamefont {Fertig}},
  \bibinfo {author} {\bibfnamefont {P.}~\bibnamefont {Hamilton}}, \bibinfo
  {author} {\bibfnamefont {J.~M.}\ \bibnamefont {Brown}}, \bibinfo {author}
  {\bibfnamefont {B.}~\bibnamefont {Estey}}, \bibinfo {author} {\bibfnamefont
  {H.}~\bibnamefont {Müller}},\ and\ \bibinfo {author} {\bibfnamefont {R.~L.}\
  \bibnamefont {Compton}},\ }\href {https://doi.org/10.1063/5.0010070}
  {\bibfield  {journal} {\bibinfo  {journal} {Applied Physics Letters}\
  }\textbf {\bibinfo {volume} {116}},\ \bibinfo {pages} {234002} (\bibinfo
  {year} {2020})}\BibitemShut {NoStop}%
\bibitem [{\citenamefont {Poli}\ \emph {et~al.}(2011)\citenamefont {Poli},
  \citenamefont {Wang}, \citenamefont {Tarallo}, \citenamefont {Alberti},
  \citenamefont {Prevedelli},\ and\ \citenamefont {Tino}}]{Poli}%
  \BibitemOpen
  \bibfield  {author} {\bibinfo {author} {\bibfnamefont {N.}~\bibnamefont
  {Poli}}, \bibinfo {author} {\bibfnamefont {F.-Y.}\ \bibnamefont {Wang}},
  \bibinfo {author} {\bibfnamefont {M.~G.}\ \bibnamefont {Tarallo}}, \bibinfo
  {author} {\bibfnamefont {A.}~\bibnamefont {Alberti}}, \bibinfo {author}
  {\bibfnamefont {M.}~\bibnamefont {Prevedelli}},\ and\ \bibinfo {author}
  {\bibfnamefont {G.~M.}\ \bibnamefont {Tino}},\ }\href
  {https://doi.org/10.1103/PhysRevLett.106.038501} {\bibfield  {journal}
  {\bibinfo  {journal} {Phys. Rev. Lett.}\ }\textbf {\bibinfo {volume} {106}},\
  \bibinfo {pages} {038501} (\bibinfo {year} {2011})}\BibitemShut {NoStop}%
\bibitem [{\citenamefont {Bidel}\ \emph {et~al.}(2013)\citenamefont {Bidel},
  \citenamefont {Carraz}, \citenamefont {Charrière}, \citenamefont {Cadoret},
  \citenamefont {Zahzam},\ and\ \citenamefont {Bresson}}]{bidel}%
  \BibitemOpen
  \bibfield  {author} {\bibinfo {author} {\bibfnamefont {Y.}~\bibnamefont
  {Bidel}}, \bibinfo {author} {\bibfnamefont {O.}~\bibnamefont {Carraz}},
  \bibinfo {author} {\bibfnamefont {R.}~\bibnamefont {Charrière}}, \bibinfo
  {author} {\bibfnamefont {M.}~\bibnamefont {Cadoret}}, \bibinfo {author}
  {\bibfnamefont {N.}~\bibnamefont {Zahzam}},\ and\ \bibinfo {author}
  {\bibfnamefont {A.}~\bibnamefont {Bresson}},\ }\href
  {https://doi.org/10.1063/1.4801756} {\bibfield  {journal} {\bibinfo
  {journal} {Applied Physics Letters}\ }\textbf {\bibinfo {volume} {102}},\
  \bibinfo {pages} {144107} (\bibinfo {year} {2013})}\BibitemShut {NoStop}%
\bibitem [{\citenamefont {Garrido~Alzar}(2019)}]{garrido}%
  \BibitemOpen
  \bibfield  {author} {\bibinfo {author} {\bibfnamefont {C.~L.}\ \bibnamefont
  {Garrido~Alzar}},\ }\href@noop {} {\bibfield  {journal} {\bibinfo  {journal}
  {AVS Quantum Science}\ }\textbf {\bibinfo {volume} {1}},\ \bibinfo {pages}
  {014702} (\bibinfo {year} {2019})}\BibitemShut {NoStop}%
\bibitem [{\citenamefont {Bize}\ \emph {et~al.}(2005)\citenamefont {Bize},
  \citenamefont {Laurent}, \citenamefont {Abgrall}, \citenamefont {Marion},
  \citenamefont {Maksimovic}, \citenamefont {Cacciapuoti}, \citenamefont
  {Grünert}, \citenamefont {Vian}, \citenamefont {dos Santos}, \citenamefont
  {Rosenbusch}, \citenamefont {Lemonde}, \citenamefont {Santarelli},
  \citenamefont {Wolf}, \citenamefont {Clairon}, \citenamefont {Luiten},
  \citenamefont {Tobar},\ and\ \citenamefont {Salomon}}]{Bize_2005}%
  \BibitemOpen
  \bibfield  {author} {\bibinfo {author} {\bibfnamefont {S.}~\bibnamefont
  {Bize}}, \bibinfo {author} {\bibfnamefont {P.}~\bibnamefont {Laurent}},
  \bibinfo {author} {\bibfnamefont {M.}~\bibnamefont {Abgrall}}, \bibinfo
  {author} {\bibfnamefont {H.}~\bibnamefont {Marion}}, \bibinfo {author}
  {\bibfnamefont {I.}~\bibnamefont {Maksimovic}}, \bibinfo {author}
  {\bibfnamefont {L.}~\bibnamefont {Cacciapuoti}}, \bibinfo {author}
  {\bibfnamefont {J.}~\bibnamefont {Grünert}}, \bibinfo {author}
  {\bibfnamefont {C.}~\bibnamefont {Vian}}, \bibinfo {author} {\bibfnamefont
  {F.~P.}\ \bibnamefont {dos Santos}}, \bibinfo {author} {\bibfnamefont
  {P.}~\bibnamefont {Rosenbusch}}, \bibinfo {author} {\bibfnamefont
  {P.}~\bibnamefont {Lemonde}}, \bibinfo {author} {\bibfnamefont
  {G.}~\bibnamefont {Santarelli}}, \bibinfo {author} {\bibfnamefont
  {P.}~\bibnamefont {Wolf}}, \bibinfo {author} {\bibfnamefont {A.}~\bibnamefont
  {Clairon}}, \bibinfo {author} {\bibfnamefont {A.}~\bibnamefont {Luiten}},
  \bibinfo {author} {\bibfnamefont {M.}~\bibnamefont {Tobar}},\ and\ \bibinfo
  {author} {\bibfnamefont {C.}~\bibnamefont {Salomon}},\ }\href
  {https://doi.org/10.1088/0953-4075/38/9/002} {\bibfield  {journal} {\bibinfo
  {journal} {Journal of Physics B: Atomic, Molecular and Optical Physics}\
  }\textbf {\bibinfo {volume} {38}},\ \bibinfo {pages} {S449} (\bibinfo {year}
  {2005})}\BibitemShut {NoStop}%
\bibitem [{\citenamefont {Leopardi}\ \emph {et~al.}(2021)\citenamefont
  {Leopardi}, \citenamefont {Beloy}, \citenamefont {Bothwell}, \citenamefont
  {Brewer}, \citenamefont {Bromley}, \citenamefont {Chen}, \citenamefont
  {Diddams}, \citenamefont {Fasano}, \citenamefont {Hassan}, \citenamefont
  {Hume}, \citenamefont {Kedar}, \citenamefont {Kennedy}, \citenamefont
  {Khader}, \citenamefont {Leibrandt}, \citenamefont {Ludlow}, \citenamefont
  {McGrew}, \citenamefont {Milner}, \citenamefont {Nicolodi}, \citenamefont
  {Oelker}, \citenamefont {Parker}, \citenamefont {Robinson}, \citenamefont
  {Romisch}, \citenamefont {Sherman}, \citenamefont {Sonderhouse},
  \citenamefont {Swann}, \citenamefont {Yao}, \citenamefont {Ye}, \citenamefont
  {Zhang},\ and\ \citenamefont {Fortier}}]{ludlownist}%
  \BibitemOpen
  \bibfield  {author} {\bibinfo {author} {\bibfnamefont {H.}~\bibnamefont
  {Leopardi}}, \bibinfo {author} {\bibfnamefont {K.}~\bibnamefont {Beloy}},
  \bibinfo {author} {\bibfnamefont {T.}~\bibnamefont {Bothwell}}, \bibinfo
  {author} {\bibfnamefont {S.}~\bibnamefont {Brewer}}, \bibinfo {author}
  {\bibfnamefont {S.}~\bibnamefont {Bromley}}, \bibinfo {author} {\bibfnamefont
  {J.-S.}\ \bibnamefont {Chen}}, \bibinfo {author} {\bibfnamefont
  {S.}~\bibnamefont {Diddams}}, \bibinfo {author} {\bibfnamefont
  {R.}~\bibnamefont {Fasano}}, \bibinfo {author} {\bibfnamefont
  {Y.}~\bibnamefont {Hassan}}, \bibinfo {author} {\bibfnamefont
  {D.}~\bibnamefont {Hume}}, \bibinfo {author} {\bibfnamefont {D.}~\bibnamefont
  {Kedar}}, \bibinfo {author} {\bibfnamefont {C.}~\bibnamefont {Kennedy}},
  \bibinfo {author} {\bibfnamefont {I.}~\bibnamefont {Khader}}, \bibinfo
  {author} {\bibfnamefont {D.}~\bibnamefont {Leibrandt}}, \bibinfo {author}
  {\bibfnamefont {A.}~\bibnamefont {Ludlow}}, \bibinfo {author} {\bibfnamefont
  {W.}~\bibnamefont {McGrew}}, \bibinfo {author} {\bibfnamefont
  {W.}~\bibnamefont {Milner}}, \bibinfo {author} {\bibfnamefont
  {D.}~\bibnamefont {Nicolodi}}, \bibinfo {author} {\bibfnamefont
  {E.}~\bibnamefont {Oelker}}, \bibinfo {author} {\bibfnamefont
  {T.}~\bibnamefont {Parker}}, \bibinfo {author} {\bibfnamefont
  {J.}~\bibnamefont {Robinson}}, \bibinfo {author} {\bibfnamefont
  {S.}~\bibnamefont {Romisch}}, \bibinfo {author} {\bibfnamefont
  {J.}~\bibnamefont {Sherman}}, \bibinfo {author} {\bibfnamefont
  {L.}~\bibnamefont {Sonderhouse}}, \bibinfo {author} {\bibfnamefont
  {W.}~\bibnamefont {Swann}}, \bibinfo {author} {\bibfnamefont
  {J.}~\bibnamefont {Yao}}, \bibinfo {author} {\bibfnamefont {J.}~\bibnamefont
  {Ye}}, \bibinfo {author} {\bibfnamefont {X.}~\bibnamefont {Zhang}},\ and\
  \bibinfo {author} {\bibfnamefont {T.}~\bibnamefont {Fortier}},\ }\href
  {https://tsapps.nist.gov/publication/get_pdf.cfm?pub_id=931072} {\bibfield
  {journal} {\bibinfo  {journal} {Metrologia}\ } (\bibinfo {year}
  {2021})}\BibitemShut {NoStop}%
\bibitem [{\citenamefont {McGrew}\ \emph {et~al.}(2018)\citenamefont {McGrew},
  \citenamefont {Zhang}, \citenamefont {Fasano}, \citenamefont {Sch{\"a}ffer},
  \citenamefont {Beloy}, \citenamefont {Nicolodi}, \citenamefont {Brown},
  \citenamefont {Hinkley}, \citenamefont {Milani}, \citenamefont {Schioppo},
  \citenamefont {Yoon},\ and\ \citenamefont {Ludlow}}]{McGrew2018}%
  \BibitemOpen
  \bibfield  {author} {\bibinfo {author} {\bibfnamefont {W.~F.}\ \bibnamefont
  {McGrew}}, \bibinfo {author} {\bibfnamefont {X.}~\bibnamefont {Zhang}},
  \bibinfo {author} {\bibfnamefont {R.~J.}\ \bibnamefont {Fasano}}, \bibinfo
  {author} {\bibfnamefont {S.~A.}\ \bibnamefont {Sch{\"a}ffer}}, \bibinfo
  {author} {\bibfnamefont {K.}~\bibnamefont {Beloy}}, \bibinfo {author}
  {\bibfnamefont {D.}~\bibnamefont {Nicolodi}}, \bibinfo {author}
  {\bibfnamefont {R.~C.}\ \bibnamefont {Brown}}, \bibinfo {author}
  {\bibfnamefont {N.}~\bibnamefont {Hinkley}}, \bibinfo {author} {\bibfnamefont
  {G.}~\bibnamefont {Milani}}, \bibinfo {author} {\bibfnamefont
  {M.}~\bibnamefont {Schioppo}}, \bibinfo {author} {\bibfnamefont {T.~H.}\
  \bibnamefont {Yoon}},\ and\ \bibinfo {author} {\bibfnamefont {A.~D.}\
  \bibnamefont {Ludlow}},\ }\href {https://doi.org/10.1038/s41586-018-0738-2}
  {\bibfield  {journal} {\bibinfo  {journal} {Nature}\ }\textbf {\bibinfo
  {volume} {564}},\ \bibinfo {pages} {87} (\bibinfo {year} {2018})}\BibitemShut
  {NoStop}%
\bibitem [{\citenamefont {Udem}\ \emph {et~al.}(2002)\citenamefont {Udem},
  \citenamefont {Holzwarth},\ and\ \citenamefont {H{\"a}nsch}}]{hanschreview}%
  \BibitemOpen
  \bibfield  {author} {\bibinfo {author} {\bibfnamefont {T.}~\bibnamefont
  {Udem}}, \bibinfo {author} {\bibfnamefont {R.}~\bibnamefont {Holzwarth}},\
  and\ \bibinfo {author} {\bibfnamefont {T.~W.}\ \bibnamefont {H{\"a}nsch}},\
  }\href {https://doi.org/10.1038/416233a} {\bibfield  {journal} {\bibinfo
  {journal} {Nature}\ }\textbf {\bibinfo {volume} {416}},\ \bibinfo {pages}
  {233} (\bibinfo {year} {2002})}\BibitemShut {NoStop}%
\bibitem [{\citenamefont {Ludlow}\ \emph {et~al.}(2015)\citenamefont {Ludlow},
  \citenamefont {Boyd}, \citenamefont {Ye}, \citenamefont {Peik},\ and\
  \citenamefont {Schmidt}}]{ludlowreview}%
  \BibitemOpen
  \bibfield  {author} {\bibinfo {author} {\bibfnamefont {A.~D.}\ \bibnamefont
  {Ludlow}}, \bibinfo {author} {\bibfnamefont {M.~M.}\ \bibnamefont {Boyd}},
  \bibinfo {author} {\bibfnamefont {J.}~\bibnamefont {Ye}}, \bibinfo {author}
  {\bibfnamefont {E.}~\bibnamefont {Peik}},\ and\ \bibinfo {author}
  {\bibfnamefont {P.~O.}\ \bibnamefont {Schmidt}},\ }\href
  {https://doi.org/10.1103/RevModPhys.87.637} {\bibfield  {journal} {\bibinfo
  {journal} {Rev. Mod. Phys.}\ }\textbf {\bibinfo {volume} {87}},\ \bibinfo
  {pages} {637} (\bibinfo {year} {2015})}\BibitemShut {NoStop}%
\bibitem [{\citenamefont {Kaufman}\ \emph {et~al.}(2015)\citenamefont
  {Kaufman}, \citenamefont {Lester}, \citenamefont {Foss-Feig}, \citenamefont
  {Wall}, \citenamefont {Rey},\ and\ \citenamefont {Regal}}]{Kaufman2015}%
  \BibitemOpen
  \bibfield  {author} {\bibinfo {author} {\bibfnamefont {A.~M.}\ \bibnamefont
  {Kaufman}}, \bibinfo {author} {\bibfnamefont {B.~J.}\ \bibnamefont {Lester}},
  \bibinfo {author} {\bibfnamefont {M.}~\bibnamefont {Foss-Feig}}, \bibinfo
  {author} {\bibfnamefont {M.~L.}\ \bibnamefont {Wall}}, \bibinfo {author}
  {\bibfnamefont {A.~M.}\ \bibnamefont {Rey}},\ and\ \bibinfo {author}
  {\bibfnamefont {C.~A.}\ \bibnamefont {Regal}},\ }\href
  {https://doi.org/10.1038/nature16073} {\bibfield  {journal} {\bibinfo
  {journal} {Nature}\ }\textbf {\bibinfo {volume} {527}},\ \bibinfo {pages}
  {208} (\bibinfo {year} {2015})}\BibitemShut {NoStop}%
\bibitem [{\citenamefont {Wu}\ \emph {et~al.}(2019{\natexlab{a}})\citenamefont
  {Wu}, \citenamefont {Kumar}, \citenamefont {Giraldo},\ and\ \citenamefont
  {Weiss}}]{Wu2019}%
  \BibitemOpen
  \bibfield  {author} {\bibinfo {author} {\bibfnamefont {T.-Y.}\ \bibnamefont
  {Wu}}, \bibinfo {author} {\bibfnamefont {A.}~\bibnamefont {Kumar}}, \bibinfo
  {author} {\bibfnamefont {F.}~\bibnamefont {Giraldo}},\ and\ \bibinfo {author}
  {\bibfnamefont {D.~S.}\ \bibnamefont {Weiss}},\ }\href
  {https://doi.org/10.1038/s41567-019-0478-8} {\bibfield  {journal} {\bibinfo
  {journal} {Nature Physics}\ }\textbf {\bibinfo {volume} {15}},\ \bibinfo
  {pages} {538} (\bibinfo {year} {2019}{\natexlab{a}})}\BibitemShut {NoStop}%
\bibitem [{\citenamefont {Picken}\ \emph {et~al.}(2018)\citenamefont {Picken},
  \citenamefont {Legaie}, \citenamefont {McDonnell},\ and\ \citenamefont
  {Pritchard}}]{Picken_2018}%
  \BibitemOpen
  \bibfield  {author} {\bibinfo {author} {\bibfnamefont {C.~J.}\ \bibnamefont
  {Picken}}, \bibinfo {author} {\bibfnamefont {R.}~\bibnamefont {Legaie}},
  \bibinfo {author} {\bibfnamefont {K.}~\bibnamefont {McDonnell}},\ and\
  \bibinfo {author} {\bibfnamefont {J.~D.}\ \bibnamefont {Pritchard}},\ }\href
  {https://doi.org/10.1088/2058-9565/aaf019} {\bibfield  {journal} {\bibinfo
  {journal} {Quantum Science and Technology}\ }\textbf {\bibinfo {volume}
  {4}},\ \bibinfo {pages} {015011} (\bibinfo {year} {2018})}\BibitemShut
  {NoStop}%
\bibitem [{\citenamefont {Davis}\ \emph {et~al.}(1995)\citenamefont {Davis},
  \citenamefont {Mewes}, \citenamefont {Andrews}, \citenamefont {van Druten},
  \citenamefont {Durfee}, \citenamefont {Kurn},\ and\ \citenamefont
  {Ketterle}}]{ketterle}%
  \BibitemOpen
  \bibfield  {author} {\bibinfo {author} {\bibfnamefont {K.~B.}\ \bibnamefont
  {Davis}}, \bibinfo {author} {\bibfnamefont {M.~O.}\ \bibnamefont {Mewes}},
  \bibinfo {author} {\bibfnamefont {M.~R.}\ \bibnamefont {Andrews}}, \bibinfo
  {author} {\bibfnamefont {N.~J.}\ \bibnamefont {van Druten}}, \bibinfo
  {author} {\bibfnamefont {D.~S.}\ \bibnamefont {Durfee}}, \bibinfo {author}
  {\bibfnamefont {D.~M.}\ \bibnamefont {Kurn}},\ and\ \bibinfo {author}
  {\bibfnamefont {W.}~\bibnamefont {Ketterle}},\ }\href
  {https://doi.org/10.1103/PhysRevLett.75.3969} {\bibfield  {journal} {\bibinfo
   {journal} {Phys. Rev. Lett.}\ }\textbf {\bibinfo {volume} {75}},\ \bibinfo
  {pages} {3969} (\bibinfo {year} {1995})}\BibitemShut {NoStop}%
\bibitem [{\citenamefont {Anderson}\ \emph {et~al.}(1995)\citenamefont
  {Anderson}, \citenamefont {Ensher}, \citenamefont {Matthews}, \citenamefont
  {Wieman},\ and\ \citenamefont {Cornell}}]{BECCU}%
  \BibitemOpen
  \bibfield  {author} {\bibinfo {author} {\bibfnamefont {M.~H.}\ \bibnamefont
  {Anderson}}, \bibinfo {author} {\bibfnamefont {J.~R.}\ \bibnamefont
  {Ensher}}, \bibinfo {author} {\bibfnamefont {M.~R.}\ \bibnamefont
  {Matthews}}, \bibinfo {author} {\bibfnamefont {C.~E.}\ \bibnamefont
  {Wieman}},\ and\ \bibinfo {author} {\bibfnamefont {E.~A.}\ \bibnamefont
  {Cornell}},\ }\href {https://doi.org/10.1126/science.269.5221.198} {\bibfield
   {journal} {\bibinfo  {journal} {Science}\ }\textbf {\bibinfo {volume}
  {269}},\ \bibinfo {pages} {198} (\bibinfo {year} {1995})}\BibitemShut
  {NoStop}%
\bibitem [{\citenamefont {Aveline}\ \emph {et~al.}(2020)\citenamefont
  {Aveline}, \citenamefont {Williams}, \citenamefont {Elliott}, \citenamefont
  {Dutenhoffer}, \citenamefont {Kellogg}, \citenamefont {Kohel}, \citenamefont
  {Lay}, \citenamefont {Oudrhiri}, \citenamefont {Shotwell}, \citenamefont
  {Yu},\ and\ \citenamefont {Thompson}}]{Aveline2020}%
  \BibitemOpen
  \bibfield  {author} {\bibinfo {author} {\bibfnamefont {D.~C.}\ \bibnamefont
  {Aveline}}, \bibinfo {author} {\bibfnamefont {J.~R.}\ \bibnamefont
  {Williams}}, \bibinfo {author} {\bibfnamefont {E.~R.}\ \bibnamefont
  {Elliott}}, \bibinfo {author} {\bibfnamefont {C.}~\bibnamefont
  {Dutenhoffer}}, \bibinfo {author} {\bibfnamefont {J.~R.}\ \bibnamefont
  {Kellogg}}, \bibinfo {author} {\bibfnamefont {J.~M.}\ \bibnamefont {Kohel}},
  \bibinfo {author} {\bibfnamefont {N.~E.}\ \bibnamefont {Lay}}, \bibinfo
  {author} {\bibfnamefont {K.}~\bibnamefont {Oudrhiri}}, \bibinfo {author}
  {\bibfnamefont {R.~F.}\ \bibnamefont {Shotwell}}, \bibinfo {author}
  {\bibfnamefont {N.}~\bibnamefont {Yu}},\ and\ \bibinfo {author}
  {\bibfnamefont {R.~J.}\ \bibnamefont {Thompson}},\ }\href
  {https://doi.org/10.1038/s41586-020-2346-1} {\bibfield  {journal} {\bibinfo
  {journal} {Nature}\ }\textbf {\bibinfo {volume} {582}},\ \bibinfo {pages}
  {193} (\bibinfo {year} {2020})}\BibitemShut {NoStop}%
\bibitem [{\citenamefont {Greiner}\ \emph {et~al.}(2002)\citenamefont
  {Greiner}, \citenamefont {Mandel}, \citenamefont {Esslinger}, \citenamefont
  {H{\"a}nsch},\ and\ \citenamefont {Bloch}}]{Greiner2002}%
  \BibitemOpen
  \bibfield  {author} {\bibinfo {author} {\bibfnamefont {M.}~\bibnamefont
  {Greiner}}, \bibinfo {author} {\bibfnamefont {O.}~\bibnamefont {Mandel}},
  \bibinfo {author} {\bibfnamefont {T.}~\bibnamefont {Esslinger}}, \bibinfo
  {author} {\bibfnamefont {T.~W.}\ \bibnamefont {H{\"a}nsch}},\ and\ \bibinfo
  {author} {\bibfnamefont {I.}~\bibnamefont {Bloch}},\ }\href
  {https://doi.org/10.1038/415039a} {\bibfield  {journal} {\bibinfo  {journal}
  {Nature}\ }\textbf {\bibinfo {volume} {415}},\ \bibinfo {pages} {39}
  (\bibinfo {year} {2002})}\BibitemShut {NoStop}%
\bibitem [{\citenamefont {Sherson}\ \emph {et~al.}(2010)\citenamefont
  {Sherson}, \citenamefont {Weitenberg}, \citenamefont {Endres}, \citenamefont
  {Cheneau}, \citenamefont {Bloch},\ and\ \citenamefont {Kuhr}}]{kuhr1}%
  \BibitemOpen
  \bibfield  {author} {\bibinfo {author} {\bibfnamefont {J.~F.}\ \bibnamefont
  {Sherson}}, \bibinfo {author} {\bibfnamefont {C.}~\bibnamefont {Weitenberg}},
  \bibinfo {author} {\bibfnamefont {M.}~\bibnamefont {Endres}}, \bibinfo
  {author} {\bibfnamefont {M.}~\bibnamefont {Cheneau}}, \bibinfo {author}
  {\bibfnamefont {I.}~\bibnamefont {Bloch}},\ and\ \bibinfo {author}
  {\bibfnamefont {S.}~\bibnamefont {Kuhr}},\ }\href
  {https://doi.org/10.1038/nature09378} {\bibfield  {journal} {\bibinfo
  {journal} {Nature}\ }\textbf {\bibinfo {volume} {467}},\ \bibinfo {pages}
  {68} (\bibinfo {year} {2010})}\BibitemShut {NoStop}%
\bibitem [{\citenamefont {Fukuhara}\ \emph {et~al.}(2013)\citenamefont
  {Fukuhara}, \citenamefont {Kantian}, \citenamefont {Endres}, \citenamefont
  {Cheneau}, \citenamefont {Schau{\ss}}, \citenamefont {Hild}, \citenamefont
  {Bellem}, \citenamefont {Schollw{\"o}ck}, \citenamefont {Giamarchi},
  \citenamefont {Gross}, \citenamefont {Bloch},\ and\ \citenamefont
  {Kuhr}}]{kuhr2}%
  \BibitemOpen
  \bibfield  {author} {\bibinfo {author} {\bibfnamefont {T.}~\bibnamefont
  {Fukuhara}}, \bibinfo {author} {\bibfnamefont {A.}~\bibnamefont {Kantian}},
  \bibinfo {author} {\bibfnamefont {M.}~\bibnamefont {Endres}}, \bibinfo
  {author} {\bibfnamefont {M.}~\bibnamefont {Cheneau}}, \bibinfo {author}
  {\bibfnamefont {P.}~\bibnamefont {Schau{\ss}}}, \bibinfo {author}
  {\bibfnamefont {S.}~\bibnamefont {Hild}}, \bibinfo {author} {\bibfnamefont
  {D.}~\bibnamefont {Bellem}}, \bibinfo {author} {\bibfnamefont
  {U.}~\bibnamefont {Schollw{\"o}ck}}, \bibinfo {author} {\bibfnamefont
  {T.}~\bibnamefont {Giamarchi}}, \bibinfo {author} {\bibfnamefont
  {C.}~\bibnamefont {Gross}}, \bibinfo {author} {\bibfnamefont
  {I.}~\bibnamefont {Bloch}},\ and\ \bibinfo {author} {\bibfnamefont
  {S.}~\bibnamefont {Kuhr}},\ }\href {https://doi.org/10.1038/nphys2561}
  {\bibfield  {journal} {\bibinfo  {journal} {Nature Physics}\ }\textbf
  {\bibinfo {volume} {9}},\ \bibinfo {pages} {235} (\bibinfo {year}
  {2013})}\BibitemShut {NoStop}%
\bibitem [{\citenamefont {Pfau}(2002)}]{atomlaser}%
  \BibitemOpen
  \bibfield  {author} {\bibinfo {author} {\bibfnamefont {T.}~\bibnamefont
  {Pfau}},\ }\href {https://doi.org/10.1126/science.1073984} {\bibfield
  {journal} {\bibinfo  {journal} {Science}\ }\textbf {\bibinfo {volume}
  {296}},\ \bibinfo {pages} {2155} (\bibinfo {year} {2002})}\BibitemShut
  {NoStop}%
\bibitem [{\citenamefont {Bell}\ \emph {et~al.}(1998)\citenamefont {Bell},
  \citenamefont {Pfau}, \citenamefont {Drodofsky}, \citenamefont {Stuhler},
  \citenamefont {Schulze}, \citenamefont {Brezger}, \citenamefont {Nowak},\
  and\ \citenamefont {Mlynek}}]{atomlithography}%
  \BibitemOpen
  \bibfield  {author} {\bibinfo {author} {\bibfnamefont {A.}~\bibnamefont
  {Bell}}, \bibinfo {author} {\bibfnamefont {T.}~\bibnamefont {Pfau}}, \bibinfo
  {author} {\bibfnamefont {U.}~\bibnamefont {Drodofsky}}, \bibinfo {author}
  {\bibfnamefont {J.}~\bibnamefont {Stuhler}}, \bibinfo {author} {\bibfnamefont
  {T.}~\bibnamefont {Schulze}}, \bibinfo {author} {\bibfnamefont
  {B.}~\bibnamefont {Brezger}}, \bibinfo {author} {\bibfnamefont
  {S.}~\bibnamefont {Nowak}},\ and\ \bibinfo {author} {\bibfnamefont
  {J.}~\bibnamefont {Mlynek}},\ }\href
  {https://doi.org/https://doi.org/10.1016/S0167-9317(98)00138-5} {\bibfield
  {journal} {\bibinfo  {journal} {Microelectronic Engineering}\ }\textbf
  {\bibinfo {volume} {41-42}},\ \bibinfo {pages} {587} (\bibinfo {year}
  {1998})}\BibitemShut {NoStop}%
\bibitem [{\citenamefont {Arnold}\ \emph {et~al.}(2002)\citenamefont {Arnold},
  \citenamefont {MacCormick},\ and\ \citenamefont {Boshier}}]{arnold2002}%
  \BibitemOpen
  \bibfield  {author} {\bibinfo {author} {\bibfnamefont {A.~S.}\ \bibnamefont
  {Arnold}}, \bibinfo {author} {\bibfnamefont {C.}~\bibnamefont {MacCormick}},\
  and\ \bibinfo {author} {\bibfnamefont {M.~G.}\ \bibnamefont {Boshier}},\
  }\href {https://doi.org/10.1103/physreva.65.031601} {\bibfield  {journal}
  {\bibinfo  {journal} {Physical Review A}\ }\textbf {\bibinfo {volume} {65}},\
  \bibinfo {pages} {031601} (\bibinfo {year} {2002})}\BibitemShut {NoStop}%
\bibitem [{\citenamefont {Deppner}\ \emph {et~al.}(2021)\citenamefont
  {Deppner}, \citenamefont {Herr}, \citenamefont {Cornelius}, \citenamefont
  {Stromberger}, \citenamefont {Sternke}, \citenamefont {Grzeschik},
  \citenamefont {Grote}, \citenamefont {Rudolph}, \citenamefont {Herrmann},
  \citenamefont {Krutzik}, \citenamefont {Wenzlawski}, \citenamefont {Corgier},
  \citenamefont {Charron}, \citenamefont {Gu{\'{e}}ry-Odelin}, \citenamefont
  {Gaaloul}, \citenamefont {L\"{a}mmerzahl}, \citenamefont {Peters},
  \citenamefont {Windpassinger},\ and\ \citenamefont {Rasel}}]{Deppner2021}%
  \BibitemOpen
  \bibfield  {author} {\bibinfo {author} {\bibfnamefont {C.}~\bibnamefont
  {Deppner}}, \bibinfo {author} {\bibfnamefont {W.}~\bibnamefont {Herr}},
  \bibinfo {author} {\bibfnamefont {M.}~\bibnamefont {Cornelius}}, \bibinfo
  {author} {\bibfnamefont {P.}~\bibnamefont {Stromberger}}, \bibinfo {author}
  {\bibfnamefont {T.}~\bibnamefont {Sternke}}, \bibinfo {author} {\bibfnamefont
  {C.}~\bibnamefont {Grzeschik}}, \bibinfo {author} {\bibfnamefont
  {A.}~\bibnamefont {Grote}}, \bibinfo {author} {\bibfnamefont
  {J.}~\bibnamefont {Rudolph}}, \bibinfo {author} {\bibfnamefont
  {S.}~\bibnamefont {Herrmann}}, \bibinfo {author} {\bibfnamefont
  {M.}~\bibnamefont {Krutzik}}, \bibinfo {author} {\bibfnamefont
  {A.}~\bibnamefont {Wenzlawski}}, \bibinfo {author} {\bibfnamefont
  {R.}~\bibnamefont {Corgier}}, \bibinfo {author} {\bibfnamefont
  {E.}~\bibnamefont {Charron}}, \bibinfo {author} {\bibfnamefont
  {D.}~\bibnamefont {Gu{\'{e}}ry-Odelin}}, \bibinfo {author} {\bibfnamefont
  {N.}~\bibnamefont {Gaaloul}}, \bibinfo {author} {\bibfnamefont
  {C.}~\bibnamefont {L\"{a}mmerzahl}}, \bibinfo {author} {\bibfnamefont
  {A.}~\bibnamefont {Peters}}, \bibinfo {author} {\bibfnamefont
  {P.}~\bibnamefont {Windpassinger}},\ and\ \bibinfo {author} {\bibfnamefont
  {E.~M.}\ \bibnamefont {Rasel}},\ }\href
  {https://doi.org/10.1103/physrevlett.127.100401} {\bibfield  {journal}
  {\bibinfo  {journal} {Physical Review Letters}\ }\textbf {\bibinfo {volume}
  {127}},\ \bibinfo {pages} {100401} (\bibinfo {year} {2021})}\BibitemShut
  {NoStop}%
\bibitem [{\citenamefont {Rushton}\ \emph {et~al.}(2014)\citenamefont
  {Rushton}, \citenamefont {Aldous},\ and\ \citenamefont
  {Himsworth}}]{Rushton2014}%
  \BibitemOpen
  \bibfield  {author} {\bibinfo {author} {\bibfnamefont {J.~A.}\ \bibnamefont
  {Rushton}}, \bibinfo {author} {\bibfnamefont {M.}~\bibnamefont {Aldous}},\
  and\ \bibinfo {author} {\bibfnamefont {M.~D.}\ \bibnamefont {Himsworth}},\
  }\href {https://doi.org/10.1063/1.4904066} {\bibfield  {journal} {\bibinfo
  {journal} {Review of Scientific Instruments}\ }\textbf {\bibinfo {volume}
  {85}},\ \bibinfo {pages} {121501} (\bibinfo {year} {2014})}\BibitemShut
  {NoStop}%
\bibitem [{\citenamefont {Salomon}\ \emph {et~al.}(2001)\citenamefont
  {Salomon}, \citenamefont {Dimarcq}, \citenamefont {Abgrall}, \citenamefont
  {Clairon}, \citenamefont {Laurent}, \citenamefont {Lemonde}, \citenamefont
  {Santarelli}, \citenamefont {Uhrich}, \citenamefont {Bernier}, \citenamefont
  {Busca}, \citenamefont {Jornod}, \citenamefont {Thomann}, \citenamefont
  {Samain}, \citenamefont {Wolf}, \citenamefont {Gonzalez}, \citenamefont
  {Guillemot}, \citenamefont {Leon}, \citenamefont {Nouel}, \citenamefont
  {Sirmain},\ and\ \citenamefont {Feltham}}]{SALOMONreview}%
  \BibitemOpen
  \bibfield  {author} {\bibinfo {author} {\bibfnamefont {C.}~\bibnamefont
  {Salomon}}, \bibinfo {author} {\bibfnamefont {N.}~\bibnamefont {Dimarcq}},
  \bibinfo {author} {\bibfnamefont {M.}~\bibnamefont {Abgrall}}, \bibinfo
  {author} {\bibfnamefont {A.}~\bibnamefont {Clairon}}, \bibinfo {author}
  {\bibfnamefont {P.}~\bibnamefont {Laurent}}, \bibinfo {author} {\bibfnamefont
  {P.}~\bibnamefont {Lemonde}}, \bibinfo {author} {\bibfnamefont
  {G.}~\bibnamefont {Santarelli}}, \bibinfo {author} {\bibfnamefont
  {P.}~\bibnamefont {Uhrich}}, \bibinfo {author} {\bibfnamefont
  {L.}~\bibnamefont {Bernier}}, \bibinfo {author} {\bibfnamefont
  {G.}~\bibnamefont {Busca}}, \bibinfo {author} {\bibfnamefont
  {A.}~\bibnamefont {Jornod}}, \bibinfo {author} {\bibfnamefont
  {P.}~\bibnamefont {Thomann}}, \bibinfo {author} {\bibfnamefont
  {E.}~\bibnamefont {Samain}}, \bibinfo {author} {\bibfnamefont
  {P.}~\bibnamefont {Wolf}}, \bibinfo {author} {\bibfnamefont {F.}~\bibnamefont
  {Gonzalez}}, \bibinfo {author} {\bibfnamefont {P.}~\bibnamefont {Guillemot}},
  \bibinfo {author} {\bibfnamefont {S.}~\bibnamefont {Leon}}, \bibinfo {author}
  {\bibfnamefont {F.}~\bibnamefont {Nouel}}, \bibinfo {author} {\bibfnamefont
  {C.}~\bibnamefont {Sirmain}},\ and\ \bibinfo {author} {\bibfnamefont
  {S.}~\bibnamefont {Feltham}},\ }\href
  {https://doi.org/https://doi.org/10.1016/S1296-2147(01)01274-4} {\bibfield
  {journal} {\bibinfo  {journal} {Comptes Rendus de l'Académie des Sciences -
  Series IV - Physics}\ }\textbf {\bibinfo {volume} {2}},\ \bibinfo {pages}
  {1313} (\bibinfo {year} {2001})}\BibitemShut {NoStop}%
\bibitem [{\citenamefont {Bongs}\ \emph {et~al.}(2016)\citenamefont {Bongs},
  \citenamefont {Boyer}, \citenamefont {Cruise}, \citenamefont {Freise},
  \citenamefont {Holynski}, \citenamefont {Hughes}, \citenamefont {Kaushik},
  \citenamefont {Lien}, \citenamefont {Niggebaum}, \citenamefont {Perea-Ortiz},
  \citenamefont {Petrov}, \citenamefont {Plant}, \citenamefont {Singh},
  \citenamefont {Stabrawa}, \citenamefont {Paul}, \citenamefont {Sorel},
  \citenamefont {Cumming}, \citenamefont {Marsh}, \citenamefont {Bowtell},
  \citenamefont {Bason}, \citenamefont {Beardsley}, \citenamefont {Campion},
  \citenamefont {Brookes}, \citenamefont {Fernholz}, \citenamefont {Fromhold},
  \citenamefont {Hackermuller}, \citenamefont {Krüger}, \citenamefont {Li},
  \citenamefont {Maclean}, \citenamefont {Mellor}, \citenamefont {Novikov},
  \citenamefont {Orucevic}, \citenamefont {Rushforth}, \citenamefont {Welch},
  \citenamefont {Benson}, \citenamefont {Wildman}, \citenamefont {Freegarde},
  \citenamefont {Himsworth}, \citenamefont {Ruostekoski}, \citenamefont
  {Smith}, \citenamefont {Tropper}, \citenamefont {Griffin}, \citenamefont
  {Arnold}, \citenamefont {Riis}, \citenamefont {Hastie}, \citenamefont
  {Paboeuf}, \citenamefont {Parrotta}, \citenamefont {Garraway}, \citenamefont
  {Pasquazi}, \citenamefont {Peccianti}, \citenamefont {Hensinger},
  \citenamefont {Potter}, \citenamefont {Nizamani}, \citenamefont {Bostock},
  \citenamefont {Blanco}, \citenamefont {Sinuco-Leon}, \citenamefont {Hill},
  \citenamefont {Williams}, \citenamefont {Gill}, \citenamefont {Hempler},
  \citenamefont {Malcolm}, \citenamefont {Cross}, \citenamefont {Kock},
  \citenamefont {Maddox},\ and\ \citenamefont {John}}]{UKHub}%
  \BibitemOpen
  \bibfield  {author} {\bibinfo {author} {\bibfnamefont {K.}~\bibnamefont
  {Bongs}}, \bibinfo {author} {\bibfnamefont {V.}~\bibnamefont {Boyer}},
  \bibinfo {author} {\bibfnamefont {M.~A.}\ \bibnamefont {Cruise}}, \bibinfo
  {author} {\bibfnamefont {A.}~\bibnamefont {Freise}}, \bibinfo {author}
  {\bibfnamefont {M.}~\bibnamefont {Holynski}}, \bibinfo {author}
  {\bibfnamefont {J.}~\bibnamefont {Hughes}}, \bibinfo {author} {\bibfnamefont
  {A.}~\bibnamefont {Kaushik}}, \bibinfo {author} {\bibfnamefont {Y.-H.}\
  \bibnamefont {Lien}}, \bibinfo {author} {\bibfnamefont {A.}~\bibnamefont
  {Niggebaum}}, \bibinfo {author} {\bibfnamefont {M.}~\bibnamefont
  {Perea-Ortiz}}, \bibinfo {author} {\bibfnamefont {P.}~\bibnamefont {Petrov}},
  \bibinfo {author} {\bibfnamefont {S.}~\bibnamefont {Plant}}, \bibinfo
  {author} {\bibfnamefont {Y.}~\bibnamefont {Singh}}, \bibinfo {author}
  {\bibfnamefont {A.}~\bibnamefont {Stabrawa}}, \bibinfo {author}
  {\bibfnamefont {D.~J.}\ \bibnamefont {Paul}}, \bibinfo {author}
  {\bibfnamefont {M.}~\bibnamefont {Sorel}}, \bibinfo {author} {\bibfnamefont
  {D.~R.~S.}\ \bibnamefont {Cumming}}, \bibinfo {author} {\bibfnamefont
  {J.~H.}\ \bibnamefont {Marsh}}, \bibinfo {author} {\bibfnamefont {R.~W.}\
  \bibnamefont {Bowtell}}, \bibinfo {author} {\bibfnamefont {M.~G.}\
  \bibnamefont {Bason}}, \bibinfo {author} {\bibfnamefont {R.~P.}\ \bibnamefont
  {Beardsley}}, \bibinfo {author} {\bibfnamefont {R.~P.}\ \bibnamefont
  {Campion}}, \bibinfo {author} {\bibfnamefont {M.~J.}\ \bibnamefont
  {Brookes}}, \bibinfo {author} {\bibfnamefont {T.}~\bibnamefont {Fernholz}},
  \bibinfo {author} {\bibfnamefont {T.~M.}\ \bibnamefont {Fromhold}}, \bibinfo
  {author} {\bibfnamefont {L.}~\bibnamefont {Hackermuller}}, \bibinfo {author}
  {\bibfnamefont {P.}~\bibnamefont {Krüger}}, \bibinfo {author} {\bibfnamefont
  {X.}~\bibnamefont {Li}}, \bibinfo {author} {\bibfnamefont {J.~O.}\
  \bibnamefont {Maclean}}, \bibinfo {author} {\bibfnamefont {C.~J.}\
  \bibnamefont {Mellor}}, \bibinfo {author} {\bibfnamefont {S.~V.}\
  \bibnamefont {Novikov}}, \bibinfo {author} {\bibfnamefont {F.}~\bibnamefont
  {Orucevic}}, \bibinfo {author} {\bibfnamefont {A.~W.}\ \bibnamefont
  {Rushforth}}, \bibinfo {author} {\bibfnamefont {N.}~\bibnamefont {Welch}},
  \bibinfo {author} {\bibfnamefont {T.~M.}\ \bibnamefont {Benson}}, \bibinfo
  {author} {\bibfnamefont {R.~D.}\ \bibnamefont {Wildman}}, \bibinfo {author}
  {\bibfnamefont {T.}~\bibnamefont {Freegarde}}, \bibinfo {author}
  {\bibfnamefont {M.}~\bibnamefont {Himsworth}}, \bibinfo {author}
  {\bibfnamefont {J.}~\bibnamefont {Ruostekoski}}, \bibinfo {author}
  {\bibfnamefont {P.}~\bibnamefont {Smith}}, \bibinfo {author} {\bibfnamefont
  {A.}~\bibnamefont {Tropper}}, \bibinfo {author} {\bibfnamefont {P.~F.}\
  \bibnamefont {Griffin}}, \bibinfo {author} {\bibfnamefont {A.~S.}\
  \bibnamefont {Arnold}}, \bibinfo {author} {\bibfnamefont {E.}~\bibnamefont
  {Riis}}, \bibinfo {author} {\bibfnamefont {J.~E.}\ \bibnamefont {Hastie}},
  \bibinfo {author} {\bibfnamefont {D.}~\bibnamefont {Paboeuf}}, \bibinfo
  {author} {\bibfnamefont {D.~C.}\ \bibnamefont {Parrotta}}, \bibinfo {author}
  {\bibfnamefont {B.~M.}\ \bibnamefont {Garraway}}, \bibinfo {author}
  {\bibfnamefont {A.}~\bibnamefont {Pasquazi}}, \bibinfo {author}
  {\bibfnamefont {M.}~\bibnamefont {Peccianti}}, \bibinfo {author}
  {\bibfnamefont {W.}~\bibnamefont {Hensinger}}, \bibinfo {author}
  {\bibfnamefont {E.}~\bibnamefont {Potter}}, \bibinfo {author} {\bibfnamefont
  {A.~H.}\ \bibnamefont {Nizamani}}, \bibinfo {author} {\bibfnamefont
  {H.}~\bibnamefont {Bostock}}, \bibinfo {author} {\bibfnamefont {A.~R.}\
  \bibnamefont {Blanco}}, \bibinfo {author} {\bibfnamefont {G.}~\bibnamefont
  {Sinuco-Leon}}, \bibinfo {author} {\bibfnamefont {I.~R.}\ \bibnamefont
  {Hill}}, \bibinfo {author} {\bibfnamefont {R.~A.}\ \bibnamefont {Williams}},
  \bibinfo {author} {\bibfnamefont {P.}~\bibnamefont {Gill}}, \bibinfo {author}
  {\bibfnamefont {N.}~\bibnamefont {Hempler}}, \bibinfo {author} {\bibfnamefont
  {G.~P.~A.}\ \bibnamefont {Malcolm}}, \bibinfo {author} {\bibfnamefont
  {T.}~\bibnamefont {Cross}}, \bibinfo {author} {\bibfnamefont {B.~O.}\
  \bibnamefont {Kock}}, \bibinfo {author} {\bibfnamefont {S.}~\bibnamefont
  {Maddox}},\ and\ \bibinfo {author} {\bibfnamefont {P.}~\bibnamefont {John}},\
  }in\ \href {https://doi.org/10.1117/12.2232143} {\emph {\bibinfo {booktitle}
  {Quantum Optics}}},\ Vol.\ \bibinfo {volume} {9900},\ \bibinfo {editor}
  {edited by\ \bibinfo {editor} {\bibfnamefont {J.}~\bibnamefont {Stuhler}}\
  and\ \bibinfo {editor} {\bibfnamefont {A.~J.}\ \bibnamefont {Shields}}},\
  \bibinfo {organization} {International Society for Optics and Photonics}\
  (\bibinfo  {publisher} {SPIE},\ \bibinfo {year} {2016})\ pp.\ \bibinfo
  {pages} {66 -- 72}\BibitemShut {NoStop}%
\bibitem [{\citenamefont {M{\'e}noret}\ \emph {et~al.}(2018)\citenamefont
  {M{\'e}noret}, \citenamefont {Vermeulen}, \citenamefont {Le~Moigne},
  \citenamefont {Bonvalot}, \citenamefont {Bouyer}, \citenamefont {Landragin},\
  and\ \citenamefont {Desruelle}}]{muquans}%
  \BibitemOpen
  \bibfield  {author} {\bibinfo {author} {\bibfnamefont {V.}~\bibnamefont
  {M{\'e}noret}}, \bibinfo {author} {\bibfnamefont {P.}~\bibnamefont
  {Vermeulen}}, \bibinfo {author} {\bibfnamefont {N.}~\bibnamefont
  {Le~Moigne}}, \bibinfo {author} {\bibfnamefont {S.}~\bibnamefont {Bonvalot}},
  \bibinfo {author} {\bibfnamefont {P.}~\bibnamefont {Bouyer}}, \bibinfo
  {author} {\bibfnamefont {A.}~\bibnamefont {Landragin}},\ and\ \bibinfo
  {author} {\bibfnamefont {B.}~\bibnamefont {Desruelle}},\ }\href
  {https://doi.org/10.1038/s41598-018-30608-1} {\bibfield  {journal} {\bibinfo
  {journal} {Scientific Reports}\ }\textbf {\bibinfo {volume} {8}},\ \bibinfo
  {pages} {12300} (\bibinfo {year} {2018})}\BibitemShut {NoStop}%
\bibitem [{\citenamefont {Perez}\ \emph {et~al.}(2014)\citenamefont {Perez},
  \citenamefont {Salim}, \citenamefont {Farkas}, \citenamefont {Duggan},
  \citenamefont {Ivory},\ and\ \citenamefont {Anderson}}]{ColdQuanta}%
  \BibitemOpen
  \bibfield  {author} {\bibinfo {author} {\bibfnamefont {M.~A.}\ \bibnamefont
  {Perez}}, \bibinfo {author} {\bibfnamefont {E.}~\bibnamefont {Salim}},
  \bibinfo {author} {\bibfnamefont {D.}~\bibnamefont {Farkas}}, \bibinfo
  {author} {\bibfnamefont {J.}~\bibnamefont {Duggan}}, \bibinfo {author}
  {\bibfnamefont {M.}~\bibnamefont {Ivory}},\ and\ \bibinfo {author}
  {\bibfnamefont {D.}~\bibnamefont {Anderson}},\ }in\ \href
  {https://doi.org/10.1117/12.2064311} {\emph {\bibinfo {booktitle} {Optical
  Trapping and Optical Micromanipulation XI}}},\ Vol.\ \bibinfo {volume}
  {9164},\ \bibinfo {editor} {edited by\ \bibinfo {editor} {\bibfnamefont
  {K.}~\bibnamefont {Dholakia}}\ and\ \bibinfo {editor} {\bibfnamefont {G.~C.}\
  \bibnamefont {Spalding}}},\ \bibinfo {organization} {International Society
  for Optics and Photonics}\ (\bibinfo  {publisher} {SPIE},\ \bibinfo {year}
  {2014})\ pp.\ \bibinfo {pages} {25 -- 29}\BibitemShut {NoStop}%
\bibitem [{\citenamefont {Kitching}(2018)}]{Kitching2018}%
  \BibitemOpen
  \bibfield  {author} {\bibinfo {author} {\bibfnamefont {J.}~\bibnamefont
  {Kitching}},\ }\href {https://doi.org/10.1063/1.5026238} {\bibfield
  {journal} {\bibinfo  {journal} {Applied Physics Reviews}\ }\textbf {\bibinfo
  {volume} {5}},\ \bibinfo {pages} {031302} (\bibinfo {year}
  {2018})}\BibitemShut {NoStop}%
\bibitem [{\citenamefont {Knappe}\ \emph {et~al.}(2004)\citenamefont {Knappe},
  \citenamefont {Shah}, \citenamefont {Schwindt}, \citenamefont {Hollberg},
  \citenamefont {Kitching}, \citenamefont {Liew},\ and\ \citenamefont
  {Moreland}}]{Knappe2004}%
  \BibitemOpen
  \bibfield  {author} {\bibinfo {author} {\bibfnamefont {S.}~\bibnamefont
  {Knappe}}, \bibinfo {author} {\bibfnamefont {V.}~\bibnamefont {Shah}},
  \bibinfo {author} {\bibfnamefont {P.~D.~D.}\ \bibnamefont {Schwindt}},
  \bibinfo {author} {\bibfnamefont {L.}~\bibnamefont {Hollberg}}, \bibinfo
  {author} {\bibfnamefont {J.}~\bibnamefont {Kitching}}, \bibinfo {author}
  {\bibfnamefont {L.-A.}\ \bibnamefont {Liew}},\ and\ \bibinfo {author}
  {\bibfnamefont {J.}~\bibnamefont {Moreland}},\ }\href
  {https://doi.org/10.1063/1.1787942} {\bibfield  {journal} {\bibinfo
  {journal} {Applied Physics Letters}\ }\textbf {\bibinfo {volume} {85}},\
  \bibinfo {pages} {1460} (\bibinfo {year} {2004})}\BibitemShut {NoStop}%
\bibitem [{\citenamefont {Liew}\ \emph {et~al.}(2004)\citenamefont {Liew},
  \citenamefont {Knappe}, \citenamefont {Moreland}, \citenamefont {Robinson},
  \citenamefont {Hollberg},\ and\ \citenamefont {Kitching}}]{moreland}%
  \BibitemOpen
  \bibfield  {author} {\bibinfo {author} {\bibfnamefont {L.-A.}\ \bibnamefont
  {Liew}}, \bibinfo {author} {\bibfnamefont {S.}~\bibnamefont {Knappe}},
  \bibinfo {author} {\bibfnamefont {J.}~\bibnamefont {Moreland}}, \bibinfo
  {author} {\bibfnamefont {H.}~\bibnamefont {Robinson}}, \bibinfo {author}
  {\bibfnamefont {L.}~\bibnamefont {Hollberg}},\ and\ \bibinfo {author}
  {\bibfnamefont {J.}~\bibnamefont {Kitching}},\ }\href
  {https://doi.org/10.1063/1.1691490} {\bibfield  {journal} {\bibinfo
  {journal} {Applied Physics Letters}\ }\textbf {\bibinfo {volume} {84}},\
  \bibinfo {pages} {2694} (\bibinfo {year} {2004})}\BibitemShut {NoStop}%
\bibitem [{\citenamefont {Vicarini}\ \emph {et~al.}(2019)\citenamefont
  {Vicarini}, \citenamefont {Abdel~Hafiz}, \citenamefont {Maurice},
  \citenamefont {Passilly}, \citenamefont {Kroemer}, \citenamefont {Ribetto},
  \citenamefont {Gaff}, \citenamefont {Gorecki}, \citenamefont {Galliou},\ and\
  \citenamefont {Boudot}}]{boudotmemscell}%
  \BibitemOpen
  \bibfield  {author} {\bibinfo {author} {\bibfnamefont {R.}~\bibnamefont
  {Vicarini}}, \bibinfo {author} {\bibfnamefont {M.}~\bibnamefont
  {Abdel~Hafiz}}, \bibinfo {author} {\bibfnamefont {V.}~\bibnamefont
  {Maurice}}, \bibinfo {author} {\bibfnamefont {N.}~\bibnamefont {Passilly}},
  \bibinfo {author} {\bibfnamefont {E.}~\bibnamefont {Kroemer}}, \bibinfo
  {author} {\bibfnamefont {L.}~\bibnamefont {Ribetto}}, \bibinfo {author}
  {\bibfnamefont {V.}~\bibnamefont {Gaff}}, \bibinfo {author} {\bibfnamefont
  {C.}~\bibnamefont {Gorecki}}, \bibinfo {author} {\bibfnamefont
  {S.}~\bibnamefont {Galliou}},\ and\ \bibinfo {author} {\bibfnamefont
  {R.}~\bibnamefont {Boudot}},\ }\href
  {https://doi.org/10.1109/TUFFC.2019.2933051} {\bibfield  {journal} {\bibinfo
  {journal} {IEEE Transactions on Ultrasonics, Ferroelectrics and Frequency
  Control}\ }\textbf {\bibinfo {volume} {66}},\ \bibinfo {pages} {1962 }
  (\bibinfo {year} {2019})}\BibitemShut {NoStop}%
\bibitem [{\citenamefont {Hasegawa}\ \emph {et~al.}(2011)\citenamefont
  {Hasegawa}, \citenamefont {Chutani}, \citenamefont {Gorecki}, \citenamefont
  {Boudot}, \citenamefont {Dziuban}, \citenamefont {Giordano}, \citenamefont
  {Clatot},\ and\ \citenamefont {Mauri}}]{Hasegawa2011_buffer_gas}%
  \BibitemOpen
  \bibfield  {author} {\bibinfo {author} {\bibfnamefont {M.}~\bibnamefont
  {Hasegawa}}, \bibinfo {author} {\bibfnamefont {R.}~\bibnamefont {Chutani}},
  \bibinfo {author} {\bibfnamefont {C.}~\bibnamefont {Gorecki}}, \bibinfo
  {author} {\bibfnamefont {R.}~\bibnamefont {Boudot}}, \bibinfo {author}
  {\bibfnamefont {P.}~\bibnamefont {Dziuban}}, \bibinfo {author} {\bibfnamefont
  {V.}~\bibnamefont {Giordano}}, \bibinfo {author} {\bibfnamefont
  {S.}~\bibnamefont {Clatot}},\ and\ \bibinfo {author} {\bibfnamefont
  {L.}~\bibnamefont {Mauri}},\ }\href
  {https://doi.org/10.1016/j.sna.2011.02.039} {\bibfield  {journal} {\bibinfo
  {journal} {Sensors and Actuators A: Physical}\ }\textbf {\bibinfo {volume}
  {167}},\ \bibinfo {pages} {594} (\bibinfo {year} {2011})}\BibitemShut
  {NoStop}%
\bibitem [{\citenamefont {Hummon}\ \emph {et~al.}(2018)\citenamefont {Hummon},
  \citenamefont {Kang}, \citenamefont {Bopp}, \citenamefont {Li}, \citenamefont
  {Westly}, \citenamefont {Kim}, \citenamefont {Fredrick}, \citenamefont
  {Diddams}, \citenamefont {Srinivasan}, \citenamefont {Aksyuk},\ and\
  \citenamefont {Kitching}}]{Hummon:18}%
  \BibitemOpen
  \bibfield  {author} {\bibinfo {author} {\bibfnamefont {M.~T.}\ \bibnamefont
  {Hummon}}, \bibinfo {author} {\bibfnamefont {S.}~\bibnamefont {Kang}},
  \bibinfo {author} {\bibfnamefont {D.~G.}\ \bibnamefont {Bopp}}, \bibinfo
  {author} {\bibfnamefont {Q.}~\bibnamefont {Li}}, \bibinfo {author}
  {\bibfnamefont {D.~A.}\ \bibnamefont {Westly}}, \bibinfo {author}
  {\bibfnamefont {S.}~\bibnamefont {Kim}}, \bibinfo {author} {\bibfnamefont
  {C.}~\bibnamefont {Fredrick}}, \bibinfo {author} {\bibfnamefont {S.~A.}\
  \bibnamefont {Diddams}}, \bibinfo {author} {\bibfnamefont {K.}~\bibnamefont
  {Srinivasan}}, \bibinfo {author} {\bibfnamefont {V.}~\bibnamefont {Aksyuk}},\
  and\ \bibinfo {author} {\bibfnamefont {J.~E.}\ \bibnamefont {Kitching}},\
  }\href {https://doi.org/10.1364/OPTICA.5.000443} {\bibfield  {journal}
  {\bibinfo  {journal} {Optica}\ }\textbf {\bibinfo {volume} {5}},\ \bibinfo
  {pages} {443} (\bibinfo {year} {2018})}\BibitemShut {NoStop}%
\bibitem [{\citenamefont {Raab}\ \emph {et~al.}(1987)\citenamefont {Raab},
  \citenamefont {Prentiss}, \citenamefont {Cable}, \citenamefont {Chu},\ and\
  \citenamefont {Pritchard}}]{Raab}%
  \BibitemOpen
  \bibfield  {author} {\bibinfo {author} {\bibfnamefont {E.~L.}\ \bibnamefont
  {Raab}}, \bibinfo {author} {\bibfnamefont {M.}~\bibnamefont {Prentiss}},
  \bibinfo {author} {\bibfnamefont {A.}~\bibnamefont {Cable}}, \bibinfo
  {author} {\bibfnamefont {S.}~\bibnamefont {Chu}},\ and\ \bibinfo {author}
  {\bibfnamefont {D.~E.}\ \bibnamefont {Pritchard}},\ }\href
  {https://doi.org/10.1103/PhysRevLett.59.2631} {\bibfield  {journal} {\bibinfo
   {journal} {Phys. Rev. Lett.}\ }\textbf {\bibinfo {volume} {59}},\ \bibinfo
  {pages} {2631} (\bibinfo {year} {1987})}\BibitemShut {NoStop}%
\bibitem [{\citenamefont {Phillips}\ and\ \citenamefont
  {Metcalf}(1982)}]{2Dcooling}%
  \BibitemOpen
  \bibfield  {author} {\bibinfo {author} {\bibfnamefont {W.~D.}\ \bibnamefont
  {Phillips}}\ and\ \bibinfo {author} {\bibfnamefont {H.}~\bibnamefont
  {Metcalf}},\ }\href {https://doi.org/10.1103/PhysRevLett.48.596} {\bibfield
  {journal} {\bibinfo  {journal} {Phys. Rev. Lett.}\ }\textbf {\bibinfo
  {volume} {48}},\ \bibinfo {pages} {596} (\bibinfo {year} {1982})}\BibitemShut
  {NoStop}%
\bibitem [{\citenamefont {Lett}\ \emph {et~al.}(1989)\citenamefont {Lett},
  \citenamefont {Phillips}, \citenamefont {Rolston}, \citenamefont {Tanner},
  \citenamefont {Watts},\ and\ \citenamefont {Westbrook}}]{Lett:89}%
  \BibitemOpen
  \bibfield  {author} {\bibinfo {author} {\bibfnamefont {P.~D.}\ \bibnamefont
  {Lett}}, \bibinfo {author} {\bibfnamefont {W.~D.}\ \bibnamefont {Phillips}},
  \bibinfo {author} {\bibfnamefont {S.~L.}\ \bibnamefont {Rolston}}, \bibinfo
  {author} {\bibfnamefont {C.~E.}\ \bibnamefont {Tanner}}, \bibinfo {author}
  {\bibfnamefont {R.~N.}\ \bibnamefont {Watts}},\ and\ \bibinfo {author}
  {\bibfnamefont {C.~I.}\ \bibnamefont {Westbrook}},\ }\href
  {https://doi.org/10.1364/JOSAB.6.002084} {\bibfield  {journal} {\bibinfo
  {journal} {J. Opt. Soc. Am. B}\ }\textbf {\bibinfo {volume} {6}},\ \bibinfo
  {pages} {2084} (\bibinfo {year} {1989})}\BibitemShut {NoStop}%
\bibitem [{\citenamefont {Ungar}\ \emph {et~al.}(1989)\citenamefont {Ungar},
  \citenamefont {Weiss}, \citenamefont {Riis},\ and\ \citenamefont
  {Chu}}]{Ungar:89}%
  \BibitemOpen
  \bibfield  {author} {\bibinfo {author} {\bibfnamefont {P.~J.}\ \bibnamefont
  {Ungar}}, \bibinfo {author} {\bibfnamefont {D.~S.}\ \bibnamefont {Weiss}},
  \bibinfo {author} {\bibfnamefont {E.}~\bibnamefont {Riis}},\ and\ \bibinfo
  {author} {\bibfnamefont {S.}~\bibnamefont {Chu}},\ }\href
  {https://doi.org/10.1364/JOSAB.6.002058} {\bibfield  {journal} {\bibinfo
  {journal} {J. Opt. Soc. Am. B}\ }\textbf {\bibinfo {volume} {6}},\ \bibinfo
  {pages} {2058} (\bibinfo {year} {1989})}\BibitemShut {NoStop}%
\bibitem [{\citenamefont {Steck}(2001)}]{Steck}%
  \BibitemOpen
  \bibfield  {author} {\bibinfo {author} {\bibfnamefont {D.~A.}\ \bibnamefont
  {Steck}},\ }\href {http://steck.us/alkalidata} {\bibinfo {title} {Rubidium 87
  d line data}},\ \bibinfo {howpublished} {Available Online} (\bibinfo {year}
  {2001})\BibitemShut {NoStop}%
\bibitem [{\citenamefont {Lett}\ \emph {et~al.}(1988)\citenamefont {Lett},
  \citenamefont {Watts}, \citenamefont {Westbrook}, \citenamefont {Phillips},
  \citenamefont {Gould},\ and\ \citenamefont {Metcalf}}]{molasses1}%
  \BibitemOpen
  \bibfield  {author} {\bibinfo {author} {\bibfnamefont {P.~D.}\ \bibnamefont
  {Lett}}, \bibinfo {author} {\bibfnamefont {R.~N.}\ \bibnamefont {Watts}},
  \bibinfo {author} {\bibfnamefont {C.~I.}\ \bibnamefont {Westbrook}}, \bibinfo
  {author} {\bibfnamefont {W.~D.}\ \bibnamefont {Phillips}}, \bibinfo {author}
  {\bibfnamefont {P.~L.}\ \bibnamefont {Gould}},\ and\ \bibinfo {author}
  {\bibfnamefont {H.~J.}\ \bibnamefont {Metcalf}},\ }\href
  {https://doi.org/10.1103/PhysRevLett.61.169} {\bibfield  {journal} {\bibinfo
  {journal} {Phys. Rev. Lett.}\ }\textbf {\bibinfo {volume} {61}},\ \bibinfo
  {pages} {169} (\bibinfo {year} {1988})}\BibitemShut {NoStop}%
\bibitem [{\citenamefont {Castin}\ \emph {et~al.}(1989)\citenamefont {Castin},
  \citenamefont {Wallis},\ and\ \citenamefont {Dalibard}}]{molasses2}%
  \BibitemOpen
  \bibfield  {author} {\bibinfo {author} {\bibfnamefont {Y.}~\bibnamefont
  {Castin}}, \bibinfo {author} {\bibfnamefont {H.}~\bibnamefont {Wallis}},\
  and\ \bibinfo {author} {\bibfnamefont {J.}~\bibnamefont {Dalibard}},\ }\href
  {https://doi.org/10.1364/JOSAB.6.002046} {\bibfield  {journal} {\bibinfo
  {journal} {J. Opt. Soc. Am. B}\ }\textbf {\bibinfo {volume} {6}},\ \bibinfo
  {pages} {2046} (\bibinfo {year} {1989})}\BibitemShut {NoStop}%
\bibitem [{\citenamefont {Weiss}\ \emph {et~al.}(1989)\citenamefont {Weiss},
  \citenamefont {Riis}, \citenamefont {Shevy}, \citenamefont {Ungar},\ and\
  \citenamefont {Chu}}]{Weiss:89}%
  \BibitemOpen
  \bibfield  {author} {\bibinfo {author} {\bibfnamefont {D.~S.}\ \bibnamefont
  {Weiss}}, \bibinfo {author} {\bibfnamefont {E.}~\bibnamefont {Riis}},
  \bibinfo {author} {\bibfnamefont {Y.}~\bibnamefont {Shevy}}, \bibinfo
  {author} {\bibfnamefont {P.~J.}\ \bibnamefont {Ungar}},\ and\ \bibinfo
  {author} {\bibfnamefont {S.}~\bibnamefont {Chu}},\ }\href
  {https://doi.org/10.1364/JOSAB.6.002072} {\bibfield  {journal} {\bibinfo
  {journal} {J. Opt. Soc. Am. B}\ }\textbf {\bibinfo {volume} {6}},\ \bibinfo
  {pages} {2072} (\bibinfo {year} {1989})}\BibitemShut {NoStop}%
\bibitem [{\citenamefont {Monroe}\ \emph {et~al.}(1990)\citenamefont {Monroe},
  \citenamefont {Swann}, \citenamefont {Robinson},\ and\ \citenamefont
  {Wieman}}]{monroe}%
  \BibitemOpen
  \bibfield  {author} {\bibinfo {author} {\bibfnamefont {C.}~\bibnamefont
  {Monroe}}, \bibinfo {author} {\bibfnamefont {W.}~\bibnamefont {Swann}},
  \bibinfo {author} {\bibfnamefont {H.}~\bibnamefont {Robinson}},\ and\
  \bibinfo {author} {\bibfnamefont {C.}~\bibnamefont {Wieman}},\ }\href
  {https://doi.org/10.1103/PhysRevLett.65.1571} {\bibfield  {journal} {\bibinfo
   {journal} {Phys. Rev. Lett.}\ }\textbf {\bibinfo {volume} {65}},\ \bibinfo
  {pages} {1571} (\bibinfo {year} {1990})}\BibitemShut {NoStop}%
\bibitem [{\citenamefont {Wieman}\ \emph {et~al.}(1995)\citenamefont {Wieman},
  \citenamefont {Flowers},\ and\ \citenamefont {Gilbert}}]{wieman}%
  \BibitemOpen
  \bibfield  {author} {\bibinfo {author} {\bibfnamefont {C.}~\bibnamefont
  {Wieman}}, \bibinfo {author} {\bibfnamefont {G.}~\bibnamefont {Flowers}},\
  and\ \bibinfo {author} {\bibfnamefont {S.}~\bibnamefont {Gilbert}},\ }\href
  {https://doi.org/10.1119/1.18072} {\bibfield  {journal} {\bibinfo  {journal}
  {American Journal of Physics}\ }\textbf {\bibinfo {volume} {63}},\ \bibinfo
  {pages} {317} (\bibinfo {year} {1995})}\BibitemShut {NoStop}%
\bibitem [{\citenamefont {McGilligan}\ \emph {et~al.}(2015)\citenamefont
  {McGilligan}, \citenamefont {Griffin}, \citenamefont {Riis},\ and\
  \citenamefont {Arnold}}]{McGilliganphase}%
  \BibitemOpen
  \bibfield  {author} {\bibinfo {author} {\bibfnamefont {J.~P.}\ \bibnamefont
  {McGilligan}}, \bibinfo {author} {\bibfnamefont {P.~F.}\ \bibnamefont
  {Griffin}}, \bibinfo {author} {\bibfnamefont {E.}~\bibnamefont {Riis}},\ and\
  \bibinfo {author} {\bibfnamefont {A.~S.}\ \bibnamefont {Arnold}},\ }\href
  {https://doi.org/10.1364/OE.23.008948} {\bibfield  {journal} {\bibinfo
  {journal} {Opt. Express}\ }\textbf {\bibinfo {volume} {23}},\ \bibinfo
  {pages} {8948} (\bibinfo {year} {2015})}\BibitemShut {NoStop}%
\bibitem [{\citenamefont {Arnold}(2022)}]{aidaninprep}%
  \BibitemOpen
  \bibfield  {author} {\bibinfo {author} {\bibfnamefont {A.~S.}\ \bibnamefont
  {Arnold}},\ }\href@noop {} {\bibfield  {journal} {\bibinfo  {journal} {In
  Preparation}\ } (\bibinfo {year} {2022})}\BibitemShut {NoStop}%
\bibitem [{\citenamefont {Lindquist}\ \emph {et~al.}(1992)\citenamefont
  {Lindquist}, \citenamefont {Stephens},\ and\ \citenamefont
  {Wieman}}]{wieman1992}%
  \BibitemOpen
  \bibfield  {author} {\bibinfo {author} {\bibfnamefont {K.}~\bibnamefont
  {Lindquist}}, \bibinfo {author} {\bibfnamefont {M.}~\bibnamefont
  {Stephens}},\ and\ \bibinfo {author} {\bibfnamefont {C.}~\bibnamefont
  {Wieman}},\ }\href {https://doi.org/10.1103/PhysRevA.46.4082} {\bibfield
  {journal} {\bibinfo  {journal} {Physical Review A}\ }\textbf {\bibinfo
  {volume} {46}},\ \bibinfo {pages} {4082} (\bibinfo {year}
  {1992})}\BibitemShut {NoStop}%
\bibitem [{\citenamefont {Arnold}(1999)}]{aidanthesis}%
  \BibitemOpen
  \bibfield  {author} {\bibinfo {author} {\bibfnamefont {A.~S.}\ \bibnamefont
  {Arnold}},\ }\href@noop {} {\bibfield  {journal} {\bibinfo  {journal} {PhD
  Dissertation - University of Sussex}\ } (\bibinfo {year} {1999})}\BibitemShut
  {NoStop}%
\bibitem [{\citenamefont {Griffin}\ \emph {et~al.}(2005)\citenamefont
  {Griffin}, \citenamefont {Weatherill},\ and\ \citenamefont
  {Adams}}]{grifflaserheating}%
  \BibitemOpen
  \bibfield  {author} {\bibinfo {author} {\bibfnamefont {P.~F.}\ \bibnamefont
  {Griffin}}, \bibinfo {author} {\bibfnamefont {K.~J.}\ \bibnamefont
  {Weatherill}},\ and\ \bibinfo {author} {\bibfnamefont {C.~S.}\ \bibnamefont
  {Adams}},\ }\href {https://doi.org/10.1063/1.2038167} {\bibfield  {journal}
  {\bibinfo  {journal} {Review of Scientific Instruments}\ }\textbf {\bibinfo
  {volume} {76}},\ \bibinfo {pages} {093102} (\bibinfo {year}
  {2005})}\BibitemShut {NoStop}%
\bibitem [{\citenamefont {Roach}\ and\ \citenamefont
  {Henclewood}(2004)}]{amdpaper}%
  \BibitemOpen
  \bibfield  {author} {\bibinfo {author} {\bibfnamefont {T.~M.}\ \bibnamefont
  {Roach}}\ and\ \bibinfo {author} {\bibfnamefont {D.}~\bibnamefont
  {Henclewood}},\ }\href {https://doi.org/10.1116/1.1806440} {\bibfield
  {journal} {\bibinfo  {journal} {Journal of Vacuum Science and Technology A}\
  }\textbf {\bibinfo {volume} {22}},\ \bibinfo {pages} {2384} (\bibinfo {year}
  {2004})}\BibitemShut {NoStop}%
\bibitem [{\citenamefont {Fortagh}\ \emph {et~al.}(1998)\citenamefont
  {Fortagh}, \citenamefont {Grossmann}, \citenamefont {Hänsch},\ and\
  \citenamefont {Zimmermann}}]{fastAMD}%
  \BibitemOpen
  \bibfield  {author} {\bibinfo {author} {\bibfnamefont {J.}~\bibnamefont
  {Fortagh}}, \bibinfo {author} {\bibfnamefont {A.}~\bibnamefont {Grossmann}},
  \bibinfo {author} {\bibfnamefont {T.~W.}\ \bibnamefont {Hänsch}},\ and\
  \bibinfo {author} {\bibfnamefont {C.}~\bibnamefont {Zimmermann}},\ }\href
  {https://doi.org/10.1063/1.369018} {\bibfield  {journal} {\bibinfo  {journal}
  {Journal of Applied Physics}\ }\textbf {\bibinfo {volume} {84}},\ \bibinfo
  {pages} {6499} (\bibinfo {year} {1998})}\BibitemShut {NoStop}%
\bibitem [{\citenamefont {Hoth}\ \emph {et~al.}(2013)\citenamefont {Hoth},
  \citenamefont {Donley},\ and\ \citenamefont {Kitching}}]{Hoth13}%
  \BibitemOpen
  \bibfield  {author} {\bibinfo {author} {\bibfnamefont {G.~W.}\ \bibnamefont
  {Hoth}}, \bibinfo {author} {\bibfnamefont {E.~A.}\ \bibnamefont {Donley}},\
  and\ \bibinfo {author} {\bibfnamefont {J.}~\bibnamefont {Kitching}},\ }\href
  {https://doi.org/10.1364/ol.38.000661} {\bibfield  {journal} {\bibinfo
  {journal} {Optics Letters}\ }\textbf {\bibinfo {volume} {38}},\ \bibinfo
  {pages} {661} (\bibinfo {year} {2013})}\BibitemShut {NoStop}%
\bibitem [{\citenamefont {McGilligan}\ \emph {et~al.}(2017)\citenamefont
  {McGilligan}, \citenamefont {Griffin}, \citenamefont {Elvin}, \citenamefont
  {Ingleby}, \citenamefont {Riis},\ and\ \citenamefont
  {Arnold}}]{McGilligan2017}%
  \BibitemOpen
  \bibfield  {author} {\bibinfo {author} {\bibfnamefont {J.~P.}\ \bibnamefont
  {McGilligan}}, \bibinfo {author} {\bibfnamefont {P.~F.}\ \bibnamefont
  {Griffin}}, \bibinfo {author} {\bibfnamefont {R.}~\bibnamefont {Elvin}},
  \bibinfo {author} {\bibfnamefont {S.~J.}\ \bibnamefont {Ingleby}}, \bibinfo
  {author} {\bibfnamefont {E.}~\bibnamefont {Riis}},\ and\ \bibinfo {author}
  {\bibfnamefont {A.~S.}\ \bibnamefont {Arnold}},\ }\href
  {https://doi.org/10.1038/s41598-017-00254-0} {\bibfield  {journal} {\bibinfo
  {journal} {Scientific Reports}\ }\textbf {\bibinfo {volume} {7}},\ \bibinfo
  {pages} {384} (\bibinfo {year} {2017})}\BibitemShut {NoStop}%
\bibitem [{\citenamefont {Ren}\ \emph {et~al.}(2020)\citenamefont {Ren},
  \citenamefont {Li}, \citenamefont {Qu}, \citenamefont {Wang}, \citenamefont
  {Li}, \citenamefont {Lü}, \citenamefont {Chen},\ and\ \citenamefont
  {Liu}}]{pressurecriteria}%
  \BibitemOpen
  \bibfield  {author} {\bibinfo {author} {\bibfnamefont {W.}~\bibnamefont
  {Ren}}, \bibinfo {author} {\bibfnamefont {T.}~\bibnamefont {Li}}, \bibinfo
  {author} {\bibfnamefont {Q.}~\bibnamefont {Qu}}, \bibinfo {author}
  {\bibfnamefont {B.}~\bibnamefont {Wang}}, \bibinfo {author} {\bibfnamefont
  {L.}~\bibnamefont {Li}}, \bibinfo {author} {\bibfnamefont {D.}~\bibnamefont
  {Lü}}, \bibinfo {author} {\bibfnamefont {W.}~\bibnamefont {Chen}},\ and\
  \bibinfo {author} {\bibfnamefont {L.}~\bibnamefont {Liu}},\ }\href
  {https://doi.org/10.1093/nsr/nwaa215} {\bibfield  {journal} {\bibinfo
  {journal} {National Science Review}\ }\textbf {\bibinfo {volume} {7}},\
  \bibinfo {pages} {1828} (\bibinfo {year} {2020})}\BibitemShut {NoStop}%
\bibitem [{\citenamefont {Flemming}\ \emph {et~al.}(1997)\citenamefont
  {Flemming}, \citenamefont {Tuboy}, \citenamefont {Milori}, \citenamefont
  {Marcassa}, \citenamefont {Zilio},\ and\ \citenamefont
  {Bagnato}}]{FLEMMING1997269}%
  \BibitemOpen
  \bibfield  {author} {\bibinfo {author} {\bibfnamefont {J.}~\bibnamefont
  {Flemming}}, \bibinfo {author} {\bibfnamefont {A.}~\bibnamefont {Tuboy}},
  \bibinfo {author} {\bibfnamefont {D.}~\bibnamefont {Milori}}, \bibinfo
  {author} {\bibfnamefont {L.}~\bibnamefont {Marcassa}}, \bibinfo {author}
  {\bibfnamefont {S.}~\bibnamefont {Zilio}},\ and\ \bibinfo {author}
  {\bibfnamefont {V.}~\bibnamefont {Bagnato}},\ }\href
  {https://doi.org/https://doi.org/10.1016/S0030-4018(96)00660-8} {\bibfield
  {journal} {\bibinfo  {journal} {Optics Communications}\ }\textbf {\bibinfo
  {volume} {135}},\ \bibinfo {pages} {269} (\bibinfo {year}
  {1997})}\BibitemShut {NoStop}%
\bibitem [{\citenamefont {Bongs}\ \emph {et~al.}(2019)\citenamefont {Bongs},
  \citenamefont {Holynski}, \citenamefont {Vovrosh}, \citenamefont {Bouyer},
  \citenamefont {Condon}, \citenamefont {Rasel}, \citenamefont {Schubert},
  \citenamefont {Schleich},\ and\ \citenamefont {Roura}}]{Bongs2019}%
  \BibitemOpen
  \bibfield  {author} {\bibinfo {author} {\bibfnamefont {K.}~\bibnamefont
  {Bongs}}, \bibinfo {author} {\bibfnamefont {M.}~\bibnamefont {Holynski}},
  \bibinfo {author} {\bibfnamefont {J.}~\bibnamefont {Vovrosh}}, \bibinfo
  {author} {\bibfnamefont {P.}~\bibnamefont {Bouyer}}, \bibinfo {author}
  {\bibfnamefont {G.}~\bibnamefont {Condon}}, \bibinfo {author} {\bibfnamefont
  {E.}~\bibnamefont {Rasel}}, \bibinfo {author} {\bibfnamefont
  {C.}~\bibnamefont {Schubert}}, \bibinfo {author} {\bibfnamefont {W.~P.}\
  \bibnamefont {Schleich}},\ and\ \bibinfo {author} {\bibfnamefont
  {A.}~\bibnamefont {Roura}},\ }\href
  {https://doi.org/10.1038/s42254-019-0117-4} {\bibfield  {journal} {\bibinfo
  {journal} {Nature Reviews Physics}\ }\textbf {\bibinfo {volume} {1}},\
  \bibinfo {pages} {731} (\bibinfo {year} {2019})}\BibitemShut {NoStop}%
\bibitem [{\citenamefont {Lee}\ \emph {et~al.}(2021{\natexlab{a}})\citenamefont
  {Lee}, \citenamefont {Lee}, \citenamefont {Park}, \citenamefont {Hong},
  \citenamefont {Heo}, \citenamefont {Seo}, \citenamefont {Jeong},
  \citenamefont {Kwon},\ and\ \citenamefont {Moon}}]{LEE2021106698}%
  \BibitemOpen
  \bibfield  {author} {\bibinfo {author} {\bibfnamefont {S.}~\bibnamefont
  {Lee}}, \bibinfo {author} {\bibfnamefont {S.-B.}\ \bibnamefont {Lee}},
  \bibinfo {author} {\bibfnamefont {S.~E.}\ \bibnamefont {Park}}, \bibinfo
  {author} {\bibfnamefont {H.-G.}\ \bibnamefont {Hong}}, \bibinfo {author}
  {\bibfnamefont {M.-S.}\ \bibnamefont {Heo}}, \bibinfo {author} {\bibfnamefont
  {S.}~\bibnamefont {Seo}}, \bibinfo {author} {\bibfnamefont {J.}~\bibnamefont
  {Jeong}}, \bibinfo {author} {\bibfnamefont {T.~Y.}\ \bibnamefont {Kwon}},\
  and\ \bibinfo {author} {\bibfnamefont {G.}~\bibnamefont {Moon}},\ }\href
  {https://doi.org/https://doi.org/10.1016/j.optlaseng.2021.106698} {\bibfield
  {journal} {\bibinfo  {journal} {Optics and Lasers in Engineering}\ }\textbf
  {\bibinfo {volume} {146}},\ \bibinfo {pages} {106698} (\bibinfo {year}
  {2021}{\natexlab{a}})}\BibitemShut {NoStop}%
\bibitem [{\citenamefont {Earl}\ \emph {et~al.}(2022)\citenamefont {Earl},
  \citenamefont {Vovrosh}, \citenamefont {Wright}, \citenamefont {Roberts},
  \citenamefont {Winch}, \citenamefont {Perea-Ortiz}, \citenamefont {Lamb},
  \citenamefont {Hayati}, \citenamefont {Griffin}, \citenamefont {Metje},
  \citenamefont {Bongs},\ and\ \citenamefont {Holynski}}]{atoms10010032}%
  \BibitemOpen
  \bibfield  {author} {\bibinfo {author} {\bibfnamefont {L.}~\bibnamefont
  {Earl}}, \bibinfo {author} {\bibfnamefont {J.}~\bibnamefont {Vovrosh}},
  \bibinfo {author} {\bibfnamefont {M.}~\bibnamefont {Wright}}, \bibinfo
  {author} {\bibfnamefont {D.}~\bibnamefont {Roberts}}, \bibinfo {author}
  {\bibfnamefont {J.}~\bibnamefont {Winch}}, \bibinfo {author} {\bibfnamefont
  {M.}~\bibnamefont {Perea-Ortiz}}, \bibinfo {author} {\bibfnamefont
  {A.}~\bibnamefont {Lamb}}, \bibinfo {author} {\bibfnamefont {F.}~\bibnamefont
  {Hayati}}, \bibinfo {author} {\bibfnamefont {P.}~\bibnamefont {Griffin}},
  \bibinfo {author} {\bibfnamefont {N.}~\bibnamefont {Metje}}, \bibinfo
  {author} {\bibfnamefont {K.}~\bibnamefont {Bongs}},\ and\ \bibinfo {author}
  {\bibfnamefont {M.}~\bibnamefont {Holynski}},\ }\bibfield  {journal}
  {\bibinfo  {journal} {Atoms}\ }\textbf {\bibinfo {volume} {10}},\ \href
  {https://doi.org/10.3390/atoms10010032} {10.3390/atoms10010032} (\bibinfo
  {year} {2022})\BibitemShut {NoStop}%
\bibitem [{\citenamefont {Strangfeld}\ \emph {et~al.}(2022)\citenamefont
  {Strangfeld}, \citenamefont {Wiegand}, \citenamefont {Kluge}, \citenamefont
  {Schoch},\ and\ \citenamefont {Krutzik}}]{Strangfeld:22}%
  \BibitemOpen
  \bibfield  {author} {\bibinfo {author} {\bibfnamefont {A.}~\bibnamefont
  {Strangfeld}}, \bibinfo {author} {\bibfnamefont {B.}~\bibnamefont {Wiegand}},
  \bibinfo {author} {\bibfnamefont {J.}~\bibnamefont {Kluge}}, \bibinfo
  {author} {\bibfnamefont {M.}~\bibnamefont {Schoch}},\ and\ \bibinfo {author}
  {\bibfnamefont {M.}~\bibnamefont {Krutzik}},\ }\href
  {https://doi.org/10.1364/OE.453942} {\bibfield  {journal} {\bibinfo
  {journal} {Opt. Express}\ }\textbf {\bibinfo {volume} {30}},\ \bibinfo
  {pages} {12039} (\bibinfo {year} {2022})}\BibitemShut {NoStop}%
\bibitem [{\citenamefont {Elliott}\ \emph {et~al.}(2018)\citenamefont
  {Elliott}, \citenamefont {Krutzik}, \citenamefont {Williams}, \citenamefont
  {Thompson},\ and\ \citenamefont {Aveline}}]{Elliott2018}%
  \BibitemOpen
  \bibfield  {author} {\bibinfo {author} {\bibfnamefont {E.~R.}\ \bibnamefont
  {Elliott}}, \bibinfo {author} {\bibfnamefont {M.~C.}\ \bibnamefont
  {Krutzik}}, \bibinfo {author} {\bibfnamefont {J.~R.}\ \bibnamefont
  {Williams}}, \bibinfo {author} {\bibfnamefont {R.~J.}\ \bibnamefont
  {Thompson}},\ and\ \bibinfo {author} {\bibfnamefont {D.~C.}\ \bibnamefont
  {Aveline}},\ }\href {https://doi.org/10.1038/s41526-018-0049-9} {\bibfield
  {journal} {\bibinfo  {journal} {npj Microgravity}\ }\textbf {\bibinfo
  {volume} {4}},\ \bibinfo {pages} {16} (\bibinfo {year} {2018})}\BibitemShut
  {NoStop}%
\bibitem [{\citenamefont {Frye}\ \emph {et~al.}(2021)\citenamefont {Frye},
  \citenamefont {Abend}, \citenamefont {Bartosch}, \citenamefont {Bawamia},
  \citenamefont {Becker}, \citenamefont {Blume}, \citenamefont {Braxmaier},
  \citenamefont {Chiow}, \citenamefont {Efremov}, \citenamefont {Ertmer},
  \citenamefont {Fierlinger}, \citenamefont {Franz}, \citenamefont {Gaaloul},
  \citenamefont {Grosse}, \citenamefont {Grzeschik}, \citenamefont {Hellmig},
  \citenamefont {Henderson}, \citenamefont {Herr}, \citenamefont {Israelsson},
  \citenamefont {Kohel}, \citenamefont {Krutzik}, \citenamefont {K{\"u}rbis},
  \citenamefont {L{\"a}mmerzahl}, \citenamefont {List}, \citenamefont
  {L{\"u}dtke}, \citenamefont {Lundblad}, \citenamefont {Marburger},
  \citenamefont {Meister}, \citenamefont {Mihm}, \citenamefont {M{\"u}ller},
  \citenamefont {M{\"u}ntinga}, \citenamefont {Nepal}, \citenamefont
  {Oberschulte}, \citenamefont {Papakonstantinou}, \citenamefont {Jaka},
  \citenamefont {Peters}, \citenamefont {Prat}, \citenamefont {Rasel},
  \citenamefont {Roura}, \citenamefont {Sbroscia}, \citenamefont {Schleich},
  \citenamefont {Schubert}, \citenamefont {Seidel}, \citenamefont {Sommer},
  \citenamefont {Spindeldreier}, \citenamefont {Stamper-Kurn}, \citenamefont
  {Stuhl}, \citenamefont {Warner}, \citenamefont {Wendrich}, \citenamefont
  {Wenzlawski}, \citenamefont {Wicht}, \citenamefont {Windpassinger},
  \citenamefont {Yu},\ and\ \citenamefont {W\"{o}ner}}]{Frye2021}%
  \BibitemOpen
  \bibfield  {author} {\bibinfo {author} {\bibfnamefont {K.}~\bibnamefont
  {Frye}}, \bibinfo {author} {\bibfnamefont {S.}~\bibnamefont {Abend}},
  \bibinfo {author} {\bibfnamefont {W.}~\bibnamefont {Bartosch}}, \bibinfo
  {author} {\bibfnamefont {A.}~\bibnamefont {Bawamia}}, \bibinfo {author}
  {\bibfnamefont {D.}~\bibnamefont {Becker}}, \bibinfo {author} {\bibfnamefont
  {H.}~\bibnamefont {Blume}}, \bibinfo {author} {\bibfnamefont
  {C.}~\bibnamefont {Braxmaier}}, \bibinfo {author} {\bibfnamefont {S.-W.}\
  \bibnamefont {Chiow}}, \bibinfo {author} {\bibfnamefont {M.~A.}\ \bibnamefont
  {Efremov}}, \bibinfo {author} {\bibfnamefont {W.}~\bibnamefont {Ertmer}},
  \bibinfo {author} {\bibfnamefont {P.}~\bibnamefont {Fierlinger}}, \bibinfo
  {author} {\bibfnamefont {T.}~\bibnamefont {Franz}}, \bibinfo {author}
  {\bibfnamefont {N.}~\bibnamefont {Gaaloul}}, \bibinfo {author} {\bibfnamefont
  {J.}~\bibnamefont {Grosse}}, \bibinfo {author} {\bibfnamefont
  {C.}~\bibnamefont {Grzeschik}}, \bibinfo {author} {\bibfnamefont
  {O.}~\bibnamefont {Hellmig}}, \bibinfo {author} {\bibfnamefont {V.~A.}\
  \bibnamefont {Henderson}}, \bibinfo {author} {\bibfnamefont {W.}~\bibnamefont
  {Herr}}, \bibinfo {author} {\bibfnamefont {U.}~\bibnamefont {Israelsson}},
  \bibinfo {author} {\bibfnamefont {J.}~\bibnamefont {Kohel}}, \bibinfo
  {author} {\bibfnamefont {M.}~\bibnamefont {Krutzik}}, \bibinfo {author}
  {\bibfnamefont {C.}~\bibnamefont {K{\"u}rbis}}, \bibinfo {author}
  {\bibfnamefont {C.}~\bibnamefont {L{\"a}mmerzahl}}, \bibinfo {author}
  {\bibfnamefont {M.}~\bibnamefont {List}}, \bibinfo {author} {\bibfnamefont
  {D.}~\bibnamefont {L{\"u}dtke}}, \bibinfo {author} {\bibfnamefont
  {N.}~\bibnamefont {Lundblad}}, \bibinfo {author} {\bibfnamefont {J.~P.}\
  \bibnamefont {Marburger}}, \bibinfo {author} {\bibfnamefont {M.}~\bibnamefont
  {Meister}}, \bibinfo {author} {\bibfnamefont {M.}~\bibnamefont {Mihm}},
  \bibinfo {author} {\bibfnamefont {H.}~\bibnamefont {M{\"u}ller}}, \bibinfo
  {author} {\bibfnamefont {H.}~\bibnamefont {M{\"u}ntinga}}, \bibinfo {author}
  {\bibfnamefont {A.~M.}\ \bibnamefont {Nepal}}, \bibinfo {author}
  {\bibfnamefont {T.}~\bibnamefont {Oberschulte}}, \bibinfo {author}
  {\bibfnamefont {A.}~\bibnamefont {Papakonstantinou}}, \bibinfo {author}
  {\bibfnamefont {P.}~\bibnamefont {Jaka}}, \bibinfo {author} {\bibfnamefont
  {A.}~\bibnamefont {Peters}}, \bibinfo {author} {\bibfnamefont
  {A.}~\bibnamefont {Prat}}, \bibinfo {author} {\bibfnamefont {E.~M.}\
  \bibnamefont {Rasel}}, \bibinfo {author} {\bibfnamefont {A.}~\bibnamefont
  {Roura}}, \bibinfo {author} {\bibfnamefont {M.}~\bibnamefont {Sbroscia}},
  \bibinfo {author} {\bibfnamefont {W.~P.}\ \bibnamefont {Schleich}}, \bibinfo
  {author} {\bibfnamefont {C.}~\bibnamefont {Schubert}}, \bibinfo {author}
  {\bibfnamefont {S.~T.}\ \bibnamefont {Seidel}}, \bibinfo {author}
  {\bibfnamefont {J.}~\bibnamefont {Sommer}}, \bibinfo {author} {\bibfnamefont
  {C.}~\bibnamefont {Spindeldreier}}, \bibinfo {author} {\bibfnamefont
  {D.}~\bibnamefont {Stamper-Kurn}}, \bibinfo {author} {\bibfnamefont {B.~K.}\
  \bibnamefont {Stuhl}}, \bibinfo {author} {\bibfnamefont {M.}~\bibnamefont
  {Warner}}, \bibinfo {author} {\bibfnamefont {T.}~\bibnamefont {Wendrich}},
  \bibinfo {author} {\bibfnamefont {A.}~\bibnamefont {Wenzlawski}}, \bibinfo
  {author} {\bibfnamefont {A.}~\bibnamefont {Wicht}}, \bibinfo {author}
  {\bibfnamefont {P.}~\bibnamefont {Windpassinger}}, \bibinfo {author}
  {\bibfnamefont {N.}~\bibnamefont {Yu}},\ and\ \bibinfo {author}
  {\bibfnamefont {L.}~\bibnamefont {W\"{o}ner}},\ }\href
  {https://doi.org/10.1140/epjqt/s40507-020-00090-8} {\bibfield  {journal}
  {\bibinfo  {journal} {EPJ Quantum Technology}\ }\textbf {\bibinfo {volume}
  {8}},\ \bibinfo {pages} {1} (\bibinfo {year} {2021})}\BibitemShut {NoStop}%
\bibitem [{\citenamefont {Dinkelaker}\ \emph {et~al.}(2017)\citenamefont
  {Dinkelaker}, \citenamefont {Schiemangk}, \citenamefont {Schkolnik},
  \citenamefont {Kenyon}, \citenamefont {Lampmann}, \citenamefont {Wenzlawski},
  \citenamefont {Windpassinger}, \citenamefont {Hellmig}, \citenamefont
  {Wendrich}, \citenamefont {Rasel}, \citenamefont {Giunta}, \citenamefont
  {Deutsch}, \citenamefont {K\"{u}rbis}, \citenamefont {Smol}, \citenamefont
  {Wicht}, \citenamefont {Krutzik},\ and\ \citenamefont
  {Peters}}]{Dinkelaker:17}%
  \BibitemOpen
  \bibfield  {author} {\bibinfo {author} {\bibfnamefont {A.~N.}\ \bibnamefont
  {Dinkelaker}}, \bibinfo {author} {\bibfnamefont {M.}~\bibnamefont
  {Schiemangk}}, \bibinfo {author} {\bibfnamefont {V.}~\bibnamefont
  {Schkolnik}}, \bibinfo {author} {\bibfnamefont {A.}~\bibnamefont {Kenyon}},
  \bibinfo {author} {\bibfnamefont {K.}~\bibnamefont {Lampmann}}, \bibinfo
  {author} {\bibfnamefont {A.}~\bibnamefont {Wenzlawski}}, \bibinfo {author}
  {\bibfnamefont {P.}~\bibnamefont {Windpassinger}}, \bibinfo {author}
  {\bibfnamefont {O.}~\bibnamefont {Hellmig}}, \bibinfo {author} {\bibfnamefont
  {T.}~\bibnamefont {Wendrich}}, \bibinfo {author} {\bibfnamefont {E.~M.}\
  \bibnamefont {Rasel}}, \bibinfo {author} {\bibfnamefont {M.}~\bibnamefont
  {Giunta}}, \bibinfo {author} {\bibfnamefont {C.}~\bibnamefont {Deutsch}},
  \bibinfo {author} {\bibfnamefont {C.}~\bibnamefont {K\"{u}rbis}}, \bibinfo
  {author} {\bibfnamefont {R.}~\bibnamefont {Smol}}, \bibinfo {author}
  {\bibfnamefont {A.}~\bibnamefont {Wicht}}, \bibinfo {author} {\bibfnamefont
  {M.}~\bibnamefont {Krutzik}},\ and\ \bibinfo {author} {\bibfnamefont
  {A.}~\bibnamefont {Peters}},\ }\href {https://doi.org/10.1364/AO.56.001388}
  {\bibfield  {journal} {\bibinfo  {journal} {Appl. Opt.}\ }\textbf {\bibinfo
  {volume} {56}},\ \bibinfo {pages} {1388} (\bibinfo {year}
  {2017})}\BibitemShut {NoStop}%
\bibitem [{\citenamefont {Luvsandamdin}\ \emph {et~al.}(2014)\citenamefont
  {Luvsandamdin}, \citenamefont {K\"{u}rbis}, \citenamefont {Schiemangk},
  \citenamefont {Sahm}, \citenamefont {Wicht}, \citenamefont {Peters},
  \citenamefont {Erbert},\ and\ \citenamefont {Tr\"{a}nkle}}]{Luvsandamdin:14}%
  \BibitemOpen
  \bibfield  {author} {\bibinfo {author} {\bibfnamefont {E.}~\bibnamefont
  {Luvsandamdin}}, \bibinfo {author} {\bibfnamefont {C.}~\bibnamefont
  {K\"{u}rbis}}, \bibinfo {author} {\bibfnamefont {M.}~\bibnamefont
  {Schiemangk}}, \bibinfo {author} {\bibfnamefont {A.}~\bibnamefont {Sahm}},
  \bibinfo {author} {\bibfnamefont {A.}~\bibnamefont {Wicht}}, \bibinfo
  {author} {\bibfnamefont {A.}~\bibnamefont {Peters}}, \bibinfo {author}
  {\bibfnamefont {G.}~\bibnamefont {Erbert}},\ and\ \bibinfo {author}
  {\bibfnamefont {G.}~\bibnamefont {Tr\"{a}nkle}},\ }\href
  {https://doi.org/10.1364/OE.22.007790} {\bibfield  {journal} {\bibinfo
  {journal} {Opt. Express}\ }\textbf {\bibinfo {volume} {22}},\ \bibinfo
  {pages} {7790} (\bibinfo {year} {2014})}\BibitemShut {NoStop}%
\bibitem [{\citenamefont {Ashkin}(1978)}]{ashkin1}%
  \BibitemOpen
  \bibfield  {author} {\bibinfo {author} {\bibfnamefont {A.}~\bibnamefont
  {Ashkin}},\ }\href {https://doi.org/10.1103/PhysRevLett.40.729} {\bibfield
  {journal} {\bibinfo  {journal} {Phys. Rev. Lett.}\ }\textbf {\bibinfo
  {volume} {40}},\ \bibinfo {pages} {729} (\bibinfo {year} {1978})}\BibitemShut
  {NoStop}%
\bibitem [{\citenamefont {Katori}(2011)}]{katori}%
  \BibitemOpen
  \bibfield  {author} {\bibinfo {author} {\bibfnamefont {H.}~\bibnamefont
  {Katori}},\ }\href {https://doi.org/10.1038/nphoton.2011.45} {\bibfield
  {journal} {\bibinfo  {journal} {Nature Photonics}\ }\textbf {\bibinfo
  {volume} {5}},\ \bibinfo {pages} {203} (\bibinfo {year} {2011})}\BibitemShut
  {NoStop}%
\bibitem [{\citenamefont {Kippenberg}\ \emph {et~al.}(2004)\citenamefont
  {Kippenberg}, \citenamefont {Spillane},\ and\ \citenamefont
  {Vahala}}]{kippenberg}%
  \BibitemOpen
  \bibfield  {author} {\bibinfo {author} {\bibfnamefont {T.~J.}\ \bibnamefont
  {Kippenberg}}, \bibinfo {author} {\bibfnamefont {S.~M.}\ \bibnamefont
  {Spillane}},\ and\ \bibinfo {author} {\bibfnamefont {K.~J.}\ \bibnamefont
  {Vahala}},\ }\href {https://doi.org/10.1103/PhysRevLett.93.083904} {\bibfield
   {journal} {\bibinfo  {journal} {Phys. Rev. Lett.}\ }\textbf {\bibinfo
  {volume} {93}},\ \bibinfo {pages} {083904} (\bibinfo {year}
  {2004})}\BibitemShut {NoStop}%
\bibitem [{\citenamefont {Heavner}\ \emph {et~al.}(2005)\citenamefont
  {Heavner}, \citenamefont {Jefferts}, \citenamefont {Donley}, \citenamefont
  {Shirley},\ and\ \citenamefont {Parker}}]{Heavner_2005}%
  \BibitemOpen
  \bibfield  {author} {\bibinfo {author} {\bibfnamefont {T.~P.}\ \bibnamefont
  {Heavner}}, \bibinfo {author} {\bibfnamefont {S.~R.}\ \bibnamefont
  {Jefferts}}, \bibinfo {author} {\bibfnamefont {E.~A.}\ \bibnamefont
  {Donley}}, \bibinfo {author} {\bibfnamefont {J.~H.}\ \bibnamefont
  {Shirley}},\ and\ \bibinfo {author} {\bibfnamefont {T.~E.}\ \bibnamefont
  {Parker}},\ }\href {https://doi.org/10.1088/0026-1394/42/5/012} {\bibfield
  {journal} {\bibinfo  {journal} {Metrologia}\ }\textbf {\bibinfo {volume}
  {42}},\ \bibinfo {pages} {411} (\bibinfo {year} {2005})}\BibitemShut
  {NoStop}%
\bibitem [{\citenamefont {MacAdam}\ \emph {et~al.}(1992)\citenamefont
  {MacAdam}, \citenamefont {Steinbach},\ and\ \citenamefont
  {Wieman}}]{diodelaserwieman}%
  \BibitemOpen
  \bibfield  {author} {\bibinfo {author} {\bibfnamefont {K.~B.}\ \bibnamefont
  {MacAdam}}, \bibinfo {author} {\bibfnamefont {A.}~\bibnamefont {Steinbach}},\
  and\ \bibinfo {author} {\bibfnamefont {C.}~\bibnamefont {Wieman}},\ }\href
  {https://doi.org/10.1119/1.16955} {\bibfield  {journal} {\bibinfo  {journal}
  {American Journal of Physics}\ }\textbf {\bibinfo {volume} {60}},\ \bibinfo
  {pages} {1098} (\bibinfo {year} {1992})}\BibitemShut {NoStop}%
\bibitem [{\citenamefont {Wieman}\ and\ \citenamefont
  {Hollberg}(1991)}]{diodelaserhollberg}%
  \BibitemOpen
  \bibfield  {author} {\bibinfo {author} {\bibfnamefont {C.~E.}\ \bibnamefont
  {Wieman}}\ and\ \bibinfo {author} {\bibfnamefont {L.}~\bibnamefont
  {Hollberg}},\ }\href {https://doi.org/10.1063/1.1142305} {\bibfield
  {journal} {\bibinfo  {journal} {Review of Scientific Instruments}\ }\textbf
  {\bibinfo {volume} {62}},\ \bibinfo {pages} {1} (\bibinfo {year}
  {1991})}\BibitemShut {NoStop}%
\bibitem [{\citenamefont {Fleming}\ and\ \citenamefont
  {Mooradian}(1981)}]{ecdlmooradian}%
  \BibitemOpen
  \bibfield  {author} {\bibinfo {author} {\bibfnamefont {M.}~\bibnamefont
  {Fleming}}\ and\ \bibinfo {author} {\bibfnamefont {A.}~\bibnamefont
  {Mooradian}},\ }\href {https://doi.org/10.1109/JQE.1981.1070634} {\bibfield
  {journal} {\bibinfo  {journal} {IEEE Journal of Quantum Electronics}\
  }\textbf {\bibinfo {volume} {17}},\ \bibinfo {pages} {44} (\bibinfo {year}
  {1981})}\BibitemShut {NoStop}%
\bibitem [{\citenamefont {Ricci}\ \emph {et~al.}(1995)\citenamefont {Ricci},
  \citenamefont {Weidem\"{u}ller}, \citenamefont {Esslinger}, \citenamefont
  {Hemmerich}, \citenamefont {Zimmermann}, \citenamefont {Vuletic},
  \citenamefont {K\"{o}nig},\ and\ \citenamefont {H\"{a}nsch}}]{Ricci1995}%
  \BibitemOpen
  \bibfield  {author} {\bibinfo {author} {\bibfnamefont {L.}~\bibnamefont
  {Ricci}}, \bibinfo {author} {\bibfnamefont {M.}~\bibnamefont
  {Weidem\"{u}ller}}, \bibinfo {author} {\bibfnamefont {T.}~\bibnamefont
  {Esslinger}}, \bibinfo {author} {\bibfnamefont {A.}~\bibnamefont
  {Hemmerich}}, \bibinfo {author} {\bibfnamefont {C.}~\bibnamefont
  {Zimmermann}}, \bibinfo {author} {\bibfnamefont {V.}~\bibnamefont {Vuletic}},
  \bibinfo {author} {\bibfnamefont {W.}~\bibnamefont {K\"{o}nig}},\ and\
  \bibinfo {author} {\bibfnamefont {T.}~\bibnamefont {H\"{a}nsch}},\ }\href
  {https://doi.org/10.1016/0030-4018(95)00146-y} {\bibfield  {journal}
  {\bibinfo  {journal} {Optics Communications}\ }\textbf {\bibinfo {volume}
  {117}},\ \bibinfo {pages} {541} (\bibinfo {year} {1995})}\BibitemShut
  {NoStop}%
\bibitem [{\citenamefont {Arnold}\ \emph {et~al.}(1998)\citenamefont {Arnold},
  \citenamefont {Wilson},\ and\ \citenamefont {Boshier}}]{ecdlarnold}%
  \BibitemOpen
  \bibfield  {author} {\bibinfo {author} {\bibfnamefont {A.~S.}\ \bibnamefont
  {Arnold}}, \bibinfo {author} {\bibfnamefont {J.~S.}\ \bibnamefont {Wilson}},\
  and\ \bibinfo {author} {\bibfnamefont {M.~G.}\ \bibnamefont {Boshier}},\
  }\href {https://doi.org/10.1063/1.1148756} {\bibfield  {journal} {\bibinfo
  {journal} {Review of Scientific Instruments}\ }\textbf {\bibinfo {volume}
  {69}},\ \bibinfo {pages} {1236} (\bibinfo {year} {1998})}\BibitemShut
  {NoStop}%
\bibitem [{\citenamefont {Cook}\ \emph {et~al.}(2012)\citenamefont {Cook},
  \citenamefont {Martin}, \citenamefont {Brown-Heft}, \citenamefont {Garman},\
  and\ \citenamefont {Steck}}]{Cook2012}%
  \BibitemOpen
  \bibfield  {author} {\bibinfo {author} {\bibfnamefont {E.~C.}\ \bibnamefont
  {Cook}}, \bibinfo {author} {\bibfnamefont {P.~J.}\ \bibnamefont {Martin}},
  \bibinfo {author} {\bibfnamefont {T.~L.}\ \bibnamefont {Brown-Heft}},
  \bibinfo {author} {\bibfnamefont {J.~C.}\ \bibnamefont {Garman}},\ and\
  \bibinfo {author} {\bibfnamefont {D.~A.}\ \bibnamefont {Steck}},\ }\href
  {https://doi.org/10.1063/1.3698003} {\bibfield  {journal} {\bibinfo
  {journal} {Review of Scientific Instruments}\ }\textbf {\bibinfo {volume}
  {83}},\ \bibinfo {pages} {043101} (\bibinfo {year} {2012})}\BibitemShut
  {NoStop}%
\bibitem [{\citenamefont {Baillard}\ \emph {et~al.}(2006)\citenamefont
  {Baillard}, \citenamefont {Gauguet}, \citenamefont {Bize}, \citenamefont
  {Lemonde}, \citenamefont {Laurent}, \citenamefont {Clairon},\ and\
  \citenamefont {Rosenbusch}}]{BAILLARD2006609}%
  \BibitemOpen
  \bibfield  {author} {\bibinfo {author} {\bibfnamefont {X.}~\bibnamefont
  {Baillard}}, \bibinfo {author} {\bibfnamefont {A.}~\bibnamefont {Gauguet}},
  \bibinfo {author} {\bibfnamefont {S.}~\bibnamefont {Bize}}, \bibinfo {author}
  {\bibfnamefont {P.}~\bibnamefont {Lemonde}}, \bibinfo {author} {\bibfnamefont
  {P.}~\bibnamefont {Laurent}}, \bibinfo {author} {\bibfnamefont
  {A.}~\bibnamefont {Clairon}},\ and\ \bibinfo {author} {\bibfnamefont
  {P.}~\bibnamefont {Rosenbusch}},\ }\href
  {https://doi.org/https://doi.org/10.1016/j.optcom.2006.05.011} {\bibfield
  {journal} {\bibinfo  {journal} {Optics Communications}\ }\textbf {\bibinfo
  {volume} {266}},\ \bibinfo {pages} {609} (\bibinfo {year}
  {2006})}\BibitemShut {NoStop}%
\bibitem [{\citenamefont {Daffurn}\ \emph {et~al.}(2021)\citenamefont
  {Daffurn}, \citenamefont {Offer},\ and\ \citenamefont
  {Arnold}}]{Daffurn2021}%
  \BibitemOpen
  \bibfield  {author} {\bibinfo {author} {\bibfnamefont {A.}~\bibnamefont
  {Daffurn}}, \bibinfo {author} {\bibfnamefont {R.~F.}\ \bibnamefont {Offer}},\
  and\ \bibinfo {author} {\bibfnamefont {A.~S.}\ \bibnamefont {Arnold}},\
  }\href {https://doi.org/10.1364/ao.426844} {\bibfield  {journal} {\bibinfo
  {journal} {Applied Optics}\ }\textbf {\bibinfo {volume} {60}},\ \bibinfo
  {pages} {5832} (\bibinfo {year} {2021})}\BibitemShut {NoStop}%
\bibitem [{\citenamefont {Sakaguchi}\ \emph {et~al.}(1988)\citenamefont
  {Sakaguchi}, \citenamefont {Koyama},\ and\ \citenamefont {Iga}}]{vcsel1}%
  \BibitemOpen
  \bibfield  {author} {\bibinfo {author} {\bibfnamefont {T.}~\bibnamefont
  {Sakaguchi}}, \bibinfo {author} {\bibfnamefont {F.}~\bibnamefont {Koyama}},\
  and\ \bibinfo {author} {\bibfnamefont {K.}~\bibnamefont {Iga}},\ }\href
  {https://doi.org/10.1049/el:19880632} {\bibfield  {journal} {\bibinfo
  {journal} {Electronics Letters}\ }\textbf {\bibinfo {volume} {24}},\ \bibinfo
  {pages} {928} (\bibinfo {year} {1988})}\BibitemShut {NoStop}%
\bibitem [{\citenamefont {Serkland}\ \emph {et~al.}(2006)\citenamefont
  {Serkland}, \citenamefont {Peake}, \citenamefont {Geib}, \citenamefont
  {Lutwak}, \citenamefont {Garvey}, \citenamefont {Varghese},\ and\
  \citenamefont {Mescher}}]{Serkland_06}%
  \BibitemOpen
  \bibfield  {author} {\bibinfo {author} {\bibfnamefont {D.~K.}\ \bibnamefont
  {Serkland}}, \bibinfo {author} {\bibfnamefont {G.~M.}\ \bibnamefont {Peake}},
  \bibinfo {author} {\bibfnamefont {K.~M.}\ \bibnamefont {Geib}}, \bibinfo
  {author} {\bibfnamefont {R.}~\bibnamefont {Lutwak}}, \bibinfo {author}
  {\bibfnamefont {R.~M.}\ \bibnamefont {Garvey}}, \bibinfo {author}
  {\bibfnamefont {M.}~\bibnamefont {Varghese}},\ and\ \bibinfo {author}
  {\bibfnamefont {M.}~\bibnamefont {Mescher}},\ }in\ \href@noop {} {\emph
  {\bibinfo {booktitle} {Conference on Vertical-Cavity Surface-Emitting Lasers
  X}}},\ \bibinfo {series} {Proceedings of SPIE}, Vol.\ \bibinfo {volume}
  {6132}\ (\bibinfo {address} {San Jose, CA},\ \bibinfo {year}
  {2006})\BibitemShut {NoStop}%
\bibitem [{\citenamefont {Serkland}\ \emph {et~al.}(2007)\citenamefont
  {Serkland}, \citenamefont {Geib}, \citenamefont {Peake}, \citenamefont
  {Lutwak}, \citenamefont {Rashed}, \citenamefont {Varghese}, \citenamefont
  {Tepolt},\ and\ \citenamefont {Prouty}}]{Serkland_07}%
  \BibitemOpen
  \bibfield  {author} {\bibinfo {author} {\bibfnamefont {D.~K.}\ \bibnamefont
  {Serkland}}, \bibinfo {author} {\bibfnamefont {K.~M.}\ \bibnamefont {Geib}},
  \bibinfo {author} {\bibfnamefont {G.~M.}\ \bibnamefont {Peake}}, \bibinfo
  {author} {\bibfnamefont {R.}~\bibnamefont {Lutwak}}, \bibinfo {author}
  {\bibfnamefont {A.}~\bibnamefont {Rashed}}, \bibinfo {author} {\bibfnamefont
  {M.}~\bibnamefont {Varghese}}, \bibinfo {author} {\bibfnamefont
  {G.}~\bibnamefont {Tepolt}},\ and\ \bibinfo {author} {\bibfnamefont
  {M.}~\bibnamefont {Prouty}},\ }in\ \href@noop {} {\emph {\bibinfo {booktitle}
  {11th Vertical Cavity Surface Emitting Lasers Conference}}},\ \bibinfo
  {series} {Proceedings of SPIE}, Vol.\ \bibinfo {volume} {6484}\ (\bibinfo
  {address} {San Jose, CA},\ \bibinfo {year} {2007})\BibitemShut {NoStop}%
\bibitem [{\citenamefont {Iga}(2018)}]{Iga_18}%
  \BibitemOpen
  \bibfield  {author} {\bibinfo {author} {\bibfnamefont {K.}~\bibnamefont
  {Iga}},\ }\bibfield  {journal} {\bibinfo  {journal} {Japanese Journal of
  Applied Physics}\ }\textbf {\bibinfo {volume} {57}},\ \href
  {https://doi.org/10.7567/jjap.57.08pa01} {10.7567/jjap.57.08pa01} (\bibinfo
  {year} {2018})\BibitemShut {NoStop}%
\bibitem [{\citenamefont {Mikhailov}\ \emph {et~al.}(2010)\citenamefont
  {Mikhailov}, \citenamefont {Horrom}, \citenamefont {Belcher},\ and\
  \citenamefont {Novikova}}]{Mikhailov_10}%
  \BibitemOpen
  \bibfield  {author} {\bibinfo {author} {\bibfnamefont {E.~E.}\ \bibnamefont
  {Mikhailov}}, \bibinfo {author} {\bibfnamefont {T.}~\bibnamefont {Horrom}},
  \bibinfo {author} {\bibfnamefont {N.}~\bibnamefont {Belcher}},\ and\ \bibinfo
  {author} {\bibfnamefont {I.}~\bibnamefont {Novikova}},\ }\href
  {https://doi.org/10.1364/JOSAB.27.000417} {\bibfield  {journal} {\bibinfo
  {journal} {Journal of the Optical Society of America B-Optical Physics}\
  }\textbf {\bibinfo {volume} {27}},\ \bibinfo {pages} {417} (\bibinfo {year}
  {2010})}\BibitemShut {NoStop}%
\bibitem [{\citenamefont {Kitching}\ \emph {et~al.}(2000)\citenamefont
  {Kitching}, \citenamefont {Knappe}, \citenamefont {Vukicevic}, \citenamefont
  {Hollberg}, \citenamefont {Wynands},\ and\ \citenamefont
  {Weidmann}}]{vcselkitching}%
  \BibitemOpen
  \bibfield  {author} {\bibinfo {author} {\bibfnamefont {J.}~\bibnamefont
  {Kitching}}, \bibinfo {author} {\bibfnamefont {S.}~\bibnamefont {Knappe}},
  \bibinfo {author} {\bibfnamefont {M.}~\bibnamefont {Vukicevic}}, \bibinfo
  {author} {\bibfnamefont {L.}~\bibnamefont {Hollberg}}, \bibinfo {author}
  {\bibfnamefont {R.}~\bibnamefont {Wynands}},\ and\ \bibinfo {author}
  {\bibfnamefont {W.}~\bibnamefont {Weidmann}},\ }\href
  {https://doi.org/10.1109/19.893276} {\bibfield  {journal} {\bibinfo
  {journal} {IEEE Transactions on Instrumentation and Measurement}\ }\textbf
  {\bibinfo {volume} {49}},\ \bibinfo {pages} {1313} (\bibinfo {year}
  {2000})}\BibitemShut {NoStop}%
\bibitem [{\citenamefont {Moriya}\ \emph {et~al.}(2020)\citenamefont {Moriya},
  \citenamefont {Singh}, \citenamefont {Bongs},\ and\ \citenamefont
  {Hastie}}]{Moriya_20}%
  \BibitemOpen
  \bibfield  {author} {\bibinfo {author} {\bibfnamefont {P.~H.}\ \bibnamefont
  {Moriya}}, \bibinfo {author} {\bibfnamefont {Y.}~\bibnamefont {Singh}},
  \bibinfo {author} {\bibfnamefont {K.}~\bibnamefont {Bongs}},\ and\ \bibinfo
  {author} {\bibfnamefont {J.~E.}\ \bibnamefont {Hastie}},\ }\href
  {https://doi.org/10.1364/OE.390982} {\bibfield  {journal} {\bibinfo
  {journal} {Optics Express}\ }\textbf {\bibinfo {volume} {28}},\ \bibinfo
  {pages} {15943} (\bibinfo {year} {2020})}\BibitemShut {NoStop}%
\bibitem [{\citenamefont {Scifres}\ \emph {et~al.}(1974)\citenamefont
  {Scifres}, \citenamefont {Burnham},\ and\ \citenamefont
  {Streifer}}]{Scifres_1974}%
  \BibitemOpen
  \bibfield  {author} {\bibinfo {author} {\bibfnamefont {D.~R.}\ \bibnamefont
  {Scifres}}, \bibinfo {author} {\bibfnamefont {R.~D.}\ \bibnamefont
  {Burnham}},\ and\ \bibinfo {author} {\bibfnamefont {W.}~\bibnamefont
  {Streifer}},\ }\href {https://doi.org/10.1063/1.1655440} {\bibfield
  {journal} {\bibinfo  {journal} {Applied Physics Letters}\ }\textbf {\bibinfo
  {volume} {25}},\ \bibinfo {pages} {203} (\bibinfo {year} {1974})}\BibitemShut
  {NoStop}%
\bibitem [{\citenamefont {Gaetano}\ and\ \citenamefont
  {Sorel}(2019)}]{Gaetano:19}%
  \BibitemOpen
  \bibfield  {author} {\bibinfo {author} {\bibfnamefont {E.~D.}\ \bibnamefont
  {Gaetano}}\ and\ \bibinfo {author} {\bibfnamefont {M.}~\bibnamefont
  {Sorel}},\ }\href {https://doi.org/10.1364/OL.44.001642} {\bibfield
  {journal} {\bibinfo  {journal} {Opt. Lett.}\ }\textbf {\bibinfo {volume}
  {44}},\ \bibinfo {pages} {1642} (\bibinfo {year} {2019})}\BibitemShut
  {NoStop}%
\bibitem [{\citenamefont {Poulin}\ \emph {et~al.}(1997)\citenamefont {Poulin},
  \citenamefont {Latrasse}, \citenamefont {Cyr},\ and\ \citenamefont
  {Tetu}}]{Poulin_1997}%
  \BibitemOpen
  \bibfield  {author} {\bibinfo {author} {\bibfnamefont {M.}~\bibnamefont
  {Poulin}}, \bibinfo {author} {\bibfnamefont {C.}~\bibnamefont {Latrasse}},
  \bibinfo {author} {\bibfnamefont {N.}~\bibnamefont {Cyr}},\ and\ \bibinfo
  {author} {\bibfnamefont {M.}~\bibnamefont {Tetu}},\ }\href
  {https://doi.org/10.1109/68.643293} {\bibfield  {journal} {\bibinfo
  {journal} {Ieee Photonics Technology Letters}\ }\textbf {\bibinfo {volume}
  {9}},\ \bibinfo {pages} {1631} (\bibinfo {year} {1997})}\BibitemShut
  {NoStop}%
\bibitem [{\citenamefont {Kraft}\ \emph {et~al.}(2005)\citenamefont {Kraft},
  \citenamefont {Deninger}, \citenamefont {Truck}, \citenamefont {Fortagh},
  \citenamefont {Lison},\ and\ \citenamefont {Zimmermann}}]{Kraft_05}%
  \BibitemOpen
  \bibfield  {author} {\bibinfo {author} {\bibfnamefont {S.}~\bibnamefont
  {Kraft}}, \bibinfo {author} {\bibfnamefont {A.}~\bibnamefont {Deninger}},
  \bibinfo {author} {\bibfnamefont {C.}~\bibnamefont {Truck}}, \bibinfo
  {author} {\bibfnamefont {J.}~\bibnamefont {Fortagh}}, \bibinfo {author}
  {\bibfnamefont {F.}~\bibnamefont {Lison}},\ and\ \bibinfo {author}
  {\bibfnamefont {C.}~\bibnamefont {Zimmermann}},\ }\href@noop {} {\bibfield
  {journal} {\bibinfo  {journal} {Laser Physics Letters}\ }\textbf {\bibinfo
  {volume} {2}},\ \bibinfo {pages} {71} (\bibinfo {year} {2005})}\BibitemShut
  {NoStop}%
\bibitem [{\citenamefont {Ligeret}\ \emph {et~al.}(2008)\citenamefont
  {Ligeret}, \citenamefont {Holleville}, \citenamefont {Perrin}, \citenamefont
  {Bansropun}, \citenamefont {Lecomte}, \citenamefont {Parillaud},
  \citenamefont {Calligaro}, \citenamefont {Krakowski},\ and\ \citenamefont
  {Dimarcq}}]{Ligeret_08}%
  \BibitemOpen
  \bibfield  {author} {\bibinfo {author} {\bibfnamefont {V.}~\bibnamefont
  {Ligeret}}, \bibinfo {author} {\bibfnamefont {D.}~\bibnamefont {Holleville}},
  \bibinfo {author} {\bibfnamefont {S.}~\bibnamefont {Perrin}}, \bibinfo
  {author} {\bibfnamefont {S.}~\bibnamefont {Bansropun}}, \bibinfo {author}
  {\bibfnamefont {M.}~\bibnamefont {Lecomte}}, \bibinfo {author} {\bibfnamefont
  {O.}~\bibnamefont {Parillaud}}, \bibinfo {author} {\bibfnamefont
  {M.}~\bibnamefont {Calligaro}}, \bibinfo {author} {\bibfnamefont
  {M.}~\bibnamefont {Krakowski}},\ and\ \bibinfo {author} {\bibfnamefont
  {N.}~\bibnamefont {Dimarcq}},\ }\href {https://doi.org/10.1049/el:20080059}
  {\bibfield  {journal} {\bibinfo  {journal} {Electronics Letters}\ }\textbf
  {\bibinfo {volume} {44}},\ \bibinfo {pages} {804} (\bibinfo {year}
  {2008})}\BibitemShut {NoStop}%
\bibitem [{\citenamefont {Gaetano}\ \emph {et~al.}(2020)\citenamefont
  {Gaetano}, \citenamefont {Watson}, \citenamefont {McBrearty}, \citenamefont
  {Sorel},\ and\ \citenamefont {Paul}}]{DiGaetano:20}%
  \BibitemOpen
  \bibfield  {author} {\bibinfo {author} {\bibfnamefont {E.~D.}\ \bibnamefont
  {Gaetano}}, \bibinfo {author} {\bibfnamefont {S.}~\bibnamefont {Watson}},
  \bibinfo {author} {\bibfnamefont {E.}~\bibnamefont {McBrearty}}, \bibinfo
  {author} {\bibfnamefont {M.}~\bibnamefont {Sorel}},\ and\ \bibinfo {author}
  {\bibfnamefont {D.~J.}\ \bibnamefont {Paul}},\ }\href
  {https://doi.org/10.1364/OL.394185} {\bibfield  {journal} {\bibinfo
  {journal} {Opt. Lett.}\ }\textbf {\bibinfo {volume} {45}},\ \bibinfo {pages}
  {3529} (\bibinfo {year} {2020})}\BibitemShut {NoStop}%
\bibitem [{\citenamefont {Wang}\ \emph {et~al.}(2015)\citenamefont {Wang},
  \citenamefont {Tian}, \citenamefont {Pantouvaki}, \citenamefont {Guo},
  \citenamefont {Absil}, \citenamefont {Van~Campenhout}, \citenamefont
  {Merckling},\ and\ \citenamefont {Van~Thourhout}}]{Wang_15}%
  \BibitemOpen
  \bibfield  {author} {\bibinfo {author} {\bibfnamefont {Z.~C.}\ \bibnamefont
  {Wang}}, \bibinfo {author} {\bibfnamefont {B.}~\bibnamefont {Tian}}, \bibinfo
  {author} {\bibfnamefont {M.}~\bibnamefont {Pantouvaki}}, \bibinfo {author}
  {\bibfnamefont {W.~M.}\ \bibnamefont {Guo}}, \bibinfo {author} {\bibfnamefont
  {P.}~\bibnamefont {Absil}}, \bibinfo {author} {\bibfnamefont
  {J.}~\bibnamefont {Van~Campenhout}}, \bibinfo {author} {\bibfnamefont
  {C.}~\bibnamefont {Merckling}},\ and\ \bibinfo {author} {\bibfnamefont
  {D.}~\bibnamefont {Van~Thourhout}},\ }\href
  {https://doi.org/10.1038/nphoton.2015.199} {\bibfield  {journal} {\bibinfo
  {journal} {Nature Photonics}\ }\textbf {\bibinfo {volume} {9}},\ \bibinfo
  {pages} {837} (\bibinfo {year} {2015})}\BibitemShut {NoStop}%
\bibitem [{\citenamefont {Xie}\ \emph {et~al.}(2019)\citenamefont {Xie},
  \citenamefont {Komljenovic}, \citenamefont {Huang}, \citenamefont {Tran},
  \citenamefont {Davenport}, \citenamefont {Torres}, \citenamefont {Pintus},\
  and\ \citenamefont {Bowers}}]{Xie:19}%
  \BibitemOpen
  \bibfield  {author} {\bibinfo {author} {\bibfnamefont {W.}~\bibnamefont
  {Xie}}, \bibinfo {author} {\bibfnamefont {T.}~\bibnamefont {Komljenovic}},
  \bibinfo {author} {\bibfnamefont {J.}~\bibnamefont {Huang}}, \bibinfo
  {author} {\bibfnamefont {M.}~\bibnamefont {Tran}}, \bibinfo {author}
  {\bibfnamefont {M.}~\bibnamefont {Davenport}}, \bibinfo {author}
  {\bibfnamefont {A.}~\bibnamefont {Torres}}, \bibinfo {author} {\bibfnamefont
  {P.}~\bibnamefont {Pintus}},\ and\ \bibinfo {author} {\bibfnamefont
  {J.}~\bibnamefont {Bowers}},\ }\href {https://doi.org/10.1364/OE.27.003642}
  {\bibfield  {journal} {\bibinfo  {journal} {Opt. Express}\ }\textbf {\bibinfo
  {volume} {27}},\ \bibinfo {pages} {3642} (\bibinfo {year}
  {2019})}\BibitemShut {NoStop}%
\bibitem [{\citenamefont {Xiang}\ \emph {et~al.}(2020)\citenamefont {Xiang},
  \citenamefont {Jin}, \citenamefont {Guo}, \citenamefont {Peters},
  \citenamefont {Kennedy}, \citenamefont {Selvidge}, \citenamefont {Morton},\
  and\ \citenamefont {Bowers}}]{Xiang_20}%
  \BibitemOpen
  \bibfield  {author} {\bibinfo {author} {\bibfnamefont {C.}~\bibnamefont
  {Xiang}}, \bibinfo {author} {\bibfnamefont {W.}~\bibnamefont {Jin}}, \bibinfo
  {author} {\bibfnamefont {J.}~\bibnamefont {Guo}}, \bibinfo {author}
  {\bibfnamefont {J.~D.}\ \bibnamefont {Peters}}, \bibinfo {author}
  {\bibfnamefont {M.~J.}\ \bibnamefont {Kennedy}}, \bibinfo {author}
  {\bibfnamefont {J.}~\bibnamefont {Selvidge}}, \bibinfo {author}
  {\bibfnamefont {P.~A.}\ \bibnamefont {Morton}},\ and\ \bibinfo {author}
  {\bibfnamefont {J.~E.}\ \bibnamefont {Bowers}},\ }\href
  {https://doi.org/10.1364/OPTICA.384026} {\bibfield  {journal} {\bibinfo
  {journal} {Optica}\ }\textbf {\bibinfo {volume} {7}},\ \bibinfo {pages} {20}
  (\bibinfo {year} {2020})}\BibitemShut {NoStop}%
\bibitem [{\citenamefont {Ohira}\ \emph {et~al.}(2010)\citenamefont {Ohira},
  \citenamefont {Kobayashi}, \citenamefont {Iizuka}, \citenamefont {Yoshida},
  \citenamefont {Ezaki}, \citenamefont {Uemura}, \citenamefont {Kojima},
  \citenamefont {Nakamura}, \citenamefont {Furuyama},\ and\ \citenamefont
  {Shibata}}]{Ohira_10}%
  \BibitemOpen
  \bibfield  {author} {\bibinfo {author} {\bibfnamefont {K.}~\bibnamefont
  {Ohira}}, \bibinfo {author} {\bibfnamefont {K.}~\bibnamefont {Kobayashi}},
  \bibinfo {author} {\bibfnamefont {N.}~\bibnamefont {Iizuka}}, \bibinfo
  {author} {\bibfnamefont {H.}~\bibnamefont {Yoshida}}, \bibinfo {author}
  {\bibfnamefont {M.}~\bibnamefont {Ezaki}}, \bibinfo {author} {\bibfnamefont
  {H.}~\bibnamefont {Uemura}}, \bibinfo {author} {\bibfnamefont
  {A.}~\bibnamefont {Kojima}}, \bibinfo {author} {\bibfnamefont
  {K.}~\bibnamefont {Nakamura}}, \bibinfo {author} {\bibfnamefont
  {H.}~\bibnamefont {Furuyama}},\ and\ \bibinfo {author} {\bibfnamefont
  {H.}~\bibnamefont {Shibata}},\ }\href {https://doi.org/10.1364/OE.18.015440}
  {\bibfield  {journal} {\bibinfo  {journal} {Optics Express}\ }\textbf
  {\bibinfo {volume} {18}},\ \bibinfo {pages} {15440} (\bibinfo {year}
  {2010})}\BibitemShut {NoStop}%
\bibitem [{\citenamefont {Gallacher}\ \emph {et~al.}(2019)\citenamefont
  {Gallacher}, \citenamefont {Sinclair}, \citenamefont {Millar}, \citenamefont
  {Sharp}, \citenamefont {Mirando}, \citenamefont {Ternent}, \citenamefont
  {Mills}, \citenamefont {Casey}, \citenamefont {Paul},\ and\ \citenamefont
  {Ieee}}]{Gallacher_19}%
  \BibitemOpen
  \bibfield  {author} {\bibinfo {author} {\bibfnamefont {K.}~\bibnamefont
  {Gallacher}}, \bibinfo {author} {\bibfnamefont {M.}~\bibnamefont {Sinclair}},
  \bibinfo {author} {\bibfnamefont {R.~W.}\ \bibnamefont {Millar}}, \bibinfo
  {author} {\bibfnamefont {O.}~\bibnamefont {Sharp}}, \bibinfo {author}
  {\bibfnamefont {F.}~\bibnamefont {Mirando}}, \bibinfo {author} {\bibfnamefont
  {G.}~\bibnamefont {Ternent}}, \bibinfo {author} {\bibfnamefont
  {G.}~\bibnamefont {Mills}}, \bibinfo {author} {\bibfnamefont
  {B.}~\bibnamefont {Casey}}, \bibinfo {author} {\bibfnamefont {D.~J.}\
  \bibnamefont {Paul}},\ and\ \bibinfo {author} {\bibnamefont {Ieee}},\ }in\
  \href {https://doi.org/10.1364/CLEO_SI.2019.STu4O.7} {\emph {\bibinfo
  {booktitle} {Conference on Lasers and Electro-Optics (CLEO)}}},\ \bibinfo
  {series and number} {Conference on Lasers and Electro-Optics}\ (\bibinfo
  {address} {San Jose, CA},\ \bibinfo {year} {2019})\BibitemShut {NoStop}%
\bibitem [{\citenamefont {Romero-Garcia}\ \emph {et~al.}(2014)\citenamefont
  {Romero-Garcia}, \citenamefont {Marzban}, \citenamefont {Merget},
  \citenamefont {Shen},\ and\ \citenamefont {Witzens}}]{Romero-Garcia_14}%
  \BibitemOpen
  \bibfield  {author} {\bibinfo {author} {\bibfnamefont {S.}~\bibnamefont
  {Romero-Garcia}}, \bibinfo {author} {\bibfnamefont {B.}~\bibnamefont
  {Marzban}}, \bibinfo {author} {\bibfnamefont {F.}~\bibnamefont {Merget}},
  \bibinfo {author} {\bibfnamefont {B.}~\bibnamefont {Shen}},\ and\ \bibinfo
  {author} {\bibfnamefont {J.}~\bibnamefont {Witzens}},\ }\bibfield  {journal}
  {\bibinfo  {journal} {Ieee Journal of Selected Topics in Quantum
  Electronics}\ }\textbf {\bibinfo {volume} {20}},\ \href
  {https://doi.org/10.1109/JSTQE.2013.2292523} {10.1109/JSTQE.2013.2292523}
  (\bibinfo {year} {2014})\BibitemShut {NoStop}%
\bibitem [{\citenamefont {Boller}\ \emph {et~al.}(2020)\citenamefont {Boller},
  \citenamefont {van Rees}, \citenamefont {Fan}, \citenamefont {Mak},
  \citenamefont {Lammerink}, \citenamefont {Franken}, \citenamefont {van~der
  Slot}, \citenamefont {Marpaung}, \citenamefont {Fallnich}, \citenamefont
  {Epping}, \citenamefont {Oldenbeuving}, \citenamefont {Geskus}, \citenamefont
  {Dekker}, \citenamefont {Visscher}, \citenamefont {Grootjans}, \citenamefont
  {Roeloffzen}, \citenamefont {Hoekman}, \citenamefont {Klein}, \citenamefont
  {Leinse},\ and\ \citenamefont {Heideman}}]{Boller_20}%
  \BibitemOpen
  \bibfield  {author} {\bibinfo {author} {\bibfnamefont {K.-J.}\ \bibnamefont
  {Boller}}, \bibinfo {author} {\bibfnamefont {A.}~\bibnamefont {van Rees}},
  \bibinfo {author} {\bibfnamefont {Y.}~\bibnamefont {Fan}}, \bibinfo {author}
  {\bibfnamefont {J.}~\bibnamefont {Mak}}, \bibinfo {author} {\bibfnamefont
  {R.~E.~M.}\ \bibnamefont {Lammerink}}, \bibinfo {author} {\bibfnamefont
  {C.~A.~A.}\ \bibnamefont {Franken}}, \bibinfo {author} {\bibfnamefont
  {P.~J.~M.}\ \bibnamefont {van~der Slot}}, \bibinfo {author} {\bibfnamefont
  {D.~A.~I.}\ \bibnamefont {Marpaung}}, \bibinfo {author} {\bibfnamefont
  {C.}~\bibnamefont {Fallnich}}, \bibinfo {author} {\bibfnamefont {J.~P.}\
  \bibnamefont {Epping}}, \bibinfo {author} {\bibfnamefont {R.~M.}\
  \bibnamefont {Oldenbeuving}}, \bibinfo {author} {\bibfnamefont
  {D.}~\bibnamefont {Geskus}}, \bibinfo {author} {\bibfnamefont
  {R.}~\bibnamefont {Dekker}}, \bibinfo {author} {\bibfnamefont
  {I.}~\bibnamefont {Visscher}}, \bibinfo {author} {\bibfnamefont
  {R.}~\bibnamefont {Grootjans}}, \bibinfo {author} {\bibfnamefont {C.~G.~H.}\
  \bibnamefont {Roeloffzen}}, \bibinfo {author} {\bibfnamefont
  {M.}~\bibnamefont {Hoekman}}, \bibinfo {author} {\bibfnamefont {E.~J.}\
  \bibnamefont {Klein}}, \bibinfo {author} {\bibfnamefont {A.}~\bibnamefont
  {Leinse}},\ and\ \bibinfo {author} {\bibfnamefont {R.~G.}\ \bibnamefont
  {Heideman}},\ }\bibfield  {journal} {\bibinfo  {journal} {Photonics}\
  }\textbf {\bibinfo {volume} {7}},\ \href
  {https://doi.org/10.3390/photonics7010004} {10.3390/photonics7010004}
  (\bibinfo {year} {2020})\BibitemShut {NoStop}%
\bibitem [{\citenamefont {Jin}\ \emph {et~al.}(2021)\citenamefont {Jin},
  \citenamefont {Yang}, \citenamefont {Chang}, \citenamefont {Shen},
  \citenamefont {Wang}, \citenamefont {Leal}, \citenamefont {Wu}, \citenamefont
  {Gao}, \citenamefont {Feshali}, \citenamefont {Paniccia}, \citenamefont
  {Vahala},\ and\ \citenamefont {Bowers}}]{Jin_21}%
  \BibitemOpen
  \bibfield  {author} {\bibinfo {author} {\bibfnamefont {W.}~\bibnamefont
  {Jin}}, \bibinfo {author} {\bibfnamefont {Q.-F.}\ \bibnamefont {Yang}},
  \bibinfo {author} {\bibfnamefont {L.}~\bibnamefont {Chang}}, \bibinfo
  {author} {\bibfnamefont {B.}~\bibnamefont {Shen}}, \bibinfo {author}
  {\bibfnamefont {H.}~\bibnamefont {Wang}}, \bibinfo {author} {\bibfnamefont
  {M.~A.}\ \bibnamefont {Leal}}, \bibinfo {author} {\bibfnamefont
  {L.}~\bibnamefont {Wu}}, \bibinfo {author} {\bibfnamefont {M.}~\bibnamefont
  {Gao}}, \bibinfo {author} {\bibfnamefont {A.}~\bibnamefont {Feshali}},
  \bibinfo {author} {\bibfnamefont {M.}~\bibnamefont {Paniccia}}, \bibinfo
  {author} {\bibfnamefont {K.~J.}\ \bibnamefont {Vahala}},\ and\ \bibinfo
  {author} {\bibfnamefont {J.~E.}\ \bibnamefont {Bowers}},\ }\href
  {https://doi.org/10.1038/s41566-021-00761-7} {\bibfield  {journal} {\bibinfo
  {journal} {Nature Photonics}\ }\textbf {\bibinfo {volume} {15}},\ \bibinfo
  {pages} {346} (\bibinfo {year} {2021})}\BibitemShut {NoStop}%
\bibitem [{\citenamefont {Blumenthal}(2020)}]{Blumenthal_20}%
  \BibitemOpen
  \bibfield  {author} {\bibinfo {author} {\bibfnamefont {D.~J.}\ \bibnamefont
  {Blumenthal}},\ }\bibfield  {journal} {\bibinfo  {journal} {Apl Photonics}\
  }\textbf {\bibinfo {volume} {5}},\ \href {https://doi.org/10.1063/1.5131683}
  {10.1063/1.5131683} (\bibinfo {year} {2020})\BibitemShut {NoStop}%
\bibitem [{\citenamefont {Belt}\ \emph {et~al.}(2017)\citenamefont {Belt},
  \citenamefont {Davenport}, \citenamefont {Bowers},\ and\ \citenamefont
  {Blumenthal}}]{Belt_17}%
  \BibitemOpen
  \bibfield  {author} {\bibinfo {author} {\bibfnamefont {M.}~\bibnamefont
  {Belt}}, \bibinfo {author} {\bibfnamefont {M.~L.}\ \bibnamefont {Davenport}},
  \bibinfo {author} {\bibfnamefont {J.~E.}\ \bibnamefont {Bowers}},\ and\
  \bibinfo {author} {\bibfnamefont {D.~J.}\ \bibnamefont {Blumenthal}},\ }\href
  {https://doi.org/10.1364/OPTICA.4.000532} {\bibfield  {journal} {\bibinfo
  {journal} {Optica}\ }\textbf {\bibinfo {volume} {4}},\ \bibinfo {pages} {532}
  (\bibinfo {year} {2017})}\BibitemShut {NoStop}%
\bibitem [{\citenamefont {Lu}\ \emph {et~al.}(2018)\citenamefont {Lu},
  \citenamefont {Fanto}, \citenamefont {Choi}, \citenamefont {Thomas},
  \citenamefont {Steidle}, \citenamefont {Mouradian}, \citenamefont {Kong},
  \citenamefont {Zhu}, \citenamefont {Moon}, \citenamefont {Berggren},
  \citenamefont {Kim}, \citenamefont {Soltani}, \citenamefont {Preble},\ and\
  \citenamefont {Englund}}]{Lu_18}%
  \BibitemOpen
  \bibfield  {author} {\bibinfo {author} {\bibfnamefont {T.~J.}\ \bibnamefont
  {Lu}}, \bibinfo {author} {\bibfnamefont {M.}~\bibnamefont {Fanto}}, \bibinfo
  {author} {\bibfnamefont {H.}~\bibnamefont {Choi}}, \bibinfo {author}
  {\bibfnamefont {P.}~\bibnamefont {Thomas}}, \bibinfo {author} {\bibfnamefont
  {J.}~\bibnamefont {Steidle}}, \bibinfo {author} {\bibfnamefont
  {S.}~\bibnamefont {Mouradian}}, \bibinfo {author} {\bibfnamefont
  {W.}~\bibnamefont {Kong}}, \bibinfo {author} {\bibfnamefont {D.}~\bibnamefont
  {Zhu}}, \bibinfo {author} {\bibfnamefont {H.}~\bibnamefont {Moon}}, \bibinfo
  {author} {\bibfnamefont {K.}~\bibnamefont {Berggren}}, \bibinfo {author}
  {\bibfnamefont {J.}~\bibnamefont {Kim}}, \bibinfo {author} {\bibfnamefont
  {M.}~\bibnamefont {Soltani}}, \bibinfo {author} {\bibfnamefont
  {S.}~\bibnamefont {Preble}},\ and\ \bibinfo {author} {\bibfnamefont
  {D.}~\bibnamefont {Englund}},\ }\href {https://doi.org/10.1364/OE.26.011147}
  {\bibfield  {journal} {\bibinfo  {journal} {Optics Express}\ }\textbf
  {\bibinfo {volume} {26}},\ \bibinfo {pages} {11147} (\bibinfo {year}
  {2018})}\BibitemShut {NoStop}%
\bibitem [{\citenamefont {Aslan}\ \emph {et~al.}(2010)\citenamefont {Aslan},
  \citenamefont {Webster}, \citenamefont {Byard}, \citenamefont {Pereira},
  \citenamefont {Hayes}, \citenamefont {Wiederkehr},\ and\ \citenamefont
  {Mendes}}]{Aslan_10}%
  \BibitemOpen
  \bibfield  {author} {\bibinfo {author} {\bibfnamefont {M.~M.}\ \bibnamefont
  {Aslan}}, \bibinfo {author} {\bibfnamefont {N.~A.}\ \bibnamefont {Webster}},
  \bibinfo {author} {\bibfnamefont {C.~L.}\ \bibnamefont {Byard}}, \bibinfo
  {author} {\bibfnamefont {M.~B.}\ \bibnamefont {Pereira}}, \bibinfo {author}
  {\bibfnamefont {C.~M.}\ \bibnamefont {Hayes}}, \bibinfo {author}
  {\bibfnamefont {R.~S.}\ \bibnamefont {Wiederkehr}},\ and\ \bibinfo {author}
  {\bibfnamefont {S.~B.}\ \bibnamefont {Mendes}},\ }\href
  {https://doi.org/https://doi.org/10.1016/j.tsf.2010.03.011} {\bibfield
  {journal} {\bibinfo  {journal} {Thin Solid Films}\ }\textbf {\bibinfo
  {volume} {518}},\ \bibinfo {pages} {4935} (\bibinfo {year}
  {2010})}\BibitemShut {NoStop}%
\bibitem [{\citenamefont {Sinclair}\ \emph {et~al.}(2020)\citenamefont
  {Sinclair}, \citenamefont {Gallacher}, \citenamefont {Sorel}, \citenamefont
  {Bayley}, \citenamefont {McBrearty}, \citenamefont {Millar}, \citenamefont
  {Hild},\ and\ \citenamefont {Paul}}]{Sinclair_20}%
  \BibitemOpen
  \bibfield  {author} {\bibinfo {author} {\bibfnamefont {M.}~\bibnamefont
  {Sinclair}}, \bibinfo {author} {\bibfnamefont {K.}~\bibnamefont {Gallacher}},
  \bibinfo {author} {\bibfnamefont {M.}~\bibnamefont {Sorel}}, \bibinfo
  {author} {\bibfnamefont {J.~C.}\ \bibnamefont {Bayley}}, \bibinfo {author}
  {\bibfnamefont {E.}~\bibnamefont {McBrearty}}, \bibinfo {author}
  {\bibfnamefont {R.~W.}\ \bibnamefont {Millar}}, \bibinfo {author}
  {\bibfnamefont {S.}~\bibnamefont {Hild}},\ and\ \bibinfo {author}
  {\bibfnamefont {D.~J.}\ \bibnamefont {Paul}},\ }\href
  {https://doi.org/10.1364/OE.381224} {\bibfield  {journal} {\bibinfo
  {journal} {Optics Express}\ }\textbf {\bibinfo {volume} {28}},\ \bibinfo
  {pages} {4010} (\bibinfo {year} {2020})}\BibitemShut {NoStop}%
\bibitem [{\citenamefont {West}\ \emph {et~al.}(2019)\citenamefont {West},
  \citenamefont {Loh}, \citenamefont {Kharas}, \citenamefont {Sorace-Agaskar},
  \citenamefont {Mehta}, \citenamefont {Sage}, \citenamefont {Chiaverini},\
  and\ \citenamefont {Ram}}]{West_19}%
  \BibitemOpen
  \bibfield  {author} {\bibinfo {author} {\bibfnamefont {G.~N.}\ \bibnamefont
  {West}}, \bibinfo {author} {\bibfnamefont {W.}~\bibnamefont {Loh}}, \bibinfo
  {author} {\bibfnamefont {D.}~\bibnamefont {Kharas}}, \bibinfo {author}
  {\bibfnamefont {C.}~\bibnamefont {Sorace-Agaskar}}, \bibinfo {author}
  {\bibfnamefont {K.~K.}\ \bibnamefont {Mehta}}, \bibinfo {author}
  {\bibfnamefont {J.}~\bibnamefont {Sage}}, \bibinfo {author} {\bibfnamefont
  {J.}~\bibnamefont {Chiaverini}},\ and\ \bibinfo {author} {\bibfnamefont
  {R.~J.}\ \bibnamefont {Ram}},\ }\bibfield  {journal} {\bibinfo  {journal}
  {Apl Photonics}\ }\textbf {\bibinfo {volume} {4}},\ \href
  {https://doi.org/10.1063/1.5052502} {10.1063/1.5052502} (\bibinfo {year}
  {2019})\BibitemShut {NoStop}%
\bibitem [{\citenamefont {Munoz}\ \emph {et~al.}(2019)\citenamefont {Munoz},
  \citenamefont {Van~Dijk}, \citenamefont {Geuzebroek}, \citenamefont
  {Geiselmann}, \citenamefont {Dominguez}, \citenamefont {Stassen},
  \citenamefont {Domenech}, \citenamefont {Zervas}, \citenamefont {Leinse},
  \citenamefont {Roeloffzen}, \citenamefont {Gargallo}, \citenamefont {Banos},
  \citenamefont {Fernandez}, \citenamefont {Cabanes}, \citenamefont {Bru},\
  and\ \citenamefont {Pastor}}]{Munoz_19}%
  \BibitemOpen
  \bibfield  {author} {\bibinfo {author} {\bibfnamefont {P.}~\bibnamefont
  {Munoz}}, \bibinfo {author} {\bibfnamefont {P.~W.~L.}\ \bibnamefont
  {Van~Dijk}}, \bibinfo {author} {\bibfnamefont {D.}~\bibnamefont
  {Geuzebroek}}, \bibinfo {author} {\bibfnamefont {M.}~\bibnamefont
  {Geiselmann}}, \bibinfo {author} {\bibfnamefont {C.}~\bibnamefont
  {Dominguez}}, \bibinfo {author} {\bibfnamefont {A.}~\bibnamefont {Stassen}},
  \bibinfo {author} {\bibfnamefont {J.~D.}\ \bibnamefont {Domenech}}, \bibinfo
  {author} {\bibfnamefont {M.}~\bibnamefont {Zervas}}, \bibinfo {author}
  {\bibfnamefont {A.}~\bibnamefont {Leinse}}, \bibinfo {author} {\bibfnamefont
  {C.~G.~H.}\ \bibnamefont {Roeloffzen}}, \bibinfo {author} {\bibfnamefont
  {B.}~\bibnamefont {Gargallo}}, \bibinfo {author} {\bibfnamefont
  {R.}~\bibnamefont {Banos}}, \bibinfo {author} {\bibfnamefont
  {J.}~\bibnamefont {Fernandez}}, \bibinfo {author} {\bibfnamefont {G.~M.}\
  \bibnamefont {Cabanes}}, \bibinfo {author} {\bibfnamefont {L.~A.}\
  \bibnamefont {Bru}},\ and\ \bibinfo {author} {\bibfnamefont {D.}~\bibnamefont
  {Pastor}},\ }\href {https://doi.org/10.1109/JSTQE.2019.2902903} {\bibfield
  {journal} {\bibinfo  {journal} {IEEE Journal of Selected Topics in Quantum
  Electronics}\ }\textbf {\bibinfo {volume} {25}},\ \bibinfo {pages} {1}
  (\bibinfo {year} {2019})}\BibitemShut {NoStop}%
\bibitem [{\citenamefont {Chang}\ \emph {et~al.}(2017)\citenamefont {Chang},
  \citenamefont {Pfeiffer}, \citenamefont {Volet}, \citenamefont {Zervas},
  \citenamefont {Peters}, \citenamefont {Manganelli}, \citenamefont {Stanton},
  \citenamefont {Li}, \citenamefont {Kippenberg},\ and\ \citenamefont
  {Bowers}}]{Chang_17}%
  \BibitemOpen
  \bibfield  {author} {\bibinfo {author} {\bibfnamefont {L.}~\bibnamefont
  {Chang}}, \bibinfo {author} {\bibfnamefont {M.~H.~P.}\ \bibnamefont
  {Pfeiffer}}, \bibinfo {author} {\bibfnamefont {N.}~\bibnamefont {Volet}},
  \bibinfo {author} {\bibfnamefont {M.}~\bibnamefont {Zervas}}, \bibinfo
  {author} {\bibfnamefont {J.~D.}\ \bibnamefont {Peters}}, \bibinfo {author}
  {\bibfnamefont {C.~L.}\ \bibnamefont {Manganelli}}, \bibinfo {author}
  {\bibfnamefont {E.~J.}\ \bibnamefont {Stanton}}, \bibinfo {author}
  {\bibfnamefont {Y.~F.}\ \bibnamefont {Li}}, \bibinfo {author} {\bibfnamefont
  {T.~J.}\ \bibnamefont {Kippenberg}},\ and\ \bibinfo {author} {\bibfnamefont
  {J.~E.}\ \bibnamefont {Bowers}},\ }\href
  {https://doi.org/10.1364/OL.42.000803} {\bibfield  {journal} {\bibinfo
  {journal} {Optics Letters}\ }\textbf {\bibinfo {volume} {42}},\ \bibinfo
  {pages} {803} (\bibinfo {year} {2017})}\BibitemShut {NoStop}%
\bibitem [{\citenamefont {Joo}\ \emph {et~al.}(2018)\citenamefont {Joo},
  \citenamefont {Park},\ and\ \citenamefont {Kim}}]{Joo_18}%
  \BibitemOpen
  \bibfield  {author} {\bibinfo {author} {\bibfnamefont {J.}~\bibnamefont
  {Joo}}, \bibinfo {author} {\bibfnamefont {J.}~\bibnamefont {Park}},\ and\
  \bibinfo {author} {\bibfnamefont {G.}~\bibnamefont {Kim}},\ }\href
  {https://doi.org/10.1109/LPT.2018.2814616} {\bibfield  {journal} {\bibinfo
  {journal} {Ieee Photonics Technology Letters}\ }\textbf {\bibinfo {volume}
  {30}},\ \bibinfo {pages} {740} (\bibinfo {year} {2018})}\BibitemShut
  {NoStop}%
\bibitem [{\citenamefont {Zheng}\ \emph {et~al.}(2019)\citenamefont {Zheng},
  \citenamefont {Domenech}, \citenamefont {Pan}, \citenamefont {Zou},
  \citenamefont {Yan},\ and\ \citenamefont {Perez}}]{Zheng_19}%
  \BibitemOpen
  \bibfield  {author} {\bibinfo {author} {\bibfnamefont {D.}~\bibnamefont
  {Zheng}}, \bibinfo {author} {\bibfnamefont {J.~D.}\ \bibnamefont {Domenech}},
  \bibinfo {author} {\bibfnamefont {W.}~\bibnamefont {Pan}}, \bibinfo {author}
  {\bibfnamefont {X.~H.}\ \bibnamefont {Zou}}, \bibinfo {author} {\bibfnamefont
  {L.~S.}\ \bibnamefont {Yan}},\ and\ \bibinfo {author} {\bibfnamefont
  {D.}~\bibnamefont {Perez}},\ }\href {https://doi.org/10.1364/OL.44.002629}
  {\bibfield  {journal} {\bibinfo  {journal} {Optics Letters}\ }\textbf
  {\bibinfo {volume} {44}},\ \bibinfo {pages} {2629} (\bibinfo {year}
  {2019})}\BibitemShut {NoStop}%
\bibitem [{\citenamefont {Sun}\ \emph {et~al.}(2016)\citenamefont {Sun},
  \citenamefont {Alam}, \citenamefont {Aitchison},\ and\ \citenamefont
  {Mojahedi}}]{Sun16}%
  \BibitemOpen
  \bibfield  {author} {\bibinfo {author} {\bibfnamefont {X.}~\bibnamefont
  {Sun}}, \bibinfo {author} {\bibfnamefont {M.~Z.}\ \bibnamefont {Alam}},
  \bibinfo {author} {\bibfnamefont {J.~S.}\ \bibnamefont {Aitchison}},\ and\
  \bibinfo {author} {\bibfnamefont {M.}~\bibnamefont {Mojahedi}},\ }\href
  {https://doi.org/10.1364/OL.41.003229} {\bibfield  {journal} {\bibinfo
  {journal} {Optics Letters}\ }\textbf {\bibinfo {volume} {41}},\ \bibinfo
  {pages} {3229} (\bibinfo {year} {2016})}\BibitemShut {NoStop}%
\bibitem [{\citenamefont {Zhan}\ \emph {et~al.}(2021)\citenamefont {Zhan},
  \citenamefont {Brock}, \citenamefont {Veilleux},\ and\ \citenamefont
  {Dagenais}}]{Zhan21}%
  \BibitemOpen
  \bibfield  {author} {\bibinfo {author} {\bibfnamefont {J.~H.}\ \bibnamefont
  {Zhan}}, \bibinfo {author} {\bibfnamefont {J.}~\bibnamefont {Brock}},
  \bibinfo {author} {\bibfnamefont {S.}~\bibnamefont {Veilleux}},\ and\
  \bibinfo {author} {\bibfnamefont {M.}~\bibnamefont {Dagenais}},\ }\href
  {https://doi.org/10.1364/OE.420499} {\bibfield  {journal} {\bibinfo
  {journal} {Optics Express}\ }\textbf {\bibinfo {volume} {29}},\ \bibinfo
  {pages} {14476} (\bibinfo {year} {2021})}\BibitemShut {NoStop}%
\bibitem [{\citenamefont {Sacher}\ \emph {et~al.}(2014)\citenamefont {Sacher},
  \citenamefont {Huang}, \citenamefont {Ding}, \citenamefont {Taylor},
  \citenamefont {Jayatilleka}, \citenamefont {Lo},\ and\ \citenamefont
  {Poon}}]{Sacher_14}%
  \BibitemOpen
  \bibfield  {author} {\bibinfo {author} {\bibfnamefont {W.~D.}\ \bibnamefont
  {Sacher}}, \bibinfo {author} {\bibfnamefont {Y.}~\bibnamefont {Huang}},
  \bibinfo {author} {\bibfnamefont {L.}~\bibnamefont {Ding}}, \bibinfo {author}
  {\bibfnamefont {B.~J.~F.}\ \bibnamefont {Taylor}}, \bibinfo {author}
  {\bibfnamefont {H.}~\bibnamefont {Jayatilleka}}, \bibinfo {author}
  {\bibfnamefont {G.~Q.}\ \bibnamefont {Lo}},\ and\ \bibinfo {author}
  {\bibfnamefont {J.~K.~S.}\ \bibnamefont {Poon}},\ }\href
  {https://doi.org/10.1364/OE.22.010938} {\bibfield  {journal} {\bibinfo
  {journal} {Optics Express}\ }\textbf {\bibinfo {volume} {22}},\ \bibinfo
  {pages} {10938} (\bibinfo {year} {2014})}\BibitemShut {NoStop}%
\bibitem [{\citenamefont {Mak}\ \emph {et~al.}(2018)\citenamefont {Mak},
  \citenamefont {Sacher}, \citenamefont {Ying}, \citenamefont {Luo},
  \citenamefont {Lo},\ and\ \citenamefont {Poon}}]{Mak_18}%
  \BibitemOpen
  \bibfield  {author} {\bibinfo {author} {\bibfnamefont {J.~C.~C.}\
  \bibnamefont {Mak}}, \bibinfo {author} {\bibfnamefont {W.~D.}\ \bibnamefont
  {Sacher}}, \bibinfo {author} {\bibfnamefont {H.}~\bibnamefont {Ying}},
  \bibinfo {author} {\bibfnamefont {X.}~\bibnamefont {Luo}}, \bibinfo {author}
  {\bibfnamefont {P.~G.-Q.}\ \bibnamefont {Lo}},\ and\ \bibinfo {author}
  {\bibfnamefont {J.~K.~S.}\ \bibnamefont {Poon}},\ }\href
  {https://doi.org/10.1364/OE.26.030623} {\bibfield  {journal} {\bibinfo
  {journal} {Optics Express}\ }\textbf {\bibinfo {volume} {26}},\ \bibinfo
  {pages} {30623} (\bibinfo {year} {2018})}\BibitemShut {NoStop}%
\bibitem [{\citenamefont {Yan}\ \emph {et~al.}(2020)\citenamefont {Yan},
  \citenamefont {Yang}, \citenamefont {Liu}, \citenamefont {Zhang},
  \citenamefont {Xia}, \citenamefont {Kang}, \citenamefont {Yang},
  \citenamefont {Qin}, \citenamefont {Deng},\ and\ \citenamefont
  {Bi}}]{Yan_20}%
  \BibitemOpen
  \bibfield  {author} {\bibinfo {author} {\bibfnamefont {W.}~\bibnamefont
  {Yan}}, \bibinfo {author} {\bibfnamefont {Y.~C.}\ \bibnamefont {Yang}},
  \bibinfo {author} {\bibfnamefont {S.~Y.}\ \bibnamefont {Liu}}, \bibinfo
  {author} {\bibfnamefont {Y.}~\bibnamefont {Zhang}}, \bibinfo {author}
  {\bibfnamefont {S.}~\bibnamefont {Xia}}, \bibinfo {author} {\bibfnamefont
  {T.~T.}\ \bibnamefont {Kang}}, \bibinfo {author} {\bibfnamefont {W.~H.}\
  \bibnamefont {Yang}}, \bibinfo {author} {\bibfnamefont {J.}~\bibnamefont
  {Qin}}, \bibinfo {author} {\bibfnamefont {L.~J.}\ \bibnamefont {Deng}},\ and\
  \bibinfo {author} {\bibfnamefont {L.}~\bibnamefont {Bi}},\ }\href
  {https://doi.org/10.1364/OPTICA.408458} {\bibfield  {journal} {\bibinfo
  {journal} {Optica}\ }\textbf {\bibinfo {volume} {7}},\ \bibinfo {pages}
  {1555} (\bibinfo {year} {2020})}\BibitemShut {NoStop}%
\bibitem [{\citenamefont {Tian}\ \emph {et~al.}(2021)\citenamefont {Tian},
  \citenamefont {Liu}, \citenamefont {Siddharth}, \citenamefont {Wang},
  \citenamefont {Blesin}, \citenamefont {He}, \citenamefont {Kippenberg},\ and\
  \citenamefont {Bhave}}]{Tian_21}%
  \BibitemOpen
  \bibfield  {author} {\bibinfo {author} {\bibfnamefont {H.}~\bibnamefont
  {Tian}}, \bibinfo {author} {\bibfnamefont {J.~Q.}\ \bibnamefont {Liu}},
  \bibinfo {author} {\bibfnamefont {A.}~\bibnamefont {Siddharth}}, \bibinfo
  {author} {\bibfnamefont {R.~N.}\ \bibnamefont {Wang}}, \bibinfo {author}
  {\bibfnamefont {T.}~\bibnamefont {Blesin}}, \bibinfo {author} {\bibfnamefont
  {J.~J.}\ \bibnamefont {He}}, \bibinfo {author} {\bibfnamefont {T.~J.}\
  \bibnamefont {Kippenberg}},\ and\ \bibinfo {author} {\bibfnamefont {S.~A.}\
  \bibnamefont {Bhave}},\ }\href {https://doi.org/10.1038/s41566-021-00882-z}
  {\bibfield  {journal} {\bibinfo  {journal} {Nature Photonics}\ }\textbf
  {\bibinfo {volume} {15}},\ \bibinfo {pages} {828} (\bibinfo {year}
  {2021})}\BibitemShut {NoStop}%
\bibitem [{\citenamefont {Okawachi}\ \emph {et~al.}(2011)\citenamefont
  {Okawachi}, \citenamefont {Saha}, \citenamefont {Levy}, \citenamefont {Wen},
  \citenamefont {Lipson},\ and\ \citenamefont {Gaeta}}]{Okawachi_11}%
  \BibitemOpen
  \bibfield  {author} {\bibinfo {author} {\bibfnamefont {Y.}~\bibnamefont
  {Okawachi}}, \bibinfo {author} {\bibfnamefont {K.}~\bibnamefont {Saha}},
  \bibinfo {author} {\bibfnamefont {J.~S.}\ \bibnamefont {Levy}}, \bibinfo
  {author} {\bibfnamefont {Y.~H.}\ \bibnamefont {Wen}}, \bibinfo {author}
  {\bibfnamefont {M.}~\bibnamefont {Lipson}},\ and\ \bibinfo {author}
  {\bibfnamefont {A.~L.}\ \bibnamefont {Gaeta}},\ }\href
  {https://doi.org/10.1364/OL.36.003398} {\bibfield  {journal} {\bibinfo
  {journal} {Optics Letters}\ }\textbf {\bibinfo {volume} {36}},\ \bibinfo
  {pages} {3398} (\bibinfo {year} {2011})}\BibitemShut {NoStop}%
\bibitem [{\citenamefont {Marin-Palomo}\ \emph {et~al.}(2017)\citenamefont
  {Marin-Palomo}, \citenamefont {Kemal}, \citenamefont {Karpov}, \citenamefont
  {Kordts}, \citenamefont {Pfeifle}, \citenamefont {Pfeiffer}, \citenamefont
  {Trocha}, \citenamefont {Wolf}, \citenamefont {Brasch}, \citenamefont
  {Anderson}, \citenamefont {Rosenberger}, \citenamefont {Vijayan},
  \citenamefont {Freude}, \citenamefont {Kippenberg},\ and\ \citenamefont
  {Koos}}]{Marin-Palomo_17}%
  \BibitemOpen
  \bibfield  {author} {\bibinfo {author} {\bibfnamefont {P.}~\bibnamefont
  {Marin-Palomo}}, \bibinfo {author} {\bibfnamefont {J.~N.}\ \bibnamefont
  {Kemal}}, \bibinfo {author} {\bibfnamefont {M.}~\bibnamefont {Karpov}},
  \bibinfo {author} {\bibfnamefont {A.}~\bibnamefont {Kordts}}, \bibinfo
  {author} {\bibfnamefont {J.}~\bibnamefont {Pfeifle}}, \bibinfo {author}
  {\bibfnamefont {M.~H.~P.}\ \bibnamefont {Pfeiffer}}, \bibinfo {author}
  {\bibfnamefont {P.}~\bibnamefont {Trocha}}, \bibinfo {author} {\bibfnamefont
  {S.}~\bibnamefont {Wolf}}, \bibinfo {author} {\bibfnamefont {V.}~\bibnamefont
  {Brasch}}, \bibinfo {author} {\bibfnamefont {M.~H.}\ \bibnamefont
  {Anderson}}, \bibinfo {author} {\bibfnamefont {R.}~\bibnamefont
  {Rosenberger}}, \bibinfo {author} {\bibfnamefont {K.}~\bibnamefont
  {Vijayan}}, \bibinfo {author} {\bibfnamefont {W.}~\bibnamefont {Freude}},
  \bibinfo {author} {\bibfnamefont {T.~J.}\ \bibnamefont {Kippenberg}},\ and\
  \bibinfo {author} {\bibfnamefont {C.}~\bibnamefont {Koos}},\ }\href
  {https://doi.org/10.1038/nature22387} {\bibfield  {journal} {\bibinfo
  {journal} {Nature}\ }\textbf {\bibinfo {volume} {546}},\ \bibinfo {pages}
  {274} (\bibinfo {year} {2017})}\BibitemShut {NoStop}%
\bibitem [{\citenamefont {Stern}\ \emph {et~al.}(2018)\citenamefont {Stern},
  \citenamefont {Ji}, \citenamefont {Okawachi}, \citenamefont {Gaeta},\ and\
  \citenamefont {Lipson}}]{Stern_18}%
  \BibitemOpen
  \bibfield  {author} {\bibinfo {author} {\bibfnamefont {B.}~\bibnamefont
  {Stern}}, \bibinfo {author} {\bibfnamefont {X.}~\bibnamefont {Ji}}, \bibinfo
  {author} {\bibfnamefont {Y.}~\bibnamefont {Okawachi}}, \bibinfo {author}
  {\bibfnamefont {A.~L.}\ \bibnamefont {Gaeta}},\ and\ \bibinfo {author}
  {\bibfnamefont {M.}~\bibnamefont {Lipson}},\ }\href
  {https://doi.org/10.1038/s41586-018-0598-9} {\bibfield  {journal} {\bibinfo
  {journal} {Nature}\ }\textbf {\bibinfo {volume} {562}},\ \bibinfo {pages}
  {401} (\bibinfo {year} {2018})}\BibitemShut {NoStop}%
\bibitem [{\citenamefont {Okawachi}\ \emph {et~al.}(2014)\citenamefont
  {Okawachi}, \citenamefont {Lamont}, \citenamefont {Luke}, \citenamefont
  {Carvalho}, \citenamefont {Yu}, \citenamefont {Lipson},\ and\ \citenamefont
  {Gaeta}}]{Okawachi_14}%
  \BibitemOpen
  \bibfield  {author} {\bibinfo {author} {\bibfnamefont {Y.}~\bibnamefont
  {Okawachi}}, \bibinfo {author} {\bibfnamefont {M.~R.~E.}\ \bibnamefont
  {Lamont}}, \bibinfo {author} {\bibfnamefont {K.}~\bibnamefont {Luke}},
  \bibinfo {author} {\bibfnamefont {D.~O.}\ \bibnamefont {Carvalho}}, \bibinfo
  {author} {\bibfnamefont {M.}~\bibnamefont {Yu}}, \bibinfo {author}
  {\bibfnamefont {M.}~\bibnamefont {Lipson}},\ and\ \bibinfo {author}
  {\bibfnamefont {A.~L.}\ \bibnamefont {Gaeta}},\ }\href
  {https://doi.org/10.1364/OL.39.003535} {\bibfield  {journal} {\bibinfo
  {journal} {Optics Letters}\ }\textbf {\bibinfo {volume} {39}},\ \bibinfo
  {pages} {3535} (\bibinfo {year} {2014})}\BibitemShut {NoStop}%
\bibitem [{\citenamefont {Brasch}\ \emph {et~al.}(2017)\citenamefont {Brasch},
  \citenamefont {Lucas}, \citenamefont {Jost}, \citenamefont {Geiselmann},\
  and\ \citenamefont {Kippenberg}}]{Brasch_17}%
  \BibitemOpen
  \bibfield  {author} {\bibinfo {author} {\bibfnamefont {V.}~\bibnamefont
  {Brasch}}, \bibinfo {author} {\bibfnamefont {E.}~\bibnamefont {Lucas}},
  \bibinfo {author} {\bibfnamefont {J.~D.}\ \bibnamefont {Jost}}, \bibinfo
  {author} {\bibfnamefont {M.}~\bibnamefont {Geiselmann}},\ and\ \bibinfo
  {author} {\bibfnamefont {T.~J.}\ \bibnamefont {Kippenberg}},\ }\bibfield
  {journal} {\bibinfo  {journal} {Light-Science and Applications}\ }\textbf
  {\bibinfo {volume} {6}},\ \href {https://doi.org/10.1038/lsa.2016.202}
  {10.1038/lsa.2016.202} (\bibinfo {year} {2017})\BibitemShut {NoStop}%
\bibitem [{\citenamefont {Pasquazi}\ \emph {et~al.}(2018)\citenamefont
  {Pasquazi}, \citenamefont {Peccianti}, \citenamefont {Razzari}, \citenamefont
  {Moss}, \citenamefont {Coen}, \citenamefont {Erkintalo}, \citenamefont
  {Chembo}, \citenamefont {Hansson}, \citenamefont {Wabnitz}, \citenamefont
  {Del'Haye}, \citenamefont {Xue}, \citenamefont {Weiner},\ and\ \citenamefont
  {Morandotti}}]{Pasquazi_18}%
  \BibitemOpen
  \bibfield  {author} {\bibinfo {author} {\bibfnamefont {A.}~\bibnamefont
  {Pasquazi}}, \bibinfo {author} {\bibfnamefont {M.}~\bibnamefont {Peccianti}},
  \bibinfo {author} {\bibfnamefont {L.}~\bibnamefont {Razzari}}, \bibinfo
  {author} {\bibfnamefont {D.~J.}\ \bibnamefont {Moss}}, \bibinfo {author}
  {\bibfnamefont {S.}~\bibnamefont {Coen}}, \bibinfo {author} {\bibfnamefont
  {M.}~\bibnamefont {Erkintalo}}, \bibinfo {author} {\bibfnamefont {Y.~K.}\
  \bibnamefont {Chembo}}, \bibinfo {author} {\bibfnamefont {T.}~\bibnamefont
  {Hansson}}, \bibinfo {author} {\bibfnamefont {S.}~\bibnamefont {Wabnitz}},
  \bibinfo {author} {\bibfnamefont {P.}~\bibnamefont {Del'Haye}}, \bibinfo
  {author} {\bibfnamefont {X.}~\bibnamefont {Xue}}, \bibinfo {author}
  {\bibfnamefont {A.~M.}\ \bibnamefont {Weiner}},\ and\ \bibinfo {author}
  {\bibfnamefont {R.}~\bibnamefont {Morandotti}},\ }\href
  {https://doi.org/https://doi.org/10.1016/j.physrep.2017.08.004} {\bibfield
  {journal} {\bibinfo  {journal} {Physics Reports-Review Section of Physics
  Letters}\ }\textbf {\bibinfo {volume} {729}},\ \bibinfo {pages} {1} (\bibinfo
  {year} {2018})}\BibitemShut {NoStop}%
\bibitem [{\citenamefont {Miller}\ \emph {et~al.}(2014)\citenamefont {Miller},
  \citenamefont {Luke}, \citenamefont {Okawachi}, \citenamefont {Cardenas},
  \citenamefont {Gaeta},\ and\ \citenamefont {Lipson}}]{Miller_14}%
  \BibitemOpen
  \bibfield  {author} {\bibinfo {author} {\bibfnamefont {S.}~\bibnamefont
  {Miller}}, \bibinfo {author} {\bibfnamefont {K.}~\bibnamefont {Luke}},
  \bibinfo {author} {\bibfnamefont {Y.}~\bibnamefont {Okawachi}}, \bibinfo
  {author} {\bibfnamefont {J.}~\bibnamefont {Cardenas}}, \bibinfo {author}
  {\bibfnamefont {A.~L.}\ \bibnamefont {Gaeta}},\ and\ \bibinfo {author}
  {\bibfnamefont {M.}~\bibnamefont {Lipson}},\ }\href
  {https://doi.org/10.1364/OE.22.026517} {\bibfield  {journal} {\bibinfo
  {journal} {Optics Express}\ }\textbf {\bibinfo {volume} {22}},\ \bibinfo
  {pages} {26517} (\bibinfo {year} {2014})}\BibitemShut {NoStop}%
\bibitem [{\citenamefont {Wang}\ \emph {et~al.}(2016)\citenamefont {Wang},
  \citenamefont {Chang}, \citenamefont {Volet}, \citenamefont {Pfeiffer},
  \citenamefont {Zervas}, \citenamefont {Guo}, \citenamefont {Kippenberg},\
  and\ \citenamefont {Bowers}}]{Wang_16}%
  \BibitemOpen
  \bibfield  {author} {\bibinfo {author} {\bibfnamefont {L.}~\bibnamefont
  {Wang}}, \bibinfo {author} {\bibfnamefont {L.}~\bibnamefont {Chang}},
  \bibinfo {author} {\bibfnamefont {N.}~\bibnamefont {Volet}}, \bibinfo
  {author} {\bibfnamefont {M.~H.~P.}\ \bibnamefont {Pfeiffer}}, \bibinfo
  {author} {\bibfnamefont {M.}~\bibnamefont {Zervas}}, \bibinfo {author}
  {\bibfnamefont {H.}~\bibnamefont {Guo}}, \bibinfo {author} {\bibfnamefont
  {T.~J.}\ \bibnamefont {Kippenberg}},\ and\ \bibinfo {author} {\bibfnamefont
  {J.~E.}\ \bibnamefont {Bowers}},\ }\href
  {https://doi.org/https://doi.org/10.1002/lpor.201600006} {\bibfield
  {journal} {\bibinfo  {journal} {Laser \& Photonics Reviews}\ }\textbf
  {\bibinfo {volume} {10}},\ \bibinfo {pages} {631} (\bibinfo {year}
  {2016})}\BibitemShut {NoStop}%
\bibitem [{\citenamefont {Newman}\ \emph {et~al.}(2019)\citenamefont {Newman},
  \citenamefont {Maurice}, \citenamefont {Drake}, \citenamefont {Stone},
  \citenamefont {Briles}, \citenamefont {Spencer}, \citenamefont {Fredrick},
  \citenamefont {Li}, \citenamefont {Westly}, \citenamefont {Ilic},
  \citenamefont {Shen}, \citenamefont {Suh}, \citenamefont {Yang},
  \citenamefont {Johnson}, \citenamefont {Johnson}, \citenamefont {Hollberg},
  \citenamefont {Vahala}, \citenamefont {Srinivasan}, \citenamefont {Diddams},
  \citenamefont {Kitching}, \citenamefont {Papp},\ and\ \citenamefont
  {Hummon}}]{Newman:19}%
  \BibitemOpen
  \bibfield  {author} {\bibinfo {author} {\bibfnamefont {Z.~L.}\ \bibnamefont
  {Newman}}, \bibinfo {author} {\bibfnamefont {V.}~\bibnamefont {Maurice}},
  \bibinfo {author} {\bibfnamefont {T.}~\bibnamefont {Drake}}, \bibinfo
  {author} {\bibfnamefont {J.~R.}\ \bibnamefont {Stone}}, \bibinfo {author}
  {\bibfnamefont {T.~C.}\ \bibnamefont {Briles}}, \bibinfo {author}
  {\bibfnamefont {D.~T.}\ \bibnamefont {Spencer}}, \bibinfo {author}
  {\bibfnamefont {C.}~\bibnamefont {Fredrick}}, \bibinfo {author}
  {\bibfnamefont {Q.}~\bibnamefont {Li}}, \bibinfo {author} {\bibfnamefont
  {D.}~\bibnamefont {Westly}}, \bibinfo {author} {\bibfnamefont {B.~R.}\
  \bibnamefont {Ilic}}, \bibinfo {author} {\bibfnamefont {B.}~\bibnamefont
  {Shen}}, \bibinfo {author} {\bibfnamefont {M.-G.}\ \bibnamefont {Suh}},
  \bibinfo {author} {\bibfnamefont {K.~Y.}\ \bibnamefont {Yang}}, \bibinfo
  {author} {\bibfnamefont {C.}~\bibnamefont {Johnson}}, \bibinfo {author}
  {\bibfnamefont {D.~M.~S.}\ \bibnamefont {Johnson}}, \bibinfo {author}
  {\bibfnamefont {L.}~\bibnamefont {Hollberg}}, \bibinfo {author}
  {\bibfnamefont {K.~J.}\ \bibnamefont {Vahala}}, \bibinfo {author}
  {\bibfnamefont {K.}~\bibnamefont {Srinivasan}}, \bibinfo {author}
  {\bibfnamefont {S.~A.}\ \bibnamefont {Diddams}}, \bibinfo {author}
  {\bibfnamefont {J.}~\bibnamefont {Kitching}}, \bibinfo {author}
  {\bibfnamefont {S.~B.}\ \bibnamefont {Papp}},\ and\ \bibinfo {author}
  {\bibfnamefont {M.~T.}\ \bibnamefont {Hummon}},\ }\href
  {https://doi.org/10.1364/OPTICA.6.000680} {\bibfield  {journal} {\bibinfo
  {journal} {Optica}\ }\textbf {\bibinfo {volume} {6}},\ \bibinfo {pages} {680}
  (\bibinfo {year} {2019})}\BibitemShut {NoStop}%
\bibitem [{\citenamefont {Arbabi}\ and\ \citenamefont
  {Goddard}(2013)}]{Arbabi_13}%
  \BibitemOpen
  \bibfield  {author} {\bibinfo {author} {\bibfnamefont {A.}~\bibnamefont
  {Arbabi}}\ and\ \bibinfo {author} {\bibfnamefont {L.~L.}\ \bibnamefont
  {Goddard}},\ }\href {https://doi.org/10.1364/OL.38.003878} {\bibfield
  {journal} {\bibinfo  {journal} {Optics Letters}\ }\textbf {\bibinfo {volume}
  {38}},\ \bibinfo {pages} {3878} (\bibinfo {year} {2013})}\BibitemShut
  {NoStop}%
\bibitem [{\citenamefont {Wilkes}\ \emph {et~al.}(2016)\citenamefont {Wilkes},
  \citenamefont {Qiang}, \citenamefont {Wang}, \citenamefont {Santagati},
  \citenamefont {Paesani}, \citenamefont {Zhou}, \citenamefont {Miller},
  \citenamefont {Marshall}, \citenamefont {Thompson},\ and\ \citenamefont
  {O’Brien}}]{Wilkes_16}%
  \BibitemOpen
  \bibfield  {author} {\bibinfo {author} {\bibfnamefont {C.~M.}\ \bibnamefont
  {Wilkes}}, \bibinfo {author} {\bibfnamefont {X.}~\bibnamefont {Qiang}},
  \bibinfo {author} {\bibfnamefont {J.}~\bibnamefont {Wang}}, \bibinfo {author}
  {\bibfnamefont {R.}~\bibnamefont {Santagati}}, \bibinfo {author}
  {\bibfnamefont {S.}~\bibnamefont {Paesani}}, \bibinfo {author} {\bibfnamefont
  {X.}~\bibnamefont {Zhou}}, \bibinfo {author} {\bibfnamefont {D.~A.~B.}\
  \bibnamefont {Miller}}, \bibinfo {author} {\bibfnamefont {G.~D.}\
  \bibnamefont {Marshall}}, \bibinfo {author} {\bibfnamefont {M.~G.}\
  \bibnamefont {Thompson}},\ and\ \bibinfo {author} {\bibfnamefont {J.~L.}\
  \bibnamefont {O’Brien}},\ }\href {https://doi.org/10.1364/OL.41.005318}
  {\bibfield  {journal} {\bibinfo  {journal} {Optics Letters}\ }\textbf
  {\bibinfo {volume} {41}},\ \bibinfo {pages} {5318} (\bibinfo {year}
  {2016})}\BibitemShut {NoStop}%
\bibitem [{\citenamefont {Nagai}\ and\ \citenamefont {Hane}(2018)}]{Nagai_18}%
  \BibitemOpen
  \bibfield  {author} {\bibinfo {author} {\bibfnamefont {T.}~\bibnamefont
  {Nagai}}\ and\ \bibinfo {author} {\bibfnamefont {K.}~\bibnamefont {Hane}},\
  }\href {https://doi.org/10.1364/oe.26.033906} {\bibfield  {journal} {\bibinfo
   {journal} {Optics Express}\ }\textbf {\bibinfo {volume} {26}},\ \bibinfo
  {pages} {33906} (\bibinfo {year} {2018})}\BibitemShut {NoStop}%
\bibitem [{\citenamefont {Shiramin}\ and\ \citenamefont
  {Van~Thourhout}(2017)}]{Shiramin_17}%
  \BibitemOpen
  \bibfield  {author} {\bibinfo {author} {\bibfnamefont {L.~A.}\ \bibnamefont
  {Shiramin}}\ and\ \bibinfo {author} {\bibfnamefont {D.}~\bibnamefont
  {Van~Thourhout}},\ }\bibfield  {journal} {\bibinfo  {journal} {Ieee Journal
  of Selected Topics in Quantum Electronics}\ }\textbf {\bibinfo {volume}
  {23}},\ \href {https://doi.org/10.1109/jstqe.2016.2586458}
  {10.1109/jstqe.2016.2586458} (\bibinfo {year} {2017})\BibitemShut {NoStop}%
\bibitem [{\citenamefont {Alexander}\ \emph {et~al.}(2018)\citenamefont
  {Alexander}, \citenamefont {George}, \citenamefont {Verbist}, \citenamefont
  {Neyts}, \citenamefont {Kuyken}, \citenamefont {Van~Thourhout},\ and\
  \citenamefont {Beeckman}}]{Alexander_18}%
  \BibitemOpen
  \bibfield  {author} {\bibinfo {author} {\bibfnamefont {K.}~\bibnamefont
  {Alexander}}, \bibinfo {author} {\bibfnamefont {J.~P.}\ \bibnamefont
  {George}}, \bibinfo {author} {\bibfnamefont {J.}~\bibnamefont {Verbist}},
  \bibinfo {author} {\bibfnamefont {K.}~\bibnamefont {Neyts}}, \bibinfo
  {author} {\bibfnamefont {B.}~\bibnamefont {Kuyken}}, \bibinfo {author}
  {\bibfnamefont {D.}~\bibnamefont {Van~Thourhout}},\ and\ \bibinfo {author}
  {\bibfnamefont {J.}~\bibnamefont {Beeckman}},\ }\bibfield  {journal}
  {\bibinfo  {journal} {Nature Communications}\ }\textbf {\bibinfo {volume}
  {9}},\ \href {https://doi.org/10.1038/s41467-018-05846-6}
  {10.1038/s41467-018-05846-6} (\bibinfo {year} {2018})\BibitemShut {NoStop}%
\bibitem [{\citenamefont {Jin}\ \emph {et~al.}(2016)\citenamefont {Jin},
  \citenamefont {Xu}, \citenamefont {Zhang},\ and\ \citenamefont
  {Li}}]{Jin_16}%
  \BibitemOpen
  \bibfield  {author} {\bibinfo {author} {\bibfnamefont {S.}~\bibnamefont
  {Jin}}, \bibinfo {author} {\bibfnamefont {L.}~\bibnamefont {Xu}}, \bibinfo
  {author} {\bibfnamefont {H.}~\bibnamefont {Zhang}},\ and\ \bibinfo {author}
  {\bibfnamefont {Y.}~\bibnamefont {Li}},\ }\href
  {https://doi.org/10.1109/lpt.2015.2507136} {\bibfield  {journal} {\bibinfo
  {journal} {Ieee Photonics Technology Letters}\ }\textbf {\bibinfo {volume}
  {28}},\ \bibinfo {pages} {736} (\bibinfo {year} {2016})}\BibitemShut
  {NoStop}%
\bibitem [{\citenamefont {Stadler}\ and\ \citenamefont
  {Mizumoto}(2014)}]{Stadler_14}%
  \BibitemOpen
  \bibfield  {author} {\bibinfo {author} {\bibfnamefont {B.~J.~H.}\
  \bibnamefont {Stadler}}\ and\ \bibinfo {author} {\bibfnamefont
  {T.}~\bibnamefont {Mizumoto}},\ }\bibfield  {journal} {\bibinfo  {journal}
  {Ieee Photonics Journal}\ }\textbf {\bibinfo {volume} {6}},\ \href
  {https://doi.org/10.1109/JPHOT.2013.2293618} {10.1109/JPHOT.2013.2293618}
  (\bibinfo {year} {2014})\BibitemShut {NoStop}%
\bibitem [{\citenamefont {Shoji}\ \emph {et~al.}(2008)\citenamefont {Shoji},
  \citenamefont {Mizumoto}, \citenamefont {Yokoi}, \citenamefont {Hsieh},\ and\
  \citenamefont {Osgood}}]{Shoji_08}%
  \BibitemOpen
  \bibfield  {author} {\bibinfo {author} {\bibfnamefont {Y.}~\bibnamefont
  {Shoji}}, \bibinfo {author} {\bibfnamefont {T.}~\bibnamefont {Mizumoto}},
  \bibinfo {author} {\bibfnamefont {H.}~\bibnamefont {Yokoi}}, \bibinfo
  {author} {\bibfnamefont {I.~W.}\ \bibnamefont {Hsieh}},\ and\ \bibinfo
  {author} {\bibfnamefont {R.~M.}\ \bibnamefont {Osgood}},\ }\bibfield
  {journal} {\bibinfo  {journal} {Applied Physics Letters}\ }\textbf {\bibinfo
  {volume} {92}},\ \href {https://doi.org/10.1063/1.2884855}
  {10.1063/1.2884855} (\bibinfo {year} {2008})\BibitemShut {NoStop}%
\bibitem [{\citenamefont {Dai}\ \emph {et~al.}(2012)\citenamefont {Dai},
  \citenamefont {Bauters},\ and\ \citenamefont {Bowers}}]{Dai_12}%
  \BibitemOpen
  \bibfield  {author} {\bibinfo {author} {\bibfnamefont {D.~X.}\ \bibnamefont
  {Dai}}, \bibinfo {author} {\bibfnamefont {J.}~\bibnamefont {Bauters}},\ and\
  \bibinfo {author} {\bibfnamefont {J.~E.}\ \bibnamefont {Bowers}},\ }\bibfield
   {journal} {\bibinfo  {journal} {Light-Science and Applications}\ }\textbf
  {\bibinfo {volume} {1}},\ \href {https://doi.org/10.1038/lsa.2012.1}
  {10.1038/lsa.2012.1} (\bibinfo {year} {2012})\BibitemShut {NoStop}%
\bibitem [{\citenamefont {Zhang}\ \emph {et~al.}(2019)\citenamefont {Zhang},
  \citenamefont {Du}, \citenamefont {Wang}, \citenamefont {Fakhrul},
  \citenamefont {Liu}, \citenamefont {Deng}, \citenamefont {Huang},
  \citenamefont {Pintus}, \citenamefont {Bowers}, \citenamefont {Ross},
  \citenamefont {Hu},\ and\ \citenamefont {Bi}}]{Zhang_19}%
  \BibitemOpen
  \bibfield  {author} {\bibinfo {author} {\bibfnamefont {Y.}~\bibnamefont
  {Zhang}}, \bibinfo {author} {\bibfnamefont {Q.~Y.}\ \bibnamefont {Du}},
  \bibinfo {author} {\bibfnamefont {C.~T.}\ \bibnamefont {Wang}}, \bibinfo
  {author} {\bibfnamefont {T.}~\bibnamefont {Fakhrul}}, \bibinfo {author}
  {\bibfnamefont {S.~Y.}\ \bibnamefont {Liu}}, \bibinfo {author} {\bibfnamefont
  {L.~J.}\ \bibnamefont {Deng}}, \bibinfo {author} {\bibfnamefont {D.~N.}\
  \bibnamefont {Huang}}, \bibinfo {author} {\bibfnamefont {P.}~\bibnamefont
  {Pintus}}, \bibinfo {author} {\bibfnamefont {J.}~\bibnamefont {Bowers}},
  \bibinfo {author} {\bibfnamefont {C.~A.}\ \bibnamefont {Ross}}, \bibinfo
  {author} {\bibfnamefont {J.~J.}\ \bibnamefont {Hu}},\ and\ \bibinfo {author}
  {\bibfnamefont {L.}~\bibnamefont {Bi}},\ }\href
  {https://doi.org/10.1364/OPTICA.6.000473} {\bibfield  {journal} {\bibinfo
  {journal} {Optica}\ }\textbf {\bibinfo {volume} {6}},\ \bibinfo {pages} {473}
  (\bibinfo {year} {2019})}\BibitemShut {NoStop}%
\bibitem [{\citenamefont {Goto}\ \emph {et~al.}(2012)\citenamefont {Goto},
  \citenamefont {Onbasli},\ and\ \citenamefont {Ross}}]{Goto_12}%
  \BibitemOpen
  \bibfield  {author} {\bibinfo {author} {\bibfnamefont {T.}~\bibnamefont
  {Goto}}, \bibinfo {author} {\bibfnamefont {M.~C.}\ \bibnamefont {Onbasli}},\
  and\ \bibinfo {author} {\bibfnamefont {C.~A.}\ \bibnamefont {Ross}},\ }\href
  {https://doi.org/10.1364/OE.20.028507} {\bibfield  {journal} {\bibinfo
  {journal} {Optics Express}\ }\textbf {\bibinfo {volume} {20}},\ \bibinfo
  {pages} {28507} (\bibinfo {year} {2012})}\BibitemShut {NoStop}%
\bibitem [{\citenamefont {Onbasli}\ \emph {et~al.}(2016)\citenamefont
  {Onbasli}, \citenamefont {Beran}, \citenamefont {Zahradnik}, \citenamefont
  {Kucera}, \citenamefont {Antos}, \citenamefont {Mistrik}, \citenamefont
  {Dionne}, \citenamefont {Veis},\ and\ \citenamefont {Ross}}]{Onbasli_16}%
  \BibitemOpen
  \bibfield  {author} {\bibinfo {author} {\bibfnamefont {M.~C.}\ \bibnamefont
  {Onbasli}}, \bibinfo {author} {\bibfnamefont {L.}~\bibnamefont {Beran}},
  \bibinfo {author} {\bibfnamefont {M.}~\bibnamefont {Zahradnik}}, \bibinfo
  {author} {\bibfnamefont {M.}~\bibnamefont {Kucera}}, \bibinfo {author}
  {\bibfnamefont {R.}~\bibnamefont {Antos}}, \bibinfo {author} {\bibfnamefont
  {J.}~\bibnamefont {Mistrik}}, \bibinfo {author} {\bibfnamefont {G.~F.}\
  \bibnamefont {Dionne}}, \bibinfo {author} {\bibfnamefont {M.}~\bibnamefont
  {Veis}},\ and\ \bibinfo {author} {\bibfnamefont {C.~A.}\ \bibnamefont
  {Ross}},\ }\bibfield  {journal} {\bibinfo  {journal} {Scientific Reports}\
  }\textbf {\bibinfo {volume} {6}},\ \href {https://doi.org/10.1038/srep23640}
  {10.1038/srep23640} (\bibinfo {year} {2016})\BibitemShut {NoStop}%
\bibitem [{\citenamefont {Hua}\ \emph {et~al.}(2016)\citenamefont {Hua},
  \citenamefont {Wen}, \citenamefont {Jiang}, \citenamefont {Hua},
  \citenamefont {Jiang},\ and\ \citenamefont {Xiao}}]{Hua_16}%
  \BibitemOpen
  \bibfield  {author} {\bibinfo {author} {\bibfnamefont {S.~Y.}\ \bibnamefont
  {Hua}}, \bibinfo {author} {\bibfnamefont {J.~M.}\ \bibnamefont {Wen}},
  \bibinfo {author} {\bibfnamefont {X.~S.}\ \bibnamefont {Jiang}}, \bibinfo
  {author} {\bibfnamefont {Q.}~\bibnamefont {Hua}}, \bibinfo {author}
  {\bibfnamefont {L.}~\bibnamefont {Jiang}},\ and\ \bibinfo {author}
  {\bibfnamefont {M.}~\bibnamefont {Xiao}},\ }\href
  {https://doi.org/10.1038/ncomms13657} {\bibfield  {journal} {\bibinfo
  {journal} {Nature Communications}\ }\textbf {\bibinfo {volume} {7}},\
  \bibinfo {pages} {13657} (\bibinfo {year} {2016})}\BibitemShut {NoStop}%
\bibitem [{\citenamefont {Kittlaus}\ \emph {et~al.}(2018)\citenamefont
  {Kittlaus}, \citenamefont {Otterstrom}, \citenamefont {Kharel}, \citenamefont
  {Gertler},\ and\ \citenamefont {Rakich}}]{Kittlaus_18}%
  \BibitemOpen
  \bibfield  {author} {\bibinfo {author} {\bibfnamefont {E.~A.}\ \bibnamefont
  {Kittlaus}}, \bibinfo {author} {\bibfnamefont {N.~T.}\ \bibnamefont
  {Otterstrom}}, \bibinfo {author} {\bibfnamefont {P.}~\bibnamefont {Kharel}},
  \bibinfo {author} {\bibfnamefont {S.}~\bibnamefont {Gertler}},\ and\ \bibinfo
  {author} {\bibfnamefont {P.~T.}\ \bibnamefont {Rakich}},\ }\href
  {https://doi.org/10.1038/s41566-018-0254-9} {\bibfield  {journal} {\bibinfo
  {journal} {Nature Photonics}\ }\textbf {\bibinfo {volume} {12}},\ \bibinfo
  {pages} {613} (\bibinfo {year} {2018})}\BibitemShut {NoStop}%
\bibitem [{\citenamefont {Kittlaus}\ \emph {et~al.}(2021)\citenamefont
  {Kittlaus}, \citenamefont {Jones}, \citenamefont {Rakich}, \citenamefont
  {Otterstrom}, \citenamefont {Muller},\ and\ \citenamefont
  {Rais-Zadeh}}]{Kittlaus_21}%
  \BibitemOpen
  \bibfield  {author} {\bibinfo {author} {\bibfnamefont {E.~A.}\ \bibnamefont
  {Kittlaus}}, \bibinfo {author} {\bibfnamefont {W.~M.}\ \bibnamefont {Jones}},
  \bibinfo {author} {\bibfnamefont {P.~T.}\ \bibnamefont {Rakich}}, \bibinfo
  {author} {\bibfnamefont {N.~T.}\ \bibnamefont {Otterstrom}}, \bibinfo
  {author} {\bibfnamefont {R.~E.}\ \bibnamefont {Muller}},\ and\ \bibinfo
  {author} {\bibfnamefont {M.}~\bibnamefont {Rais-Zadeh}},\ }\href
  {https://doi.org/10.1038/s41566-020-00711-9} {\bibfield  {journal} {\bibinfo
  {journal} {Nature Photonics}\ }\textbf {\bibinfo {volume} {15}},\ \bibinfo
  {pages} {43} (\bibinfo {year} {2021})}\BibitemShut {NoStop}%
\bibitem [{\citenamefont {Dai}(2018)}]{Dai18}%
  \BibitemOpen
  \bibfield  {author} {\bibinfo {author} {\bibfnamefont {D.}~\bibnamefont
  {Dai}},\ }\href {https://doi.org/10.1109/JPROC.2018.2822787} {\bibfield
  {journal} {\bibinfo  {journal} {Proceedings of the IEEE}\ }\textbf {\bibinfo
  {volume} {106}},\ \bibinfo {pages} {2117} (\bibinfo {year}
  {2018})}\BibitemShut {NoStop}%
\bibitem [{\citenamefont {Gallacher}\ \emph {et~al.}(2022)\citenamefont
  {Gallacher}, \citenamefont {Griffin}, \citenamefont {Riis}, \citenamefont
  {Sorel},\ and\ \citenamefont {Paul}}]{Gallacher_22}%
  \BibitemOpen
  \bibfield  {author} {\bibinfo {author} {\bibfnamefont {K.}~\bibnamefont
  {Gallacher}}, \bibinfo {author} {\bibfnamefont {P.~F.}\ \bibnamefont
  {Griffin}}, \bibinfo {author} {\bibfnamefont {E.}~\bibnamefont {Riis}},
  \bibinfo {author} {\bibfnamefont {M.}~\bibnamefont {Sorel}},\ and\ \bibinfo
  {author} {\bibfnamefont {D.~J.}\ \bibnamefont {Paul}},\ }\bibfield  {journal}
  {\bibinfo  {journal} {Apl Photonics}\ }\textbf {\bibinfo {volume} {7}},\
  \href {https://doi.org/10.1063/5.0077738} {10.1063/5.0077738} (\bibinfo
  {year} {2022})\BibitemShut {NoStop}%
\bibitem [{\citenamefont {Yang}\ \emph {et~al.}(2007)\citenamefont {Yang},
  \citenamefont {Conkey}, \citenamefont {Wu}, \citenamefont {Yin},
  \citenamefont {Hawkins},\ and\ \citenamefont {Schmidt}}]{Yang2007}%
  \BibitemOpen
  \bibfield  {author} {\bibinfo {author} {\bibfnamefont {W.}~\bibnamefont
  {Yang}}, \bibinfo {author} {\bibfnamefont {D.~B.}\ \bibnamefont {Conkey}},
  \bibinfo {author} {\bibfnamefont {B.}~\bibnamefont {Wu}}, \bibinfo {author}
  {\bibfnamefont {D.}~\bibnamefont {Yin}}, \bibinfo {author} {\bibfnamefont
  {A.~R.}\ \bibnamefont {Hawkins}},\ and\ \bibinfo {author} {\bibfnamefont
  {H.}~\bibnamefont {Schmidt}},\ }\href
  {https://doi.org/10.1038/nphoton.2007.74} {\bibfield  {journal} {\bibinfo
  {journal} {Nature Photonics}\ }\textbf {\bibinfo {volume} {1}},\ \bibinfo
  {pages} {331} (\bibinfo {year} {2007})}\BibitemShut {NoStop}%
\bibitem [{\citenamefont {Zektzer}\ \emph {et~al.}(2021)\citenamefont
  {Zektzer}, \citenamefont {Mazurski}, \citenamefont {Barash},\ and\
  \citenamefont {Levy}}]{Zektzer21}%
  \BibitemOpen
  \bibfield  {author} {\bibinfo {author} {\bibfnamefont {R.}~\bibnamefont
  {Zektzer}}, \bibinfo {author} {\bibfnamefont {N.}~\bibnamefont {Mazurski}},
  \bibinfo {author} {\bibfnamefont {Y.}~\bibnamefont {Barash}},\ and\ \bibinfo
  {author} {\bibfnamefont {U.}~\bibnamefont {Levy}},\ }\bibfield  {journal}
  {\bibinfo  {journal} {Nature Photonics}\ }\href
  {https://doi.org/10.1038/s41566-021-00853-4} {10.1038/s41566-021-00853-4}
  (\bibinfo {year} {2021})\BibitemShut {NoStop}%
\bibitem [{\citenamefont {Kim}\ \emph {et~al.}(2018)\citenamefont {Kim},
  \citenamefont {Westly}, \citenamefont {Roxworthy}, \citenamefont {Li},
  \citenamefont {Yulaev}, \citenamefont {Srinivasan},\ and\ \citenamefont
  {Aksyuk}}]{Kim2018}%
  \BibitemOpen
  \bibfield  {author} {\bibinfo {author} {\bibfnamefont {S.}~\bibnamefont
  {Kim}}, \bibinfo {author} {\bibfnamefont {D.~A.}\ \bibnamefont {Westly}},
  \bibinfo {author} {\bibfnamefont {B.~J.}\ \bibnamefont {Roxworthy}}, \bibinfo
  {author} {\bibfnamefont {Q.}~\bibnamefont {Li}}, \bibinfo {author}
  {\bibfnamefont {A.}~\bibnamefont {Yulaev}}, \bibinfo {author} {\bibfnamefont
  {K.}~\bibnamefont {Srinivasan}},\ and\ \bibinfo {author} {\bibfnamefont
  {V.~A.}\ \bibnamefont {Aksyuk}},\ }\href
  {https://doi.org/10.1038/s41377-018-0073-2} {\bibfield  {journal} {\bibinfo
  {journal} {Light: Science {\&} Applications}\ }\textbf {\bibinfo {volume}
  {7}},\ \bibinfo {pages} {72} (\bibinfo {year} {2018})}\BibitemShut {NoStop}%
\bibitem [{\citenamefont {Yulaev}\ \emph {et~al.}(2021)\citenamefont {Yulaev},
  \citenamefont {Zhu}, \citenamefont {Ropp}, \citenamefont {Westly},
  \citenamefont {Simelgor}, \citenamefont {Zhang}, \citenamefont {Lezec},
  \citenamefont {Agrawal},\ and\ \citenamefont {Aksyuk}}]{Yulaev:21}%
  \BibitemOpen
  \bibfield  {author} {\bibinfo {author} {\bibfnamefont {A.}~\bibnamefont
  {Yulaev}}, \bibinfo {author} {\bibfnamefont {W.}~\bibnamefont {Zhu}},
  \bibinfo {author} {\bibfnamefont {C.}~\bibnamefont {Ropp}}, \bibinfo {author}
  {\bibfnamefont {D.~A.}\ \bibnamefont {Westly}}, \bibinfo {author}
  {\bibfnamefont {G.}~\bibnamefont {Simelgor}}, \bibinfo {author}
  {\bibfnamefont {C.}~\bibnamefont {Zhang}}, \bibinfo {author} {\bibfnamefont
  {H.~J.}\ \bibnamefont {Lezec}}, \bibinfo {author} {\bibfnamefont
  {A.}~\bibnamefont {Agrawal}},\ and\ \bibinfo {author} {\bibfnamefont {V.~A.}\
  \bibnamefont {Aksyuk}},\ }in\ \href {https://doi.org/10.1364/OFC.2021.F2B.1}
  {\emph {\bibinfo {booktitle} {Optical Fiber Communication Conference (OFC)
  2021}}}\ (\bibinfo  {publisher} {Optica Publishing Group},\ \bibinfo {year}
  {2021})\ p.\ \bibinfo {pages} {F2B.1}\BibitemShut {NoStop}%
\bibitem [{\citenamefont {McGehee}\ \emph {et~al.}(2021)\citenamefont
  {McGehee}, \citenamefont {Zhu}, \citenamefont {Barker}, \citenamefont
  {Westly}, \citenamefont {Yulaev}, \citenamefont {Klimov}, \citenamefont
  {Agrawal}, \citenamefont {Eckel}, \citenamefont {Aksyuk},\ and\ \citenamefont
  {McClelland}}]{McGehee_2021}%
  \BibitemOpen
  \bibfield  {author} {\bibinfo {author} {\bibfnamefont {W.~R.}\ \bibnamefont
  {McGehee}}, \bibinfo {author} {\bibfnamefont {W.}~\bibnamefont {Zhu}},
  \bibinfo {author} {\bibfnamefont {D.~S.}\ \bibnamefont {Barker}}, \bibinfo
  {author} {\bibfnamefont {D.}~\bibnamefont {Westly}}, \bibinfo {author}
  {\bibfnamefont {A.}~\bibnamefont {Yulaev}}, \bibinfo {author} {\bibfnamefont
  {N.}~\bibnamefont {Klimov}}, \bibinfo {author} {\bibfnamefont
  {A.}~\bibnamefont {Agrawal}}, \bibinfo {author} {\bibfnamefont
  {S.}~\bibnamefont {Eckel}}, \bibinfo {author} {\bibfnamefont
  {V.}~\bibnamefont {Aksyuk}},\ and\ \bibinfo {author} {\bibfnamefont {J.~J.}\
  \bibnamefont {McClelland}},\ }\href
  {https://doi.org/10.1088/1367-2630/abdce3} {\bibfield  {journal} {\bibinfo
  {journal} {New Journal of Physics}\ }\textbf {\bibinfo {volume} {23}},\
  \bibinfo {pages} {013021} (\bibinfo {year} {2021})}\BibitemShut {NoStop}%
\bibitem [{\citenamefont {Hunter}\ \emph {et~al.}(2018)\citenamefont {Hunter},
  \citenamefont {Jim\'{e}nez-Mart\'{i}nez}, \citenamefont {Herbsommer},
  \citenamefont {Ramaswamy}, \citenamefont {Li},\ and\ \citenamefont
  {Riis}}]{Hunter:18}%
  \BibitemOpen
  \bibfield  {author} {\bibinfo {author} {\bibfnamefont {D.}~\bibnamefont
  {Hunter}}, \bibinfo {author} {\bibfnamefont {R.}~\bibnamefont
  {Jim\'{e}nez-Mart\'{i}nez}}, \bibinfo {author} {\bibfnamefont
  {J.}~\bibnamefont {Herbsommer}}, \bibinfo {author} {\bibfnamefont
  {S.}~\bibnamefont {Ramaswamy}}, \bibinfo {author} {\bibfnamefont
  {W.}~\bibnamefont {Li}},\ and\ \bibinfo {author} {\bibfnamefont
  {E.}~\bibnamefont {Riis}},\ }\href {https://doi.org/10.1364/OE.26.030523}
  {\bibfield  {journal} {\bibinfo  {journal} {Opt. Express}\ }\textbf {\bibinfo
  {volume} {26}},\ \bibinfo {pages} {30523} (\bibinfo {year}
  {2018})}\BibitemShut {NoStop}%
\bibitem [{\citenamefont {Donley}\ \emph {et~al.}(2009)\citenamefont {Donley},
  \citenamefont {Long}, \citenamefont {Liebisch}, \citenamefont {Hodby},
  \citenamefont {Fisher},\ and\ \citenamefont {Kitching}}]{donleyrotation}%
  \BibitemOpen
  \bibfield  {author} {\bibinfo {author} {\bibfnamefont {E.~A.}\ \bibnamefont
  {Donley}}, \bibinfo {author} {\bibfnamefont {J.~L.}\ \bibnamefont {Long}},
  \bibinfo {author} {\bibfnamefont {T.~C.}\ \bibnamefont {Liebisch}}, \bibinfo
  {author} {\bibfnamefont {E.~R.}\ \bibnamefont {Hodby}}, \bibinfo {author}
  {\bibfnamefont {T.~A.}\ \bibnamefont {Fisher}},\ and\ \bibinfo {author}
  {\bibfnamefont {J.}~\bibnamefont {Kitching}},\ }\href
  {https://doi.org/10.1103/PhysRevA.79.013420} {\bibfield  {journal} {\bibinfo
  {journal} {Phys. Rev. A}\ }\textbf {\bibinfo {volume} {79}},\ \bibinfo
  {pages} {013420} (\bibinfo {year} {2009})}\BibitemShut {NoStop}%
\bibitem [{\citenamefont {Degen}\ \emph {et~al.}(2017)\citenamefont {Degen},
  \citenamefont {Reinhard},\ and\ \citenamefont
  {Cappellaro}}]{RevModPhys.89.035002}%
  \BibitemOpen
  \bibfield  {author} {\bibinfo {author} {\bibfnamefont {C.~L.}\ \bibnamefont
  {Degen}}, \bibinfo {author} {\bibfnamefont {F.}~\bibnamefont {Reinhard}},\
  and\ \bibinfo {author} {\bibfnamefont {P.}~\bibnamefont {Cappellaro}},\
  }\href {https://doi.org/10.1103/RevModPhys.89.035002} {\bibfield  {journal}
  {\bibinfo  {journal} {Rev. Mod. Phys.}\ }\textbf {\bibinfo {volume} {89}},\
  \bibinfo {pages} {035002} (\bibinfo {year} {2017})}\BibitemShut {NoStop}%
\bibitem [{\citenamefont {Smith}\ and\ \citenamefont {Hughes}(2004)}]{hughes}%
  \BibitemOpen
  \bibfield  {author} {\bibinfo {author} {\bibfnamefont {D.~A.}\ \bibnamefont
  {Smith}}\ and\ \bibinfo {author} {\bibfnamefont {I.~G.}\ \bibnamefont
  {Hughes}},\ }\href {https://doi.org/10.1119/1.1652039} {\bibfield  {journal}
  {\bibinfo  {journal} {American Journal of Physics}\ }\textbf {\bibinfo
  {volume} {72}},\ \bibinfo {pages} {631} (\bibinfo {year} {2004})}\BibitemShut
  {NoStop}%
\bibitem [{\citenamefont {Siddons}\ \emph {et~al.}(2008)\citenamefont
  {Siddons}, \citenamefont {Adams}, \citenamefont {Ge},\ and\ \citenamefont
  {Hughes}}]{Siddons_2008}%
  \BibitemOpen
  \bibfield  {author} {\bibinfo {author} {\bibfnamefont {P.}~\bibnamefont
  {Siddons}}, \bibinfo {author} {\bibfnamefont {C.~S.}\ \bibnamefont {Adams}},
  \bibinfo {author} {\bibfnamefont {C.}~\bibnamefont {Ge}},\ and\ \bibinfo
  {author} {\bibfnamefont {I.~G.}\ \bibnamefont {Hughes}},\ }\href
  {https://doi.org/10.1088/0953-4075/41/15/155004} {\bibfield  {journal}
  {\bibinfo  {journal} {Journal of Physics B: Atomic, Molecular and Optical
  Physics}\ }\textbf {\bibinfo {volume} {41}},\ \bibinfo {pages} {155004}
  (\bibinfo {year} {2008})}\BibitemShut {NoStop}%
\bibitem [{\citenamefont {Affolderbach}\ and\ \citenamefont
  {Mileti}(2005)}]{AFFOLDERBACH2005291}%
  \BibitemOpen
  \bibfield  {author} {\bibinfo {author} {\bibfnamefont {C.}~\bibnamefont
  {Affolderbach}}\ and\ \bibinfo {author} {\bibfnamefont {G.}~\bibnamefont
  {Mileti}},\ }\href
  {https://doi.org/https://doi.org/10.1016/j.optlaseng.2004.02.009} {\bibfield
  {journal} {\bibinfo  {journal} {Optics and Lasers in Engineering}\ }\textbf
  {\bibinfo {volume} {43}},\ \bibinfo {pages} {291} (\bibinfo {year}
  {2005})}\BibitemShut {NoStop}%
\bibitem [{\citenamefont {Ye}\ \emph {et~al.}(1996)\citenamefont {Ye},
  \citenamefont {Swartz}, \citenamefont {Jungner},\ and\ \citenamefont
  {Hall}}]{Ye:96}%
  \BibitemOpen
  \bibfield  {author} {\bibinfo {author} {\bibfnamefont {J.}~\bibnamefont
  {Ye}}, \bibinfo {author} {\bibfnamefont {S.}~\bibnamefont {Swartz}}, \bibinfo
  {author} {\bibfnamefont {P.}~\bibnamefont {Jungner}},\ and\ \bibinfo {author}
  {\bibfnamefont {J.~L.}\ \bibnamefont {Hall}},\ }\href
  {https://doi.org/10.1364/OL.21.001280} {\bibfield  {journal} {\bibinfo
  {journal} {Opt. Lett.}\ }\textbf {\bibinfo {volume} {21}},\ \bibinfo {pages}
  {1280} (\bibinfo {year} {1996})}\BibitemShut {NoStop}%
\bibitem [{\citenamefont {Knapkiewicz}(2019{\natexlab{a}})}]{pawel}%
  \BibitemOpen
  \bibfield  {author} {\bibinfo {author} {\bibfnamefont {P.}~\bibnamefont
  {Knapkiewicz}},\ }\href {https://doi.org/10.1088/1361-6641/aafecc} {\bibfield
   {journal} {\bibinfo  {journal} {Semiconductor Science and Technology}\
  }\textbf {\bibinfo {volume} {34}},\ \bibinfo {pages} {035005} (\bibinfo
  {year} {2019}{\natexlab{a}})}\BibitemShut {NoStop}%
\bibitem [{\citenamefont {Eklund}\ \emph {et~al.}(2008)\citenamefont {Eklund},
  \citenamefont {Shkel}, \citenamefont {Knappe}, \citenamefont {Donley},\ and\
  \citenamefont {Kitching}}]{EKLUND2008175}%
  \BibitemOpen
  \bibfield  {author} {\bibinfo {author} {\bibfnamefont {E.~J.}\ \bibnamefont
  {Eklund}}, \bibinfo {author} {\bibfnamefont {A.~M.}\ \bibnamefont {Shkel}},
  \bibinfo {author} {\bibfnamefont {S.}~\bibnamefont {Knappe}}, \bibinfo
  {author} {\bibfnamefont {E.}~\bibnamefont {Donley}},\ and\ \bibinfo {author}
  {\bibfnamefont {J.}~\bibnamefont {Kitching}},\ }\href
  {https://doi.org/https://doi.org/10.1016/j.sna.2007.10.006} {\bibfield
  {journal} {\bibinfo  {journal} {Sensors and Actuators A: Physical}\ }\textbf
  {\bibinfo {volume} {143}},\ \bibinfo {pages} {175} (\bibinfo {year}
  {2008})}\BibitemShut {NoStop}%
\bibitem [{\citenamefont {Shah}\ \emph {et~al.}(2007)\citenamefont {Shah},
  \citenamefont {Knappe}, \citenamefont {Schwindt},\ and\ \citenamefont
  {Kitching}}]{kitchingmag}%
  \BibitemOpen
  \bibfield  {author} {\bibinfo {author} {\bibfnamefont {V.}~\bibnamefont
  {Shah}}, \bibinfo {author} {\bibfnamefont {S.}~\bibnamefont {Knappe}},
  \bibinfo {author} {\bibfnamefont {P.~D.~D.}\ \bibnamefont {Schwindt}},\ and\
  \bibinfo {author} {\bibfnamefont {J.}~\bibnamefont {Kitching}},\ }\href
  {https://doi.org/10.1038/nphoton.2007.201} {\bibfield  {journal} {\bibinfo
  {journal} {Nature Photonics}\ }\textbf {\bibinfo {volume} {1}},\ \bibinfo
  {pages} {649} (\bibinfo {year} {2007})}\BibitemShut {NoStop}%
\bibitem [{\citenamefont {Schwartz}\ and\ \citenamefont
  {Schaible}(1979)}]{RIE}%
  \BibitemOpen
  \bibfield  {author} {\bibinfo {author} {\bibfnamefont {G.~C.}\ \bibnamefont
  {Schwartz}}\ and\ \bibinfo {author} {\bibfnamefont {P.~M.}\ \bibnamefont
  {Schaible}},\ }\href {https://doi.org/10.1116/1.569962} {\bibfield  {journal}
  {\bibinfo  {journal} {Journal of Vacuum Science and Technology}\ }\textbf
  {\bibinfo {volume} {16}},\ \bibinfo {pages} {410} (\bibinfo {year}
  {1979})}\BibitemShut {NoStop}%
\bibitem [{\citenamefont {Laermer}\ \emph {et~al.}(2020)\citenamefont
  {Laermer}, \citenamefont {Franssila}, \citenamefont {Sainiemi},\ and\
  \citenamefont {Kolari}}]{DRIE}%
  \BibitemOpen
  \bibfield  {author} {\bibinfo {author} {\bibfnamefont {F.}~\bibnamefont
  {Laermer}}, \bibinfo {author} {\bibfnamefont {S.}~\bibnamefont {Franssila}},
  \bibinfo {author} {\bibfnamefont {L.}~\bibnamefont {Sainiemi}},\ and\
  \bibinfo {author} {\bibfnamefont {K.}~\bibnamefont {Kolari}},\ }in\ \href
  {https://doi.org/https://doi.org/10.1016/B978-0-12-817786-0.00016-5} {\emph
  {\bibinfo {booktitle} {Handbook of Silicon Based MEMS Materials and
  Technologies (Third Edition)}}},\ \bibinfo {series and number} {Micro and
  Nano Technologies},\ \bibinfo {editor} {edited by\ \bibinfo {editor}
  {\bibfnamefont {M.}~\bibnamefont {Tilli}}, \bibinfo {editor} {\bibfnamefont
  {M.}~\bibnamefont {Paulasto-Krockel}}, \bibinfo {editor} {\bibfnamefont
  {M.}~\bibnamefont {Petzold}}, \bibinfo {editor} {\bibfnamefont
  {H.}~\bibnamefont {Theuss}}, \bibinfo {editor} {\bibfnamefont
  {T.}~\bibnamefont {Motooka}},\ and\ \bibinfo {editor} {\bibfnamefont
  {V.}~\bibnamefont {Lindroos}}}\ (\bibinfo  {publisher} {Elsevier},\ \bibinfo
  {year} {2020})\ \bibinfo {edition} {third edition}\ ed.,\ pp.\ \bibinfo
  {pages} {417--446}\BibitemShut {NoStop}%
\bibitem [{\citenamefont {Vladimirsky}(1999)}]{VLADIMIRSKY1999205}%
  \BibitemOpen
  \bibfield  {author} {\bibinfo {author} {\bibfnamefont {Y.}~\bibnamefont
  {Vladimirsky}},\ }in\ \href
  {https://doi.org/https://doi.org/10.1016/B978-012617560-8/50032-3} {\emph
  {\bibinfo {booktitle} {Vacuum Ultraviolet Spectroscopy}}},\ \bibinfo {editor}
  {edited by\ \bibinfo {editor} {\bibfnamefont {J.}~\bibnamefont {Samson}}\
  and\ \bibinfo {editor} {\bibfnamefont {D.}~\bibnamefont {Ederer}}}\ (\bibinfo
   {publisher} {Academic Press},\ \bibinfo {address} {Burlington},\ \bibinfo
  {year} {1999})\ pp.\ \bibinfo {pages} {205--223}\BibitemShut {NoStop}%
\bibitem [{\citenamefont {Gosálvez}\ \emph {et~al.}(2010)\citenamefont
  {Gosálvez}, \citenamefont {Zubel},\ and\ \citenamefont
  {Viinikka}}]{wetetch}%
  \BibitemOpen
  \bibfield  {author} {\bibinfo {author} {\bibfnamefont {M.~A.}\ \bibnamefont
  {Gosálvez}}, \bibinfo {author} {\bibfnamefont {I.}~\bibnamefont {Zubel}},\
  and\ \bibinfo {author} {\bibfnamefont {E.}~\bibnamefont {Viinikka}},\ }in\
  \href {https://doi.org/https://doi.org/10.1016/B978-0-12-817786-0.00017-7}
  {\emph {\bibinfo {booktitle} {Handbook of Silicon Based MEMS Materials and
  Technologies (Third Edition)}}},\ \bibinfo {series and number} {Micro and
  Nano Technologies},\ \bibinfo {editor} {edited by\ \bibinfo {editor}
  {\bibfnamefont {M.}~\bibnamefont {Tilli}}, \bibinfo {editor} {\bibfnamefont
  {M.}~\bibnamefont {Paulasto-Krockel}}, \bibinfo {editor} {\bibfnamefont
  {M.}~\bibnamefont {Petzold}}, \bibinfo {editor} {\bibfnamefont
  {H.}~\bibnamefont {Theuss}}, \bibinfo {editor} {\bibfnamefont
  {T.}~\bibnamefont {Motooka}},\ and\ \bibinfo {editor} {\bibfnamefont
  {V.}~\bibnamefont {Lindroos}}}\ (\bibinfo  {publisher} {Elsevier},\ \bibinfo
  {year} {2010})\ \bibinfo {edition} {third edition}\ ed.,\ pp.\ \bibinfo
  {pages} {447--480}\BibitemShut {NoStop}%
\bibitem [{\citenamefont {Golshan}\ \emph {et~al.}(2017)\citenamefont
  {Golshan}, \citenamefont {Baharudin}, \citenamefont {Aoyama}, \citenamefont
  {Ariffin}, \citenamefont {Ismail},\ and\ \citenamefont
  {Ehsan}}]{siliconmachining}%
  \BibitemOpen
  \bibfield  {author} {\bibinfo {author} {\bibfnamefont {A.}~\bibnamefont
  {Golshan}}, \bibinfo {author} {\bibfnamefont {B.}~\bibnamefont {Baharudin}},
  \bibinfo {author} {\bibfnamefont {H.}~\bibnamefont {Aoyama}}, \bibinfo
  {author} {\bibfnamefont {M.}~\bibnamefont {Ariffin}}, \bibinfo {author}
  {\bibfnamefont {M.}~\bibnamefont {Ismail}},\ and\ \bibinfo {author}
  {\bibfnamefont {A.}~\bibnamefont {Ehsan}},\ }\href
  {https://doi.org/https://doi.org/10.1016/j.proeng.2017.04.085} {\bibfield
  {journal} {\bibinfo  {journal} {Procedia Engineering}\ }\textbf {\bibinfo
  {volume} {184}},\ \bibinfo {pages} {192} (\bibinfo {year}
  {2017})}\BibitemShut {NoStop}%
\bibitem [{\citenamefont {Li}(2008)}]{lasermachining}%
  \BibitemOpen
  \bibinfo {editor} {\bibfnamefont {D.}~\bibnamefont {Li}},\ ed.,\ \bibinfo
  {title} {Laser micromachining},\ in\ \href
  {https://doi.org/10.1007/978-0-387-48998-8_797} {\emph {\bibinfo {booktitle}
  {Encyclopedia of Microfluidics and Nanofluidics}}}\ (\bibinfo  {publisher}
  {Springer US},\ \bibinfo {address} {Boston, MA},\ \bibinfo {year} {2008})\
  pp.\ \bibinfo {pages} {981--981}\BibitemShut {NoStop}%
\bibitem [{\citenamefont {Wang}\ and\ \citenamefont
  {Yang}(2021)}]{lasercuttingreview}%
  \BibitemOpen
  \bibfield  {author} {\bibinfo {author} {\bibfnamefont {H.-J.}\ \bibnamefont
  {Wang}}\ and\ \bibinfo {author} {\bibfnamefont {T.}~\bibnamefont {Yang}},\
  }\href {https://doi.org/https://doi.org/10.1016/j.jeurceramsoc.2021.04.019}
  {\bibfield  {journal} {\bibinfo  {journal} {Journal of the European Ceramic
  Society}\ }\textbf {\bibinfo {volume} {41}},\ \bibinfo {pages} {4997}
  (\bibinfo {year} {2021})}\BibitemShut {NoStop}%
\bibitem [{\citenamefont {Lee}\ and\ \citenamefont {Lin}(2015)}]{laserdicing}%
  \BibitemOpen
  \bibfield  {author} {\bibinfo {author} {\bibfnamefont {Y.-C.}\ \bibnamefont
  {Lee}}\ and\ \bibinfo {author} {\bibfnamefont {J.-T.}\ \bibnamefont {Lin}}\
  }(\bibinfo {year} {2015})\ pp.\ \bibinfo {pages} {348--350}\BibitemShut
  {NoStop}%
\bibitem [{\citenamefont {Dyer}\ \emph {et~al.}(2022)\citenamefont {Dyer},
  \citenamefont {Griffin}, \citenamefont {Arnold}, \citenamefont {Mirando},
  \citenamefont {Burt}, \citenamefont {Riis},\ and\ \citenamefont
  {McGilligan}}]{dyer}%
  \BibitemOpen
  \bibfield  {author} {\bibinfo {author} {\bibfnamefont {S.}~\bibnamefont
  {Dyer}}, \bibinfo {author} {\bibfnamefont {P.~F.}\ \bibnamefont {Griffin}},
  \bibinfo {author} {\bibfnamefont {A.~S.}\ \bibnamefont {Arnold}}, \bibinfo
  {author} {\bibfnamefont {F.}~\bibnamefont {Mirando}}, \bibinfo {author}
  {\bibfnamefont {D.~P.}\ \bibnamefont {Burt}}, \bibinfo {author}
  {\bibfnamefont {E.}~\bibnamefont {Riis}},\ and\ \bibinfo {author}
  {\bibfnamefont {J.~P.}\ \bibnamefont {McGilligan}},\ }\href
  {https://doi.org/10.48550/.2207.12904} {\bibinfo {title} {Micro-machined deep
  silicon atomic vapor cells}} (\bibinfo {year} {2022}),\ \Eprint
  {https://arxiv.org/abs/2207.12904} {arXiv:2207.12904} \BibitemShut {NoStop}%
\bibitem [{\citenamefont {Wallis}\ and\ \citenamefont
  {Pomerantz}(1969)}]{wallis}%
  \BibitemOpen
  \bibfield  {author} {\bibinfo {author} {\bibfnamefont {G.}~\bibnamefont
  {Wallis}}\ and\ \bibinfo {author} {\bibfnamefont {D.~I.}\ \bibnamefont
  {Pomerantz}},\ }\href {https://doi.org/10.1063/1.1657121} {\bibfield
  {journal} {\bibinfo  {journal} {Journal of Applied Physics}\ }\textbf
  {\bibinfo {volume} {40}},\ \bibinfo {pages} {3946} (\bibinfo {year}
  {1969})}\BibitemShut {NoStop}%
\bibitem [{\citenamefont {Vesborg}\ \emph {et~al.}(2010)\citenamefont
  {Vesborg}, \citenamefont {Olsen}, \citenamefont {Henriksen}, \citenamefont
  {Chorkendorff},\ and\ \citenamefont {Hansen}}]{heatinganodic}%
  \BibitemOpen
  \bibfield  {author} {\bibinfo {author} {\bibfnamefont {P.~C.~K.}\
  \bibnamefont {Vesborg}}, \bibinfo {author} {\bibfnamefont {J.~L.}\
  \bibnamefont {Olsen}}, \bibinfo {author} {\bibfnamefont {T.~R.}\ \bibnamefont
  {Henriksen}}, \bibinfo {author} {\bibfnamefont {I.}~\bibnamefont
  {Chorkendorff}},\ and\ \bibinfo {author} {\bibfnamefont {O.}~\bibnamefont
  {Hansen}},\ }\href {https://doi.org/10.1063/1.3277117} {\bibfield  {journal}
  {\bibinfo  {journal} {Review of Scientific Instruments}\ }\textbf {\bibinfo
  {volume} {81}},\ \bibinfo {pages} {016111} (\bibinfo {year}
  {2010})}\BibitemShut {NoStop}%
\bibitem [{\citenamefont {Maurice}(2016)}]{vincentthesis}%
  \BibitemOpen
  \bibfield  {author} {\bibinfo {author} {\bibfnamefont {V.}~\bibnamefont
  {Maurice}},\ }\emph {\bibinfo {title} {Design, microfabrication and
  characterization of alkali vapor cells for miniature atomic frequency
  references}},\ \href {http://www.theses.fr/2016BESA2001} {Ph.D. thesis},\
  \bibinfo  {school} {Université de Franche-Comté} (\bibinfo {year} {2016}),\
  \bibinfo {note} {thèse de doctorat dirigée par C. Gorecki, N. Passilly, and
  R. Boudot Sciences pour l'ingénieur Besançon 2016}\BibitemShut {NoStop}%
\bibitem [{\citenamefont {Knowles}\ and\ \citenamefont {van
  Helvoort}(2006)}]{anodicbondtextbook}%
  \BibitemOpen
  \bibfield  {author} {\bibinfo {author} {\bibfnamefont {K.~M.}\ \bibnamefont
  {Knowles}}\ and\ \bibinfo {author} {\bibfnamefont {A.~T.~J.}\ \bibnamefont
  {van Helvoort}},\ }\href {https://doi.org/10.1179/174328006X102501}
  {\bibfield  {journal} {\bibinfo  {journal} {International Materials Reviews}\
  }\textbf {\bibinfo {volume} {51}},\ \bibinfo {pages} {273} (\bibinfo {year}
  {2006})}\BibitemShut {NoStop}%
\bibitem [{\citenamefont {Li}\ \emph {et~al.}(2015)\citenamefont {Li},
  \citenamefont {Xiao}, \citenamefont {Hou}, \citenamefont {Wang},
  \citenamefont {Chen},\ and\ \citenamefont {Wu}}]{anodicaccelerometer}%
  \BibitemOpen
  \bibfield  {author} {\bibinfo {author} {\bibfnamefont {Q.}~\bibnamefont
  {Li}}, \bibinfo {author} {\bibfnamefont {D.}~\bibnamefont {Xiao}}, \bibinfo
  {author} {\bibfnamefont {Z.}~\bibnamefont {Hou}}, \bibinfo {author}
  {\bibfnamefont {X.}~\bibnamefont {Wang}}, \bibinfo {author} {\bibfnamefont
  {Z.}~\bibnamefont {Chen}},\ and\ \bibinfo {author} {\bibfnamefont
  {X.}~\bibnamefont {Wu}},\ }in\ \href
  {https://doi.org/10.1109/ICSENS.2015.7370430} {\emph {\bibinfo {booktitle}
  {IEEE SENSORS}}}\ (\bibinfo {year} {2015})\ pp.\ \bibinfo {pages}
  {1--4}\BibitemShut {NoStop}%
\bibitem [{\citenamefont {Plaza}\ \emph {et~al.}(2004)\citenamefont {Plaza},
  \citenamefont {Lopez-Bosque}, \citenamefont {Gracia}, \citenamefont {Cane},
  \citenamefont {Wollenstein}, \citenamefont {Kuhner}, \citenamefont
  {Plescher},\ and\ \citenamefont {Bottner}}]{anodicgas}%
  \BibitemOpen
  \bibfield  {author} {\bibinfo {author} {\bibfnamefont {J.}~\bibnamefont
  {Plaza}}, \bibinfo {author} {\bibfnamefont {M.}~\bibnamefont {Lopez-Bosque}},
  \bibinfo {author} {\bibfnamefont {I.}~\bibnamefont {Gracia}}, \bibinfo
  {author} {\bibfnamefont {C.}~\bibnamefont {Cane}}, \bibinfo {author}
  {\bibfnamefont {J.}~\bibnamefont {Wollenstein}}, \bibinfo {author}
  {\bibfnamefont {G.}~\bibnamefont {Kuhner}}, \bibinfo {author} {\bibfnamefont
  {G.}~\bibnamefont {Plescher}},\ and\ \bibinfo {author} {\bibfnamefont
  {H.}~\bibnamefont {Bottner}},\ }\href
  {https://doi.org/10.1109/JSEN.2004.823681} {\bibfield  {journal} {\bibinfo
  {journal} {IEEE Sensors Journal}\ }\textbf {\bibinfo {volume} {4}},\ \bibinfo
  {pages} {195} (\bibinfo {year} {2004})}\BibitemShut {NoStop}%
\bibitem [{\citenamefont {Veenstra}\ \emph {et~al.}(2001)\citenamefont
  {Veenstra}, \citenamefont {Berenschot}, \citenamefont {Gardeniers},
  \citenamefont {Sanders}, \citenamefont {Elwenspoek},\ and\ \citenamefont
  {van~den Berg}}]{anodicpump}%
  \BibitemOpen
  \bibfield  {author} {\bibinfo {author} {\bibfnamefont {T.~T.}\ \bibnamefont
  {Veenstra}}, \bibinfo {author} {\bibfnamefont {J.~W.}\ \bibnamefont
  {Berenschot}}, \bibinfo {author} {\bibfnamefont {J.~G.~E.}\ \bibnamefont
  {Gardeniers}}, \bibinfo {author} {\bibfnamefont {R.~G.~P.}\ \bibnamefont
  {Sanders}}, \bibinfo {author} {\bibfnamefont {M.~C.}\ \bibnamefont
  {Elwenspoek}},\ and\ \bibinfo {author} {\bibfnamefont {A.}~\bibnamefont
  {van~den Berg}},\ }\href {https://doi.org/10.1149/1.1339873} {\bibfield
  {journal} {\bibinfo  {journal} {Journal of The Electrochemical Society}\
  }\textbf {\bibinfo {volume} {148}},\ \bibinfo {pages} {G68} (\bibinfo {year}
  {2001})}\BibitemShut {NoStop}%
\bibitem [{\citenamefont {Cui}(2008)}]{Cui2008}%
  \BibitemOpen
  \bibfield  {author} {\bibinfo {author} {\bibfnamefont {Z.}~\bibnamefont
  {Cui}},\ }\bibinfo {title} {Anodic bonding},\ in\ \href
  {https://doi.org/10.1007/978-0-387-48998-8_41} {\emph {\bibinfo {booktitle}
  {Encyclopedia of Microfluidics and Nanofluidics}}},\ \bibinfo {editor}
  {edited by\ \bibinfo {editor} {\bibfnamefont {D.}~\bibnamefont {Li}}}\
  (\bibinfo  {publisher} {Springer US},\ \bibinfo {address} {Boston, MA},\
  \bibinfo {year} {2008})\ pp.\ \bibinfo {pages} {50--54}\BibitemShut {NoStop}%
\bibitem [{\citenamefont {Pl{\"o}{\ss}l}(1999)}]{directbond2}%
  \BibitemOpen
  \bibfield  {author} {\bibinfo {author} {\bibfnamefont {A.~D.}\ \bibnamefont
  {Pl{\"o}{\ss}l}},\ }\href@noop {} {\bibfield  {journal} {\bibinfo  {journal}
  {Materials Science \& Engineering R-reports}\ }\textbf {\bibinfo {volume}
  {25}},\ \bibinfo {pages} {1} (\bibinfo {year} {1999})}\BibitemShut {NoStop}%
\bibitem [{\citenamefont {Gösele}\ \emph {et~al.}(1995)\citenamefont
  {Gösele}, \citenamefont {Stenzel}, \citenamefont {Martini}, \citenamefont
  {Steinkirchner}, \citenamefont {Conrad},\ and\ \citenamefont
  {Scheerschmidt}}]{directbondUHV}%
  \BibitemOpen
  \bibfield  {author} {\bibinfo {author} {\bibfnamefont {U.}~\bibnamefont
  {Gösele}}, \bibinfo {author} {\bibfnamefont {H.}~\bibnamefont {Stenzel}},
  \bibinfo {author} {\bibfnamefont {T.}~\bibnamefont {Martini}}, \bibinfo
  {author} {\bibfnamefont {J.}~\bibnamefont {Steinkirchner}}, \bibinfo {author}
  {\bibfnamefont {D.}~\bibnamefont {Conrad}},\ and\ \bibinfo {author}
  {\bibfnamefont {K.}~\bibnamefont {Scheerschmidt}},\ }\href
  {https://doi.org/10.1063/1.115335} {\bibfield  {journal} {\bibinfo  {journal}
  {Applied Physics Letters}\ }\textbf {\bibinfo {volume} {67}},\ \bibinfo
  {pages} {3614} (\bibinfo {year} {1995})}\BibitemShut {NoStop}%
\bibitem [{\citenamefont {Karlen}\ \emph {et~al.}(2020)\citenamefont {Karlen},
  \citenamefont {Haesler}, \citenamefont {Overstolz}, \citenamefont
  {Bergonzi},\ and\ \citenamefont {Lecomte}}]{karlencucu}%
  \BibitemOpen
  \bibfield  {author} {\bibinfo {author} {\bibfnamefont {S.}~\bibnamefont
  {Karlen}}, \bibinfo {author} {\bibfnamefont {J.}~\bibnamefont {Haesler}},
  \bibinfo {author} {\bibfnamefont {T.}~\bibnamefont {Overstolz}}, \bibinfo
  {author} {\bibfnamefont {G.}~\bibnamefont {Bergonzi}},\ and\ \bibinfo
  {author} {\bibfnamefont {S.}~\bibnamefont {Lecomte}},\ }\href
  {https://doi.org/10.1109/JMEMS.2019.2949349} {\bibfield  {journal} {\bibinfo
  {journal} {Journal of Microelectromechanical Systems}\ }\textbf {\bibinfo
  {volume} {29}},\ \bibinfo {pages} {95} (\bibinfo {year} {2020})}\BibitemShut
  {NoStop}%
\bibitem [{\citenamefont {Karlen}(2018)}]{karlenthesis}%
  \BibitemOpen
  \bibfield  {author} {\bibinfo {author} {\bibfnamefont {S.}~\bibnamefont
  {Karlen}},\ }\href@noop {} {\bibfield  {journal} {\bibinfo  {journal} {PhD
  Dissertation - Universite de Neuch\^{a}tel}\ } (\bibinfo {year}
  {2018})}\BibitemShut {NoStop}%
\bibitem [{\citenamefont {Wolffenbuttel}\ and\ \citenamefont
  {Wise}(1994)}]{eutectic}%
  \BibitemOpen
  \bibfield  {author} {\bibinfo {author} {\bibfnamefont {R.}~\bibnamefont
  {Wolffenbuttel}}\ and\ \bibinfo {author} {\bibfnamefont {K.}~\bibnamefont
  {Wise}},\ }\href
  {https://doi.org/https://doi.org/10.1016/0924-4247(93)00653-L} {\bibfield
  {journal} {\bibinfo  {journal} {Sensors and Actuators A: Physical}\ }\textbf
  {\bibinfo {volume} {43}},\ \bibinfo {pages} {223} (\bibinfo {year}
  {1994})}\BibitemShut {NoStop}%
\bibitem [{\citenamefont {Chutani}\ \emph {et~al.}(2014)\citenamefont
  {Chutani}, \citenamefont {Hasegawa}, \citenamefont {Maurice}, \citenamefont
  {Passilly},\ and\ \citenamefont {Gorecki}}]{CHUTANIdrie}%
  \BibitemOpen
  \bibfield  {author} {\bibinfo {author} {\bibfnamefont {R.}~\bibnamefont
  {Chutani}}, \bibinfo {author} {\bibfnamefont {M.}~\bibnamefont {Hasegawa}},
  \bibinfo {author} {\bibfnamefont {V.}~\bibnamefont {Maurice}}, \bibinfo
  {author} {\bibfnamefont {N.}~\bibnamefont {Passilly}},\ and\ \bibinfo
  {author} {\bibfnamefont {C.}~\bibnamefont {Gorecki}},\ }\href
  {https://doi.org/https://doi.org/10.1016/j.sna.2013.12.031} {\bibfield
  {journal} {\bibinfo  {journal} {Sensors and Actuators A: Physical}\ }\textbf
  {\bibinfo {volume} {208}},\ \bibinfo {pages} {66} (\bibinfo {year}
  {2014})}\BibitemShut {NoStop}%
\bibitem [{\citenamefont {P{\'{e}}tremand}\ \emph {et~al.}(2012)\citenamefont
  {P{\'{e}}tremand}, \citenamefont {Affolderbach}, \citenamefont {Straessle},
  \citenamefont {Pellaton}, \citenamefont {Briand}, \citenamefont {Mileti},\
  and\ \citenamefont {de~Rooij}}]{P_tremand_2012}%
  \BibitemOpen
  \bibfield  {author} {\bibinfo {author} {\bibfnamefont {Y.}~\bibnamefont
  {P{\'{e}}tremand}}, \bibinfo {author} {\bibfnamefont {C.}~\bibnamefont
  {Affolderbach}}, \bibinfo {author} {\bibfnamefont {R.}~\bibnamefont
  {Straessle}}, \bibinfo {author} {\bibfnamefont {M.}~\bibnamefont {Pellaton}},
  \bibinfo {author} {\bibfnamefont {D.}~\bibnamefont {Briand}}, \bibinfo
  {author} {\bibfnamefont {G.}~\bibnamefont {Mileti}},\ and\ \bibinfo {author}
  {\bibfnamefont {N.~F.}\ \bibnamefont {de~Rooij}},\ }\href
  {https://doi.org/10.1088/0960-1317/22/2/025013} {\bibfield  {journal}
  {\bibinfo  {journal} {Journal of Micromechanics and Microengineering}\
  }\textbf {\bibinfo {volume} {22}},\ \bibinfo {pages} {025013} (\bibinfo
  {year} {2012})}\BibitemShut {NoStop}%
\bibitem [{\citenamefont {Newman}\ \emph {et~al.}(2021)\citenamefont {Newman},
  \citenamefont {Maurice}, \citenamefont {Fredrick}, \citenamefont {Fortier},
  \citenamefont {Leopardi}, \citenamefont {Hollberg}, \citenamefont {Diddams},
  \citenamefont {Kitching},\ and\ \citenamefont
  {Hummon}}]{newman2021highperformance}%
  \BibitemOpen
  \bibfield  {author} {\bibinfo {author} {\bibfnamefont {Z.~L.}\ \bibnamefont
  {Newman}}, \bibinfo {author} {\bibfnamefont {V.}~\bibnamefont {Maurice}},
  \bibinfo {author} {\bibfnamefont {C.}~\bibnamefont {Fredrick}}, \bibinfo
  {author} {\bibfnamefont {T.}~\bibnamefont {Fortier}}, \bibinfo {author}
  {\bibfnamefont {H.}~\bibnamefont {Leopardi}}, \bibinfo {author}
  {\bibfnamefont {L.}~\bibnamefont {Hollberg}}, \bibinfo {author}
  {\bibfnamefont {S.~A.}\ \bibnamefont {Diddams}}, \bibinfo {author}
  {\bibfnamefont {J.}~\bibnamefont {Kitching}},\ and\ \bibinfo {author}
  {\bibfnamefont {M.~T.}\ \bibnamefont {Hummon}},\ }\href
  {https://doi.org/10.1364/OL.435603} {\bibfield  {journal} {\bibinfo
  {journal} {Opt. Lett.}\ }\textbf {\bibinfo {volume} {46}},\ \bibinfo {pages}
  {4702} (\bibinfo {year} {2021})}\BibitemShut {NoStop}%
\bibitem [{\citenamefont {Chutani}\ \emph {et~al.}(2015)\citenamefont
  {Chutani}, \citenamefont {Maurice}, \citenamefont {Passilly}, \citenamefont
  {Gorecki}, \citenamefont {Boudot}, \citenamefont {Abdel~Hafiz}, \citenamefont
  {Abb{\'e}}, \citenamefont {Galliou}, \citenamefont {Rauch},\ and\
  \citenamefont {de~Clercq}}]{chutani}%
  \BibitemOpen
  \bibfield  {author} {\bibinfo {author} {\bibfnamefont {R.}~\bibnamefont
  {Chutani}}, \bibinfo {author} {\bibfnamefont {V.}~\bibnamefont {Maurice}},
  \bibinfo {author} {\bibfnamefont {N.}~\bibnamefont {Passilly}}, \bibinfo
  {author} {\bibfnamefont {C.}~\bibnamefont {Gorecki}}, \bibinfo {author}
  {\bibfnamefont {R.}~\bibnamefont {Boudot}}, \bibinfo {author} {\bibfnamefont
  {M.}~\bibnamefont {Abdel~Hafiz}}, \bibinfo {author} {\bibfnamefont
  {P.}~\bibnamefont {Abb{\'e}}}, \bibinfo {author} {\bibfnamefont
  {S.}~\bibnamefont {Galliou}}, \bibinfo {author} {\bibfnamefont {J.-Y.}\
  \bibnamefont {Rauch}},\ and\ \bibinfo {author} {\bibfnamefont
  {E.}~\bibnamefont {de~Clercq}},\ }\href {https://doi.org/10.1038/srep14001}
  {\bibfield  {journal} {\bibinfo  {journal} {Scientific Reports}\ }\textbf
  {\bibinfo {volume} {5}},\ \bibinfo {pages} {14001} (\bibinfo {year}
  {2015})}\BibitemShut {NoStop}%
\bibitem [{\citenamefont {Schmidt}\ and\ \citenamefont
  {Hawkins}(2010)}]{hollowcoreschmidt}%
  \BibitemOpen
  \bibfield  {author} {\bibinfo {author} {\bibfnamefont {H.}~\bibnamefont
  {Schmidt}}\ and\ \bibinfo {author} {\bibfnamefont {A.}~\bibnamefont
  {Hawkins}},\ }\href {https://doi.org/https://doi.org/10.1002/lpor.200900040}
  {\bibfield  {journal} {\bibinfo  {journal} {Laser and Photonics Reviews}\
  }\textbf {\bibinfo {volume} {4}},\ \bibinfo {pages} {720} (\bibinfo {year}
  {2010})}\BibitemShut {NoStop}%
\bibitem [{\citenamefont {Nishino}\ \emph {et~al.}(2019)\citenamefont
  {Nishino}, \citenamefont {Hara}, \citenamefont {Yano}, \citenamefont {Toda},
  \citenamefont {Kanamori}, \citenamefont {Kajita}, \citenamefont {Ido},\ and\
  \citenamefont {Ono}}]{nishino}%
  \BibitemOpen
  \bibfield  {author} {\bibinfo {author} {\bibfnamefont {H.}~\bibnamefont
  {Nishino}}, \bibinfo {author} {\bibfnamefont {M.}~\bibnamefont {Hara}},
  \bibinfo {author} {\bibfnamefont {Y.}~\bibnamefont {Yano}}, \bibinfo {author}
  {\bibfnamefont {M.}~\bibnamefont {Toda}}, \bibinfo {author} {\bibfnamefont
  {Y.}~\bibnamefont {Kanamori}}, \bibinfo {author} {\bibfnamefont
  {M.}~\bibnamefont {Kajita}}, \bibinfo {author} {\bibfnamefont
  {T.}~\bibnamefont {Ido}},\ and\ \bibinfo {author} {\bibfnamefont
  {T.}~\bibnamefont {Ono}},\ }\href {https://doi.org/10.7567/1882-0786/ab2a3c}
  {\bibfield  {journal} {\bibinfo  {journal} {Applied Physics Express}\
  }\textbf {\bibinfo {volume} {12}},\ \bibinfo {pages} {072012} (\bibinfo
  {year} {2019})}\BibitemShut {NoStop}%
\bibitem [{\citenamefont {Backlund}\ and\ \citenamefont
  {Rosengren}(1992)}]{Backlund_1992}%
  \BibitemOpen
  \bibfield  {author} {\bibinfo {author} {\bibfnamefont {Y.}~\bibnamefont
  {Backlund}}\ and\ \bibinfo {author} {\bibfnamefont {L.}~\bibnamefont
  {Rosengren}},\ }\href {https://doi.org/10.1088/0960-1317/2/2/002} {\bibfield
  {journal} {\bibinfo  {journal} {Journal of Micromechanics and
  Microengineering}\ }\textbf {\bibinfo {volume} {2}},\ \bibinfo {pages} {75}
  (\bibinfo {year} {1992})}\BibitemShut {NoStop}%
\bibitem [{\citenamefont {Rola}\ \emph {et~al.}(2014)\citenamefont {Rola},
  \citenamefont {Ptasi{\'{n}}ski}, \citenamefont {Zakrzewski},\ and\
  \citenamefont {Zubel}}]{Rola2014}%
  \BibitemOpen
  \bibfield  {author} {\bibinfo {author} {\bibfnamefont {K.~P.}\ \bibnamefont
  {Rola}}, \bibinfo {author} {\bibfnamefont {K.}~\bibnamefont
  {Ptasi{\'{n}}ski}}, \bibinfo {author} {\bibfnamefont {A.}~\bibnamefont
  {Zakrzewski}},\ and\ \bibinfo {author} {\bibfnamefont {I.}~\bibnamefont
  {Zubel}},\ }\href {https://doi.org/10.1007/s00542-013-1859-z} {\bibfield
  {journal} {\bibinfo  {journal} {Microsystem Technologies}\ }\textbf {\bibinfo
  {volume} {20}},\ \bibinfo {pages} {221} (\bibinfo {year} {2014})}\BibitemShut
  {NoStop}%
\bibitem [{\citenamefont {Nishino}\ \emph
  {et~al.}(2021{\natexlab{a}})\citenamefont {Nishino}, \citenamefont {Yano},
  \citenamefont {Hara}, \citenamefont {Toda}, \citenamefont {Kajita},
  \citenamefont {Ido},\ and\ \citenamefont {Ono}}]{Nishino:212}%
  \BibitemOpen
  \bibfield  {author} {\bibinfo {author} {\bibfnamefont {H.}~\bibnamefont
  {Nishino}}, \bibinfo {author} {\bibfnamefont {Y.}~\bibnamefont {Yano}},
  \bibinfo {author} {\bibfnamefont {M.}~\bibnamefont {Hara}}, \bibinfo {author}
  {\bibfnamefont {M.}~\bibnamefont {Toda}}, \bibinfo {author} {\bibfnamefont
  {M.}~\bibnamefont {Kajita}}, \bibinfo {author} {\bibfnamefont
  {T.}~\bibnamefont {Ido}},\ and\ \bibinfo {author} {\bibfnamefont
  {T.}~\bibnamefont {Ono}},\ }\href {https://doi.org/10.1364/OL.424354}
  {\bibfield  {journal} {\bibinfo  {journal} {Opt. Lett.}\ }\textbf {\bibinfo
  {volume} {46}},\ \bibinfo {pages} {2272} (\bibinfo {year}
  {2021}{\natexlab{a}})}\BibitemShut {NoStop}%
\bibitem [{\citenamefont {Slepkov}\ \emph {et~al.}(2010)\citenamefont
  {Slepkov}, \citenamefont {Bhagwat}, \citenamefont {Venkataraman},
  \citenamefont {Londero},\ and\ \citenamefont {Gaeta}}]{hollowcore}%
  \BibitemOpen
  \bibfield  {author} {\bibinfo {author} {\bibfnamefont {A.~D.}\ \bibnamefont
  {Slepkov}}, \bibinfo {author} {\bibfnamefont {A.~R.}\ \bibnamefont
  {Bhagwat}}, \bibinfo {author} {\bibfnamefont {V.}~\bibnamefont
  {Venkataraman}}, \bibinfo {author} {\bibfnamefont {P.}~\bibnamefont
  {Londero}},\ and\ \bibinfo {author} {\bibfnamefont {A.~L.}\ \bibnamefont
  {Gaeta}},\ }\href {https://doi.org/10.1103/PhysRevA.81.053825} {\bibfield
  {journal} {\bibinfo  {journal} {Phys. Rev. A}\ }\textbf {\bibinfo {volume}
  {81}},\ \bibinfo {pages} {053825} (\bibinfo {year} {2010})}\BibitemShut
  {NoStop}%
\bibitem [{\citenamefont {Yin}\ \emph {et~al.}(2004)\citenamefont {Yin},
  \citenamefont {Schmidt}, \citenamefont {Barber},\ and\ \citenamefont
  {Hawkins}}]{Yin:04}%
  \BibitemOpen
  \bibfield  {author} {\bibinfo {author} {\bibfnamefont {D.}~\bibnamefont
  {Yin}}, \bibinfo {author} {\bibfnamefont {H.}~\bibnamefont {Schmidt}},
  \bibinfo {author} {\bibfnamefont {J.}~\bibnamefont {Barber}},\ and\ \bibinfo
  {author} {\bibfnamefont {A.}~\bibnamefont {Hawkins}},\ }\href
  {https://doi.org/10.1364/OPEX.12.002710} {\bibfield  {journal} {\bibinfo
  {journal} {Opt. Express}\ }\textbf {\bibinfo {volume} {12}},\ \bibinfo
  {pages} {2710} (\bibinfo {year} {2004})}\BibitemShut {NoStop}%
\bibitem [{\citenamefont {Perrella}\ \emph {et~al.}(2013)\citenamefont
  {Perrella}, \citenamefont {Light}, \citenamefont {Anstie}, \citenamefont
  {Baynes}, \citenamefont {Benabid},\ and\ \citenamefont
  {Luiten}}]{Perrella:13}%
  \BibitemOpen
  \bibfield  {author} {\bibinfo {author} {\bibfnamefont {C.}~\bibnamefont
  {Perrella}}, \bibinfo {author} {\bibfnamefont {P.~S.}\ \bibnamefont {Light}},
  \bibinfo {author} {\bibfnamefont {J.~D.}\ \bibnamefont {Anstie}}, \bibinfo
  {author} {\bibfnamefont {F.~N.}\ \bibnamefont {Baynes}}, \bibinfo {author}
  {\bibfnamefont {F.}~\bibnamefont {Benabid}},\ and\ \bibinfo {author}
  {\bibfnamefont {A.~N.}\ \bibnamefont {Luiten}},\ }\href
  {https://doi.org/10.1364/OL.38.002122} {\bibfield  {journal} {\bibinfo
  {journal} {Opt. Lett.}\ }\textbf {\bibinfo {volume} {38}},\ \bibinfo {pages}
  {2122} (\bibinfo {year} {2013})}\BibitemShut {NoStop}%
\bibitem [{\citenamefont {Perrella}\ \emph {et~al.}(2012)\citenamefont
  {Perrella}, \citenamefont {Light}, \citenamefont {Stace}, \citenamefont
  {Benabid},\ and\ \citenamefont {Luiten}}]{Perellahollowcore}%
  \BibitemOpen
  \bibfield  {author} {\bibinfo {author} {\bibfnamefont {C.}~\bibnamefont
  {Perrella}}, \bibinfo {author} {\bibfnamefont {P.~S.}\ \bibnamefont {Light}},
  \bibinfo {author} {\bibfnamefont {T.~M.}\ \bibnamefont {Stace}}, \bibinfo
  {author} {\bibfnamefont {F.}~\bibnamefont {Benabid}},\ and\ \bibinfo {author}
  {\bibfnamefont {A.~N.}\ \bibnamefont {Luiten}},\ }\href
  {https://doi.org/10.1103/PhysRevA.85.012518} {\bibfield  {journal} {\bibinfo
  {journal} {Phys. Rev. A}\ }\textbf {\bibinfo {volume} {85}},\ \bibinfo
  {pages} {012518} (\bibinfo {year} {2012})}\BibitemShut {NoStop}%
\bibitem [{\citenamefont {Wang}\ \emph {et~al.}(2018)\citenamefont {Wang},
  \citenamefont {Yi}, \citenamefont {Mawdsley}, \citenamefont {Kim},
  \citenamefont {Wang},\ and\ \citenamefont {Han}}]{csmc}%
  \BibitemOpen
  \bibfield  {author} {\bibinfo {author} {\bibfnamefont {C.}~\bibnamefont
  {Wang}}, \bibinfo {author} {\bibfnamefont {X.}~\bibnamefont {Yi}}, \bibinfo
  {author} {\bibfnamefont {J.}~\bibnamefont {Mawdsley}}, \bibinfo {author}
  {\bibfnamefont {M.}~\bibnamefont {Kim}}, \bibinfo {author} {\bibfnamefont
  {Z.}~\bibnamefont {Wang}},\ and\ \bibinfo {author} {\bibfnamefont
  {R.}~\bibnamefont {Han}},\ }\href {https://doi.org/10.1038/s41928-018-0102-4}
  {\bibfield  {journal} {\bibinfo  {journal} {Nature Electronics}\ }\textbf
  {\bibinfo {volume} {1}},\ \bibinfo {pages} {421} (\bibinfo {year}
  {2018})}\BibitemShut {NoStop}%
\bibitem [{\citenamefont {Stern}\ \emph {et~al.}(2013)\citenamefont {Stern},
  \citenamefont {Desiatov}, \citenamefont {Goykhman},\ and\ \citenamefont
  {Levy}}]{Stern2013}%
  \BibitemOpen
  \bibfield  {author} {\bibinfo {author} {\bibfnamefont {L.}~\bibnamefont
  {Stern}}, \bibinfo {author} {\bibfnamefont {B.}~\bibnamefont {Desiatov}},
  \bibinfo {author} {\bibfnamefont {I.}~\bibnamefont {Goykhman}},\ and\
  \bibinfo {author} {\bibfnamefont {U.}~\bibnamefont {Levy}},\ }\href
  {https://doi.org/10.1038/ncomms2554} {\bibfield  {journal} {\bibinfo
  {journal} {Nature Communications}\ }\textbf {\bibinfo {volume} {4}},\
  \bibinfo {pages} {1548} (\bibinfo {year} {2013})}\BibitemShut {NoStop}%
\bibitem [{\citenamefont {Stern}\ \emph {et~al.}(2017)\citenamefont {Stern},
  \citenamefont {Desiatov}, \citenamefont {Mazurski},\ and\ \citenamefont
  {Levy}}]{Stern2017}%
  \BibitemOpen
  \bibfield  {author} {\bibinfo {author} {\bibfnamefont {L.}~\bibnamefont
  {Stern}}, \bibinfo {author} {\bibfnamefont {B.}~\bibnamefont {Desiatov}},
  \bibinfo {author} {\bibfnamefont {N.}~\bibnamefont {Mazurski}},\ and\
  \bibinfo {author} {\bibfnamefont {U.}~\bibnamefont {Levy}},\ }\href
  {https://doi.org/10.1038/ncomms14461} {\bibfield  {journal} {\bibinfo
  {journal} {Nature Communications}\ }\textbf {\bibinfo {volume} {8}},\
  \bibinfo {pages} {14461} (\bibinfo {year} {2017})}\BibitemShut {NoStop}%
\bibitem [{\citenamefont {Zektzer}\ \emph {et~al.}(2020)\citenamefont
  {Zektzer}, \citenamefont {Hummon}, \citenamefont {Stern}, \citenamefont
  {Sebbag}, \citenamefont {Barash}, \citenamefont {Mazurski}, \citenamefont
  {Kitching},\ and\ \citenamefont {Levy}}]{hummonstern}%
  \BibitemOpen
  \bibfield  {author} {\bibinfo {author} {\bibfnamefont {R.}~\bibnamefont
  {Zektzer}}, \bibinfo {author} {\bibfnamefont {M.~T.}\ \bibnamefont {Hummon}},
  \bibinfo {author} {\bibfnamefont {L.}~\bibnamefont {Stern}}, \bibinfo
  {author} {\bibfnamefont {Y.}~\bibnamefont {Sebbag}}, \bibinfo {author}
  {\bibfnamefont {Y.}~\bibnamefont {Barash}}, \bibinfo {author} {\bibfnamefont
  {N.}~\bibnamefont {Mazurski}}, \bibinfo {author} {\bibfnamefont
  {J.}~\bibnamefont {Kitching}},\ and\ \bibinfo {author} {\bibfnamefont
  {U.}~\bibnamefont {Levy}},\ }\href
  {https://doi.org/https://doi.org/10.1002/lpor.201900414} {\bibfield
  {journal} {\bibinfo  {journal} {Laser and Photonics Reviews}\ }\textbf
  {\bibinfo {volume} {14}},\ \bibinfo {pages} {1900414} (\bibinfo {year}
  {2020})}\BibitemShut {NoStop}%
\bibitem [{\citenamefont {Bridge}\ \emph {et~al.}(2009)\citenamefont {Bridge},
  \citenamefont {Millen}, \citenamefont {Adams},\ and\ \citenamefont
  {Jones}}]{c-adams}%
  \BibitemOpen
  \bibfield  {author} {\bibinfo {author} {\bibfnamefont {E.~M.}\ \bibnamefont
  {Bridge}}, \bibinfo {author} {\bibfnamefont {J.}~\bibnamefont {Millen}},
  \bibinfo {author} {\bibfnamefont {C.~S.}\ \bibnamefont {Adams}},\ and\
  \bibinfo {author} {\bibfnamefont {M.~P.~A.}\ \bibnamefont {Jones}},\ }\href
  {https://doi.org/10.1063/1.3036980} {\bibfield  {journal} {\bibinfo
  {journal} {Review of Scientific Instruments}\ }\textbf {\bibinfo {volume}
  {80}},\ \bibinfo {pages} {013101} (\bibinfo {year} {2009})}\BibitemShut
  {NoStop}%
\bibitem [{\citenamefont {Kitching}\ \emph {et~al.}(2002)\citenamefont
  {Kitching}, \citenamefont {Knappe},\ and\ \citenamefont
  {Hollberg}}]{kitching2002}%
  \BibitemOpen
  \bibfield  {author} {\bibinfo {author} {\bibfnamefont {J.}~\bibnamefont
  {Kitching}}, \bibinfo {author} {\bibfnamefont {S.}~\bibnamefont {Knappe}},\
  and\ \bibinfo {author} {\bibfnamefont {L.}~\bibnamefont {Hollberg}},\ }\href
  {https://doi.org/10.1063/1.1494115} {\bibfield  {journal} {\bibinfo
  {journal} {Applied Physics Letters}\ }\textbf {\bibinfo {volume} {81}},\
  \bibinfo {pages} {553} (\bibinfo {year} {2002})}\BibitemShut {NoStop}%
\bibitem [{\citenamefont {Bopp}\ \emph {et~al.}(2020)\citenamefont {Bopp},
  \citenamefont {Maurice},\ and\ \citenamefont {Kitching}}]{Bopp_2020}%
  \BibitemOpen
  \bibfield  {author} {\bibinfo {author} {\bibfnamefont {D.~G.}\ \bibnamefont
  {Bopp}}, \bibinfo {author} {\bibfnamefont {V.~M.}\ \bibnamefont {Maurice}},\
  and\ \bibinfo {author} {\bibfnamefont {J.~E.}\ \bibnamefont {Kitching}},\
  }\href {https://doi.org/10.1088/2515-7647/abcbe5} {\bibfield  {journal}
  {\bibinfo  {journal} {Journal of Physics: Photonics}\ }\textbf {\bibinfo
  {volume} {3}},\ \bibinfo {pages} {015002} (\bibinfo {year}
  {2020})}\BibitemShut {NoStop}%
\bibitem [{\citenamefont {Douahi}\ \emph
  {et~al.}(2007{\natexlab{a}})\citenamefont {Douahi}, \citenamefont {Nieradko},
  \citenamefont {Beugnot}, \citenamefont {Dziuban}, \citenamefont {Maillote},
  \citenamefont {Guerandel}, \citenamefont {Moraja}, \citenamefont {Gorecki},\
  and\ \citenamefont {Giordano}}]{gorecki}%
  \BibitemOpen
  \bibfield  {author} {\bibinfo {author} {\bibfnamefont {A.}~\bibnamefont
  {Douahi}}, \bibinfo {author} {\bibfnamefont {L.}~\bibnamefont {Nieradko}},
  \bibinfo {author} {\bibfnamefont {J.}~\bibnamefont {Beugnot}}, \bibinfo
  {author} {\bibfnamefont {J.}~\bibnamefont {Dziuban}}, \bibinfo {author}
  {\bibfnamefont {H.}~\bibnamefont {Maillote}}, \bibinfo {author}
  {\bibfnamefont {S.}~\bibnamefont {Guerandel}}, \bibinfo {author}
  {\bibfnamefont {M.}~\bibnamefont {Moraja}}, \bibinfo {author} {\bibfnamefont
  {C.}~\bibnamefont {Gorecki}},\ and\ \bibinfo {author} {\bibfnamefont
  {V.}~\bibnamefont {Giordano}},\ }\href
  {https://digital-library.theiet.org/content/journals/10.1049/el_20070147}
  {\bibfield  {journal} {\bibinfo  {journal} {Electronics Letters}\ }\textbf
  {\bibinfo {volume} {43}},\ \bibinfo {pages} {279} (\bibinfo {year}
  {2007}{\natexlab{a}})}\BibitemShut {NoStop}%
\bibitem [{\citenamefont {Douahi}\ \emph
  {et~al.}(2007{\natexlab{b}})\citenamefont {Douahi}, \citenamefont {Nieradko},
  \citenamefont {Beugnot}, \citenamefont {Dziuban}, \citenamefont {Maillote},
  \citenamefont {Boudot}, \citenamefont {Guerandel}, \citenamefont {Moraja},
  \citenamefont {Gorecki},\ and\ \citenamefont {Giordano}}]{boudotpill}%
  \BibitemOpen
  \bibfield  {author} {\bibinfo {author} {\bibfnamefont {A.}~\bibnamefont
  {Douahi}}, \bibinfo {author} {\bibfnamefont {L.}~\bibnamefont {Nieradko}},
  \bibinfo {author} {\bibfnamefont {J.}~\bibnamefont {Beugnot}}, \bibinfo
  {author} {\bibfnamefont {J.}~\bibnamefont {Dziuban}}, \bibinfo {author}
  {\bibfnamefont {H.}~\bibnamefont {Maillote}}, \bibinfo {author}
  {\bibfnamefont {R.}~\bibnamefont {Boudot}}, \bibinfo {author} {\bibfnamefont
  {S.}~\bibnamefont {Guerandel}}, \bibinfo {author} {\bibfnamefont
  {M.}~\bibnamefont {Moraja}}, \bibinfo {author} {\bibfnamefont
  {C.}~\bibnamefont {Gorecki}},\ and\ \bibinfo {author} {\bibfnamefont
  {V.}~\bibnamefont {Giordano}},\ }in\ \href
  {https://doi.org/10.1109/FREQ.2007.4319031} {\emph {\bibinfo {booktitle}
  {2007 IEEE International Frequency Control Symposium Joint with the 21st
  European Frequency and Time Forum}}}\ (\bibinfo {year} {2007})\ pp.\ \bibinfo
  {pages} {58--61}\BibitemShut {NoStop}%
\bibitem [{\citenamefont {Bregazzi}\ \emph
  {et~al.}(2021{\natexlab{a}})\citenamefont {Bregazzi}, \citenamefont {Dyer},
  \citenamefont {Griffin}, \citenamefont {Burt}, \citenamefont {Arnold},
  \citenamefont {Riis},\ and\ \citenamefont {McGilligan}}]{mcgilliganSPIE}%
  \BibitemOpen
  \bibfield  {author} {\bibinfo {author} {\bibfnamefont {A.}~\bibnamefont
  {Bregazzi}}, \bibinfo {author} {\bibfnamefont {S.}~\bibnamefont {Dyer}},
  \bibinfo {author} {\bibfnamefont {P.~F.}\ \bibnamefont {Griffin}}, \bibinfo
  {author} {\bibfnamefont {D.~P.}\ \bibnamefont {Burt}}, \bibinfo {author}
  {\bibfnamefont {A.~S.}\ \bibnamefont {Arnold}}, \bibinfo {author}
  {\bibfnamefont {E.}~\bibnamefont {Riis}},\ and\ \bibinfo {author}
  {\bibfnamefont {J.~P.}\ \bibnamefont {McGilligan}},\ }in\ \href
  {https://doi.org/10.1117/12.2601340} {\emph {\bibinfo {booktitle} {Quantum
  Technology: Driving Commercialisation of an Enabling Science II}}},\ Vol.\
  \bibinfo {volume} {11881},\ \bibinfo {editor} {edited by\ \bibinfo {editor}
  {\bibfnamefont {M.~J.}\ \bibnamefont {Padgett}}, \bibinfo {editor}
  {\bibfnamefont {K.}~\bibnamefont {Bongs}}, \bibinfo {editor} {\bibfnamefont
  {A.}~\bibnamefont {Fedrizzi}},\ and\ \bibinfo {editor} {\bibfnamefont
  {A.}~\bibnamefont {Politi}}},\ \bibinfo {organization} {International Society
  for Optics and Photonics}\ (\bibinfo  {publisher} {SPIE},\ \bibinfo {year}
  {2021})\ pp.\ \bibinfo {pages} {100 -- 110}\BibitemShut {NoStop}%
\bibitem [{\citenamefont {Nishino}\ \emph
  {et~al.}(2021{\natexlab{b}})\citenamefont {Nishino}, \citenamefont {Furuya},\
  and\ \citenamefont {Ono}}]{Nishino:21}%
  \BibitemOpen
  \bibfield  {author} {\bibinfo {author} {\bibfnamefont {H.}~\bibnamefont
  {Nishino}}, \bibinfo {author} {\bibfnamefont {Y.}~\bibnamefont {Furuya}},\
  and\ \bibinfo {author} {\bibfnamefont {T.}~\bibnamefont {Ono}},\ }\href
  {https://doi.org/10.1364/OE.442859} {\bibfield  {journal} {\bibinfo
  {journal} {Opt. Express}\ }\textbf {\bibinfo {volume} {29}},\ \bibinfo
  {pages} {44316} (\bibinfo {year} {2021}{\natexlab{b}})}\BibitemShut {NoStop}%
\bibitem [{\citenamefont {Henmi}\ \emph {et~al.}(1994)\citenamefont {Henmi},
  \citenamefont {Shoji}, \citenamefont {Shoji}, \citenamefont {Yoshimi},\ and\
  \citenamefont {Esashi}}]{HENMI1994243}%
  \BibitemOpen
  \bibfield  {author} {\bibinfo {author} {\bibfnamefont {H.}~\bibnamefont
  {Henmi}}, \bibinfo {author} {\bibfnamefont {S.}~\bibnamefont {Shoji}},
  \bibinfo {author} {\bibfnamefont {Y.}~\bibnamefont {Shoji}}, \bibinfo
  {author} {\bibfnamefont {K.}~\bibnamefont {Yoshimi}},\ and\ \bibinfo {author}
  {\bibfnamefont {M.}~\bibnamefont {Esashi}},\ }\href
  {https://doi.org/https://doi.org/10.1016/0924-4247(94)80003-0} {\bibfield
  {journal} {\bibinfo  {journal} {Sensors and Actuators A: Physical}\ }\textbf
  {\bibinfo {volume} {43}},\ \bibinfo {pages} {243} (\bibinfo {year}
  {1994})}\BibitemShut {NoStop}%
\bibitem [{\citenamefont {Maurice}\ \emph
  {et~al.}(2017{\natexlab{a}})\citenamefont {Maurice}, \citenamefont
  {Rutkowski}, \citenamefont {Kroemer}, \citenamefont {Bargiel}, \citenamefont
  {Passilly}, \citenamefont {Boudot}, \citenamefont {Chutani}, \citenamefont
  {Galliou}, \citenamefont {Gorecki}, \citenamefont {Mauri},\ and\
  \citenamefont {Moraja}}]{boudotlineofsightchannel}%
  \BibitemOpen
  \bibfield  {author} {\bibinfo {author} {\bibfnamefont {V.}~\bibnamefont
  {Maurice}}, \bibinfo {author} {\bibfnamefont {J.}~\bibnamefont {Rutkowski}},
  \bibinfo {author} {\bibfnamefont {E.}~\bibnamefont {Kroemer}}, \bibinfo
  {author} {\bibfnamefont {S.}~\bibnamefont {Bargiel}}, \bibinfo {author}
  {\bibfnamefont {N.}~\bibnamefont {Passilly}}, \bibinfo {author}
  {\bibfnamefont {R.}~\bibnamefont {Boudot}}, \bibinfo {author} {\bibfnamefont
  {R.}~\bibnamefont {Chutani}}, \bibinfo {author} {\bibfnamefont
  {S.}~\bibnamefont {Galliou}}, \bibinfo {author} {\bibfnamefont
  {C.}~\bibnamefont {Gorecki}}, \bibinfo {author} {\bibfnamefont
  {L.}~\bibnamefont {Mauri}},\ and\ \bibinfo {author} {\bibfnamefont
  {M.}~\bibnamefont {Moraja}},\ }in\ \href
  {https://doi.org/10.1109/FCS.2017.8088984} {\emph {\bibinfo {booktitle} {2017
  Joint Conference of the European Frequency and Time Forum and IEEE
  International Frequency Control Symposium (EFTF/IFCS)}}}\ (\bibinfo {year}
  {2017})\ pp.\ \bibinfo {pages} {636--637}\BibitemShut {NoStop}%
\bibitem [{\citenamefont {Li}\ \emph {et~al.}(2019)\citenamefont {Li},
  \citenamefont {Chai}, \citenamefont {Wei}, \citenamefont {Yang},
  \citenamefont {Daruwalla}, \citenamefont {Ayazi},\ and\ \citenamefont
  {Raman}}]{Li2019}%
  \BibitemOpen
  \bibfield  {author} {\bibinfo {author} {\bibfnamefont {C.}~\bibnamefont
  {Li}}, \bibinfo {author} {\bibfnamefont {X.}~\bibnamefont {Chai}}, \bibinfo
  {author} {\bibfnamefont {B.}~\bibnamefont {Wei}}, \bibinfo {author}
  {\bibfnamefont {J.}~\bibnamefont {Yang}}, \bibinfo {author} {\bibfnamefont
  {A.}~\bibnamefont {Daruwalla}}, \bibinfo {author} {\bibfnamefont
  {F.}~\bibnamefont {Ayazi}},\ and\ \bibinfo {author} {\bibfnamefont
  {C.}~\bibnamefont {Raman}},\ }\href
  {https://doi.org/10.1038/s41467-019-09647-3} {\bibfield  {journal} {\bibinfo
  {journal} {Nature Communications}\ }\textbf {\bibinfo {volume} {10}},\
  \bibinfo {pages} {1831} (\bibinfo {year} {2019})}\BibitemShut {NoStop}%
\bibitem [{\citenamefont {Maurice}\ \emph
  {et~al.}(2017{\natexlab{b}})\citenamefont {Maurice}, \citenamefont
  {Rutkowski}, \citenamefont {Kroemer}, \citenamefont {Bargiel}, \citenamefont
  {Passilly}, \citenamefont {Boudot}, \citenamefont {Gorecki}, \citenamefont
  {Mauri},\ and\ \citenamefont {Moraja}}]{paste}%
  \BibitemOpen
  \bibfield  {author} {\bibinfo {author} {\bibfnamefont {V.}~\bibnamefont
  {Maurice}}, \bibinfo {author} {\bibfnamefont {J.}~\bibnamefont {Rutkowski}},
  \bibinfo {author} {\bibfnamefont {E.}~\bibnamefont {Kroemer}}, \bibinfo
  {author} {\bibfnamefont {S.}~\bibnamefont {Bargiel}}, \bibinfo {author}
  {\bibfnamefont {N.}~\bibnamefont {Passilly}}, \bibinfo {author}
  {\bibfnamefont {R.}~\bibnamefont {Boudot}}, \bibinfo {author} {\bibfnamefont
  {C.}~\bibnamefont {Gorecki}}, \bibinfo {author} {\bibfnamefont
  {L.}~\bibnamefont {Mauri}},\ and\ \bibinfo {author} {\bibfnamefont
  {M.}~\bibnamefont {Moraja}},\ }\href {https://doi.org/10.1063/1.4981772}
  {\bibfield  {journal} {\bibinfo  {journal} {Applied Physics Letters}\
  }\textbf {\bibinfo {volume} {110}},\ \bibinfo {pages} {164103} (\bibinfo
  {year} {2017}{\natexlab{b}})}\BibitemShut {NoStop}%
\bibitem [{\citenamefont {Stern}\ \emph {et~al.}(2019)\citenamefont {Stern},
  \citenamefont {Bopp}, \citenamefont {Schima}, \citenamefont {Maurice},\ and\
  \citenamefont {Kitching}}]{BoppStern}%
  \BibitemOpen
  \bibfield  {author} {\bibinfo {author} {\bibfnamefont {L.}~\bibnamefont
  {Stern}}, \bibinfo {author} {\bibfnamefont {D.~G.}\ \bibnamefont {Bopp}},
  \bibinfo {author} {\bibfnamefont {S.~A.}\ \bibnamefont {Schima}}, \bibinfo
  {author} {\bibfnamefont {V.~N.}\ \bibnamefont {Maurice}},\ and\ \bibinfo
  {author} {\bibfnamefont {J.~E.}\ \bibnamefont {Kitching}},\ }\href
  {https://doi.org/10.1038/s41467-019-11145-5} {\bibfield  {journal} {\bibinfo
  {journal} {Nature Communications}\ }\textbf {\bibinfo {volume} {10}},\
  \bibinfo {pages} {3156} (\bibinfo {year} {2019})}\BibitemShut {NoStop}%
\bibitem [{\citenamefont {Vicarini}\ \emph {et~al.}(2018)\citenamefont
  {Vicarini}, \citenamefont {Maurice}, \citenamefont {{Abdel Hafiz}},
  \citenamefont {Rutkowski}, \citenamefont {Gorecki}, \citenamefont {Passilly},
  \citenamefont {Ribetto}, \citenamefont {Gaff}, \citenamefont {Volant},
  \citenamefont {Galliou},\ and\ \citenamefont {Boudot}}]{VICARINI201899}%
  \BibitemOpen
  \bibfield  {author} {\bibinfo {author} {\bibfnamefont {R.}~\bibnamefont
  {Vicarini}}, \bibinfo {author} {\bibfnamefont {V.}~\bibnamefont {Maurice}},
  \bibinfo {author} {\bibfnamefont {M.}~\bibnamefont {{Abdel Hafiz}}}, \bibinfo
  {author} {\bibfnamefont {J.}~\bibnamefont {Rutkowski}}, \bibinfo {author}
  {\bibfnamefont {C.}~\bibnamefont {Gorecki}}, \bibinfo {author} {\bibfnamefont
  {N.}~\bibnamefont {Passilly}}, \bibinfo {author} {\bibfnamefont
  {L.}~\bibnamefont {Ribetto}}, \bibinfo {author} {\bibfnamefont
  {V.}~\bibnamefont {Gaff}}, \bibinfo {author} {\bibfnamefont {V.}~\bibnamefont
  {Volant}}, \bibinfo {author} {\bibfnamefont {S.}~\bibnamefont {Galliou}},\
  and\ \bibinfo {author} {\bibfnamefont {R.}~\bibnamefont {Boudot}},\ }\href
  {https://doi.org/https://doi.org/10.1016/j.sna.2018.07.032} {\bibfield
  {journal} {\bibinfo  {journal} {Sensors and Actuators A: Physical}\ }\textbf
  {\bibinfo {volume} {280}},\ \bibinfo {pages} {99} (\bibinfo {year}
  {2018})}\BibitemShut {NoStop}%
\bibitem [{\citenamefont {Kohn}\ \emph {et~al.}(2020)\citenamefont {Kohn},
  \citenamefont {Bigelow}, \citenamefont {Spanjers}, \citenamefont {Stuhl},
  \citenamefont {Kasch}, \citenamefont {Olson}, \citenamefont {Imhof},
  \citenamefont {Hostutler},\ and\ \citenamefont {Squires}}]{AFRL}%
  \BibitemOpen
  \bibfield  {author} {\bibinfo {author} {\bibfnamefont {R.~N.}\ \bibnamefont
  {Kohn}}, \bibinfo {author} {\bibfnamefont {M.~S.}\ \bibnamefont {Bigelow}},
  \bibinfo {author} {\bibfnamefont {M.}~\bibnamefont {Spanjers}}, \bibinfo
  {author} {\bibfnamefont {B.~K.}\ \bibnamefont {Stuhl}}, \bibinfo {author}
  {\bibfnamefont {B.~L.}\ \bibnamefont {Kasch}}, \bibinfo {author}
  {\bibfnamefont {S.~E.}\ \bibnamefont {Olson}}, \bibinfo {author}
  {\bibfnamefont {E.~A.}\ \bibnamefont {Imhof}}, \bibinfo {author}
  {\bibfnamefont {D.~A.}\ \bibnamefont {Hostutler}},\ and\ \bibinfo {author}
  {\bibfnamefont {M.~B.}\ \bibnamefont {Squires}},\ }\href
  {https://doi.org/10.1063/1.5128120} {\bibfield  {journal} {\bibinfo
  {journal} {Review of Scientific Instruments}\ }\textbf {\bibinfo {volume}
  {91}},\ \bibinfo {pages} {035108} (\bibinfo {year} {2020})}\BibitemShut
  {NoStop}%
\bibitem [{\citenamefont {Kang}\ \emph {et~al.}(2019)\citenamefont {Kang},
  \citenamefont {Moore}, \citenamefont {McGilligan}, \citenamefont {Mott},
  \citenamefont {Mis}, \citenamefont {Roper}, \citenamefont {Donley},\ and\
  \citenamefont {Kitching}}]{Kang:19}%
  \BibitemOpen
  \bibfield  {author} {\bibinfo {author} {\bibfnamefont {S.}~\bibnamefont
  {Kang}}, \bibinfo {author} {\bibfnamefont {K.~R.}\ \bibnamefont {Moore}},
  \bibinfo {author} {\bibfnamefont {J.~P.}\ \bibnamefont {McGilligan}},
  \bibinfo {author} {\bibfnamefont {R.}~\bibnamefont {Mott}}, \bibinfo {author}
  {\bibfnamefont {A.}~\bibnamefont {Mis}}, \bibinfo {author} {\bibfnamefont
  {C.}~\bibnamefont {Roper}}, \bibinfo {author} {\bibfnamefont {E.~A.}\
  \bibnamefont {Donley}},\ and\ \bibinfo {author} {\bibfnamefont
  {J.}~\bibnamefont {Kitching}},\ }\href {https://doi.org/10.1364/OL.44.003002}
  {\bibfield  {journal} {\bibinfo  {journal} {Optics Letters}\ }\textbf
  {\bibinfo {volume} {44}},\ \bibinfo {pages} {3002} (\bibinfo {year}
  {2019})}\BibitemShut {NoStop}%
\bibitem [{\citenamefont {McGilligan}\ \emph
  {et~al.}(2020{\natexlab{a}})\citenamefont {McGilligan}, \citenamefont
  {Moore}, \citenamefont {Kang}, \citenamefont {Mott}, \citenamefont {Mis},
  \citenamefont {Roper}, \citenamefont {Donley},\ and\ \citenamefont
  {Kitching}}]{mcgilliganAIB}%
  \BibitemOpen
  \bibfield  {author} {\bibinfo {author} {\bibfnamefont {J.~P.}\ \bibnamefont
  {McGilligan}}, \bibinfo {author} {\bibfnamefont {K.~R.}\ \bibnamefont
  {Moore}}, \bibinfo {author} {\bibfnamefont {S.}~\bibnamefont {Kang}},
  \bibinfo {author} {\bibfnamefont {R.}~\bibnamefont {Mott}}, \bibinfo {author}
  {\bibfnamefont {A.}~\bibnamefont {Mis}}, \bibinfo {author} {\bibfnamefont
  {C.}~\bibnamefont {Roper}}, \bibinfo {author} {\bibfnamefont {E.~A.}\
  \bibnamefont {Donley}},\ and\ \bibinfo {author} {\bibfnamefont
  {J.}~\bibnamefont {Kitching}},\ }\href
  {https://doi.org/10.1103/PhysRevApplied.13.044038} {\bibfield  {journal}
  {\bibinfo  {journal} {Phys. Rev. Applied}\ }\textbf {\bibinfo {volume}
  {13}},\ \bibinfo {pages} {044038} (\bibinfo {year}
  {2020}{\natexlab{a}})}\BibitemShut {NoStop}%
\bibitem [{\citenamefont {Karlen}\ \emph {et~al.}(2018)\citenamefont {Karlen},
  \citenamefont {Overstolz}, \citenamefont {Gobet}, \citenamefont {Haesler},
  \citenamefont {Droz},\ and\ \citenamefont {Lecomte}}]{microdisks}%
  \BibitemOpen
  \bibfield  {author} {\bibinfo {author} {\bibfnamefont {S.}~\bibnamefont
  {Karlen}}, \bibinfo {author} {\bibfnamefont {T.}~\bibnamefont {Overstolz}},
  \bibinfo {author} {\bibfnamefont {J.}~\bibnamefont {Gobet}}, \bibinfo
  {author} {\bibfnamefont {J.}~\bibnamefont {Haesler}}, \bibinfo {author}
  {\bibfnamefont {F.}~\bibnamefont {Droz}},\ and\ \bibinfo {author}
  {\bibfnamefont {S.}~\bibnamefont {Lecomte}},\ }in\ \href
  {https://doi.org/10.1109/EFTF.2018.8409005} {\emph {\bibinfo {booktitle}
  {2018 European Frequency and Time Forum (EFTF)}}}\ (\bibinfo {year} {2018})\
  pp.\ \bibinfo {pages} {91--93}\BibitemShut {NoStop}%
\bibitem [{\citenamefont {Dugrain}\ \emph {et~al.}(2014)\citenamefont
  {Dugrain}, \citenamefont {Rosenbusch},\ and\ \citenamefont
  {Reichel}}]{fastamd2}%
  \BibitemOpen
  \bibfield  {author} {\bibinfo {author} {\bibfnamefont {V.}~\bibnamefont
  {Dugrain}}, \bibinfo {author} {\bibfnamefont {P.}~\bibnamefont
  {Rosenbusch}},\ and\ \bibinfo {author} {\bibfnamefont {J.}~\bibnamefont
  {Reichel}},\ }\href {https://doi.org/10.1063/1.4892996} {\bibfield  {journal}
  {\bibinfo  {journal} {Review of Scientific Instruments}\ }\textbf {\bibinfo
  {volume} {85}},\ \bibinfo {pages} {083112} (\bibinfo {year}
  {2014})}\BibitemShut {NoStop}%
\bibitem [{\citenamefont {Bogi}\ \emph {et~al.}(2009)\citenamefont {Bogi},
  \citenamefont {Marinelli}, \citenamefont {Burchianti}, \citenamefont
  {Mariotti}, \citenamefont {Moi}, \citenamefont {Gozzini}, \citenamefont
  {Marmugi},\ and\ \citenamefont {Lucchesini}}]{Bogi:09}%
  \BibitemOpen
  \bibfield  {author} {\bibinfo {author} {\bibfnamefont {A.}~\bibnamefont
  {Bogi}}, \bibinfo {author} {\bibfnamefont {C.}~\bibnamefont {Marinelli}},
  \bibinfo {author} {\bibfnamefont {A.}~\bibnamefont {Burchianti}}, \bibinfo
  {author} {\bibfnamefont {E.}~\bibnamefont {Mariotti}}, \bibinfo {author}
  {\bibfnamefont {L.}~\bibnamefont {Moi}}, \bibinfo {author} {\bibfnamefont
  {S.}~\bibnamefont {Gozzini}}, \bibinfo {author} {\bibfnamefont
  {L.}~\bibnamefont {Marmugi}},\ and\ \bibinfo {author} {\bibfnamefont
  {A.}~\bibnamefont {Lucchesini}},\ }\href
  {https://doi.org/10.1364/OL.34.002643} {\bibfield  {journal} {\bibinfo
  {journal} {Opt. Lett.}\ }\textbf {\bibinfo {volume} {34}},\ \bibinfo {pages}
  {2643} (\bibinfo {year} {2009})}\BibitemShut {NoStop}%
\bibitem [{\citenamefont {Marmugi}\ \emph {et~al.}(2012)\citenamefont
  {Marmugi}, \citenamefont {Gozzini}, \citenamefont {Lucchesini}, \citenamefont
  {Bogi}, \citenamefont {Burchianti},\ and\ \citenamefont
  {Marinelli}}]{Marmugi:12}%
  \BibitemOpen
  \bibfield  {author} {\bibinfo {author} {\bibfnamefont {L.}~\bibnamefont
  {Marmugi}}, \bibinfo {author} {\bibfnamefont {S.}~\bibnamefont {Gozzini}},
  \bibinfo {author} {\bibfnamefont {A.}~\bibnamefont {Lucchesini}}, \bibinfo
  {author} {\bibfnamefont {A.}~\bibnamefont {Bogi}}, \bibinfo {author}
  {\bibfnamefont {A.}~\bibnamefont {Burchianti}},\ and\ \bibinfo {author}
  {\bibfnamefont {C.}~\bibnamefont {Marinelli}},\ }\href
  {https://doi.org/10.1364/JOSAB.29.002729} {\bibfield  {journal} {\bibinfo
  {journal} {J. Opt. Soc. Am. B}\ }\textbf {\bibinfo {volume} {29}},\ \bibinfo
  {pages} {2729} (\bibinfo {year} {2012})}\BibitemShut {NoStop}%
\bibitem [{\citenamefont {Torralbo-Campo}\ \emph {et~al.}(2015)\citenamefont
  {Torralbo-Campo}, \citenamefont {Bruce}, \citenamefont {Smirne},\ and\
  \citenamefont {Cassettari}}]{Torralbo-Campo2015}%
  \BibitemOpen
  \bibfield  {author} {\bibinfo {author} {\bibfnamefont {L.}~\bibnamefont
  {Torralbo-Campo}}, \bibinfo {author} {\bibfnamefont {G.~D.}\ \bibnamefont
  {Bruce}}, \bibinfo {author} {\bibfnamefont {G.}~\bibnamefont {Smirne}},\ and\
  \bibinfo {author} {\bibfnamefont {D.}~\bibnamefont {Cassettari}},\ }\href
  {https://doi.org/10.1038/srep14729} {\bibfield  {journal} {\bibinfo
  {journal} {Scientific Reports}\ }\textbf {\bibinfo {volume} {5}},\ \bibinfo
  {pages} {14729} (\bibinfo {year} {2015})}\BibitemShut {NoStop}%
\bibitem [{\citenamefont {Talker}\ \emph {et~al.}(2021)\citenamefont {Talker},
  \citenamefont {Arora}, \citenamefont {Zektzer}, \citenamefont {Sebbag},
  \citenamefont {Dikopltsev},\ and\ \citenamefont {Levy}}]{MEMSLIAD}%
  \BibitemOpen
  \bibfield  {author} {\bibinfo {author} {\bibfnamefont {E.}~\bibnamefont
  {Talker}}, \bibinfo {author} {\bibfnamefont {P.}~\bibnamefont {Arora}},
  \bibinfo {author} {\bibfnamefont {R.}~\bibnamefont {Zektzer}}, \bibinfo
  {author} {\bibfnamefont {Y.}~\bibnamefont {Sebbag}}, \bibinfo {author}
  {\bibfnamefont {M.}~\bibnamefont {Dikopltsev}},\ and\ \bibinfo {author}
  {\bibfnamefont {U.}~\bibnamefont {Levy}},\ }\href
  {https://doi.org/10.1103/PhysRevApplied.15.L051001} {\bibfield  {journal}
  {\bibinfo  {journal} {Phys. Rev. Applied}\ }\textbf {\bibinfo {volume}
  {15}},\ \bibinfo {pages} {L051001} (\bibinfo {year} {2021})}\BibitemShut
  {NoStop}%
\bibitem [{\citenamefont {Klempt}\ \emph {et~al.}(2006)\citenamefont {Klempt},
  \citenamefont {van Zoest}, \citenamefont {Henninger}, \citenamefont {Topic},
  \citenamefont {Rasel}, \citenamefont {Ertmer},\ and\ \citenamefont
  {Arlt}}]{MOTLIAD}%
  \BibitemOpen
  \bibfield  {author} {\bibinfo {author} {\bibfnamefont {C.}~\bibnamefont
  {Klempt}}, \bibinfo {author} {\bibfnamefont {T.}~\bibnamefont {van Zoest}},
  \bibinfo {author} {\bibfnamefont {T.}~\bibnamefont {Henninger}}, \bibinfo
  {author} {\bibfnamefont {O.}~\bibnamefont {Topic}}, \bibinfo {author}
  {\bibfnamefont {E.}~\bibnamefont {Rasel}}, \bibinfo {author} {\bibfnamefont
  {W.}~\bibnamefont {Ertmer}},\ and\ \bibinfo {author} {\bibfnamefont
  {J.}~\bibnamefont {Arlt}},\ }\href
  {https://doi.org/10.1103/PhysRevA.73.013410} {\bibfield  {journal} {\bibinfo
  {journal} {Phys. Rev. A}\ }\textbf {\bibinfo {volume} {73}},\ \bibinfo
  {pages} {013410} (\bibinfo {year} {2006})}\BibitemShut {NoStop}%
\bibitem [{\citenamefont {Kang}\ \emph {et~al.}(2017)\citenamefont {Kang},
  \citenamefont {Mott}, \citenamefont {Gilmore}, \citenamefont {Sorenson},
  \citenamefont {Rakher}, \citenamefont {Donley}, \citenamefont {Kitching},\
  and\ \citenamefont {Roper}}]{kang-roper}%
  \BibitemOpen
  \bibfield  {author} {\bibinfo {author} {\bibfnamefont {S.}~\bibnamefont
  {Kang}}, \bibinfo {author} {\bibfnamefont {R.~P.}\ \bibnamefont {Mott}},
  \bibinfo {author} {\bibfnamefont {K.~A.}\ \bibnamefont {Gilmore}}, \bibinfo
  {author} {\bibfnamefont {L.~D.}\ \bibnamefont {Sorenson}}, \bibinfo {author}
  {\bibfnamefont {M.~T.}\ \bibnamefont {Rakher}}, \bibinfo {author}
  {\bibfnamefont {E.~A.}\ \bibnamefont {Donley}}, \bibinfo {author}
  {\bibfnamefont {J.}~\bibnamefont {Kitching}},\ and\ \bibinfo {author}
  {\bibfnamefont {C.~S.}\ \bibnamefont {Roper}},\ }\href
  {https://doi.org/10.1063/1.4986197} {\bibfield  {journal} {\bibinfo
  {journal} {Applied Physics Letters}\ }\textbf {\bibinfo {volume} {110}},\
  \bibinfo {pages} {244101} (\bibinfo {year} {2017})}\BibitemShut {NoStop}%
\bibitem [{\citenamefont {Lee}\ \emph {et~al.}(1996)\citenamefont {Lee},
  \citenamefont {Kim}, \citenamefont {Noh},\ and\ \citenamefont
  {Jhe}}]{Lee:96}%
  \BibitemOpen
  \bibfield  {author} {\bibinfo {author} {\bibfnamefont {K.~I.}\ \bibnamefont
  {Lee}}, \bibinfo {author} {\bibfnamefont {J.~A.}\ \bibnamefont {Kim}},
  \bibinfo {author} {\bibfnamefont {H.~R.}\ \bibnamefont {Noh}},\ and\ \bibinfo
  {author} {\bibfnamefont {W.}~\bibnamefont {Jhe}},\ }\href
  {https://doi.org/10.1364/OL.21.001177} {\bibfield  {journal} {\bibinfo
  {journal} {Opt. Lett.}\ }\textbf {\bibinfo {volume} {21}},\ \bibinfo {pages}
  {1177} (\bibinfo {year} {1996})}\BibitemShut {NoStop}%
\bibitem [{\citenamefont {Ravenhall}\ \emph {et~al.}(2021)\citenamefont
  {Ravenhall}, \citenamefont {Yuen},\ and\ \citenamefont {Foot}}]{footpyramid}%
  \BibitemOpen
  \bibfield  {author} {\bibinfo {author} {\bibfnamefont {S.}~\bibnamefont
  {Ravenhall}}, \bibinfo {author} {\bibfnamefont {B.}~\bibnamefont {Yuen}},\
  and\ \bibinfo {author} {\bibfnamefont {C.}~\bibnamefont {Foot}},\ }\href
  {https://doi.org/10.1364/OE.423662} {\bibfield  {journal} {\bibinfo
  {journal} {Opt. Express}\ }\textbf {\bibinfo {volume} {29}},\ \bibinfo
  {pages} {21143} (\bibinfo {year} {2021})}\BibitemShut {NoStop}%
\bibitem [{\citenamefont {Bodart}\ \emph {et~al.}(2010)\citenamefont {Bodart},
  \citenamefont {Merlet}, \citenamefont {Malossi}, \citenamefont {Dos~Santos},
  \citenamefont {Bouyer},\ and\ \citenamefont {Landragin}}]{landraginpyramid}%
  \BibitemOpen
  \bibfield  {author} {\bibinfo {author} {\bibfnamefont {Q.}~\bibnamefont
  {Bodart}}, \bibinfo {author} {\bibfnamefont {S.}~\bibnamefont {Merlet}},
  \bibinfo {author} {\bibfnamefont {N.}~\bibnamefont {Malossi}}, \bibinfo
  {author} {\bibfnamefont {F.~P.}\ \bibnamefont {Dos~Santos}}, \bibinfo
  {author} {\bibfnamefont {P.}~\bibnamefont {Bouyer}},\ and\ \bibinfo {author}
  {\bibfnamefont {A.}~\bibnamefont {Landragin}},\ }\href
  {https://doi.org/10.1063/1.3373917} {\bibfield  {journal} {\bibinfo
  {journal} {Applied Physics Letters}\ }\textbf {\bibinfo {volume} {96}},\
  \bibinfo {pages} {134101} (\bibinfo {year} {2010})}\BibitemShut {NoStop}%
\bibitem [{\citenamefont {Wu}\ \emph {et~al.}(2019{\natexlab{b}})\citenamefont
  {Wu}, \citenamefont {Pagel}, \citenamefont {Malek}, \citenamefont {Nguyen},
  \citenamefont {Zi}, \citenamefont {Scheirer},\ and\ \citenamefont
  {Müller}}]{mueller}%
  \BibitemOpen
  \bibfield  {author} {\bibinfo {author} {\bibfnamefont {X.}~\bibnamefont
  {Wu}}, \bibinfo {author} {\bibfnamefont {Z.}~\bibnamefont {Pagel}}, \bibinfo
  {author} {\bibfnamefont {B.~S.}\ \bibnamefont {Malek}}, \bibinfo {author}
  {\bibfnamefont {T.~H.}\ \bibnamefont {Nguyen}}, \bibinfo {author}
  {\bibfnamefont {F.}~\bibnamefont {Zi}}, \bibinfo {author} {\bibfnamefont
  {D.~S.}\ \bibnamefont {Scheirer}},\ and\ \bibinfo {author} {\bibfnamefont
  {H.}~\bibnamefont {Müller}},\ }\href
  {https://doi.org/10.1126/sciadv.aax0800} {\bibfield  {journal} {\bibinfo
  {journal} {Science Advances}\ }\textbf {\bibinfo {volume} {5}},\ \bibinfo
  {pages} {eaax0800} (\bibinfo {year} {2019}{\natexlab{b}})}\BibitemShut
  {NoStop}%
\bibitem [{\citenamefont {dos Santos}\ and\ \citenamefont
  {Bonvalot}(2016)}]{dosSantos2016}%
  \BibitemOpen
  \bibfield  {author} {\bibinfo {author} {\bibfnamefont {F.~P.}\ \bibnamefont
  {dos Santos}}\ and\ \bibinfo {author} {\bibfnamefont {S.}~\bibnamefont
  {Bonvalot}},\ }\bibinfo {title} {Cold-atom absolute gravimetry},\ in\ \href
  {https://doi.org/10.1007/978-3-319-02370-0_30-2} {\emph {\bibinfo {booktitle}
  {Encyclopedia of Geodesy}}},\ \bibinfo {editor} {edited by\ \bibinfo {editor}
  {\bibfnamefont {E.}~\bibnamefont {Grafarend}}}\ (\bibinfo  {publisher}
  {Springer International Publishing},\ \bibinfo {address} {Cham},\ \bibinfo
  {year} {2016})\ pp.\ \bibinfo {pages} {1--6}\BibitemShut {NoStop}%
\bibitem [{\citenamefont {Bowden}\ \emph {et~al.}(2019)\citenamefont {Bowden},
  \citenamefont {Hobson}, \citenamefont {Hill}, \citenamefont {Vianello},
  \citenamefont {Schioppo}, \citenamefont {Silva}, \citenamefont {Margolis},
  \citenamefont {Baird},\ and\ \citenamefont {Gill}}]{Bowden2019}%
  \BibitemOpen
  \bibfield  {author} {\bibinfo {author} {\bibfnamefont {W.}~\bibnamefont
  {Bowden}}, \bibinfo {author} {\bibfnamefont {R.}~\bibnamefont {Hobson}},
  \bibinfo {author} {\bibfnamefont {I.~R.}\ \bibnamefont {Hill}}, \bibinfo
  {author} {\bibfnamefont {A.}~\bibnamefont {Vianello}}, \bibinfo {author}
  {\bibfnamefont {M.}~\bibnamefont {Schioppo}}, \bibinfo {author}
  {\bibfnamefont {A.}~\bibnamefont {Silva}}, \bibinfo {author} {\bibfnamefont
  {H.~S.}\ \bibnamefont {Margolis}}, \bibinfo {author} {\bibfnamefont
  {P.~E.~G.}\ \bibnamefont {Baird}},\ and\ \bibinfo {author} {\bibfnamefont
  {P.}~\bibnamefont {Gill}},\ }\href
  {https://doi.org/10.1038/s41598-019-48168-3} {\bibfield  {journal} {\bibinfo
  {journal} {Scientific Reports}\ }\textbf {\bibinfo {volume} {9}},\ \bibinfo
  {pages} {11704} (\bibinfo {year} {2019})}\BibitemShut {NoStop}%
\bibitem [{\citenamefont {Shimizu}\ \emph {et~al.}(1991)\citenamefont
  {Shimizu}, \citenamefont {Shimizu},\ and\ \citenamefont
  {Takuma}}]{Shimizu91}%
  \BibitemOpen
  \bibfield  {author} {\bibinfo {author} {\bibfnamefont {F.}~\bibnamefont
  {Shimizu}}, \bibinfo {author} {\bibfnamefont {K.}~\bibnamefont {Shimizu}},\
  and\ \bibinfo {author} {\bibfnamefont {H.}~\bibnamefont {Takuma}},\ }\href
  {https://doi.org/10.1364/OL.16.000339} {\bibfield  {journal} {\bibinfo
  {journal} {Opt. Lett.}\ }\textbf {\bibinfo {volume} {16}},\ \bibinfo {pages}
  {339} (\bibinfo {year} {1991})}\BibitemShut {NoStop}%
\bibitem [{\citenamefont {Vangeleyn}\ \emph {et~al.}(2009)\citenamefont
  {Vangeleyn}, \citenamefont {Griffin}, \citenamefont {Riis},\ and\
  \citenamefont {Arnold}}]{Vangeleyn:09}%
  \BibitemOpen
  \bibfield  {author} {\bibinfo {author} {\bibfnamefont {M.}~\bibnamefont
  {Vangeleyn}}, \bibinfo {author} {\bibfnamefont {P.~F.}\ \bibnamefont
  {Griffin}}, \bibinfo {author} {\bibfnamefont {E.}~\bibnamefont {Riis}},\ and\
  \bibinfo {author} {\bibfnamefont {A.~S.}\ \bibnamefont {Arnold}},\ }\href
  {https://doi.org/10.1364/OE.17.013601} {\bibfield  {journal} {\bibinfo
  {journal} {Opt. Express}\ }\textbf {\bibinfo {volume} {17}},\ \bibinfo
  {pages} {13601} (\bibinfo {year} {2009})}\BibitemShut {NoStop}%
\bibitem [{\citenamefont {Pollock}\ \emph {et~al.}(2011)\citenamefont
  {Pollock}, \citenamefont {Cotter}, \citenamefont {Laliotis}, \citenamefont
  {Ramirez-Martinez},\ and\ \citenamefont {Hinds}}]{Pollock_2011}%
  \BibitemOpen
  \bibfield  {author} {\bibinfo {author} {\bibfnamefont {S.}~\bibnamefont
  {Pollock}}, \bibinfo {author} {\bibfnamefont {J.~P.}\ \bibnamefont {Cotter}},
  \bibinfo {author} {\bibfnamefont {A.}~\bibnamefont {Laliotis}}, \bibinfo
  {author} {\bibfnamefont {F.}~\bibnamefont {Ramirez-Martinez}},\ and\ \bibinfo
  {author} {\bibfnamefont {E.~A.}\ \bibnamefont {Hinds}},\ }\href
  {https://doi.org/10.1088/1367-2630/13/4/043029} {\bibfield  {journal}
  {\bibinfo  {journal} {New Journal of Physics}\ }\textbf {\bibinfo {volume}
  {13}},\ \bibinfo {pages} {043029} (\bibinfo {year} {2011})}\BibitemShut
  {NoStop}%
\bibitem [{\citenamefont {Trupke}\ \emph {et~al.}(2006)\citenamefont {Trupke},
  \citenamefont {Ramirez-Martinez}, \citenamefont {Curtis}, \citenamefont
  {Ashmore}, \citenamefont {Eriksson}, \citenamefont {Hinds}, \citenamefont
  {Moktadir}, \citenamefont {Gollasch}, \citenamefont {Kraft}, \citenamefont
  {Vijaya~Prakash},\ and\ \citenamefont {Baumberg}}]{trumpke}%
  \BibitemOpen
  \bibfield  {author} {\bibinfo {author} {\bibfnamefont {M.}~\bibnamefont
  {Trupke}}, \bibinfo {author} {\bibfnamefont {F.}~\bibnamefont
  {Ramirez-Martinez}}, \bibinfo {author} {\bibfnamefont {E.~A.}\ \bibnamefont
  {Curtis}}, \bibinfo {author} {\bibfnamefont {J.~P.}\ \bibnamefont {Ashmore}},
  \bibinfo {author} {\bibfnamefont {S.}~\bibnamefont {Eriksson}}, \bibinfo
  {author} {\bibfnamefont {E.~A.}\ \bibnamefont {Hinds}}, \bibinfo {author}
  {\bibfnamefont {Z.}~\bibnamefont {Moktadir}}, \bibinfo {author}
  {\bibfnamefont {C.}~\bibnamefont {Gollasch}}, \bibinfo {author}
  {\bibfnamefont {M.}~\bibnamefont {Kraft}}, \bibinfo {author} {\bibfnamefont
  {G.}~\bibnamefont {Vijaya~Prakash}},\ and\ \bibinfo {author} {\bibfnamefont
  {J.~J.}\ \bibnamefont {Baumberg}},\ }\href
  {https://doi.org/10.1063/1.2172412} {\bibfield  {journal} {\bibinfo
  {journal} {Applied Physics Letters}\ }\textbf {\bibinfo {volume} {88}},\
  \bibinfo {pages} {071116} (\bibinfo {year} {2006})}\BibitemShut {NoStop}%
\bibitem [{\citenamefont {Vangeleyn}\ \emph {et~al.}(2010)\citenamefont
  {Vangeleyn}, \citenamefont {Griffin}, \citenamefont {Riis},\ and\
  \citenamefont {Arnold}}]{Vangeleyn:10}%
  \BibitemOpen
  \bibfield  {author} {\bibinfo {author} {\bibfnamefont {M.}~\bibnamefont
  {Vangeleyn}}, \bibinfo {author} {\bibfnamefont {P.~F.}\ \bibnamefont
  {Griffin}}, \bibinfo {author} {\bibfnamefont {E.}~\bibnamefont {Riis}},\ and\
  \bibinfo {author} {\bibfnamefont {A.~S.}\ \bibnamefont {Arnold}},\ }\href
  {https://doi.org/10.1364/OL.35.003453} {\bibfield  {journal} {\bibinfo
  {journal} {Opt. Lett.}\ }\textbf {\bibinfo {volume} {35}},\ \bibinfo {pages}
  {3453} (\bibinfo {year} {2010})}\BibitemShut {NoStop}%
\bibitem [{\citenamefont {Nshii}\ \emph {et~al.}(2013)\citenamefont {Nshii},
  \citenamefont {Vangeleyn}, \citenamefont {Cotter}, \citenamefont {Griffin},
  \citenamefont {Hinds}, \citenamefont {Ironside}, \citenamefont {See},
  \citenamefont {Sinclair}, \citenamefont {Riis},\ and\ \citenamefont
  {Arnold}}]{Nshii2013}%
  \BibitemOpen
  \bibfield  {author} {\bibinfo {author} {\bibfnamefont {C.~C.}\ \bibnamefont
  {Nshii}}, \bibinfo {author} {\bibfnamefont {M.}~\bibnamefont {Vangeleyn}},
  \bibinfo {author} {\bibfnamefont {J.~P.}\ \bibnamefont {Cotter}}, \bibinfo
  {author} {\bibfnamefont {P.~F.}\ \bibnamefont {Griffin}}, \bibinfo {author}
  {\bibfnamefont {E.~A.}\ \bibnamefont {Hinds}}, \bibinfo {author}
  {\bibfnamefont {C.~N.}\ \bibnamefont {Ironside}}, \bibinfo {author}
  {\bibfnamefont {P.}~\bibnamefont {See}}, \bibinfo {author} {\bibfnamefont
  {A.~G.}\ \bibnamefont {Sinclair}}, \bibinfo {author} {\bibfnamefont
  {E.}~\bibnamefont {Riis}},\ and\ \bibinfo {author} {\bibfnamefont {A.~S.}\
  \bibnamefont {Arnold}},\ }\href {https://doi.org/10.1038/nnano.2013.47}
  {\bibfield  {journal} {\bibinfo  {journal} {Nature Nanotechnology}\ }\textbf
  {\bibinfo {volume} {8}},\ \bibinfo {pages} {321} (\bibinfo {year}
  {2013})}\BibitemShut {NoStop}%
\bibitem [{\citenamefont {Cotter}\ \emph {et~al.}(2016)\citenamefont {Cotter},
  \citenamefont {McGilligan}, \citenamefont {Griffin}, \citenamefont {Rabey},
  \citenamefont {Docherty}, \citenamefont {Riis}, \citenamefont {Arnold},\ and\
  \citenamefont {Hinds}}]{Cotter2016}%
  \BibitemOpen
  \bibfield  {author} {\bibinfo {author} {\bibfnamefont {J.~P.}\ \bibnamefont
  {Cotter}}, \bibinfo {author} {\bibfnamefont {J.~P.}\ \bibnamefont
  {McGilligan}}, \bibinfo {author} {\bibfnamefont {P.~F.}\ \bibnamefont
  {Griffin}}, \bibinfo {author} {\bibfnamefont {I.~M.}\ \bibnamefont {Rabey}},
  \bibinfo {author} {\bibfnamefont {K.}~\bibnamefont {Docherty}}, \bibinfo
  {author} {\bibfnamefont {E.}~\bibnamefont {Riis}}, \bibinfo {author}
  {\bibfnamefont {A.~S.}\ \bibnamefont {Arnold}},\ and\ \bibinfo {author}
  {\bibfnamefont {E.~A.}\ \bibnamefont {Hinds}},\ }\href
  {https://doi.org/10.1007/s00340-016-6415-y} {\bibfield  {journal} {\bibinfo
  {journal} {Applied Physics B}\ }\textbf {\bibinfo {volume} {122}},\ \bibinfo
  {pages} {172} (\bibinfo {year} {2016})}\BibitemShut {NoStop}%
\bibitem [{\citenamefont {McGilligan}\ \emph {et~al.}(2016)\citenamefont
  {McGilligan}, \citenamefont {Griffin}, \citenamefont {Riis},\ and\
  \citenamefont {Arnold}}]{McGilligan:16}%
  \BibitemOpen
  \bibfield  {author} {\bibinfo {author} {\bibfnamefont {J.~P.}\ \bibnamefont
  {McGilligan}}, \bibinfo {author} {\bibfnamefont {P.~F.}\ \bibnamefont
  {Griffin}}, \bibinfo {author} {\bibfnamefont {E.}~\bibnamefont {Riis}},\ and\
  \bibinfo {author} {\bibfnamefont {A.~S.}\ \bibnamefont {Arnold}},\ }\href
  {https://doi.org/10.1364/JOSAB.33.001271} {\bibfield  {journal} {\bibinfo
  {journal} {J. Opt. Soc. Am. B}\ }\textbf {\bibinfo {volume} {33}},\ \bibinfo
  {pages} {1271} (\bibinfo {year} {2016})}\BibitemShut {NoStop}%
\bibitem [{\citenamefont {Bregazzi}\ \emph
  {et~al.}(2021{\natexlab{b}})\citenamefont {Bregazzi}, \citenamefont
  {Griffin}, \citenamefont {Arnold}, \citenamefont {Burt}, \citenamefont
  {Martinez}, \citenamefont {Boudot}, \citenamefont {Kitching}, \citenamefont
  {Riis},\ and\ \citenamefont {McGilligan}}]{Bregazzi}%
  \BibitemOpen
  \bibfield  {author} {\bibinfo {author} {\bibfnamefont {A.}~\bibnamefont
  {Bregazzi}}, \bibinfo {author} {\bibfnamefont {P.~F.}\ \bibnamefont
  {Griffin}}, \bibinfo {author} {\bibfnamefont {A.~S.}\ \bibnamefont {Arnold}},
  \bibinfo {author} {\bibfnamefont {D.~P.}\ \bibnamefont {Burt}}, \bibinfo
  {author} {\bibfnamefont {G.}~\bibnamefont {Martinez}}, \bibinfo {author}
  {\bibfnamefont {R.}~\bibnamefont {Boudot}}, \bibinfo {author} {\bibfnamefont
  {J.}~\bibnamefont {Kitching}}, \bibinfo {author} {\bibfnamefont
  {E.}~\bibnamefont {Riis}},\ and\ \bibinfo {author} {\bibfnamefont {J.~P.}\
  \bibnamefont {McGilligan}},\ }\href {https://doi.org/10.1063/5.0068725}
  {\bibfield  {journal} {\bibinfo  {journal} {Applied Physics Letters}\
  }\textbf {\bibinfo {volume} {119}},\ \bibinfo {pages} {184002} (\bibinfo
  {year} {2021}{\natexlab{b}})}\BibitemShut {NoStop}%
\bibitem [{\citenamefont {Elvin}\ \emph {et~al.}(2019)\citenamefont {Elvin},
  \citenamefont {Hoth}, \citenamefont {Wright}, \citenamefont {Lewis},
  \citenamefont {McGilligan}, \citenamefont {Arnold}, \citenamefont {Griffin},\
  and\ \citenamefont {Riis}}]{Elvin:19}%
  \BibitemOpen
  \bibfield  {author} {\bibinfo {author} {\bibfnamefont {R.}~\bibnamefont
  {Elvin}}, \bibinfo {author} {\bibfnamefont {G.~W.}\ \bibnamefont {Hoth}},
  \bibinfo {author} {\bibfnamefont {M.}~\bibnamefont {Wright}}, \bibinfo
  {author} {\bibfnamefont {B.}~\bibnamefont {Lewis}}, \bibinfo {author}
  {\bibfnamefont {J.~P.}\ \bibnamefont {McGilligan}}, \bibinfo {author}
  {\bibfnamefont {A.~S.}\ \bibnamefont {Arnold}}, \bibinfo {author}
  {\bibfnamefont {P.~F.}\ \bibnamefont {Griffin}},\ and\ \bibinfo {author}
  {\bibfnamefont {E.}~\bibnamefont {Riis}},\ }\href
  {https://doi.org/10.1364/OE.378632} {\bibfield  {journal} {\bibinfo
  {journal} {Opt. Express}\ }\textbf {\bibinfo {volume} {27}},\ \bibinfo
  {pages} {38359} (\bibinfo {year} {2019})}\BibitemShut {NoStop}%
\bibitem [{\citenamefont {Franssen}\ \emph {et~al.}(2019)\citenamefont
  {Franssen}, \citenamefont {de~Raadt}, \citenamefont {van Ninhuijs},\ and\
  \citenamefont {Luiten}}]{franssen}%
  \BibitemOpen
  \bibfield  {author} {\bibinfo {author} {\bibfnamefont {J.~G.~H.}\
  \bibnamefont {Franssen}}, \bibinfo {author} {\bibfnamefont {T.~C.~H.}\
  \bibnamefont {de~Raadt}}, \bibinfo {author} {\bibfnamefont {M.~A.~W.}\
  \bibnamefont {van Ninhuijs}},\ and\ \bibinfo {author} {\bibfnamefont {O.~J.}\
  \bibnamefont {Luiten}},\ }\href
  {https://doi.org/10.1103/PhysRevAccelBeams.22.023401} {\bibfield  {journal}
  {\bibinfo  {journal} {Phys. Rev. Accel. Beams}\ }\textbf {\bibinfo {volume}
  {22}},\ \bibinfo {pages} {023401} (\bibinfo {year} {2019})}\BibitemShut
  {NoStop}%
\bibitem [{\citenamefont {Barker}\ \emph {et~al.}(2019)\citenamefont {Barker},
  \citenamefont {Norrgard}, \citenamefont {Klimov}, \citenamefont {Fedchak},
  \citenamefont {Scherschligt},\ and\ \citenamefont {Eckel}}]{eckelpressure}%
  \BibitemOpen
  \bibfield  {author} {\bibinfo {author} {\bibfnamefont {D.}~\bibnamefont
  {Barker}}, \bibinfo {author} {\bibfnamefont {E.}~\bibnamefont {Norrgard}},
  \bibinfo {author} {\bibfnamefont {N.}~\bibnamefont {Klimov}}, \bibinfo
  {author} {\bibfnamefont {J.}~\bibnamefont {Fedchak}}, \bibinfo {author}
  {\bibfnamefont {J.}~\bibnamefont {Scherschligt}},\ and\ \bibinfo {author}
  {\bibfnamefont {S.}~\bibnamefont {Eckel}},\ }\href
  {https://doi.org/10.1103/PhysRevApplied.11.064023} {\bibfield  {journal}
  {\bibinfo  {journal} {Phys. Rev. Applied}\ }\textbf {\bibinfo {volume}
  {11}},\ \bibinfo {pages} {064023} (\bibinfo {year} {2019})}\BibitemShut
  {NoStop}%
\bibitem [{\citenamefont {Imhof}\ \emph {et~al.}(2017)\citenamefont {Imhof},
  \citenamefont {Stuhl}, \citenamefont {Kasch}, \citenamefont {Kroese},
  \citenamefont {Olson},\ and\ \citenamefont {Squires}}]{imhofgmot}%
  \BibitemOpen
  \bibfield  {author} {\bibinfo {author} {\bibfnamefont {E.}~\bibnamefont
  {Imhof}}, \bibinfo {author} {\bibfnamefont {B.~K.}\ \bibnamefont {Stuhl}},
  \bibinfo {author} {\bibfnamefont {B.}~\bibnamefont {Kasch}}, \bibinfo
  {author} {\bibfnamefont {B.}~\bibnamefont {Kroese}}, \bibinfo {author}
  {\bibfnamefont {S.~E.}\ \bibnamefont {Olson}},\ and\ \bibinfo {author}
  {\bibfnamefont {M.~B.}\ \bibnamefont {Squires}},\ }\href
  {https://doi.org/10.1103/PhysRevA.96.033636} {\bibfield  {journal} {\bibinfo
  {journal} {Phys. Rev. A}\ }\textbf {\bibinfo {volume} {96}},\ \bibinfo
  {pages} {033636} (\bibinfo {year} {2017})}\BibitemShut {NoStop}%
\bibitem [{\citenamefont {Lee}\ \emph {et~al.}(2021{\natexlab{b}})\citenamefont
  {Lee}, \citenamefont {Ding}, \citenamefont {Christensen}, \citenamefont
  {Rosenthal}, \citenamefont {Ison}, \citenamefont {Gillund}, \citenamefont
  {Bossert}, \citenamefont {Fuerschbach}, \citenamefont {Kindel}, \citenamefont
  {Finnegan}, \citenamefont {Wendt}, \citenamefont {Gehl}, \citenamefont
  {McGuinness}, \citenamefont {Walker}, \citenamefont {Lentine}, \citenamefont
  {Kemme}, \citenamefont {Biedermann},\ and\ \citenamefont
  {Schwindt}}]{lee2021coldatom}%
  \BibitemOpen
  \bibfield  {author} {\bibinfo {author} {\bibfnamefont {J.}~\bibnamefont
  {Lee}}, \bibinfo {author} {\bibfnamefont {R.}~\bibnamefont {Ding}}, \bibinfo
  {author} {\bibfnamefont {J.}~\bibnamefont {Christensen}}, \bibinfo {author}
  {\bibfnamefont {R.~R.}\ \bibnamefont {Rosenthal}}, \bibinfo {author}
  {\bibfnamefont {A.}~\bibnamefont {Ison}}, \bibinfo {author} {\bibfnamefont
  {D.~P.}\ \bibnamefont {Gillund}}, \bibinfo {author} {\bibfnamefont
  {D.}~\bibnamefont {Bossert}}, \bibinfo {author} {\bibfnamefont {K.~H.}\
  \bibnamefont {Fuerschbach}}, \bibinfo {author} {\bibfnamefont
  {W.}~\bibnamefont {Kindel}}, \bibinfo {author} {\bibfnamefont {P.~S.}\
  \bibnamefont {Finnegan}}, \bibinfo {author} {\bibfnamefont {J.~R.}\
  \bibnamefont {Wendt}}, \bibinfo {author} {\bibfnamefont {M.}~\bibnamefont
  {Gehl}}, \bibinfo {author} {\bibfnamefont {H.}~\bibnamefont {McGuinness}},
  \bibinfo {author} {\bibfnamefont {C.~A.}\ \bibnamefont {Walker}}, \bibinfo
  {author} {\bibfnamefont {A.}~\bibnamefont {Lentine}}, \bibinfo {author}
  {\bibfnamefont {S.~A.}\ \bibnamefont {Kemme}}, \bibinfo {author}
  {\bibfnamefont {G.}~\bibnamefont {Biedermann}},\ and\ \bibinfo {author}
  {\bibfnamefont {P.~D.~D.}\ \bibnamefont {Schwindt}},\ }\href@noop {}
  {\bibinfo {title} {A cold-atom interferometer with microfabricated gratings
  and a single seed laser}} (\bibinfo {year} {2021}{\natexlab{b}}),\ \Eprint
  {https://arxiv.org/abs/2107.04792} {arXiv:2107.04792 [physics.atom-ph]}
  \BibitemShut {NoStop}%
\bibitem [{\citenamefont {Eckel}\ \emph {et~al.}(2018)\citenamefont {Eckel},
  \citenamefont {Barker}, \citenamefont {Fedchak}, \citenamefont {Klimov},
  \citenamefont {Norrgard}, \citenamefont {Scherschligt}, \citenamefont
  {Makrides},\ and\ \citenamefont {Tiesinga}}]{eckelbarker}%
  \BibitemOpen
  \bibfield  {author} {\bibinfo {author} {\bibfnamefont {S.}~\bibnamefont
  {Eckel}}, \bibinfo {author} {\bibfnamefont {D.~S.}\ \bibnamefont {Barker}},
  \bibinfo {author} {\bibfnamefont {J.~A.}\ \bibnamefont {Fedchak}}, \bibinfo
  {author} {\bibfnamefont {N.~N.}\ \bibnamefont {Klimov}}, \bibinfo {author}
  {\bibfnamefont {E.}~\bibnamefont {Norrgard}}, \bibinfo {author}
  {\bibfnamefont {J.}~\bibnamefont {Scherschligt}}, \bibinfo {author}
  {\bibfnamefont {C.}~\bibnamefont {Makrides}},\ and\ \bibinfo {author}
  {\bibfnamefont {E.}~\bibnamefont {Tiesinga}},\ }\href
  {https://doi.org/10.1088/1681-7575/aadbe4} {\bibfield  {journal} {\bibinfo
  {journal} {Metrologia}\ }\textbf {\bibinfo {volume} {55}},\ \bibinfo {pages}
  {S182} (\bibinfo {year} {2018})}\BibitemShut {NoStop}%
\bibitem [{\citenamefont {Sitaram}\ \emph {et~al.}(2020)\citenamefont
  {Sitaram}, \citenamefont {Elgee}, \citenamefont {Campbell}, \citenamefont
  {Klimov}, \citenamefont {Eckel},\ and\ \citenamefont {Barker}}]{eckelsr}%
  \BibitemOpen
  \bibfield  {author} {\bibinfo {author} {\bibfnamefont {A.}~\bibnamefont
  {Sitaram}}, \bibinfo {author} {\bibfnamefont {P.~K.}\ \bibnamefont {Elgee}},
  \bibinfo {author} {\bibfnamefont {G.~K.}\ \bibnamefont {Campbell}}, \bibinfo
  {author} {\bibfnamefont {N.~N.}\ \bibnamefont {Klimov}}, \bibinfo {author}
  {\bibfnamefont {S.}~\bibnamefont {Eckel}},\ and\ \bibinfo {author}
  {\bibfnamefont {D.~S.}\ \bibnamefont {Barker}},\ }\href
  {https://doi.org/10.1063/5.0019551} {\bibfield  {journal} {\bibinfo
  {journal} {Review of Scientific Instruments}\ }\textbf {\bibinfo {volume}
  {91}},\ \bibinfo {pages} {103202} (\bibinfo {year} {2020})}\BibitemShut
  {NoStop}%
\bibitem [{\citenamefont {Bondza}\ \emph {et~al.}(2022)\citenamefont {Bondza},
  \citenamefont {Lisdat}, \citenamefont {Kroker},\ and\ \citenamefont
  {Leopold}}]{bondza2021twocolor}%
  \BibitemOpen
  \bibfield  {author} {\bibinfo {author} {\bibfnamefont {S.}~\bibnamefont
  {Bondza}}, \bibinfo {author} {\bibfnamefont {C.}~\bibnamefont {Lisdat}},
  \bibinfo {author} {\bibfnamefont {S.}~\bibnamefont {Kroker}},\ and\ \bibinfo
  {author} {\bibfnamefont {T.}~\bibnamefont {Leopold}},\ }\href
  {https://doi.org/10.1103/physrevapplied.17.044002} {\bibfield  {journal}
  {\bibinfo  {journal} {Physical Review Applied}\ }\textbf {\bibinfo {volume}
  {17}},\ \bibinfo {pages} {044002} (\bibinfo {year} {2022})}\BibitemShut
  {NoStop}%
\bibitem [{\citenamefont {Reichel}\ \emph {et~al.}(1999)\citenamefont
  {Reichel}, \citenamefont {H\"ansel},\ and\ \citenamefont
  {H\"ansch}}]{mirrorMOT1}%
  \BibitemOpen
  \bibfield  {author} {\bibinfo {author} {\bibfnamefont {J.}~\bibnamefont
  {Reichel}}, \bibinfo {author} {\bibfnamefont {W.}~\bibnamefont {H\"ansel}},\
  and\ \bibinfo {author} {\bibfnamefont {T.~W.}\ \bibnamefont {H\"ansch}},\
  }\href {https://doi.org/10.1103/PhysRevLett.83.3398} {\bibfield  {journal}
  {\bibinfo  {journal} {Phys. Rev. Lett.}\ }\textbf {\bibinfo {volume} {83}},\
  \bibinfo {pages} {3398} (\bibinfo {year} {1999})}\BibitemShut {NoStop}%
\bibitem [{\citenamefont {Folman}\ \emph {et~al.}(2000)\citenamefont {Folman},
  \citenamefont {Kr\"uger}, \citenamefont {Cassettari}, \citenamefont {Hessmo},
  \citenamefont {Maier},\ and\ \citenamefont {Schmiedmayer}}]{mirrorMOT2}%
  \BibitemOpen
  \bibfield  {author} {\bibinfo {author} {\bibfnamefont {R.}~\bibnamefont
  {Folman}}, \bibinfo {author} {\bibfnamefont {P.}~\bibnamefont {Kr\"uger}},
  \bibinfo {author} {\bibfnamefont {D.}~\bibnamefont {Cassettari}}, \bibinfo
  {author} {\bibfnamefont {B.}~\bibnamefont {Hessmo}}, \bibinfo {author}
  {\bibfnamefont {T.}~\bibnamefont {Maier}},\ and\ \bibinfo {author}
  {\bibfnamefont {J.}~\bibnamefont {Schmiedmayer}},\ }\href
  {https://doi.org/10.1103/PhysRevLett.84.4749} {\bibfield  {journal} {\bibinfo
   {journal} {Phys. Rev. Lett.}\ }\textbf {\bibinfo {volume} {84}},\ \bibinfo
  {pages} {4749} (\bibinfo {year} {2000})}\BibitemShut {NoStop}%
\bibitem [{\citenamefont {Wildermuth}\ \emph {et~al.}(2004)\citenamefont
  {Wildermuth}, \citenamefont {Kr\"uger}, \citenamefont {Becker}, \citenamefont
  {Brajdic}, \citenamefont {Haupt}, \citenamefont {Kasper}, \citenamefont
  {Folman},\ and\ \citenamefont {Schmiedmayer}}]{mirrorMOT3}%
  \BibitemOpen
  \bibfield  {author} {\bibinfo {author} {\bibfnamefont {S.}~\bibnamefont
  {Wildermuth}}, \bibinfo {author} {\bibfnamefont {P.}~\bibnamefont
  {Kr\"uger}}, \bibinfo {author} {\bibfnamefont {C.}~\bibnamefont {Becker}},
  \bibinfo {author} {\bibfnamefont {M.}~\bibnamefont {Brajdic}}, \bibinfo
  {author} {\bibfnamefont {S.}~\bibnamefont {Haupt}}, \bibinfo {author}
  {\bibfnamefont {A.}~\bibnamefont {Kasper}}, \bibinfo {author} {\bibfnamefont
  {R.}~\bibnamefont {Folman}},\ and\ \bibinfo {author} {\bibfnamefont
  {J.}~\bibnamefont {Schmiedmayer}},\ }\href
  {https://doi.org/10.1103/PhysRevA.69.030901} {\bibfield  {journal} {\bibinfo
  {journal} {Phys. Rev. A}\ }\textbf {\bibinfo {volume} {69}},\ \bibinfo
  {pages} {030901} (\bibinfo {year} {2004})}\BibitemShut {NoStop}%
\bibitem [{\citenamefont {Roy}\ \emph {et~al.}(2018)\citenamefont {Roy},
  \citenamefont {Rushton}, \citenamefont {Dragomir}, \citenamefont {Aldous},\
  and\ \citenamefont {Himsworth}}]{Roy2018}%
  \BibitemOpen
  \bibfield  {author} {\bibinfo {author} {\bibfnamefont {R.}~\bibnamefont
  {Roy}}, \bibinfo {author} {\bibfnamefont {J.}~\bibnamefont {Rushton}},
  \bibinfo {author} {\bibfnamefont {A.}~\bibnamefont {Dragomir}}, \bibinfo
  {author} {\bibfnamefont {M.}~\bibnamefont {Aldous}},\ and\ \bibinfo {author}
  {\bibfnamefont {M.}~\bibnamefont {Himsworth}},\ }\href
  {https://doi.org/10.1038/s41598-018-28464-0} {\bibfield  {journal} {\bibinfo
  {journal} {Scientific Reports}\ }\textbf {\bibinfo {volume} {8}},\ \bibinfo
  {pages} {10095} (\bibinfo {year} {2018})}\BibitemShut {NoStop}%
\bibitem [{\citenamefont {Henderson}\ \emph {et~al.}(2020)\citenamefont
  {Henderson}, \citenamefont {Johnson}, \citenamefont {Kale}, \citenamefont
  {Griffin}, \citenamefont {Riis},\ and\ \citenamefont
  {Arnold}}]{Henderson:20}%
  \BibitemOpen
  \bibfield  {author} {\bibinfo {author} {\bibfnamefont {V.~A.}\ \bibnamefont
  {Henderson}}, \bibinfo {author} {\bibfnamefont {M.~Y.~H.}\ \bibnamefont
  {Johnson}}, \bibinfo {author} {\bibfnamefont {Y.~B.}\ \bibnamefont {Kale}},
  \bibinfo {author} {\bibfnamefont {P.~F.}\ \bibnamefont {Griffin}}, \bibinfo
  {author} {\bibfnamefont {E.}~\bibnamefont {Riis}},\ and\ \bibinfo {author}
  {\bibfnamefont {A.~S.}\ \bibnamefont {Arnold}},\ }\href
  {https://doi.org/10.1364/OE.388897} {\bibfield  {journal} {\bibinfo
  {journal} {Opt. Express}\ }\textbf {\bibinfo {volume} {28}},\ \bibinfo
  {pages} {9072} (\bibinfo {year} {2020})}\BibitemShut {NoStop}%
\bibitem [{\citenamefont {Engelberg}\ and\ \citenamefont
  {Levy}(2020)}]{Engelberg2020}%
  \BibitemOpen
  \bibfield  {author} {\bibinfo {author} {\bibfnamefont {J.}~\bibnamefont
  {Engelberg}}\ and\ \bibinfo {author} {\bibfnamefont {U.}~\bibnamefont
  {Levy}},\ }\href {https://doi.org/10.1038/s41467-020-15972-9} {\bibfield
  {journal} {\bibinfo  {journal} {Nature Communications}\ }\textbf {\bibinfo
  {volume} {11}},\ \bibinfo {pages} {1991} (\bibinfo {year}
  {2020})}\BibitemShut {NoStop}%
\bibitem [{\citenamefont {Walther}\ \emph {et~al.}(2012)\citenamefont
  {Walther}, \citenamefont {Helgert}, \citenamefont {Rockstuhl}, \citenamefont
  {Setzpfandt}, \citenamefont {Eilenberger}, \citenamefont {Kley},
  \citenamefont {Lederer}, \citenamefont {Tünnermann},\ and\ \citenamefont
  {Pertsch}}]{beammetashaping}%
  \BibitemOpen
  \bibfield  {author} {\bibinfo {author} {\bibfnamefont {B.}~\bibnamefont
  {Walther}}, \bibinfo {author} {\bibfnamefont {C.}~\bibnamefont {Helgert}},
  \bibinfo {author} {\bibfnamefont {C.}~\bibnamefont {Rockstuhl}}, \bibinfo
  {author} {\bibfnamefont {F.}~\bibnamefont {Setzpfandt}}, \bibinfo {author}
  {\bibfnamefont {F.}~\bibnamefont {Eilenberger}}, \bibinfo {author}
  {\bibfnamefont {E.-B.}\ \bibnamefont {Kley}}, \bibinfo {author}
  {\bibfnamefont {F.}~\bibnamefont {Lederer}}, \bibinfo {author} {\bibfnamefont
  {A.}~\bibnamefont {Tünnermann}},\ and\ \bibinfo {author} {\bibfnamefont
  {T.}~\bibnamefont {Pertsch}},\ }\href
  {https://doi.org/https://doi.org/10.1002/adma.201290300} {\bibfield
  {journal} {\bibinfo  {journal} {Advanced Materials}\ }\textbf {\bibinfo
  {volume} {24}},\ \bibinfo {pages} {6251} (\bibinfo {year}
  {2012})}\BibitemShut {NoStop}%
\bibitem [{\citenamefont {Grady}\ \emph {et~al.}(2013)\citenamefont {Grady},
  \citenamefont {Heyes}, \citenamefont {Chowdhury}, \citenamefont {Zeng},
  \citenamefont {Reiten}, \citenamefont {Azad}, \citenamefont {Taylor},
  \citenamefont {Dalvit},\ and\ \citenamefont {Chen}}]{metapolarization}%
  \BibitemOpen
  \bibfield  {author} {\bibinfo {author} {\bibfnamefont {N.~K.}\ \bibnamefont
  {Grady}}, \bibinfo {author} {\bibfnamefont {J.~E.}\ \bibnamefont {Heyes}},
  \bibinfo {author} {\bibfnamefont {D.~R.}\ \bibnamefont {Chowdhury}}, \bibinfo
  {author} {\bibfnamefont {Y.}~\bibnamefont {Zeng}}, \bibinfo {author}
  {\bibfnamefont {M.~T.}\ \bibnamefont {Reiten}}, \bibinfo {author}
  {\bibfnamefont {A.~K.}\ \bibnamefont {Azad}}, \bibinfo {author}
  {\bibfnamefont {A.~J.}\ \bibnamefont {Taylor}}, \bibinfo {author}
  {\bibfnamefont {D.~A.~R.}\ \bibnamefont {Dalvit}},\ and\ \bibinfo {author}
  {\bibfnamefont {H.-T.}\ \bibnamefont {Chen}},\ }\href
  {https://doi.org/10.1126/science.1235399} {\bibfield  {journal} {\bibinfo
  {journal} {Science}\ }\textbf {\bibinfo {volume} {340}},\ \bibinfo {pages}
  {1304} (\bibinfo {year} {2013})}\BibitemShut {NoStop}%
\bibitem [{\citenamefont {Cong}\ \emph {et~al.}(2014)\citenamefont {Cong},
  \citenamefont {Xu}, \citenamefont {Gu}, \citenamefont {Singh}, \citenamefont
  {Han},\ and\ \citenamefont {Zhang}}]{metapolarization2}%
  \BibitemOpen
  \bibfield  {author} {\bibinfo {author} {\bibfnamefont {L.}~\bibnamefont
  {Cong}}, \bibinfo {author} {\bibfnamefont {N.}~\bibnamefont {Xu}}, \bibinfo
  {author} {\bibfnamefont {J.}~\bibnamefont {Gu}}, \bibinfo {author}
  {\bibfnamefont {R.}~\bibnamefont {Singh}}, \bibinfo {author} {\bibfnamefont
  {J.}~\bibnamefont {Han}},\ and\ \bibinfo {author} {\bibfnamefont
  {W.}~\bibnamefont {Zhang}},\ }\href
  {https://doi.org/https://doi.org/10.1002/lpor.201300205} {\bibfield
  {journal} {\bibinfo  {journal} {Laser and Photonics Reviews}\ }\textbf
  {\bibinfo {volume} {8}},\ \bibinfo {pages} {626} (\bibinfo {year}
  {2014})}\BibitemShut {NoStop}%
\bibitem [{\citenamefont {Zhu}\ \emph {et~al.}(2020)\citenamefont {Zhu},
  \citenamefont {Liu}, \citenamefont {Sain}, \citenamefont {Wang},
  \citenamefont {Schlickriede}, \citenamefont {Tang}, \citenamefont {Deng},
  \citenamefont {Li}, \citenamefont {Yang}, \citenamefont {Holynski},
  \citenamefont {Zhang}, \citenamefont {Zentgraf}, \citenamefont {Bongs},
  \citenamefont {Lien},\ and\ \citenamefont {Li}}]{metasurface}%
  \BibitemOpen
  \bibfield  {author} {\bibinfo {author} {\bibfnamefont {L.}~\bibnamefont
  {Zhu}}, \bibinfo {author} {\bibfnamefont {X.}~\bibnamefont {Liu}}, \bibinfo
  {author} {\bibfnamefont {B.}~\bibnamefont {Sain}}, \bibinfo {author}
  {\bibfnamefont {M.}~\bibnamefont {Wang}}, \bibinfo {author} {\bibfnamefont
  {C.}~\bibnamefont {Schlickriede}}, \bibinfo {author} {\bibfnamefont
  {Y.}~\bibnamefont {Tang}}, \bibinfo {author} {\bibfnamefont {J.}~\bibnamefont
  {Deng}}, \bibinfo {author} {\bibfnamefont {K.}~\bibnamefont {Li}}, \bibinfo
  {author} {\bibfnamefont {J.}~\bibnamefont {Yang}}, \bibinfo {author}
  {\bibfnamefont {M.}~\bibnamefont {Holynski}}, \bibinfo {author}
  {\bibfnamefont {S.}~\bibnamefont {Zhang}}, \bibinfo {author} {\bibfnamefont
  {T.}~\bibnamefont {Zentgraf}}, \bibinfo {author} {\bibfnamefont
  {K.}~\bibnamefont {Bongs}}, \bibinfo {author} {\bibfnamefont {Y.-H.}\
  \bibnamefont {Lien}},\ and\ \bibinfo {author} {\bibfnamefont
  {G.}~\bibnamefont {Li}},\ }\href {https://doi.org/10.1126/sciadv.abb6667}
  {\bibfield  {journal} {\bibinfo  {journal} {Science Advances}\ }\textbf
  {\bibinfo {volume} {6}},\ \bibinfo {pages} {eabb6667} (\bibinfo {year}
  {2020})}\BibitemShut {NoStop}%
\bibitem [{\citenamefont {Lin}\ and\ \citenamefont
  {Li}(2019)}]{metamultifocus}%
  \BibitemOpen
  \bibfield  {author} {\bibinfo {author} {\bibfnamefont {R.}~\bibnamefont
  {Lin}}\ and\ \bibinfo {author} {\bibfnamefont {X.}~\bibnamefont {Li}},\
  }\href {https://doi.org/10.1364/OL.44.002819} {\bibfield  {journal} {\bibinfo
   {journal} {Opt. Lett.}\ }\textbf {\bibinfo {volume} {44}},\ \bibinfo {pages}
  {2819} (\bibinfo {year} {2019})}\BibitemShut {NoStop}%
\bibitem [{\citenamefont {Hsu}\ \emph {et~al.}(2020)\citenamefont {Hsu},
  \citenamefont {Thiele}, \citenamefont {Zhu}, \citenamefont {Brown},
  \citenamefont {Papp}, \citenamefont {Agrawal},\ and\ \citenamefont
  {Regal}}]{regalmeta}%
  \BibitemOpen
  \bibfield  {author} {\bibinfo {author} {\bibfnamefont {T.-W.}\ \bibnamefont
  {Hsu}}, \bibinfo {author} {\bibfnamefont {T.}~\bibnamefont {Thiele}},
  \bibinfo {author} {\bibfnamefont {W.}~\bibnamefont {Zhu}}, \bibinfo {author}
  {\bibfnamefont {M.~O.}\ \bibnamefont {Brown}}, \bibinfo {author}
  {\bibfnamefont {S.~B.}\ \bibnamefont {Papp}}, \bibinfo {author}
  {\bibfnamefont {A.}~\bibnamefont {Agrawal}},\ and\ \bibinfo {author}
  {\bibfnamefont {C.~A.}\ \bibnamefont {Regal}},\ }in\ \href
  {https://doi.org/10.1364/CLEO_AT.2020.JW2A.12} {\emph {\bibinfo {booktitle}
  {Conference on Lasers and Electro-Optics}}}\ (\bibinfo  {publisher} {Optica
  Publishing Group},\ \bibinfo {year} {2020})\ p.\ \bibinfo {pages}
  {JW2A.12}\BibitemShut {NoStop}%
\bibitem [{\citenamefont {Hsu}\ \emph {et~al.}(2021)\citenamefont {Hsu},
  \citenamefont {Zhu}, \citenamefont {Thiele}, \citenamefont {Brown},
  \citenamefont {Papp}, \citenamefont {Agrawal},\ and\ \citenamefont
  {Regal}}]{regalmeta2}%
  \BibitemOpen
  \bibfield  {author} {\bibinfo {author} {\bibfnamefont {T.-W.}\ \bibnamefont
  {Hsu}}, \bibinfo {author} {\bibfnamefont {W.}~\bibnamefont {Zhu}}, \bibinfo
  {author} {\bibfnamefont {T.}~\bibnamefont {Thiele}}, \bibinfo {author}
  {\bibfnamefont {M.~O.}\ \bibnamefont {Brown}}, \bibinfo {author}
  {\bibfnamefont {S.~B.}\ \bibnamefont {Papp}}, \bibinfo {author}
  {\bibfnamefont {A.}~\bibnamefont {Agrawal}},\ and\ \bibinfo {author}
  {\bibfnamefont {C.~A.}\ \bibnamefont {Regal}},\ }\href@noop {} {\bibinfo
  {title} {Single atom trapping in a metasurface lens optical tweezer}}
  (\bibinfo {year} {2021}),\ \Eprint {https://arxiv.org/abs/2110.11559}
  {arXiv:2110.11559 [physics.atom-ph]} \BibitemShut {NoStop}%
\bibitem [{\citenamefont {Little}\ \emph {et~al.}(2021)\citenamefont {Little},
  \citenamefont {Hoth}, \citenamefont {Christensen}, \citenamefont {Walker},
  \citenamefont {De~Smet}, \citenamefont {Biedermann}, \citenamefont {Lee},\
  and\ \citenamefont {Schwindt}}]{hothlittle}%
  \BibitemOpen
  \bibfield  {author} {\bibinfo {author} {\bibfnamefont {B.~J.}\ \bibnamefont
  {Little}}, \bibinfo {author} {\bibfnamefont {G.~W.}\ \bibnamefont {Hoth}},
  \bibinfo {author} {\bibfnamefont {J.}~\bibnamefont {Christensen}}, \bibinfo
  {author} {\bibfnamefont {C.}~\bibnamefont {Walker}}, \bibinfo {author}
  {\bibfnamefont {D.~J.}\ \bibnamefont {De~Smet}}, \bibinfo {author}
  {\bibfnamefont {G.~W.}\ \bibnamefont {Biedermann}}, \bibinfo {author}
  {\bibfnamefont {J.}~\bibnamefont {Lee}},\ and\ \bibinfo {author}
  {\bibfnamefont {P.~D.~D.}\ \bibnamefont {Schwindt}},\ }\href
  {https://doi.org/10.1116/5.0053885} {\bibfield  {journal} {\bibinfo
  {journal} {AVS Quantum Science}\ }\textbf {\bibinfo {volume} {3}},\ \bibinfo
  {pages} {035001} (\bibinfo {year} {2021})}\BibitemShut {NoStop}%
\bibitem [{\citenamefont {Schwindt}\ \emph {et~al.}(2016)\citenamefont
  {Schwindt}, \citenamefont {Jau}, \citenamefont {Partner}, \citenamefont
  {Casias}, \citenamefont {Wagner}, \citenamefont {Moorman}, \citenamefont
  {Manginell}, \citenamefont {Kellogg},\ and\ \citenamefont
  {Prestage}}]{prestage}%
  \BibitemOpen
  \bibfield  {author} {\bibinfo {author} {\bibfnamefont {P.~D.~D.}\
  \bibnamefont {Schwindt}}, \bibinfo {author} {\bibfnamefont {Y.-Y.}\
  \bibnamefont {Jau}}, \bibinfo {author} {\bibfnamefont {H.}~\bibnamefont
  {Partner}}, \bibinfo {author} {\bibfnamefont {A.}~\bibnamefont {Casias}},
  \bibinfo {author} {\bibfnamefont {A.~R.}\ \bibnamefont {Wagner}}, \bibinfo
  {author} {\bibfnamefont {M.}~\bibnamefont {Moorman}}, \bibinfo {author}
  {\bibfnamefont {R.~P.}\ \bibnamefont {Manginell}}, \bibinfo {author}
  {\bibfnamefont {J.~R.}\ \bibnamefont {Kellogg}},\ and\ \bibinfo {author}
  {\bibfnamefont {J.~D.}\ \bibnamefont {Prestage}},\ }\href
  {https://doi.org/10.1063/1.4948739} {\bibfield  {journal} {\bibinfo
  {journal} {Review of Scientific Instruments}\ }\textbf {\bibinfo {volume}
  {87}},\ \bibinfo {pages} {053112} (\bibinfo {year} {2016})}\BibitemShut
  {NoStop}%
\bibitem [{\citenamefont {Takeda}\ \emph {et~al.}(2011)\citenamefont {Takeda},
  \citenamefont {Kurisu}, \citenamefont {Yamamoto}, \citenamefont {Nakagawa},\
  and\ \citenamefont {Ishizawa}}]{takeda}%
  \BibitemOpen
  \bibfield  {author} {\bibinfo {author} {\bibfnamefont {M.}~\bibnamefont
  {Takeda}}, \bibinfo {author} {\bibfnamefont {H.}~\bibnamefont {Kurisu}},
  \bibinfo {author} {\bibfnamefont {S.}~\bibnamefont {Yamamoto}}, \bibinfo
  {author} {\bibfnamefont {H.}~\bibnamefont {Nakagawa}},\ and\ \bibinfo
  {author} {\bibfnamefont {K.}~\bibnamefont {Ishizawa}},\ }\href
  {https://doi.org/https://doi.org/10.1016/j.apsusc.2011.09.092} {\bibfield
  {journal} {\bibinfo  {journal} {Applied Surface Science}\ }\textbf {\bibinfo
  {volume} {258}},\ \bibinfo {pages} {1405} (\bibinfo {year}
  {2011})}\BibitemShut {NoStop}%
\bibitem [{\citenamefont {McGilligan}\ \emph
  {et~al.}(2020{\natexlab{b}})\citenamefont {McGilligan}, \citenamefont
  {Moore}, \citenamefont {Dellis}, \citenamefont {Martinez}, \citenamefont
  {de~Clercq}, \citenamefont {Griffin}, \citenamefont {Arnold}, \citenamefont
  {Riis}, \citenamefont {Boudot},\ and\ \citenamefont
  {Kitching}}]{McGilligan2020}%
  \BibitemOpen
  \bibfield  {author} {\bibinfo {author} {\bibfnamefont {J.~P.}\ \bibnamefont
  {McGilligan}}, \bibinfo {author} {\bibfnamefont {K.~R.}\ \bibnamefont
  {Moore}}, \bibinfo {author} {\bibfnamefont {A.}~\bibnamefont {Dellis}},
  \bibinfo {author} {\bibfnamefont {G.~D.}\ \bibnamefont {Martinez}}, \bibinfo
  {author} {\bibfnamefont {E.}~\bibnamefont {de~Clercq}}, \bibinfo {author}
  {\bibfnamefont {P.~F.}\ \bibnamefont {Griffin}}, \bibinfo {author}
  {\bibfnamefont {A.~S.}\ \bibnamefont {Arnold}}, \bibinfo {author}
  {\bibfnamefont {E.}~\bibnamefont {Riis}}, \bibinfo {author} {\bibfnamefont
  {R.}~\bibnamefont {Boudot}},\ and\ \bibinfo {author} {\bibfnamefont
  {J.}~\bibnamefont {Kitching}},\ }\href {https://doi.org/10.1063/5.0014658}
  {\bibfield  {journal} {\bibinfo  {journal} {Applied Physics Letters}\
  }\textbf {\bibinfo {volume} {117}},\ \bibinfo {pages} {054001} (\bibinfo
  {year} {2020}{\natexlab{b}})}\BibitemShut {NoStop}%
\bibitem [{\citenamefont {Vecchio}\ \emph {et~al.}(2011)\citenamefont
  {Vecchio}, \citenamefont {Venkatraman}, \citenamefont {Shea}, \citenamefont
  {Maeder},\ and\ \citenamefont {Ryser}}]{LTCCcell}%
  \BibitemOpen
  \bibfield  {author} {\bibinfo {author} {\bibfnamefont {F.}~\bibnamefont
  {Vecchio}}, \bibinfo {author} {\bibfnamefont {V.}~\bibnamefont
  {Venkatraman}}, \bibinfo {author} {\bibfnamefont {H.~R.}\ \bibnamefont
  {Shea}}, \bibinfo {author} {\bibfnamefont {T.}~\bibnamefont {Maeder}},\ and\
  \bibinfo {author} {\bibfnamefont {P.}~\bibnamefont {Ryser}},\ }\href
  {https://doi.org/https://doi.org/10.1016/j.sna.2011.03.045} {\bibfield
  {journal} {\bibinfo  {journal} {Sensors and Actuators A: Physical}\ }\textbf
  {\bibinfo {volume} {172}},\ \bibinfo {pages} {330} (\bibinfo {year}
  {2011})}\BibitemShut {NoStop}%
\bibitem [{\citenamefont {Burrow}\ \emph {et~al.}(2021)\citenamefont {Burrow},
  \citenamefont {Osborn}, \citenamefont {Boughton}, \citenamefont {Mirando},
  \citenamefont {Burt}, \citenamefont {Griffin}, \citenamefont {Arnold},\ and\
  \citenamefont {Riis}}]{burrow2021}%
  \BibitemOpen
  \bibfield  {author} {\bibinfo {author} {\bibfnamefont {O.~S.}\ \bibnamefont
  {Burrow}}, \bibinfo {author} {\bibfnamefont {P.~F.}\ \bibnamefont {Osborn}},
  \bibinfo {author} {\bibfnamefont {E.}~\bibnamefont {Boughton}}, \bibinfo
  {author} {\bibfnamefont {F.}~\bibnamefont {Mirando}}, \bibinfo {author}
  {\bibfnamefont {D.~P.}\ \bibnamefont {Burt}}, \bibinfo {author}
  {\bibfnamefont {P.~F.}\ \bibnamefont {Griffin}}, \bibinfo {author}
  {\bibfnamefont {A.~S.}\ \bibnamefont {Arnold}},\ and\ \bibinfo {author}
  {\bibfnamefont {E.}~\bibnamefont {Riis}},\ }\href
  {https://doi.org/10.1063/5.0061010} {\bibfield  {journal} {\bibinfo
  {journal} {Applied Physics Letters}\ }\textbf {\bibinfo {volume} {119}},\
  \bibinfo {pages} {124002} (\bibinfo {year} {2021})}\BibitemShut {NoStop}%
\bibitem [{\citenamefont {Cooper}\ \emph {et~al.}(2021)\citenamefont {Cooper},
  \citenamefont {Coles}, \citenamefont {Everton}, \citenamefont {Maskery},
  \citenamefont {Campion}, \citenamefont {Madkhaly}, \citenamefont {Morley},
  \citenamefont {O’Shea}, \citenamefont {Evans}, \citenamefont {Saint},
  \citenamefont {Krüger}, \citenamefont {Oručević}, \citenamefont {Tuck},
  \citenamefont {Wildman}, \citenamefont {Fromhold},\ and\ \citenamefont
  {Hackermüller}}]{COOPER2021}%
  \BibitemOpen
  \bibfield  {author} {\bibinfo {author} {\bibfnamefont {N.}~\bibnamefont
  {Cooper}}, \bibinfo {author} {\bibfnamefont {L.}~\bibnamefont {Coles}},
  \bibinfo {author} {\bibfnamefont {S.}~\bibnamefont {Everton}}, \bibinfo
  {author} {\bibfnamefont {I.}~\bibnamefont {Maskery}}, \bibinfo {author}
  {\bibfnamefont {R.}~\bibnamefont {Campion}}, \bibinfo {author} {\bibfnamefont
  {S.}~\bibnamefont {Madkhaly}}, \bibinfo {author} {\bibfnamefont
  {C.}~\bibnamefont {Morley}}, \bibinfo {author} {\bibfnamefont
  {J.}~\bibnamefont {O’Shea}}, \bibinfo {author} {\bibfnamefont
  {W.}~\bibnamefont {Evans}}, \bibinfo {author} {\bibfnamefont
  {R.}~\bibnamefont {Saint}}, \bibinfo {author} {\bibfnamefont
  {P.}~\bibnamefont {Krüger}}, \bibinfo {author} {\bibfnamefont
  {F.}~\bibnamefont {Oručević}}, \bibinfo {author} {\bibfnamefont
  {C.}~\bibnamefont {Tuck}}, \bibinfo {author} {\bibfnamefont {R.}~\bibnamefont
  {Wildman}}, \bibinfo {author} {\bibfnamefont {T.}~\bibnamefont {Fromhold}},\
  and\ \bibinfo {author} {\bibfnamefont {L.}~\bibnamefont {Hackermüller}},\
  }\href {https://doi.org/https://doi.org/10.1016/j.addma.2021.101898}
  {\bibfield  {journal} {\bibinfo  {journal} {Additive Manufacturing}\ }\textbf
  {\bibinfo {volume} {40}},\ \bibinfo {pages} {101898} (\bibinfo {year}
  {2021})}\BibitemShut {NoStop}%
\bibitem [{\citenamefont {Madkhaly}\ \emph {et~al.}(2021)\citenamefont
  {Madkhaly}, \citenamefont {Coles}, \citenamefont {Morley}, \citenamefont
  {Colquhoun}, \citenamefont {Fromhold}, \citenamefont {Cooper},\ and\
  \citenamefont {Hackerm\"uller}}]{PRXQuantum}%
  \BibitemOpen
  \bibfield  {author} {\bibinfo {author} {\bibfnamefont {S.}~\bibnamefont
  {Madkhaly}}, \bibinfo {author} {\bibfnamefont {L.}~\bibnamefont {Coles}},
  \bibinfo {author} {\bibfnamefont {C.}~\bibnamefont {Morley}}, \bibinfo
  {author} {\bibfnamefont {C.}~\bibnamefont {Colquhoun}}, \bibinfo {author}
  {\bibfnamefont {T.}~\bibnamefont {Fromhold}}, \bibinfo {author}
  {\bibfnamefont {N.}~\bibnamefont {Cooper}},\ and\ \bibinfo {author}
  {\bibfnamefont {L.}~\bibnamefont {Hackerm\"uller}},\ }\href
  {https://doi.org/10.1103/PRXQuantum.2.030326} {\bibfield  {journal} {\bibinfo
   {journal} {PRX Quantum}\ }\textbf {\bibinfo {volume} {2}},\ \bibinfo {pages}
  {030326} (\bibinfo {year} {2021})}\BibitemShut {NoStop}%
\bibitem [{\citenamefont {Dellis}\ \emph {et~al.}(2016)\citenamefont {Dellis},
  \citenamefont {Shah}, \citenamefont {Donley}, \citenamefont {Knappe},\ and\
  \citenamefont {Kitching}}]{dellis2016lowHe}%
  \BibitemOpen
  \bibfield  {author} {\bibinfo {author} {\bibfnamefont {A.~T.}\ \bibnamefont
  {Dellis}}, \bibinfo {author} {\bibfnamefont {V.}~\bibnamefont {Shah}},
  \bibinfo {author} {\bibfnamefont {E.~A.}\ \bibnamefont {Donley}}, \bibinfo
  {author} {\bibfnamefont {S.}~\bibnamefont {Knappe}},\ and\ \bibinfo {author}
  {\bibfnamefont {J.}~\bibnamefont {Kitching}},\ }\href
  {https://doi.org/10.1364/ol.41.002775} {\bibfield  {journal} {\bibinfo
  {journal} {Optics Letters}\ }\textbf {\bibinfo {volume} {41}},\ \bibinfo
  {pages} {2775} (\bibinfo {year} {2016})}\BibitemShut {NoStop}%
\bibitem [{\citenamefont {Shoji}\ \emph {et~al.}(1998)\citenamefont {Shoji},
  \citenamefont {Kikuchi},\ and\ \citenamefont {Torigoe}}]{SHOJI199895}%
  \BibitemOpen
  \bibfield  {author} {\bibinfo {author} {\bibfnamefont {S.}~\bibnamefont
  {Shoji}}, \bibinfo {author} {\bibfnamefont {H.}~\bibnamefont {Kikuchi}},\
  and\ \bibinfo {author} {\bibfnamefont {H.}~\bibnamefont {Torigoe}},\ }\href
  {https://doi.org/https://doi.org/10.1016/S0924-4247(97)01659-2} {\bibfield
  {journal} {\bibinfo  {journal} {Sensors and Actuators A: Physical}\ }\textbf
  {\bibinfo {volume} {64}},\ \bibinfo {pages} {95} (\bibinfo {year}
  {1998})}\BibitemShut {NoStop}%
\bibitem [{\citenamefont {Bunch}\ \emph {et~al.}(2008)\citenamefont {Bunch},
  \citenamefont {Verbridge}, \citenamefont {Alden}, \citenamefont {van~der
  Zande}, \citenamefont {Parpia}, \citenamefont {Craighead},\ and\
  \citenamefont {McEuen}}]{graphene}%
  \BibitemOpen
  \bibfield  {author} {\bibinfo {author} {\bibfnamefont {J.~S.}\ \bibnamefont
  {Bunch}}, \bibinfo {author} {\bibfnamefont {S.~S.}\ \bibnamefont
  {Verbridge}}, \bibinfo {author} {\bibfnamefont {J.~S.}\ \bibnamefont
  {Alden}}, \bibinfo {author} {\bibfnamefont {A.~M.}\ \bibnamefont {van~der
  Zande}}, \bibinfo {author} {\bibfnamefont {J.~M.}\ \bibnamefont {Parpia}},
  \bibinfo {author} {\bibfnamefont {H.~G.}\ \bibnamefont {Craighead}},\ and\
  \bibinfo {author} {\bibfnamefont {P.~L.}\ \bibnamefont {McEuen}},\ }\href
  {https://doi.org/10.1021/nl801457b} {\bibfield  {journal} {\bibinfo
  {journal} {Nano Letters}\ }\textbf {\bibinfo {volume} {8}},\ \bibinfo {pages}
  {2458} (\bibinfo {year} {2008})}\BibitemShut {NoStop}%
\bibitem [{\citenamefont {Karlen}\ \emph {et~al.}(2017)\citenamefont {Karlen},
  \citenamefont {Gobet}, \citenamefont {Overstolz}, \citenamefont {Haesler},\
  and\ \citenamefont {Lecomte}}]{KarlenALD}%
  \BibitemOpen
  \bibfield  {author} {\bibinfo {author} {\bibfnamefont {S.}~\bibnamefont
  {Karlen}}, \bibinfo {author} {\bibfnamefont {J.}~\bibnamefont {Gobet}},
  \bibinfo {author} {\bibfnamefont {T.}~\bibnamefont {Overstolz}}, \bibinfo
  {author} {\bibfnamefont {J.}~\bibnamefont {Haesler}},\ and\ \bibinfo {author}
  {\bibfnamefont {S.}~\bibnamefont {Lecomte}},\ }\href
  {https://doi.org/10.1364/OE.25.002187} {\bibfield  {journal} {\bibinfo
  {journal} {Opt. Express}\ }\textbf {\bibinfo {volume} {25}},\ \bibinfo
  {pages} {2187} (\bibinfo {year} {2017})}\BibitemShut {NoStop}%
\bibitem [{\citenamefont {Woetzel}\ \emph {et~al.}(2013)\citenamefont
  {Woetzel}, \citenamefont {Talkenberg}, \citenamefont {Scholtes},
  \citenamefont {IJsselsteijn}, \citenamefont {Schultze},\ and\ \citenamefont
  {Meyer}}]{WOETZEL2013158}%
  \BibitemOpen
  \bibfield  {author} {\bibinfo {author} {\bibfnamefont {S.}~\bibnamefont
  {Woetzel}}, \bibinfo {author} {\bibfnamefont {F.}~\bibnamefont {Talkenberg}},
  \bibinfo {author} {\bibfnamefont {T.}~\bibnamefont {Scholtes}}, \bibinfo
  {author} {\bibfnamefont {R.}~\bibnamefont {IJsselsteijn}}, \bibinfo {author}
  {\bibfnamefont {V.}~\bibnamefont {Schultze}},\ and\ \bibinfo {author}
  {\bibfnamefont {H.-G.}\ \bibnamefont {Meyer}},\ }\href
  {https://doi.org/https://doi.org/10.1016/j.surfcoat.2013.01.044} {\bibfield
  {journal} {\bibinfo  {journal} {Surface and Coatings Technology}\ }\textbf
  {\bibinfo {volume} {221}},\ \bibinfo {pages} {158} (\bibinfo {year}
  {2013})}\BibitemShut {NoStop}%
\bibitem [{\citenamefont {Stephens}\ \emph {et~al.}(1994)\citenamefont
  {Stephens}, \citenamefont {Rhodes},\ and\ \citenamefont
  {Wieman}}]{wiemancoatings}%
  \BibitemOpen
  \bibfield  {author} {\bibinfo {author} {\bibfnamefont {M.}~\bibnamefont
  {Stephens}}, \bibinfo {author} {\bibfnamefont {R.}~\bibnamefont {Rhodes}},\
  and\ \bibinfo {author} {\bibfnamefont {C.}~\bibnamefont {Wieman}},\ }\href
  {https://doi.org/10.1063/1.358502} {\bibfield  {journal} {\bibinfo  {journal}
  {Journal of Applied Physics}\ }\textbf {\bibinfo {volume} {76}},\ \bibinfo
  {pages} {3479} (\bibinfo {year} {1994})}\BibitemShut {NoStop}%
\bibitem [{\citenamefont {Neuman}\ \emph {et~al.}(1995)\citenamefont {Neuman},
  \citenamefont {Wang},\ and\ \citenamefont {Gallagher}}]{sapphirecm}%
  \BibitemOpen
  \bibfield  {author} {\bibinfo {author} {\bibfnamefont {J.~A.}\ \bibnamefont
  {Neuman}}, \bibinfo {author} {\bibfnamefont {P.}~\bibnamefont {Wang}},\ and\
  \bibinfo {author} {\bibfnamefont {A.}~\bibnamefont {Gallagher}},\ }\href
  {https://doi.org/10.1063/1.1145589} {\bibfield  {journal} {\bibinfo
  {journal} {Review of Scientific Instruments}\ }\textbf {\bibinfo {volume}
  {66}},\ \bibinfo {pages} {3021} (\bibinfo {year} {1995})}\BibitemShut
  {NoStop}%
\bibitem [{\citenamefont {Kurashima}\ \emph {et~al.}(2019)\citenamefont
  {Kurashima}, \citenamefont {Matsumae}, \citenamefont {Yanagimachi},
  \citenamefont {Harasaka},\ and\ \citenamefont {Takagi}}]{sapphiremm}%
  \BibitemOpen
  \bibfield  {author} {\bibinfo {author} {\bibfnamefont {Y.}~\bibnamefont
  {Kurashima}}, \bibinfo {author} {\bibfnamefont {T.}~\bibnamefont {Matsumae}},
  \bibinfo {author} {\bibfnamefont {S.}~\bibnamefont {Yanagimachi}}, \bibinfo
  {author} {\bibfnamefont {K.}~\bibnamefont {Harasaka}},\ and\ \bibinfo
  {author} {\bibfnamefont {H.}~\bibnamefont {Takagi}},\ }\href
  {https://doi.org/10.7567/1347-4065/ab3c07} {\bibfield  {journal} {\bibinfo
  {journal} {Japanese Journal of Applied Physics}\ }\textbf {\bibinfo {volume}
  {58}},\ \bibinfo {pages} {096506} (\bibinfo {year} {2019})}\BibitemShut
  {NoStop}%
\bibitem [{\citenamefont {Trocellier}\ \emph {et~al.}(2014)\citenamefont
  {Trocellier}, \citenamefont {Agarwal},\ and\ \citenamefont {Miro}}]{spinel}%
  \BibitemOpen
  \bibfield  {author} {\bibinfo {author} {\bibfnamefont {P.}~\bibnamefont
  {Trocellier}}, \bibinfo {author} {\bibfnamefont {S.}~\bibnamefont
  {Agarwal}},\ and\ \bibinfo {author} {\bibfnamefont {S.}~\bibnamefont
  {Miro}},\ }\href
  {https://doi.org/https://doi.org/10.1016/j.jnucmat.2013.10.061} {\bibfield
  {journal} {\bibinfo  {journal} {Journal of Nuclear Materials}\ }\textbf
  {\bibinfo {volume} {445}},\ \bibinfo {pages} {128} (\bibinfo {year}
  {2014})}\BibitemShut {NoStop}%
\bibitem [{\citenamefont {Lemeshev}\ \emph {et~al.}(2019)\citenamefont
  {Lemeshev}, \citenamefont {Senina}, \citenamefont {Pedchenko},\ and\
  \citenamefont {Boyko}}]{spinel2019}%
  \BibitemOpen
  \bibfield  {author} {\bibinfo {author} {\bibfnamefont {D.~O.}\ \bibnamefont
  {Lemeshev}}, \bibinfo {author} {\bibfnamefont {M.~O.}\ \bibnamefont
  {Senina}}, \bibinfo {author} {\bibfnamefont {M.~S.}\ \bibnamefont
  {Pedchenko}},\ and\ \bibinfo {author} {\bibfnamefont {A.~V.}\ \bibnamefont
  {Boyko}},\ }\href {https://doi.org/10.1088/1757-899x/525/1/012081} {\bibfield
   {journal} {\bibinfo  {journal} {IOP Conference Series: Materials Science and
  Engineering}\ }\textbf {\bibinfo {volume} {525}},\ \bibinfo {pages} {012081}
  (\bibinfo {year} {2019})}\BibitemShut {NoStop}%
\bibitem [{\citenamefont {Xiao}\ \emph {et~al.}(2020)\citenamefont {Xiao},
  \citenamefont {Yu}, \citenamefont {Li}, \citenamefont {Ruan}, \citenamefont
  {Kong}, \citenamefont {Huang}, \citenamefont {Huang}, \citenamefont {Zhou},
  \citenamefont {Su}, \citenamefont {Yao}, \citenamefont {Que}, \citenamefont
  {Liu}, \citenamefont {Zhang}, \citenamefont {Wang}, \citenamefont {Liu},
  \citenamefont {Shen}, \citenamefont {Allix}, \citenamefont {Zhang},\ and\
  \citenamefont {Tang}}]{ceramicreview}%
  \BibitemOpen
  \bibfield  {author} {\bibinfo {author} {\bibfnamefont {Z.}~\bibnamefont
  {Xiao}}, \bibinfo {author} {\bibfnamefont {S.}~\bibnamefont {Yu}}, \bibinfo
  {author} {\bibfnamefont {Y.}~\bibnamefont {Li}}, \bibinfo {author}
  {\bibfnamefont {S.}~\bibnamefont {Ruan}}, \bibinfo {author} {\bibfnamefont
  {L.~B.}\ \bibnamefont {Kong}}, \bibinfo {author} {\bibfnamefont
  {Q.}~\bibnamefont {Huang}}, \bibinfo {author} {\bibfnamefont
  {Z.}~\bibnamefont {Huang}}, \bibinfo {author} {\bibfnamefont
  {K.}~\bibnamefont {Zhou}}, \bibinfo {author} {\bibfnamefont {H.}~\bibnamefont
  {Su}}, \bibinfo {author} {\bibfnamefont {Z.}~\bibnamefont {Yao}}, \bibinfo
  {author} {\bibfnamefont {W.}~\bibnamefont {Que}}, \bibinfo {author}
  {\bibfnamefont {Y.}~\bibnamefont {Liu}}, \bibinfo {author} {\bibfnamefont
  {T.}~\bibnamefont {Zhang}}, \bibinfo {author} {\bibfnamefont
  {J.}~\bibnamefont {Wang}}, \bibinfo {author} {\bibfnamefont {P.}~\bibnamefont
  {Liu}}, \bibinfo {author} {\bibfnamefont {D.}~\bibnamefont {Shen}}, \bibinfo
  {author} {\bibfnamefont {M.}~\bibnamefont {Allix}}, \bibinfo {author}
  {\bibfnamefont {J.}~\bibnamefont {Zhang}},\ and\ \bibinfo {author}
  {\bibfnamefont {D.}~\bibnamefont {Tang}},\ }\href
  {https://doi.org/https://doi.org/10.1016/j.mser.2019.100518} {\bibfield
  {journal} {\bibinfo  {journal} {Materials Science and Engineering: R:
  Reports}\ }\textbf {\bibinfo {volume} {139}},\ \bibinfo {pages} {100518}
  (\bibinfo {year} {2020})}\BibitemShut {NoStop}%
\bibitem [{\citenamefont {Sarro}(2000)}]{SiC}%
  \BibitemOpen
  \bibfield  {author} {\bibinfo {author} {\bibfnamefont {P.~M.}\ \bibnamefont
  {Sarro}},\ }\href
  {https://doi.org/https://doi.org/10.1016/S0924-4247(99)00335-0} {\bibfield
  {journal} {\bibinfo  {journal} {Sensors and Actuators A: Physical}\ }\textbf
  {\bibinfo {volume} {82}},\ \bibinfo {pages} {210} (\bibinfo {year}
  {2000})}\BibitemShut {NoStop}%
\bibitem [{\citenamefont {Takahashi}\ \emph {et~al.}(2009)\citenamefont
  {Takahashi}, \citenamefont {Aori},\ and\ \citenamefont
  {Ikeuchi}}]{SiCbonding}%
  \BibitemOpen
  \bibfield  {author} {\bibinfo {author} {\bibfnamefont {M.}~\bibnamefont
  {Takahashi}}, \bibinfo {author} {\bibfnamefont {Y.}~\bibnamefont {Aori}},\
  and\ \bibinfo {author} {\bibfnamefont {K.}~\bibnamefont {Ikeuchi}},\ }\href
  {https://doi.org/10.2207/qjjws.27.192s} {\bibfield  {journal} {\bibinfo
  {journal} {Quarterly Journal of the Japan Welding Society}\ }\textbf
  {\bibinfo {volume} {27}},\ \bibinfo {pages} {192s} (\bibinfo {year}
  {2009})}\BibitemShut {NoStop}%
\bibitem [{\citenamefont {Huet}\ \emph {et~al.}(2012)\citenamefont {Huet},
  \citenamefont {Ammar}, \citenamefont {Morvan}, \citenamefont {Sarazin},
  \citenamefont {Pocholle}, \citenamefont {Reichel}, \citenamefont {Guerlin},\
  and\ \citenamefont {Schwartz}}]{SiCchiptrap}%
  \BibitemOpen
  \bibfield  {author} {\bibinfo {author} {\bibfnamefont {L.}~\bibnamefont
  {Huet}}, \bibinfo {author} {\bibfnamefont {M.}~\bibnamefont {Ammar}},
  \bibinfo {author} {\bibfnamefont {E.}~\bibnamefont {Morvan}}, \bibinfo
  {author} {\bibfnamefont {N.}~\bibnamefont {Sarazin}}, \bibinfo {author}
  {\bibfnamefont {J.-P.}\ \bibnamefont {Pocholle}}, \bibinfo {author}
  {\bibfnamefont {J.}~\bibnamefont {Reichel}}, \bibinfo {author} {\bibfnamefont
  {C.}~\bibnamefont {Guerlin}},\ and\ \bibinfo {author} {\bibfnamefont
  {S.}~\bibnamefont {Schwartz}},\ }\href {https://doi.org/10.1063/1.3689777}
  {\bibfield  {journal} {\bibinfo  {journal} {Applied Physics Letters}\
  }\textbf {\bibinfo {volume} {100}},\ \bibinfo {pages} {121114} (\bibinfo
  {year} {2012})}\BibitemShut {NoStop}%
\bibitem [{\citenamefont {Ring}\ \emph {et~al.}(2011)\citenamefont {Ring},
  \citenamefont {Peckys}, \citenamefont {Dukes}, \citenamefont {Baudoin},\ and\
  \citenamefont {De~Jonge}}]{SiNwindow}%
  \BibitemOpen
  \bibfield  {author} {\bibinfo {author} {\bibfnamefont {E.~A.}\ \bibnamefont
  {Ring}}, \bibinfo {author} {\bibfnamefont {D.~B.}\ \bibnamefont {Peckys}},
  \bibinfo {author} {\bibfnamefont {M.~J.}\ \bibnamefont {Dukes}}, \bibinfo
  {author} {\bibfnamefont {J.~P.}\ \bibnamefont {Baudoin}},\ and\ \bibinfo
  {author} {\bibfnamefont {N.}~\bibnamefont {De~Jonge}},\ }\href
  {https://doi.org/https://doi.org/10.1111/j.1365-2818.2011.03501.x} {\bibfield
   {journal} {\bibinfo  {journal} {Journal of Microscopy}\ }\textbf {\bibinfo
  {volume} {243}},\ \bibinfo {pages} {273} (\bibinfo {year}
  {2011})}\BibitemShut {NoStop}%
\bibitem [{\citenamefont {de~Jonge}\ \emph {et~al.}(2009)\citenamefont
  {de~Jonge}, \citenamefont {Peckys}, \citenamefont {Kremers},\ and\
  \citenamefont {Piston}}]{SiNvacuum}%
  \BibitemOpen
  \bibfield  {author} {\bibinfo {author} {\bibfnamefont {N.}~\bibnamefont
  {de~Jonge}}, \bibinfo {author} {\bibfnamefont {D.~B.}\ \bibnamefont
  {Peckys}}, \bibinfo {author} {\bibfnamefont {G.~J.}\ \bibnamefont
  {Kremers}},\ and\ \bibinfo {author} {\bibfnamefont {D.~W.}\ \bibnamefont
  {Piston}},\ }\href {https://doi.org/10.1073/pnas.0809567106} {\bibfield
  {journal} {\bibinfo  {journal} {Proceedings of the National Academy of
  Sciences}\ }\textbf {\bibinfo {volume} {106}},\ \bibinfo {pages} {2159}
  (\bibinfo {year} {2009})}\BibitemShut {NoStop}%
\bibitem [{\citenamefont {Nishiyama}\ \emph {et~al.}(2017)\citenamefont
  {Nishiyama}, \citenamefont {Ishikawa}, \citenamefont {Ohfuji}, \citenamefont
  {Marquardt}, \citenamefont {Kurnosov}, \citenamefont {Taniguchi},
  \citenamefont {Kim}, \citenamefont {Yoshida}, \citenamefont {Masuno},
  \citenamefont {Bednarcik}, \citenamefont {Kulik}, \citenamefont {Ikuhara},
  \citenamefont {Wakai},\ and\ \citenamefont {Irifune}}]{c-SiN}%
  \BibitemOpen
  \bibfield  {author} {\bibinfo {author} {\bibfnamefont {N.}~\bibnamefont
  {Nishiyama}}, \bibinfo {author} {\bibfnamefont {R.}~\bibnamefont {Ishikawa}},
  \bibinfo {author} {\bibfnamefont {H.}~\bibnamefont {Ohfuji}}, \bibinfo
  {author} {\bibfnamefont {H.}~\bibnamefont {Marquardt}}, \bibinfo {author}
  {\bibfnamefont {A.}~\bibnamefont {Kurnosov}}, \bibinfo {author}
  {\bibfnamefont {T.}~\bibnamefont {Taniguchi}}, \bibinfo {author}
  {\bibfnamefont {B.-N.}\ \bibnamefont {Kim}}, \bibinfo {author} {\bibfnamefont
  {H.}~\bibnamefont {Yoshida}}, \bibinfo {author} {\bibfnamefont
  {A.}~\bibnamefont {Masuno}}, \bibinfo {author} {\bibfnamefont
  {J.}~\bibnamefont {Bednarcik}}, \bibinfo {author} {\bibfnamefont
  {E.}~\bibnamefont {Kulik}}, \bibinfo {author} {\bibfnamefont
  {Y.}~\bibnamefont {Ikuhara}}, \bibinfo {author} {\bibfnamefont
  {F.}~\bibnamefont {Wakai}},\ and\ \bibinfo {author} {\bibfnamefont
  {T.}~\bibnamefont {Irifune}},\ }\href {https://doi.org/10.1038/srep44755}
  {\bibfield  {journal} {\bibinfo  {journal} {Scientific Reports}\ }\textbf
  {\bibinfo {volume} {7}},\ \bibinfo {pages} {44755} (\bibinfo {year}
  {2017})}\BibitemShut {NoStop}%
\bibitem [{\citenamefont {Weichel}\ \emph {et~al.}(2000)\citenamefont
  {Weichel}, \citenamefont {{de Reus}}, \citenamefont {Bouaidat}, \citenamefont
  {Rasmussen}, \citenamefont {Hansen}, \citenamefont {Birkelund},\ and\
  \citenamefont {Dirac}}]{SiNbonding}%
  \BibitemOpen
  \bibfield  {author} {\bibinfo {author} {\bibfnamefont {S.}~\bibnamefont
  {Weichel}}, \bibinfo {author} {\bibfnamefont {R.}~\bibnamefont {{de Reus}}},
  \bibinfo {author} {\bibfnamefont {S.}~\bibnamefont {Bouaidat}}, \bibinfo
  {author} {\bibfnamefont {P.}~\bibnamefont {Rasmussen}}, \bibinfo {author}
  {\bibfnamefont {O.}~\bibnamefont {Hansen}}, \bibinfo {author} {\bibfnamefont
  {K.}~\bibnamefont {Birkelund}},\ and\ \bibinfo {author} {\bibfnamefont
  {H.}~\bibnamefont {Dirac}},\ }\href
  {https://doi.org/https://doi.org/10.1016/S0924-4247(99)00372-6} {\bibfield
  {journal} {\bibinfo  {journal} {Sensors and Actuators A: Physical}\ }\textbf
  {\bibinfo {volume} {82}},\ \bibinfo {pages} {249} (\bibinfo {year}
  {2000})}\BibitemShut {NoStop}%
\bibitem [{\citenamefont {Grzebyk}(2017)}]{micropumps1}%
  \BibitemOpen
  \bibfield  {author} {\bibinfo {author} {\bibfnamefont {T.}~\bibnamefont
  {Grzebyk}},\ }\href {https://doi.org/10.1109/JMEMS.2017.2676820} {\bibfield
  {journal} {\bibinfo  {journal} {Journal of Microelectromechanical Systems}\
  }\textbf {\bibinfo {volume} {26}},\ \bibinfo {pages} {705} (\bibinfo {year}
  {2017})}\BibitemShut {NoStop}%
\bibitem [{\citenamefont {Green}\ \emph {et~al.}(2013)\citenamefont {Green},
  \citenamefont {Malhotra},\ and\ \citenamefont {Gianchandani}}]{sputterpump}%
  \BibitemOpen
  \bibfield  {author} {\bibinfo {author} {\bibfnamefont {S.~R.}\ \bibnamefont
  {Green}}, \bibinfo {author} {\bibfnamefont {R.}~\bibnamefont {Malhotra}},\
  and\ \bibinfo {author} {\bibfnamefont {Y.~B.}\ \bibnamefont {Gianchandani}},\
  }\href {https://doi.org/10.1109/JMEMS.2012.2221159} {\bibfield  {journal}
  {\bibinfo  {journal} {Journal of Microelectromechanical Systems}\ }\textbf
  {\bibinfo {volume} {22}},\ \bibinfo {pages} {309} (\bibinfo {year}
  {2013})}\BibitemShut {NoStop}%
\bibitem [{\citenamefont {Grzebyk}\ \emph {et~al.}(2014)\citenamefont
  {Grzebyk}, \citenamefont {Górecka-Drzazga},\ and\ \citenamefont
  {Dziuban}}]{dziuban}%
  \BibitemOpen
  \bibfield  {author} {\bibinfo {author} {\bibfnamefont {T.}~\bibnamefont
  {Grzebyk}}, \bibinfo {author} {\bibfnamefont {A.}~\bibnamefont
  {Górecka-Drzazga}},\ and\ \bibinfo {author} {\bibfnamefont {J.}~\bibnamefont
  {Dziuban}},\ }\href
  {https://doi.org/https://doi.org/10.1016/j.sna.2014.01.011} {\bibfield
  {journal} {\bibinfo  {journal} {Sensors and Actuators A: Physical}\ }\textbf
  {\bibinfo {volume} {208}},\ \bibinfo {pages} {113} (\bibinfo {year}
  {2014})}\BibitemShut {NoStop}%
\bibitem [{\citenamefont {Grzebyk}\ \emph {et~al.}(2011)\citenamefont
  {Grzebyk}, \citenamefont {Stasiak},\ and\ \citenamefont
  {Górecka-Drzazga}}]{deflector}%
  \BibitemOpen
  \bibfield  {author} {\bibinfo {author} {\bibfnamefont {T.}~\bibnamefont
  {Grzebyk}}, \bibinfo {author} {\bibfnamefont {P.}~\bibnamefont {Stasiak}},\
  and\ \bibinfo {author} {\bibfnamefont {A.}~\bibnamefont {Górecka-Drzazga}},\
  }\href
  {https://www.scopus.com/inward/record.uri?eid=2-s2.0-79961070455&partnerID=40&md5=77ed43caccf99167fd29002e5864602c}
  {\bibfield  {journal} {\bibinfo  {journal} {Optica Applicata}\ }\textbf
  {\bibinfo {volume} {41}},\ \bibinfo {pages} {389} (\bibinfo {year}
  {2011})}\BibitemShut {NoStop}%
\bibitem [{\citenamefont {Audi}\ and\ \citenamefont {{de
  Simon}}(1987)}]{AUDI1987629}%
  \BibitemOpen
  \bibfield  {author} {\bibinfo {author} {\bibfnamefont {M.}~\bibnamefont
  {Audi}}\ and\ \bibinfo {author} {\bibfnamefont {M.}~\bibnamefont {{de
  Simon}}},\ }\href
  {https://doi.org/https://doi.org/10.1016/0042-207X(87)90048-0} {\bibfield
  {journal} {\bibinfo  {journal} {Vacuum}\ }\textbf {\bibinfo {volume} {37}},\
  \bibinfo {pages} {629} (\bibinfo {year} {1987})}\BibitemShut {NoStop}%
\bibitem [{\citenamefont {Boudot}\ \emph {et~al.}(2020)\citenamefont {Boudot},
  \citenamefont {McGilligan}, \citenamefont {Moore}, \citenamefont {Maurice},
  \citenamefont {Martinez}, \citenamefont {Hansen}, \citenamefont {de~Clercq},\
  and\ \citenamefont {Kitching}}]{boudotmcgilligan}%
  \BibitemOpen
  \bibfield  {author} {\bibinfo {author} {\bibfnamefont {R.}~\bibnamefont
  {Boudot}}, \bibinfo {author} {\bibfnamefont {J.~P.}\ \bibnamefont
  {McGilligan}}, \bibinfo {author} {\bibfnamefont {K.~R.}\ \bibnamefont
  {Moore}}, \bibinfo {author} {\bibfnamefont {V.}~\bibnamefont {Maurice}},
  \bibinfo {author} {\bibfnamefont {G.~D.}\ \bibnamefont {Martinez}}, \bibinfo
  {author} {\bibfnamefont {A.}~\bibnamefont {Hansen}}, \bibinfo {author}
  {\bibfnamefont {E.}~\bibnamefont {de~Clercq}},\ and\ \bibinfo {author}
  {\bibfnamefont {J.}~\bibnamefont {Kitching}},\ }\href
  {https://doi.org/10.1038/s41598-020-73605-z} {\bibfield  {journal} {\bibinfo
  {journal} {Scientific Reports}\ }\textbf {\bibinfo {volume} {10}},\ \bibinfo
  {pages} {16590} (\bibinfo {year} {2020})}\BibitemShut {NoStop}%
\bibitem [{\citenamefont {Basu}\ and\ \citenamefont
  {Vel{\'{a}}squez-Garc{\'{\i}}a}(2016)}]{Basu_2016}%
  \BibitemOpen
  \bibfield  {author} {\bibinfo {author} {\bibfnamefont {A.}~\bibnamefont
  {Basu}}\ and\ \bibinfo {author} {\bibfnamefont {L.~F.}\ \bibnamefont
  {Vel{\'{a}}squez-Garc{\'{\i}}a}},\ }\href
  {https://doi.org/10.1088/0960-1317/26/12/124003} {\bibfield  {journal}
  {\bibinfo  {journal} {Journal of Micromechanics and Microengineering}\
  }\textbf {\bibinfo {volume} {26}},\ \bibinfo {pages} {124003} (\bibinfo
  {year} {2016})}\BibitemShut {NoStop}%
\bibitem [{\citenamefont {Benvenuti}\ \emph {et~al.}(2001)\citenamefont
  {Benvenuti}, \citenamefont {Chiggiato}, \citenamefont {{Costa Pinto}},
  \citenamefont {{Escudeiro Santana}}, \citenamefont {Hedley}, \citenamefont
  {Mongelluzzo}, \citenamefont {Ruzinov},\ and\ \citenamefont
  {Wevers}}]{BENVENUTI200157}%
  \BibitemOpen
  \bibfield  {author} {\bibinfo {author} {\bibfnamefont {C.}~\bibnamefont
  {Benvenuti}}, \bibinfo {author} {\bibfnamefont {P.}~\bibnamefont
  {Chiggiato}}, \bibinfo {author} {\bibfnamefont {P.}~\bibnamefont {{Costa
  Pinto}}}, \bibinfo {author} {\bibfnamefont {A.}~\bibnamefont {{Escudeiro
  Santana}}}, \bibinfo {author} {\bibfnamefont {T.}~\bibnamefont {Hedley}},
  \bibinfo {author} {\bibfnamefont {A.}~\bibnamefont {Mongelluzzo}}, \bibinfo
  {author} {\bibfnamefont {V.}~\bibnamefont {Ruzinov}},\ and\ \bibinfo {author}
  {\bibfnamefont {I.}~\bibnamefont {Wevers}},\ }\href
  {https://doi.org/https://doi.org/10.1016/S0042-207X(00)00246-3} {\bibfield
  {journal} {\bibinfo  {journal} {Vacuum}\ }\textbf {\bibinfo {volume} {60}},\
  \bibinfo {pages} {57} (\bibinfo {year} {2001})}\BibitemShut {NoStop}%
\bibitem [{\citenamefont {Benvenuti}\ \emph {et~al.}(1999)\citenamefont
  {Benvenuti}, \citenamefont {Cazeneuve}, \citenamefont {Chiggiato},
  \citenamefont {Cicoira}, \citenamefont {{Escudeiro Santana}}, \citenamefont
  {Johanek}, \citenamefont {Ruzinov},\ and\ \citenamefont
  {Fraxedas}}]{BENVENUTI1999219}%
  \BibitemOpen
  \bibfield  {author} {\bibinfo {author} {\bibfnamefont {C.}~\bibnamefont
  {Benvenuti}}, \bibinfo {author} {\bibfnamefont {J.}~\bibnamefont
  {Cazeneuve}}, \bibinfo {author} {\bibfnamefont {P.}~\bibnamefont
  {Chiggiato}}, \bibinfo {author} {\bibfnamefont {F.}~\bibnamefont {Cicoira}},
  \bibinfo {author} {\bibfnamefont {A.}~\bibnamefont {{Escudeiro Santana}}},
  \bibinfo {author} {\bibfnamefont {V.}~\bibnamefont {Johanek}}, \bibinfo
  {author} {\bibfnamefont {V.}~\bibnamefont {Ruzinov}},\ and\ \bibinfo {author}
  {\bibfnamefont {J.}~\bibnamefont {Fraxedas}},\ }\href
  {https://doi.org/https://doi.org/10.1016/S0042-207X(98)00377-7} {\bibfield
  {journal} {\bibinfo  {journal} {Vacuum}\ }\textbf {\bibinfo {volume} {53}},\
  \bibinfo {pages} {219} (\bibinfo {year} {1999})}\BibitemShut {NoStop}%
\bibitem [{\citenamefont {Scherer}\ \emph {et~al.}(2012)\citenamefont
  {Scherer}, \citenamefont {Fenner},\ and\ \citenamefont {Hensley}}]{scherer}%
  \BibitemOpen
  \bibfield  {author} {\bibinfo {author} {\bibfnamefont {D.~R.}\ \bibnamefont
  {Scherer}}, \bibinfo {author} {\bibfnamefont {D.~B.}\ \bibnamefont
  {Fenner}},\ and\ \bibinfo {author} {\bibfnamefont {J.~M.}\ \bibnamefont
  {Hensley}},\ }\href {https://doi.org/10.1116/1.4757950} {\bibfield  {journal}
  {\bibinfo  {journal} {Journal of Vacuum Science and Technology A}\ }\textbf
  {\bibinfo {volume} {30}},\ \bibinfo {pages} {061602} (\bibinfo {year}
  {2012})}\BibitemShut {NoStop}%
\bibitem [{\citenamefont {Weatherill}\ \emph {et~al.}(2009)\citenamefont
  {Weatherill}, \citenamefont {Pritchard}, \citenamefont {Griffin},
  \citenamefont {Dammalapati}, \citenamefont {Adams},\ and\ \citenamefont
  {Riis}}]{griffwindow}%
  \BibitemOpen
  \bibfield  {author} {\bibinfo {author} {\bibfnamefont {K.~J.}\ \bibnamefont
  {Weatherill}}, \bibinfo {author} {\bibfnamefont {J.~D.}\ \bibnamefont
  {Pritchard}}, \bibinfo {author} {\bibfnamefont {P.~F.}\ \bibnamefont
  {Griffin}}, \bibinfo {author} {\bibfnamefont {U.}~\bibnamefont
  {Dammalapati}}, \bibinfo {author} {\bibfnamefont {C.~S.}\ \bibnamefont
  {Adams}},\ and\ \bibinfo {author} {\bibfnamefont {E.}~\bibnamefont {Riis}},\
  }\href {https://doi.org/10.1063/1.3075547} {\bibfield  {journal} {\bibinfo
  {journal} {Review of Scientific Instruments}\ }\textbf {\bibinfo {volume}
  {80}},\ \bibinfo {pages} {026105} (\bibinfo {year} {2009})}\BibitemShut
  {NoStop}%
\bibitem [{\citenamefont {Noor}\ \emph {et~al.}(2020)\citenamefont {Noor},
  \citenamefont {Asadian},\ and\ \citenamefont {Shkel}}]{microspheres}%
  \BibitemOpen
  \bibfield  {author} {\bibinfo {author} {\bibfnamefont {R.~M.}\ \bibnamefont
  {Noor}}, \bibinfo {author} {\bibfnamefont {M.~H.}\ \bibnamefont {Asadian}},\
  and\ \bibinfo {author} {\bibfnamefont {A.~M.}\ \bibnamefont {Shkel}},\ }\href
  {https://doi.org/10.1109/JMEMS.2019.2949084} {\bibfield  {journal} {\bibinfo
  {journal} {Journal of Microelectromechanical Systems}\ }\textbf {\bibinfo
  {volume} {29}},\ \bibinfo {pages} {25} (\bibinfo {year} {2020})}\BibitemShut
  {NoStop}%
\bibitem [{bor(2020)}]{borofloat33}%
  \BibitemOpen
  \href@noop {} {\emph {\bibinfo {title} {Borofloat 33 - Thermal Properties
  Data Sheet}}},\ \bibinfo {organization} {Schott} (\bibinfo {year}
  {2020})\BibitemShut {NoStop}%
\bibitem [{\citenamefont {Singh}\ \emph {et~al.}(1972)\citenamefont {Singh},
  \citenamefont {Dilavore},\ and\ \citenamefont {Alley}}]{glassmelt}%
  \BibitemOpen
  \bibfield  {author} {\bibinfo {author} {\bibfnamefont {G.}~\bibnamefont
  {Singh}}, \bibinfo {author} {\bibfnamefont {P.}~\bibnamefont {Dilavore}},\
  and\ \bibinfo {author} {\bibfnamefont {C.~O.}\ \bibnamefont {Alley}},\ }\href
  {https://doi.org/10.1063/1.1685940} {\bibfield  {journal} {\bibinfo
  {journal} {Review of Scientific Instruments}\ }\textbf {\bibinfo {volume}
  {43}},\ \bibinfo {pages} {1388} (\bibinfo {year} {1972})}\BibitemShut
  {NoStop}%
\bibitem [{\citenamefont
  {Knapkiewicz}(2019{\natexlab{b}})}]{pawelmicromachines}%
  \BibitemOpen
  \bibfield  {author} {\bibinfo {author} {\bibfnamefont {P.}~\bibnamefont
  {Knapkiewicz}},\ }\href {https://doi.org/10.3390/mi10010025} {\bibfield
  {journal} {\bibinfo  {journal} {Micromachines}\ }\textbf {\bibinfo {volume}
  {10}},\ \bibinfo {pages} {25} (\bibinfo {year}
  {2019}{\natexlab{b}})}\BibitemShut {NoStop}%
\bibitem [{tho(2022)}]{thorlabs}%
  \BibitemOpen
  \href@noop {} {\bibinfo {title} {Thorlabs glassblown atomic reference
  cells}},\ \bibinfo {howpublished}
  {\url{https://www.thorlabs.com/newgrouppage9.cfm?objectgroup_id=1470}}
  (\bibinfo {year} {2022}),\ \bibinfo {note} {"Accessed:
  09-01-2022"}\BibitemShut {NoStop}%
\bibitem [{PGB(2022)}]{PGB}%
  \BibitemOpen
  \href@noop {} {\bibinfo {title} {Precision glassblowing atomic reference
  cells}},\ \bibinfo {howpublished}
  {\url{https://www.precisionglassblowing.com/vapor-wavelength-reference-cells/}}
  (\bibinfo {year} {2022}),\ \bibinfo {note} {accessed: 09-01-2022}\BibitemShut
  {NoStop}%
\bibitem [{Sac(2022)}]{Sacher}%
  \BibitemOpen
  \href@noop {} {\bibinfo {title} {Sacher lasertechnik atomic reference
  cells}},\ \bibinfo {howpublished}
  {\url{https://www.sacher-laser.com/home/lab-equipment/spectroscopy/reference_gas_and_vapor_cells/reference_gas_and_vapor_cells.html}}
  (\bibinfo {year} {2022}),\ \bibinfo {note} {accessed: 09-01-2022}\BibitemShut
  {NoStop}%
\bibitem [{\citenamefont {Maurice}\ \emph {et~al.}(2022)\citenamefont
  {Maurice}, \citenamefont {Carlé}, \citenamefont {Keshavarzi}, \citenamefont
  {Chutani}, \citenamefont {Queste}, \citenamefont {Gauthier-Manuel},
  \citenamefont {Cote}, \citenamefont {Vicarini}, \citenamefont {Hafiz},
  \citenamefont {Boudot},\ and\ \citenamefont {Passilly}}]{meltcellarxiv}%
  \BibitemOpen
  \bibfield  {author} {\bibinfo {author} {\bibfnamefont {V.}~\bibnamefont
  {Maurice}}, \bibinfo {author} {\bibfnamefont {C.}~\bibnamefont {Carlé}},
  \bibinfo {author} {\bibfnamefont {S.}~\bibnamefont {Keshavarzi}}, \bibinfo
  {author} {\bibfnamefont {R.}~\bibnamefont {Chutani}}, \bibinfo {author}
  {\bibfnamefont {S.}~\bibnamefont {Queste}}, \bibinfo {author} {\bibfnamefont
  {L.}~\bibnamefont {Gauthier-Manuel}}, \bibinfo {author} {\bibfnamefont
  {J.-M.}\ \bibnamefont {Cote}}, \bibinfo {author} {\bibfnamefont
  {R.}~\bibnamefont {Vicarini}}, \bibinfo {author} {\bibfnamefont {M.~A.}\
  \bibnamefont {Hafiz}}, \bibinfo {author} {\bibfnamefont {R.}~\bibnamefont
  {Boudot}},\ and\ \bibinfo {author} {\bibfnamefont {N.}~\bibnamefont
  {Passilly}},\ }\href {https://doi.org/10.48550/ARXIV.2205.10440} {\bibinfo
  {title} {Laser-actuated hermetic seals for integrated atomic devices}}
  (\bibinfo {year} {2022}),\ \Eprint {https://arxiv.org/abs/2205.10440}
  {arXiv:2205.10440} \BibitemShut {NoStop}%
\bibitem [{\citenamefont {Du}\ \emph {et~al.}(2004)\citenamefont {Du},
  \citenamefont {Squires}, \citenamefont {Imai}, \citenamefont {Czaia},
  \citenamefont {Saravanan}, \citenamefont {Bright}, \citenamefont {Reichel},
  \citenamefont {H\"ansch},\ and\ \citenamefont {Anderson}}]{atomchippinchoff}%
  \BibitemOpen
  \bibfield  {author} {\bibinfo {author} {\bibfnamefont {S.}~\bibnamefont
  {Du}}, \bibinfo {author} {\bibfnamefont {M.~B.}\ \bibnamefont {Squires}},
  \bibinfo {author} {\bibfnamefont {Y.}~\bibnamefont {Imai}}, \bibinfo {author}
  {\bibfnamefont {L.}~\bibnamefont {Czaia}}, \bibinfo {author} {\bibfnamefont
  {R.~A.}\ \bibnamefont {Saravanan}}, \bibinfo {author} {\bibfnamefont
  {V.}~\bibnamefont {Bright}}, \bibinfo {author} {\bibfnamefont
  {J.}~\bibnamefont {Reichel}}, \bibinfo {author} {\bibfnamefont {T.~W.}\
  \bibnamefont {H\"ansch}},\ and\ \bibinfo {author} {\bibfnamefont {D.~Z.}\
  \bibnamefont {Anderson}},\ }\href
  {https://doi.org/10.1103/PhysRevA.70.053606} {\bibfield  {journal} {\bibinfo
  {journal} {Phys. Rev. A}\ }\textbf {\bibinfo {volume} {70}},\ \bibinfo
  {pages} {053606} (\bibinfo {year} {2004})}\BibitemShut {NoStop}%
\bibitem [{\citenamefont {Grosse}\ \emph {et~al.}(2016)\citenamefont {Grosse},
  \citenamefont {Seidel}, \citenamefont {Becker}, \citenamefont {Lachmann},
  \citenamefont {Scharringhausen}, \citenamefont {Braxmaier},\ and\
  \citenamefont {Rasel}}]{nasapinchoff}%
  \BibitemOpen
  \bibfield  {author} {\bibinfo {author} {\bibfnamefont {J.}~\bibnamefont
  {Grosse}}, \bibinfo {author} {\bibfnamefont {S.~T.}\ \bibnamefont {Seidel}},
  \bibinfo {author} {\bibfnamefont {D.}~\bibnamefont {Becker}}, \bibinfo
  {author} {\bibfnamefont {M.~D.}\ \bibnamefont {Lachmann}}, \bibinfo {author}
  {\bibfnamefont {M.}~\bibnamefont {Scharringhausen}}, \bibinfo {author}
  {\bibfnamefont {C.}~\bibnamefont {Braxmaier}},\ and\ \bibinfo {author}
  {\bibfnamefont {E.~M.}\ \bibnamefont {Rasel}},\ }\href
  {https://doi.org/10.1116/1.4947583} {\bibfield  {journal} {\bibinfo
  {journal} {Journal of Vacuum Science and Technology A}\ }\textbf {\bibinfo
  {volume} {34}},\ \bibinfo {pages} {031606} (\bibinfo {year}
  {2016})}\BibitemShut {NoStop}%
\bibitem [{\citenamefont {Komareneni}\ \emph {et~al.}(1979)\citenamefont
  {Komareneni}, \citenamefont {Freeborn},\ and\ \citenamefont
  {Smith}}]{coldweld}%
  \BibitemOpen
  \bibfield  {author} {\bibinfo {author} {\bibfnamefont {S.}~\bibnamefont
  {Komareneni}}, \bibinfo {author} {\bibfnamefont {W.~P.}\ \bibnamefont
  {Freeborn}},\ and\ \bibinfo {author} {\bibfnamefont {C.~A.}\ \bibnamefont
  {Smith}},\ }\href@noop {} {\bibfield  {journal} {\bibinfo  {journal}
  {American Mineralogist}\ }\textbf {\bibinfo {volume} {64}},\ \bibinfo {pages}
  {650} (\bibinfo {year} {1979})}\BibitemShut {NoStop}%
\bibitem [{\citenamefont {Saint}\ \emph {et~al.}(2018)\citenamefont {Saint},
  \citenamefont {Evans}, \citenamefont {Zhou}, \citenamefont {Barrett},
  \citenamefont {Fromhold}, \citenamefont {Saleh}, \citenamefont {Maskery},
  \citenamefont {Tuck}, \citenamefont {Wildman}, \citenamefont
  {Oru{\v{c}}evi{\'{c}}},\ and\ \citenamefont {Kr{\"u}ger}}]{Saint2018}%
  \BibitemOpen
  \bibfield  {author} {\bibinfo {author} {\bibfnamefont {R.}~\bibnamefont
  {Saint}}, \bibinfo {author} {\bibfnamefont {W.}~\bibnamefont {Evans}},
  \bibinfo {author} {\bibfnamefont {Y.}~\bibnamefont {Zhou}}, \bibinfo {author}
  {\bibfnamefont {T.}~\bibnamefont {Barrett}}, \bibinfo {author} {\bibfnamefont
  {T.~M.}\ \bibnamefont {Fromhold}}, \bibinfo {author} {\bibfnamefont
  {E.}~\bibnamefont {Saleh}}, \bibinfo {author} {\bibfnamefont
  {I.}~\bibnamefont {Maskery}}, \bibinfo {author} {\bibfnamefont
  {C.}~\bibnamefont {Tuck}}, \bibinfo {author} {\bibfnamefont {R.}~\bibnamefont
  {Wildman}}, \bibinfo {author} {\bibfnamefont {F.}~\bibnamefont
  {Oru{\v{c}}evi{\'{c}}}},\ and\ \bibinfo {author} {\bibfnamefont
  {P.}~\bibnamefont {Kr{\"u}ger}},\ }\href
  {https://doi.org/10.1038/s41598-018-26455-9} {\bibfield  {journal} {\bibinfo
  {journal} {Scientific Reports}\ }\textbf {\bibinfo {volume} {8}},\ \bibinfo
  {pages} {8368} (\bibinfo {year} {2018})}\BibitemShut {NoStop}%
\bibitem [{\citenamefont {Schwindt}\ \emph {et~al.}(2007)\citenamefont
  {Schwindt}, \citenamefont {Lindseth}, \citenamefont {Knappe}, \citenamefont
  {Shah}, \citenamefont {Kitching},\ and\ \citenamefont
  {Liew}}]{microfabcoilkitching}%
  \BibitemOpen
  \bibfield  {author} {\bibinfo {author} {\bibfnamefont {P.~D.~D.}\
  \bibnamefont {Schwindt}}, \bibinfo {author} {\bibfnamefont {B.}~\bibnamefont
  {Lindseth}}, \bibinfo {author} {\bibfnamefont {S.}~\bibnamefont {Knappe}},
  \bibinfo {author} {\bibfnamefont {V.}~\bibnamefont {Shah}}, \bibinfo {author}
  {\bibfnamefont {J.}~\bibnamefont {Kitching}},\ and\ \bibinfo {author}
  {\bibfnamefont {L.-A.}\ \bibnamefont {Liew}},\ }\href
  {https://doi.org/10.1063/1.2709532} {\bibfield  {journal} {\bibinfo
  {journal} {Applied Physics Letters}\ }\textbf {\bibinfo {volume} {90}},\
  \bibinfo {pages} {081102} (\bibinfo {year} {2007})}\BibitemShut {NoStop}%
\bibitem [{\citenamefont {Huang}\ \emph {et~al.}(2019)\citenamefont {Huang},
  \citenamefont {Mazzoni}, \citenamefont {Alzar},\ and\ \citenamefont
  {Reichel}}]{jakobreichelcoil}%
  \BibitemOpen
  \bibfield  {author} {\bibinfo {author} {\bibfnamefont {M.-Z.}\ \bibnamefont
  {Huang}}, \bibinfo {author} {\bibfnamefont {T.}~\bibnamefont {Mazzoni}},
  \bibinfo {author} {\bibfnamefont {C.~L.~G.}\ \bibnamefont {Alzar}},\ and\
  \bibinfo {author} {\bibfnamefont {J.}~\bibnamefont {Reichel}},\ }in\ \href
  {https://doi.org/10.1364/QIM.2019.T5A.32} {\emph {\bibinfo {booktitle}
  {Quantum Information and Measurement (QIM) V: Quantum Technologies}}}\
  (\bibinfo  {publisher} {Optical Society of America},\ \bibinfo {year}
  {2019})\ p.\ \bibinfo {pages} {T5A.32}\BibitemShut {NoStop}%
\bibitem [{\citenamefont {Noor}\ and\ \citenamefont
  {Shkel}(2018)}]{coilheater}%
  \BibitemOpen
  \bibfield  {author} {\bibinfo {author} {\bibfnamefont {R.~M.}\ \bibnamefont
  {Noor}}\ and\ \bibinfo {author} {\bibfnamefont {A.~M.}\ \bibnamefont
  {Shkel}},\ }\href {https://doi.org/10.1109/JMEMS.2018.2874451} {\bibfield
  {journal} {\bibinfo  {journal} {Journal of Microelectromechanical Systems}\
  }\textbf {\bibinfo {volume} {27}},\ \bibinfo {pages} {1148} (\bibinfo {year}
  {2018})}\BibitemShut {NoStop}%
\bibitem [{\citenamefont {Baumg\"artner}\ \emph {et~al.}(2010)\citenamefont
  {Baumg\"artner}, \citenamefont {Sewell}, \citenamefont {Eriksson},
  \citenamefont {Llorente-Garcia}, \citenamefont {Dingjan}, \citenamefont
  {Cotter},\ and\ \citenamefont {Hinds}}]{baumgartner}%
  \BibitemOpen
  \bibfield  {author} {\bibinfo {author} {\bibfnamefont {F.}~\bibnamefont
  {Baumg\"artner}}, \bibinfo {author} {\bibfnamefont {R.~J.}\ \bibnamefont
  {Sewell}}, \bibinfo {author} {\bibfnamefont {S.}~\bibnamefont {Eriksson}},
  \bibinfo {author} {\bibfnamefont {I.}~\bibnamefont {Llorente-Garcia}},
  \bibinfo {author} {\bibfnamefont {J.}~\bibnamefont {Dingjan}}, \bibinfo
  {author} {\bibfnamefont {J.~P.}\ \bibnamefont {Cotter}},\ and\ \bibinfo
  {author} {\bibfnamefont {E.~A.}\ \bibnamefont {Hinds}},\ }\href
  {https://doi.org/10.1103/PhysRevLett.105.243003} {\bibfield  {journal}
  {\bibinfo  {journal} {Phys. Rev. Lett.}\ }\textbf {\bibinfo {volume} {105}},\
  \bibinfo {pages} {243003} (\bibinfo {year} {2010})}\BibitemShut {NoStop}%
\bibitem [{\citenamefont {Sewell}\ \emph {et~al.}(2010)\citenamefont {Sewell},
  \citenamefont {Dingjan}, \citenamefont {Baumgärtner}, \citenamefont
  {Llorente-Garc{\'{\i}}a}, \citenamefont {Eriksson}, \citenamefont {Hinds},
  \citenamefont {Lewis}, \citenamefont {Srinivasan}, \citenamefont {Moktadir},
  \citenamefont {Gollasch},\ and\ \citenamefont {Kraft}}]{Sewell_2010}%
  \BibitemOpen
  \bibfield  {author} {\bibinfo {author} {\bibfnamefont {R.~J.}\ \bibnamefont
  {Sewell}}, \bibinfo {author} {\bibfnamefont {J.}~\bibnamefont {Dingjan}},
  \bibinfo {author} {\bibfnamefont {F.}~\bibnamefont {Baumgärtner}}, \bibinfo
  {author} {\bibfnamefont {I.}~\bibnamefont {Llorente-Garc{\'{\i}}a}}, \bibinfo
  {author} {\bibfnamefont {S.}~\bibnamefont {Eriksson}}, \bibinfo {author}
  {\bibfnamefont {E.~A.}\ \bibnamefont {Hinds}}, \bibinfo {author}
  {\bibfnamefont {G.}~\bibnamefont {Lewis}}, \bibinfo {author} {\bibfnamefont
  {P.}~\bibnamefont {Srinivasan}}, \bibinfo {author} {\bibfnamefont
  {Z.}~\bibnamefont {Moktadir}}, \bibinfo {author} {\bibfnamefont {C.~O.}\
  \bibnamefont {Gollasch}},\ and\ \bibinfo {author} {\bibfnamefont
  {M.}~\bibnamefont {Kraft}},\ }\href
  {https://doi.org/10.1088/0953-4075/43/5/051003} {\bibfield  {journal}
  {\bibinfo  {journal} {Journal of Physics B: Atomic, Molecular and Optical
  Physics}\ }\textbf {\bibinfo {volume} {43}},\ \bibinfo {pages} {051003}
  (\bibinfo {year} {2010})}\BibitemShut {NoStop}%
\bibitem [{\citenamefont {Chen}\ \emph {et~al.}(2022)\citenamefont {Chen},
  \citenamefont {Huang}, \citenamefont {Xu}, \citenamefont {Zhang},
  \citenamefont {Ma}, \citenamefont {Lu}, \citenamefont {Wang}, \citenamefont
  {Chen}, \citenamefont {Zhang}, \citenamefont {Tang}, \citenamefont {Dong},
  \citenamefont {Liu}, \citenamefont {Xiang}, \citenamefont {Guo},\ and\
  \citenamefont {Zou}}]{chen2021planar}%
  \BibitemOpen
  \bibfield  {author} {\bibinfo {author} {\bibfnamefont {L.}~\bibnamefont
  {Chen}}, \bibinfo {author} {\bibfnamefont {C.-J.}\ \bibnamefont {Huang}},
  \bibinfo {author} {\bibfnamefont {X.-B.}\ \bibnamefont {Xu}}, \bibinfo
  {author} {\bibfnamefont {Y.-C.}\ \bibnamefont {Zhang}}, \bibinfo {author}
  {\bibfnamefont {D.-Q.}\ \bibnamefont {Ma}}, \bibinfo {author} {\bibfnamefont
  {Z.-T.}\ \bibnamefont {Lu}}, \bibinfo {author} {\bibfnamefont {Z.-B.}\
  \bibnamefont {Wang}}, \bibinfo {author} {\bibfnamefont {G.-J.}\ \bibnamefont
  {Chen}}, \bibinfo {author} {\bibfnamefont {J.-Z.}\ \bibnamefont {Zhang}},
  \bibinfo {author} {\bibfnamefont {H.~X.}\ \bibnamefont {Tang}}, \bibinfo
  {author} {\bibfnamefont {C.-H.}\ \bibnamefont {Dong}}, \bibinfo {author}
  {\bibfnamefont {W.}~\bibnamefont {Liu}}, \bibinfo {author} {\bibfnamefont
  {G.-Y.}\ \bibnamefont {Xiang}}, \bibinfo {author} {\bibfnamefont {G.-C.}\
  \bibnamefont {Guo}},\ and\ \bibinfo {author} {\bibfnamefont {C.-L.}\
  \bibnamefont {Zou}},\ }\href
  {https://doi.org/10.1103/PhysRevApplied.17.034031} {\bibfield  {journal}
  {\bibinfo  {journal} {Phys. Rev. Applied}\ }\textbf {\bibinfo {volume}
  {17}},\ \bibinfo {pages} {034031} (\bibinfo {year} {2022})}\BibitemShut
  {NoStop}%
\end{thebibliography}%

\end{document}